\documentclass[draft,tightenlines,nofootinbib,preprint,aps,eqsecnum,amsmath,amssymb]{revtex4}

\newcommand{\beq}{\begin{equation}}
\newcommand{\eeq}{\end{equation}}
\newcommand{\bea}{\begin{eqnarray}}
\newcommand{\eea}{\end{eqnarray}}
\newcommand{\cir}{{\buildrel \circ \over =}}
\newcommand{\sgn}{\epsilon}

\begin{document}

\title{The Einstein-Maxwell-Particle System in the York Canonical Basis
of ADM Tetrad Gravity: II) The Weak Field Approximation in the
3-Orthogonal Gauges and Hamiltonian Post-Minkowskian Gravity: the
N-Body Problem and Gravitational Waves with Asymptotic Background.}

\medskip

\author{David Alba}

\affiliation{Dipartimento di Fisica\\
Universita' di Firenze\\Polo Scientifico, via Sansone 1\\
 50019 Sesto Fiorentino, Italy\\
 E-mail ALBA@FI.INFN.IT}

\author{Luca Lusanna}

\affiliation{ Sezione INFN di Firenze\\ Polo Scientifico\\ Via Sansone 1\\
50019 Sesto Fiorentino (FI), Italy\\ Phone: 0039-055-4572334\\
FAX: 0039-055-4572364\\ E-mail: lusanna@fi.infn.it}

\today

\begin{abstract}

In this second paper we define a Post-Minkowskian (PM) weak field
approximation leading to a linearization of the Hamilton equations
of ADM tetrad gravity in the York canonical basis in a family of
non-harmonic 3-orthogonal Schwinger time gauges. The York time
${}^3K$ (the relativistic inertial gauge variable, not existing in
Newtonian gravity, parametrizing the family and connected to the
freedom in clock synchronization, i.e. to the definition of the
instantaneous 3-spaces) is put equal to an arbitrary numerical
function. The matter are point particles, with a Grassmann
regularization of self-energies, and the electro-magnetic field in
the radiation gauge: a ultraviolet cutoff allows a consistent
linearization, which is shown to be the lowest order of a
Hamiltonian Post-Minkowskian (HPM) expansion.

We solve the constraints and the Hamilton equations for the tidal
variables and we find Post-Minkowskian gravitational waves with
asymptotic background (and the correct quadrupole emission formula)
propagating on dynamically determined non-Euclidean 3-spaces. The
conserved ADM energy and the Grassmann regularizzation of
self-energies imply the correct energy balance. A generalized
transverse-traceless gauge can be identified and the main tools for
the detection of gravitational waves are reproduced in these
non-harmonic gauges. In conclusion we get a PM solution for the
gravitational field and we identify a class of PM Einstein
space-times, which will be studied in more detail in a third paper
together with the PM equations of motion for the particles and their
Post-Newtonian expansion (but in absence of the electro-magnetic
field).

Finally we make a discussion on the {\it gauge problem in general
relativity} to understand which type of experimental observations
may lead to a preferred choice for the inertial gauge variable
${}^3K$ in the PM space-times. In the third paper we will show that
this choice is connected with the problem of dark matter.

\end{abstract}

\maketitle

\vfill\eject

\section{Introduction}

In this paper we will study the linearization of the Hamiltonian
formulation of ADM tetrad gravity given in Ref.\cite{1} (quoted as
paper I in what follows) to get a Post-Minkowskian (PM) description
of gravitational waves (GW) in non-harmonic gauges in asymptotically
Minkowskian space-times with the asymptotic Minkowski metric used as
a background. We are able to reproduce all the main properties of
GW's, which are usually derived in the standard Lagrangian approach
to GW's in harmonic gauges with a Post-Newtonian (PN) expansion (see
Appendix A for a review).

\bigskip

We define a new Hamiltonian Post-Minkowskiam (HPM) approach to the
description of GW's in which we do not assume the decomposition
${}^4g_{\mu\nu} = {}^4\eta_{\mu\nu} + {}^4h_{\mu\nu}$, but we use an
asymptotic Minkowski background 4-metric ${}^4\eta_{\mu\nu(asym)}$
at spatial infinity in a certain family of asymptotically
Minkowskian space-times. As shown in paper I, this approach is based
on ADM tetrad gravity in these space-times followed by a canonical
transformation to a York canonical basis, which diagonalizes the
York-Lichnerowitz formulation of general relativity \cite{2,3}
allowing a clean separation between physical tidal degrees of
freedom and inertial gauge variables inside the gravitational field
in a 3-covariant way inside the non-Euclidean instantaneous 3-spaces
(the Cauchy surfaces for the tidal variables): while in special
relativity its shape as a sub-manifold of Minkowski space-time is
arbitrary \cite{4} (the gauge freedom in the choice of the
convention for clock synchronization), in general relativity the
3-space is dynamically determined \cite{15} except for the trace of
its extrinsic curvature, the inertial gauge variable called York
time (it describes the general relativistic remnant of the gauge
freedom in clock synchronization).

\medskip

In paper I  we have explicitly evaluated the Hamilton equations of
ADM tetrad gravity coupled to dynamical matter in the Schwinger time
gauges in the York canonical basis defined in Ref.\cite{5}. This
paper was the final development of previous researches
\cite{6,7,8,9} in canonical gravity. The matter consists of the
electro-magnetic field in the radiation gauge and of N dynamical
(not test) massive point particles \footnote{In a future paper we
will try to describe compact extended objects and their self-gravity
(see footnote 16 of Appendix A) starting from balls of perfect
fluids, whose Lagrangian description will be the extension to tetrad
gravity of the special relativistic one given in Ref\cite{10}. The
Grassmann regularization is a semi-classical way out from causality
problems like the ones arising when considering the point limit of
extended models for the electron in classical electro-dynamics.}.
The Grassmann-valued electric charges and signs of the energy
regularize both the electro-magnetic and gravitational self-energies
(see Refs.\cite{4,11,12,13,14} for the electro-magnetic case in
special relativity) in the equations of motion for the N-body
problem avoiding both the gravitational and electro-magnetic
self-energies. As a consequence the action principle for the
particles given in paper I is well posed (there are no essential
singularities on the particle world-lines).

\medskip

Dirac theory of constraints is taken into account at every step, in
particular when making gauge-fixings (instead in the standard ADM
approach and numerical gravity the gauge fixings are chosen only on
the basis of convenience). This will allow to make a clear
distinction between the instantaneous action-at-a-distance effects
implied by the super-Hamiltonian and super-momentum constraints
(like it happens in every gauge theory: with the electro-magnetic
field in the radiation gauge this is implied by the Gauss law
constraint in presence of matter) and retarded tidal effects.

\medskip

We use a family of non-harmonic 3-orthogonal Schwinger time-gauges
in which the gauge fixings imply only elliptic-type equations at a
given time for the gauge variables. As a consequence, we avoid the
wave equations for gauge variables like the lapse and shift
functions present in the harmonic gauges and requiring initial data
in the asymptotic past.

\bigskip

ADM tetrad gravity is formulated in an arbitrary admissible 3+1
splitting of the globally hyperbolic space-time, i.e. in a foliation
with instantaneous space-like 3-spaces tending to a Minkowski
space-like hyper-plane at spatial infinity in a direction
independent way: they correspond to a clock synchronization
convention and each one of them can be used as a Cauchy surface for
field equations. As shown in Refs.\cite{1,6} the absence of
super-translations implies that  the SPI group of asymptotic
symmetries is reduced to the asymptotic ADM Poincare' group
\footnote{It reduces to the special relativistic Poincare' group of
the given matter in non-inertial frames of Minkowski space-time when
the Newton constant $G$ is switched off.} and the allowed 3+1
splittings must have the instantaneous 3-spaces tending to
asymptotic special-relativistic Wigner hyper-planes orthogonal to
the ADM 4-momentum in a direction-independent way (see Ref.\cite{4}
for these non-inertial rest frames in special relativity). At
spatial infinity there are asymptotic inertial observers, carrying a
flat tetrad $\epsilon^{\mu}_A$ (${}^4\eta_{\mu\nu}\,
\epsilon^{\mu}_A\, \epsilon^{\nu}_B = {}^4\eta_{AB}$,
$\epsilon^{\mu}{}_{\alpha\beta\gamma}\, \epsilon^{\alpha}_1\,
\epsilon^{\beta}_2\, \epsilon^{\gamma}_3 = \epsilon^{\mu}_{\tau}$),
whose spatial axes can be identified with the fixed stars of star
catalogues.\medskip

\bigskip

We use radar 4-coordinates $\sigma^A = (\sigma^{\tau} = \tau ;
\sigma^r)$, $A = \tau ,r$, adapted to the admissible 3+1 splitting
of the space-time and centered on an arbitrary time-like observer
$x^{\mu}(\tau)$ (origin of the 3-coordinates $\sigma^r$): they
define a non-inertial frame centered on the observer, so that they
are {\it observer and frame- dependent}. The time variable $\tau$ is
an arbitrary  monotonically increasing function of the proper time
given by the atomic clock carried by the observer. The instantaneous
3-spaces identified by this convention for clock synchronization are
denoted $\Sigma_{\tau}$. The transformation $\sigma^A \mapsto
x^{\mu} = z^{\mu}(\tau, \sigma^r)$ to world 4-coordinates defines
the embedding $z^{\mu}(\tau, \vec \sigma)$ of the Riemannian
instantaneous 3-spaces $\Sigma_{\tau}$ into the space-time. By
choosing  world 4-coordinates centered on the time-like observer,
whose world-line is the time axis, we have $x^{\mu}(\tau) =
(x^o(\tau); 0)$: the condition $x^o(\tau) = const.$ is equivalent to
$\tau = const.$ and identifies the instantaneous 3-space
$\Sigma_{\tau}$. If the time-like observer coincides with an
asymptotic inertial observer $x^{\mu}(\tau) = x^{\mu}_o +
\epsilon^{\mu}_{\tau}\, \tau$ with $\epsilon^{\mu}_{\tau} = (1; 0)$,
$\epsilon^{\mu}_r = (0; \delta^i_r)$, $x^{\mu}_o = (x^o_o; 0)$, then
the natural embedding describing the given 3+1 splitting of
space-time is $z^{\mu}(\tau, \sigma^r) = x^{\mu}_o +
\epsilon^{\mu}_A\, \sigma^A$ and the world 4-metric is
${}^4g_{\mu\nu} = \epsilon^A_{\mu}\, \epsilon^B_{\nu}\, {}^4g_{AB}$
($\epsilon^A_{\mu}$ are flat asymptotic cotetrads,
$\epsilon^A_{\mu}\, \epsilon^{\mu}_B = \delta^A_B$,
$\epsilon^A_{\mu}\, \epsilon^{\nu}_A = \delta^{\mu}_{\nu}$).

\medskip

From now on we shall denote the curvilinear 3-coordinates $\sigma^r$
with the notation $\vec \sigma$ for the sake of simplicity. Usually
the convention of sum over repeated indices is used, except when
there are too many summations.

\bigskip

The 4-metric ${}^4g_{AB}$ has signature $\sgn\, (+---)$ with $\sgn =
\pm$ (the particle physics, $\sgn = +$, and general relativity,
$\sgn = -$, conventions). Flat indices $(\alpha )$, $\alpha = o, a$,
are raised and lowered by the flat Minkowski metric
${}^4\eta_{(\alpha )(\beta )} = \sgn\, (+---)$. We define
${}^4\eta_{(a)(b)} = - \sgn\, \delta_{(a)(b)}$ with a
positive-definite Euclidean 3-metric. On each instantaneous 3-space
$\Sigma_{\tau}$ we have that the 4-metric has a
direction-independent limit to the flat Minkowski 4-metric (the
asymptotic background) at spatial infinity ${}^4g_{AB}(\tau, \vec
\sigma) \rightarrow {}^4\eta_{AB(asym)} = \sgn\, (+---)$. This
asymptotic 4-metric allows to define both a flat d'Alambertian $\Box
= \partial_{\tau}^2 - \triangle$ and a flat Laplacian $\triangle =
\sum_r\, \partial_r^2$ on $\Sigma_{\tau}$ ($\partial_A =
{{\partial}\over {\partial\, \sigma^A}}$). We will also need the
flat distribution  $c(\vec \sigma, {\vec \sigma}^{'}) = {1\over
{\Delta}}\,\, \delta^3(\vec \sigma, {\vec \sigma}^{'}) = -
\frac{1}{4\pi}\,\frac{1}{|\vec \sigma - {\vec \sigma}^{'}|}$ with
$|\vec \sigma - {\vec \sigma}^{'}| = \sqrt{\sum_u\, (\sigma^u -
\sigma^{'\, u})^2}$, where $\delta^3(\vec \sigma, {\vec
\sigma}^{'})$ is the Dirac delta function on the 3-manifold
$\Sigma_{\tau}$.

\bigskip

In this paper we  study the linearization of the Hamilton equations
of ADM tetrad gravity in the York canonical basis with the weak
field approximation in the family of non-harmonic 3-orthogonal
gauges in which the gauge variable (a relativistic inertial effect)
York time ${}^3K(\tau, \vec \sigma)$ is an arbitrary numerical
function and the 3-coordinates in $\Sigma_{\tau}$ are chosen so that
the 3-metric is everywhere diagonal. This will be a starting point
before facing higher orders in the HPM expansion defined at the end
of Section III (at higher orders regularization problems may arise).

\medskip

Since the gravitational gauge freedom is fixed, we get well defined
Hamilton equations for the matter. In particular this {\it avoids}
the introduction of "ad hoc" Lagrangians for the motion of test
particles in the resulting gravitational field as it is usually
done.
\medskip

We will look for solutions of the Hamilton equations {\it near}
Minkowski space-time, so that the matter content of the non-flat
space-time is assumed to be restricted by a {\it ultra-violet (UV)
cutoff} $M$: this avoids the appearance of strong gravitational
fields. This UV cutoff will allow us to avoid the {\it slow motion
approximation and the Post-Newtonian (PN) expansions} for the N
point particles. Naturally our linearization will not be reliable if
we look at distances $d_i$ from the $i-th$ particle of the order (or
less) of the Schwarzschild gravitational radius $R_M = {{2\, G\,
M}\over {c^2}}$ of the particle, i.e. we must have $d_i > R_M$.
Since we have a unique treatment for the near and far zone (we have
matter everywhere) we do not need radiative multipoles for the
gravitational field but only matter multipoles.We will do a
multipolar expansion of the energy-momentum tensor in terms of
relativistic Dixon multipoles  on the 3-space $\Sigma_{\tau}$ and we
will recover the standard quadrupole emission formula.

\medskip

An important aspect of the York canonical basis is the separation of
the 12 ADM equations for the gravitational field in three
sets:\medskip

\noindent a) the four contracted Bianchi identities for the time
derivatives of the unknowns in the super-Hamiltonian and
super-momentum constraints (they imply that, if the constraints are
solved on the Cauchy surface $\Sigma_{\tau_o}$, then the solution is
respected on the subsequent 3-spaces $\Sigma_{\tau > \tau_o}$);
\medskip

\noindent b) the four equations for the time derivatives of the
primary gauge variables (the 3-coordinates and the York time): since
we have fixed these gauge variables these equations becomes
equations of elliptic type for the lapse and shift functions (the
secondary gauge variables $1 + n(\tau, \vec \sigma)$ and ${\bar
n}_{(a)}(\tau, \vec \sigma)$) of our family of gauges (instead in
the harmonic gauges the lapse and shift functions obey wave
equations);\medskip

\noindent c) the Hamilton equations for the tidal variables $R_{\bar
a}$, $\Pi_{\bar a}$, $\bar a = 1,2$, which become hyperbolic
equations  ($\partial_{\tau}^2\, R_{\bar a}(\tau, \vec \sigma) =
....$) for $R_{\bar a}$ after the elimination of the momenta
$\Pi_{\bar a}$ by inverting the first half of the Hamilton equations
(those for $\partial_{\tau}\, R_{\bar a}$).\medskip

\noindent To these equations we must add the super-Hamiltonian and
super-momentum constraints, which are equations of elliptic type for
their unknowns on the 3-space $\Sigma_{\tau}$, namely for the
conformal factor $\tilde \phi(\tau, \vec \sigma)$ of the 3-metric
(namely the 3-volume element) and the off-diagonal terms
$\sigma_{(a)(b)}{|}_{a \not= b}(\tau, \vec \sigma)$ of the shear of
the congruence of Eulerian observers associated with the 3+1
splitting.\medskip

All the previous equations and also the matter Hamilton equations
are considered on the instantaneous 3-spaces of the whole space-time
without making a separation between the {\it near} and {\it far or
radiation} zones with respect to the particles for the study of
gravitational radiation.\medskip

As a consequence, only the equations for $\partial_{\tau}^2\,
R_{\bar a}(\tau, \vec \sigma)$ will lead to linearized equations
involving the flat d'Alambertian and implying matter-dependent {\it
retarded} solutions $R_{\bar a}^{(ret)}(\tau, \vec \sigma)$ (with
the homogeneous solutions eliminated by a no-incoming radiation
condition with respect to the asymptotic Minkowski 4-metric). We
will see that the linearized wave equation implies that only the TT
(traceless-transverse) part of the matter stress tensor is relevant.
Therefore we recover TT gravitational waves also in this family of
non-harmonic  gauges. Moreover, by means of a transformation of the
3-coordinates $\sigma^r$ on the 3-space $\Sigma_{\tau}$, we can find
a generalized TT gauge with the relativistic inertial effects
connected with the York time ${}^3K$ explicitly shown.\medskip

Moreover we find the correct energy balance (the back-reaction
problem) for the emission of GW's by using the conserved ADM energy
and avoiding singular quantities like the gravitational self-forces
on the particles \cite{16,17} due to the Grassmann regularization
and objects like the Landau-Lifschitz energy-momentum pseudo-tensor
of the gravitational field.

\medskip

The solutions of the linearized super-Hamiltonian and super-momentum
constraints involve only instantaneous quantities on the 3-space
$\Sigma_{\tau}$ and the same happens for the linearized equations
for the lapse and shift functions. These are the {\it instantaneous
action-at-a-distance} effects. Moreover the linearized Bianchi
identities turn out to be automatically satisfied. Therefore, the
linearized solutions for $\tilde \phi$, $\sigma_{(a)(b)}{|}_{a \not=
b}$, $1 + n$, ${\bar n}_{(a)}$, depend upon matter-dependent
instantaneous quantities and upon the instantaneous tidal variables
$\Gamma^{(1)}_r = \sum_{\bar a}\, \gamma_{\bar ar}\, R_{\bar a}$.
When we replace $\Gamma^{(1)}_r$ with the retarded solution
$\Gamma^{(1)(ret)}_r = \sum_{\bar a}\, \gamma_{\bar ar}\,
R^{(ret)}_{\bar a}$, the previous variables become functions of
suitable combinations of {\it instantaneous} and {\it retarded}
terms all depending only on the matter. At the lowest order in
$1/c^2$ the results of the harmonic gauge used in the IAU
conventions for the solar system are recovered (if ${}^3K$ is
negligible inside the solar system).

\medskip

At this stage the extrinsic curvature tensor ${}^3K_{rs}(\tau, \vec
\sigma)$  is completely determined except for the numerical function
describing its trace ${}^3K(\tau, \vec \sigma)$. Therefore, the
final determination of the dynamical instantaneous 3-spaces
$\Sigma_{\tau}$ (whose inner 3-curvature is determined by the
gravitational waves and matter) associated to the linearized
solution requires a choice of the York time: this has to be done by
relying on the {\it observational conventions}, which hide a notion
of clock synchronization. Instead in the harmonic gauges one uses
the Euclidean instantaneous 3-spaces of the inertial observers of
Minkowski space-time, avoiding to look at the extrinsic curvature
associated to the harmonic solutions!

\medskip

Both in the previous solutions and in the resulting Hamilton
equations for matter the influence of the relativistic inertial
effects connected with the York time ${}^3K$ are explicitly shown.
It turns out that at the HPM level all the equations depend on the
following spatially non-local function of the York time: ${}^3{\cal
K} \, {\buildrel {def}\over =}\, {1\over {\triangle}}\, {}^3K$.

\bigskip

In a third paper \cite{b} we will give the PN expansion at all
orders of the matter equations of motion (in absence of the
electro-magnetic field) in the slow motion limit resulting from the
HPM linearization. There we  will evaluate the dependence on the
inertial gauge variable York time ${}^3K_{(1)}  =\, \triangle\,
{}^3{\cal K}_{(1)}$ of geometrical and physical quantities. The
relevance for astrophysics and cosmology of these results  will be
discussed in the third paper, after an analysis of the relation of
the gauge problem in general relativity with the conventions used
for the description of matter (extended bodies) in the geocentric
(GCRS), barycentric (BCRS( and celestial (ICRS) reference frames.
There we will discuss the possibility that  dark matter can be
simulated with these relativistic inertial effects in a
Post-Minkowskian (PM) extension of ICRS.

\bigskip

In Section II we review the Hamilton equations of the gravitational
field and of matter in our family of 3-orthogonal gauges, which were
given in Appendix C of paper I in the 3-orthogonal gauges defined in
that paper.

In Section III we define our linearization scheme and we define the
HPM expansion. Also the coordinate transformation connecting the
3-orthogonal gauges to the harmonic ones at the lowest HPM order is
defined.

In Section IV we solve the linearized constraints and the equations
for the lapse and shift functions in our family of 3-orthogonal
gauges. Also the ADM Poincare' generators are given till the second
order.

In Section V we give the linearization of the equations of motion
for the particles and the electro-magnetic field.

In Section VI we give the linearized second order equations of
motion for the tidal variables $R_{\bar a}$ and we show that they
imply the wave equation (with respect to the asymptotic Minkowski
4-metric) for the TT part ${}^4h^{TT}_{(1)rs}$ of the diagonal
3-metric ${}^4g_{(1)rs}$ on the 3-space $\Sigma_{\tau}$. Also a
generalized (non 3-orthogonal) TT gauge is identified

In Section VII we study the retarded solution for HPM GW with
asymptotic background and we show that the dominant term is the
emission quadrupole formula by using a multipolar expansion of the
energy-momentum tensor in terms of Dixon multipoles. Also the far
field behavior is considered. Then we study the energy balance
associated to the emission of GW's. Finally we look at the problem
of detection of GW's.

In the Conclusions, after a review of the results and of the
problems which will appear at the second HPM order, we discuss the
{\it gauge problem in general relativity}, the dependence upon the
observational conventions for the 4-coordinates of the description
of matter and the relevance of the York time ${}^3K_{(1)}$ for
explaining at least part of dark matter as relativistic inertial
effects (to be discussed in the third paper \cite{b}).

For a comparison with our formulation, in Appendix A there is a
review of the standard approach to GW's by using Einstein's
equations in harmonic gauges.

In Appendix B there is a discussion of Dixon multipoles and of the
multipolar expansion of the energy-momentum tensor.

In Appendix C there is the study of the balance for momentum and
angular momentum in GW's emission.

\vfill\eject

\section{The Hamilton Equations in the 3-Orthogonal Schwinger Time
Gauges}

As shown in paper I, the 3-orthogonal gauges of ADM tetrad gravity
are the family of Schwinger time gauges where we have

\bea
 &&\alpha_{(a)}(\tau ,\vec \sigma ) \approx 0,\qquad
 \varphi_{(a)}(\tau ,\vec \sigma ) \approx 0,\nonumber \\
 &&{}\nonumber \\
 &&\theta^i(\tau ,\vec \sigma) \approx 0,\qquad
 \pi_{\tilde
\phi}(\tau ,\vec \sigma) = {{c^3}\over {12\pi\, G}}\, {}^3K(\tau
,\vec \sigma) \approx {{c^3}\over {12\pi\, G}}\, F(\tau, \vec
\sigma),
 \label{2.1}
 \eea

\noindent in the York canonical basis ($a, r = 1,2,3$, $\bar a =
1,2$)

\beq
 \begin{minipage}[t]{4 cm}
\begin{tabular}{|ll|ll|l|l|l|} \hline
$\varphi_{(a)}$ & $\alpha_{(a)}$ & $n$ & ${\bar n}_{(a)}$ &
$\theta^r$ & $\tilde \phi$ & $R_{\bar a}$\\ \hline
$\pi_{\varphi_{(a)}} \approx0$ &
 $\pi^{(\alpha)}_{(a)} \approx 0$ & $\pi_n \approx 0$ & $\pi_{{\bar n}_{(a)}} \approx 0$
& $\pi^{(\theta )}_r$ & $\pi_{\tilde \phi}$ & $\Pi_{\bar a}$ \\
\hline
\end{tabular}
\end{minipage}
 \label{2.2}
 \eeq
\medskip

The tidal variables of the gravitational field are $R_{\bar a}$,
$\Pi_{\bar a}$. The primary gauge variables (the inertial effects
connected with the choice of 3-coordinates on $\Sigma_{\tau}$ and
with clock synchronization) are $\theta^r$ and $\pi_{\tilde \phi}$:
their conjugate variables $\pi_r^{(\theta)}$ and $\tilde \phi$ are
the unknowns in the super-momentum and super-Hamiltonian
constraints, respectively. While $\tilde \phi(\tau, \vec \sigma)$ is
the conformal factor of the 3-metric (namely the 3-volume element),
$\pi_r^{(\theta)}$ may be replaced with the off-diagonal terms
$\sigma_{(a)(b)}{|}_{a \not= b}(\tau, \vec \sigma)$ of the shear of
the congruence of Eulerian observers associated with the 3+1
splitting (the diagonal elements are connected to the tidal momenta
$\Pi_{\bar a}$). The secondary gauge (inertial) variables
(determined after a gauge fixation of the primary ones) are the
shift ($n^r = {}^3{\bar e}^r_{(a)}\, {\bar n}_{(a)}$) and lapse ($1
+ n$) functions.\medskip

The canonical variables for the particles are $\eta^r_i(\tau)$,
$\kappa_{ir}(\tau)$, $i = 1,..,N$ (we use the notation $\kappa_{ir}$
instead of the one ${\check \kappa}_{ir}$ of paper I). For the
electromagnetic field in the radiation gauge the canonical variables
are $A_{\perp r}(\tau, \vec \sigma)$, $\pi^r_{\perp}(\tau, \vec
\sigma)$ and from paper I we have the following expression for the
electro-magnetic fields $F_{rs} = \partial_r\, A_{\perp s} -
\partial_s\, A_{\perp r}$, $B_r = \epsilon_{ruv}\,
\partial_u\, A_{\perp v}$, $E_r = - F_{\tau r} = - \partial_{\tau}\,
A_{\perp r} + \partial_r\, A_{\tau}$ with $A_{\tau}$ given in
Eq.(3.32) of I.\bigskip

The members of this family of gauges differ for the value of the
York time ${}^3K(\tau, \vec \sigma) \approx F(\tau, \vec \sigma)$,
where $F(\tau, \vec \sigma)$ is a numerical function (for $F(\tau,
\vec \sigma) = const.$ we get the constant mean curvature (CMC)
gauges of ADM theory). This gauge variable, describing a
relativistic inertial effect (which does not exist in Newton theory
in Galileo space-time), is the remnant of the special relativistic
gauge freedom in choosing the convention for clock synchronization,
i.e. for the identification of the instantaneous 3-spaces
$\Sigma_{\tau}$. While in special relativity the whole extrinsic
curvature tensor ${}^3K_{rs}(\tau, \vec \sigma)$ of $\Sigma_{\tau}$
is pure gauge, in general relativity it is determined by the
dynamics with the exception of its trace \footnote{Let us remark
that positive (negative) extrinsic curvature implies that the
instantaneous 3-space, as an embedded 3-manifold, is a concave
(convex) 3-surface of the space-time}.\medskip

In this Section we will give the restriction to this family of
gauges of the Hamilton equations  and of the constraints given in
paper I by using the results of its Appendix C.

\bigskip

We will use the following notational conventions (most of them are
defined in paper I):\medskip

a) We write the conformal factor of the 3-metric in the form $\phi
(\tau ,\vec \sigma) = {\tilde \phi}^{1/6}(\tau ,\vec \sigma) =
e^{q(\tau, \vec \sigma)}$. Then we have $\phi^{-1}\, \partial_r\,
\phi = {1\over 6}\, {\tilde \phi}^{-1}\, \partial_r\, \tilde \phi =
\partial_r\, q$, $\phi^{-1}\, \partial_r^2\, \phi = \partial_r^2\, q
+ (\partial_r\, q)^2$ (this notation was not used in paper
I).\medskip

b) We use $V_{ra}$ for $V_{ra}(\theta^n)$ to simplify the notation.
We use the notation  $V_{ra}(0) = \delta_{ra}$, $V_{(i)ra} =
{{\partial\, V_{ra}(\theta^n)}\over {\partial\,
\theta^i}}{|}_{\theta^i = 0}$, $B_{(i)jw} = {{\partial\,
B_{jw}(\theta^n)}\over {\partial\, \theta^i}}{|}_{\theta^i = 0}$. As
said in Subsection IIC of paper I, we use angles $\theta^i$
corresponding to canonical coordinates of first kind on the group
manifold of SO(3), because this implies  $V_{(i)rs} = 2\, B_{(i)rs}
= \epsilon_{irs}$.\medskip

c) The set of numerical parameters $\gamma_{\bar aa}$ satisfies
\cite{5,6} $\sum_u\, \gamma_{\bar au} = 0$, $\sum_u\, \gamma_{\bar a
u}\, \gamma_{\bar b u} = \delta_{\bar a\bar b}$, $\sum_{\bar a}\,
\gamma_{\bar au}\, \gamma_{\bar av} = \delta_{uv} - {1\over 3}$. A
different York canonical basis is associated to each solution of
these equations.
\bigskip

As shown in paper I, in the York canonical basis we have the
following building blocks for the Einstein-Maxwell- Particle system
in the radiation gauge of the electro-magnetic field \footnote{The
so-called gothic inverse 4-metric ${\bf h}^{AB} = \sqrt{- {}^4g}\,
{}^4g^{AB} - {}^4\eta^{AB}$ takes the form: ${\bf h}^{\tau\tau} =
\sgn\, ({{\tilde \phi}\over {1 + n}} - 1)$, ${\bf h}^{\tau r} = -
\sgn\, {{{\tilde \phi}^{2/3}}\over {1 + n}}\, Q_r^{-1}\, {\bar
n}_{(r)}$, ${\bf h}^{rs} = - \sgn\, [{{{\tilde \phi}^{1/3}}\over {1
+ n}}\, Q_r^{-1}\, Q_s^{-1}\, ((1 + n)^2\, \delta_{rs} - {\bar
n}_{(r)}\, {\bar n}_{(s)}) - \delta_{rs}]$. At spatial infinity we
get ${\bf h}^{AB} \, \rightarrow\, 0$.} (${}^3{\bar e}_{(a)r}$ and
${}^3{\bar e}^r_{(a)}$ are cotriads and triads on $\Sigma_{\tau}$
respectively)

\begin{eqnarray*}
 {}^4g_{\tau\tau} &=& \sgn\, \Big[(1 + n)^2 - \sum_a\,
 {\bar n}_{(a)}^2\Big],\nonumber \\
 {}^4g_{\tau r} &=& - \sgn\, \sum_a\, {\bar n}_{(a)}\, {}^3{\bar e}_{(a)r} =
 - \sgn\, {\tilde \phi}^{1/3}\,  Q_r\, {\bar n}_{(r)},\nonumber \\
 {}^4g_{rs} &=& - \sgn\, {}^3g_{rs} = - \sgn\, \sum_a\, {}^3{\bar e}_{(a)r}\,
 {}^3{\bar e}_{(a)s} = - \sgn\, \phi^4\,
 {}^3{\hat g}_{rs}\, =\, - \sgn\, {\tilde \phi}^{2/3}\,  Q^2_r\,
 \delta_{rs},\nonumber \\
 &&{}\nonumber \\
 &&Q_a\, =\, e^{\Gamma_q^{(1)}},\quad \Gamma_a^{(1)} = \sum_{\bar a}^{1,2}\,
 \gamma_{\bar aa}\, R_{\bar a}, \qquad
 \tilde \phi = \phi^6 = e^{6\, q} = \sqrt{\gamma} =
 \sqrt{det\, {}^3g} = {}^3\bar e,
 \end{eqnarray*}

 \begin{eqnarray*}
 {}^3e_{(a)r} &=& {}^3{\bar e}_{(a)r} = {\tilde \phi}^{1/3}\, Q_a\,
 \delta_{ra},\qquad
 {}^3e^r_{(a)} = {}^3{\bar e}^r_{(a)} = {\tilde \phi}^{- 1/3}\, Q^{-1}_a\,
 \delta_{ra},
 \end{eqnarray*}

\begin{eqnarray*}
 \pi^{(\theta)}_i &=&  {{c^3}\over {8\pi\, G}}\, \tilde \phi\,
 \sum_{ab}\,  Q_a\, Q_b^{-1}\, \epsilon_{iab}\,
 \sigma_{(a)(b)}{|}_{a\not= b},\nonumber \\
  && \tilde \phi\, \sigma_{(a)(b)}{|}_{a \not= b} = - {{8\pi\, G}\over
 {c^3}}\,  {{\epsilon_{abi}}\over
 {Q_b\, Q_a^{-1} - Q_a\, Q_b^{-1}}}\, \pi_i^{(\theta )},\nonumber \\
 &&{}\nonumber \\
  \Pi_{\bar a} &=&  - {{c^3}\over {8\pi\, G}}\, \tilde \phi\,
 \sum_a\, \gamma_{\bar aa}\, \sigma_{(a)(a)},\qquad
  \tilde \phi\, \sigma_{(a)(a)} = - {{8\pi\, G}\over {c^3}}\,
 \sum_{\bar a}\, \gamma_{\bar aa}\, \Pi_{\bar a},
 \end{eqnarray*}

\begin{eqnarray*}
  {}^3K_{rs} &\approx&  - {{4\pi\, G}\over {c^3}}\, {\tilde \phi}^{-1/3}\,
 \Big( Q^2_r\, \delta_{rs}\,  [2\, \sum_{\bar b}\, \gamma_{\bar br}\,
 \Pi_{\bar b} -  \tilde \phi\, \pi_{\tilde \phi}] -
 2\,  Q_r\, Q_s\,   {{\epsilon_{rsi}\, \pi_i^{(\theta )}}\over {
 Q_r\, Q^{-1}_s  - Q_s\, Q^{-1}_r}} \Big) =\nonumber \\
 &=&  {\tilde \phi}^{2/3}\, \Big[ {{4\pi\, G}\over
 {c^3}}\, \pi_{\tilde \phi}\,  Q^2_r\, \delta_{rs}\,  +
 (1 - \delta_{rs})\, \sigma_{(r)(s)}\, Q_r\, Q_s -
 {{8\pi\, G}\over {c^3}}\, {\tilde \phi}^{-1}\, \sum_{\bar a}\,
 \gamma_{\bar ar}\, \Pi_{\bar a}\, Q^2_r\, \delta_{rs} \Big],
 \end{eqnarray*}

 \bea
 \sigma^2 &{\buildrel {def}\over =}& {1\over 2}\, \sum_{ab}\,
 \sigma_{(a)(b)}\, \sigma_{(a)(b)} = {1\over 2}\, \Big({{8\pi\,
 G}\over {c^3}}\Big)^2\, {\tilde \phi}^{-2}\, \Big[\sum_{\bar a}\,
 \Pi^2_{\bar a} +\nonumber \\
 &+&2\,  \Big({{(\pi_1^{(\theta)})^2}\over {(Q_2\, Q_3^{-1}
  - Q_3\, Q_2^{-1})^2}} + {{(\pi_2^{(\theta)})^2}\over {(Q_3\, Q_1^{-1}
   - Q_1\, Q_3^{-1})^2}} + {{(\pi_3^{(\theta)})^2}\over {(Q_1\, Q_2^{-1}
    - Q_2\, Q_1^{-1})^2}}\Big)\Big].
 \label{2.3}
 \eea

Eqs.(2.18) and (6.2) of I have been used.  $\theta = - \sgn\, {}^3K$
and  $\sigma_{(a)(b)}$ are the expansion and the shear of the
congruence of  Eulerian observers of $\Sigma_{\tau}$ as shown in
paper I. In paper I it is also shown that the original momenta
conjugate to the cotriads ${}^3e_{(a)r}$ before going to the York
canonical basis are ${}^3\pi^r_{(a)} \approx {\tilde \phi}^{-1/3}\,
[\delta_{ra}\, Q_a^{-1}\, (\tilde \phi\, \pi_{\tilde \phi} +
\sum_{\bar a}\, \gamma_{\bar aa}\, \Pi_{\bar a}) + \sum_i\,
Q_r^{-1}\, {{\epsilon_{ari}\, \pi_i^{(\theta)}}\over {Q_r\, Q_a^{-1}
- Q_a\, Q_r^{-1}}}]$.
\bigskip

From Eqs.(6.4) of I we have the following expressions of the mass
and momentum densities (we use the notation ${\cal M}$ and
$\kappa_{ir}$ instead of ${\check {\cal M}}$ and ${\check
\kappa}_{ir}$ of I for the quantities in the electro-magnetic
radiation gauge; see the Introduction for $c(\vec \sigma, {\vec
\sigma}^{'})$)

 \bea
 {\cal M}(\tau ,\vec \sigma ){|}_{\theta^i =0} &=& \sum_i\, \delta^3(\vec \sigma
 ,{\vec \eta}_i(\tau ))\,\, \eta_i\, \sqrt{m_i^2\, c^2 +
 {\tilde \phi}^{-2/3}\, \sum_a\, Q_a^{-2}\,  \Big(
 \kappa_{ia}(\tau ) - {{Q_i}\over c}\, A_{\perp\, a}\Big)^2\,}
 (\tau ,\vec \sigma ) +\nonumber \\
  &+& {1\over {2c}}\, \Big[{\tilde  \phi}^{-1/3}\, \Big( \sum_{rsa}\, Q_a^2\,
  \delta_{ra}\, \delta_{sa}\, \pi^r_{\perp}\,  \pi^s_{\perp} +
   {1\over 2}\, \sum_{ab}\, Q_a^{-2}\, Q_b^{-2}\,  F_{ab}\, F_{ab}\Big)\Big](\tau ,\vec
  \sigma) -\nonumber \\
   &-&  {1\over {2c}}\,
  \Big[{\tilde \phi}^{-1/3}\, \sum_{rsan}\, Q_a^2\, \delta_{ra}\, \delta_{sa}\,
  \Big(2\, \pi^r_{\perp} - \sum_m\, \delta^{rm}\, \sum_i\, Q_i\, \eta_i\, {{\partial\,
  c(\vec \sigma , {\vec \eta}_i(\tau))}\over {\partial\, \sigma^m}}\Big)
  \nonumber \\
  &&\delta^{sn}\,\sum_j\, Q_j\, \eta_j\, {{\partial\,
  c(\vec \sigma , {\vec \eta}_j(\tau))}\over {\partial\,
  \sigma^n}}\Big](\tau ,\vec \sigma),\nonumber \\
   &&{}\nonumber \\
 {\cal M}_r(\tau ,\vec \sigma )&=& \sum_{i=1}^N\, \eta_i\,
 \Big(\kappa_{ir}(\tau ) - {{Q_i}\over c}\,
 A_{\perp r}(\tau ,\vec \sigma)\Big)\, \delta^3(\vec
 \sigma ,{\vec \eta}_i(\tau )) -\nonumber \\
 &-& {1\over c}\, \sum_s\, F_{rs}(\tau ,\vec \sigma)\, \Big(
  \pi_\perp^s(\tau,  \vec{\sigma}) - \sum_n\, \delta^{sn}\, \sum_i\, Q_i\, \eta_i\,
 {{\partial\, c(\vec \sigma, {\vec \eta}_i(\tau))}\over {\partial\,
 \sigma^n}}\Big),\nonumber \\
 &&{}
   \label{2.4}
 \eea

From Eqs. (C8), (C12) and (C20) of I we get the following
expressions for the Laplace-Beltrami operator and for the intrinsic
3-curvature  $ {}^3R[\theta^n, \phi ,R_{\bar a}] = \phi^{-5}\,
\Big(- 8\,  {\hat \triangle}\, \phi + {}^3{\hat R}\, \phi \Big) =
\phi^{-6}\, \Big({\cal S} + {\cal T}\Big)$ with ${}^3\hat
R[\theta^n, R_{\bar a}] =  \phi^{-2}\, \Big({\cal S} + {\cal
T}_1\Big)$

\begin{eqnarray*}
 \hat \triangle{|}_{\theta^i = 0}\, &=&\, \sum_a\, Q_a^{-2}(\tau ,\vec \sigma)\,
 \Big[ \partial^2_a -  2\,  \sum_{\bar b}\, \gamma_{\bar ba}\, \partial_a\,
 R_{\bar b}(\tau ,\vec \sigma) \, \partial_a\Big] =\nonumber \\
 &=& \sum_a\, Q_a^{-2}(\tau ,\vec \sigma)\,
 \Big[ \partial^2_a -  2\, \partial_a\, \Gamma_a^{(1)}(\tau ,\vec
 \sigma)\, \partial_a\Big],
  \end{eqnarray*}

\bea
 {\cal S}{|}_{\theta^i = 0} &=&{\tilde \phi}^{1/3}\, \sum_a\, Q_a^{-2}\, \Big(\sum_{\bar b}\,
 \Big[\sum_{\bar c}\, (2\, \gamma_{\bar ba}\, \gamma_{\bar ca} -
 \delta_{\bar b\bar c})\, \partial_a\, R_{\bar b}\, \partial_a\, R_{\bar c}
 -\nonumber \\
 &-& 4\, \gamma_{\bar ba}\, \partial_a\, q\, \partial_a\, R_{\bar b}\Big] +
  8\, ( \partial_a\, q)^2\Big),\nonumber \\
  &&{}\nonumber \\
 {\cal T}(\tau ,\vec \sigma ){|}_{\theta^i = 0} &=&
  - 2\, \Big[{\tilde \phi}^{1/3}\, \sum_a\, Q_a^{-2}\, \Big(4\, \Big[
 \partial_a^2\, q + 2\, (\partial_a\, q)^2\Big] -\nonumber \\
 &-& \partial_a^2\, \Gamma_a^{(1)} - 2\, (5\, \partial_a\,
 q -  \partial_a\, \Gamma_a^{(1)})\, \partial_a\, \Gamma_a^{(1)}\Big)
 \Big](\tau ,\vec \sigma),\nonumber \\
 &&{}\nonumber \\
 {\cal T}_1(\tau ,\vec \sigma ){|}_{\theta^i = 0} &=&
 - 2\, \Big[{\tilde \phi}^{1/3}\, \sum_a\, Q_a^{-2}\, \Big(4\, (\partial_a\,
 q)^2 -  \partial_a^2\, \Gamma_a^{(1)} - 2\, (\partial_a\, q -
 \partial_a\, \Gamma_a^{(1)})\, \partial_a\, \Gamma_a^{(1)}\Big)
 \Big](\tau ,\vec \sigma),\nonumber \\
 &&{}\nonumber \\
 {}^3{\hat R}(\tau ,\vec \sigma){|}_{\theta^i=0} &=&
 \sum_a\, \Big(Q_a^{-2}\,  \Big[2\,
 \partial_a^2\, \Gamma^{(1)}_a - \sum_{\bar b}\, (\partial_a\, R_{\bar b})^2 -
 2\, \Big(\partial_a\, \Gamma^{(1)}_a\Big)^2\Big]
 \Big)(\tau ,\vec \sigma).\nonumber \\
 &&{}
 \label{2.5}
 \eea

\bigskip

In the 3-orthogonal gauges we have the following form of the
constraints, of the Bianchi identities and of the equations of
motion of gauge and physical variables.

\bigskip

\subsection{The Constraints}

The super-Hamiltonian and super-momentum constraints determining
$\phi = {\tilde \phi}^{1/6} = e^q$ and $\sigma_{(a)(b)}{|}_{a \not=
b} = - {{8\pi\, G}\over {c^3}}\, {\tilde \phi}^{-1}\, \sum_i\,
{{\epsilon_{abi}\, \pi_i^{(\theta)}}\over {Q_b\, Q_a^{-1} - Q_a\,
Q_b^{-1}}}$, given in Eqs. (6.5), (6.6), (C12), (C20) of I, are
[Eqs.(\ref{2.4}) and (\ref{2.5}) are also needed; we use the
notation ${\cal H}_{(a)}$ instead of ${\bar {\tilde {\cal
H}}}_{(a)}$ of I]

\bea
  {\cal H}(\tau ,\vec \sigma ){|}_{\theta^i = 0} &=&
  {{c^3}\over {2\pi\, G}}\, \phi(\tau ,\vec \sigma )\, \Big(
 ({\hat \triangle}{|}_{\theta^i = 0} -
 {1\over 8}\, {}^3\hat R{|}_{\theta^i = 0})\, \phi +
 {{2\pi\, G}\over {c^3}}\, \phi^{-1}\,
 {\cal M}{|}_{\theta^i = 0} +\nonumber \\
 &+& {{8\, \pi^2\, G^2}\over {c^6}}\, \phi^{-7}\, \sum_{\bar a}\, \Pi^2_{\bar a}
 + {1\over 8}\, \phi^{5}\, \sum_{ab, a\not= b}\, \sigma_{(a)(b)}\,
 \sigma_{(a)(b)} - {1\over {12}}\, \phi^{5}\, ({}^3K)^2
 \Big)(\tau ,\vec \sigma ) \approx 0,\nonumber \\
 \label{2.6}
 \eea

  \bea
  {\cal H}_{(a)}{|}_{\theta^i = 0}(\tau ,\vec \sigma) &=&
   - {{c^3}\over {8\pi\, G}}\,  {\tilde \phi}^{2/3}(\tau, \vec
 \sigma)\, \Big(\sum_{b \not= a}\, Q_b^{-1}\, \Big[\partial_b\,
 \sigma_{(a)(b)} +
  \Big(6\,  \partial_b\, q + \partial_b\,
  (\Gamma_a^{(1)} - \Gamma_b^{(1)})\Big)\, \sigma_{(a)(b)}\Big]
  -\nonumber \\
 &-& Q_a^{-1}\,  \Big[{2\over 3}\, \partial_a\, {}^3K
 +  {{8\pi\, G}\over {c^3}}\, {\tilde \phi}^{-1}\, \Big(\, \sum_{\bar b}\,
 (\gamma_{\bar ba}\, \partial_a\, \Pi_{\bar b}
 - \partial_a\, R_{\bar b}\, \Pi_{\bar b}) + {\cal
 M}_{(a)}\Big)\Big] \Big)(\tau ,\vec \sigma)
 \quad \approx 0.\nonumber \\
 &&{}
 \label{2.7}
 \eea

All the constraints depend on the York time ${}^3K$.

\subsection{The Contracted Bianchi Identities}

The Hamilton equations for the unknowns in the constraints, given in
Eqs.(6.7), (6.8), (C21), (C11), (C19) and (C5) of I, are ($\cir$
means evaluated by means of the equations of motion)

\bea
 \partial_{\tau}\, {\tilde \phi} (\tau ,\vec \sigma ){|}_{\theta^i = 0}
 = 6\, (\tilde \phi\, \partial_{\tau}\, q)(\tau, \vec \sigma){|}_{\theta^i = 0}
 &\cir&  \Big[ -  (1 + n)\, \tilde \phi\,
 {}^3K +   {\tilde \phi}^{2/3}\,  \sum_a\, Q_a^{-1}\,
 \Big(\partial_a\, {\bar n}_{(a)} +\nonumber \\
 &+& {\bar n}_{(a)} \,  \Big(4\, \partial_a\, q -
 \partial_a\, \Gamma_a^{(1)}\Big)
  \Big)\Big](\tau ,\vec \sigma ),\nonumber \\
 &&{}
 \label{2.8}
 \eea

 \bea
  \partial_{\tau}\, \pi_i^{(\theta )}(\tau ,\vec \sigma ){|}_{\theta^i = 0}
  &=& {{c^3}\over {8\pi\, G}}\, \tilde \phi\, \sum_{ab}\,
 \epsilon_{iab}\, Q_a\, Q_b^{-1}\, \Big[\partial_{\tau}\, \sigma_{(a)(b)} +
 \Big(6\,  \partial_{\tau}\, q +
 \partial_{\tau}\, (\Gamma_a^{(1)} - \Gamma_b^{(1)})\Big)\,
 \sigma_{(a)(b)}\Big].\nonumber \\
 &&{}
 \label{2.9}
 \eea

\noindent We can extract $\partial_{\tau}\, \sigma_{(a)(b)}$ from
Eqs.(\ref{2.9}) due to Eqs.(\ref{2.3}), i.e. ${{8\pi\, G}\over
{c^3}}\, \pi_i^{(\theta)} = \tilde \phi\, \sum_{ab}\,
\epsilon_{iab}\, Q_a\, Q_b^{-1}\, \sigma_{(a)(b)}$. By using
$V_{(i)rs} = 2\, B_{(i)rs} = \epsilon_{irs}$ and Eqs. (6.8) and
(C5), (C11), (C19), (C21) of I, we get

\bigskip

\begin{eqnarray*}
 \partial_{\tau}\, \sigma_{(a)(b)}{|}_{a \not= b, \theta^i=0} &\cir&
 - \Big[\Big(6\, \partial_{\tau}\, q +
 {{Q_b\, Q_a^{-1} + Q_a\, Q_b^{-1}}\over {Q_b\, Q_a^{-1} - Q_a\, Q_b^{-1}}}\,
 \partial_{\tau}\, (\Gamma_a^{(1)} - \Gamma_b^{(1)})\,
  \Big)\, \sigma_{(a)(b)}{|}_{a \not= b}\Big](\tau ,\vec \sigma)
 -\nonumber \\
 &-& \Big({\tilde \phi}^{-1}\, \sum_i\,
 {{\epsilon_{abi}}\over {Q_b\, Q_a^{-1} - Q_a\, Q_b^{-1}}}\Big)(\tau,
 \vec \sigma)\, \Big[F_i(\tau, \vec \sigma)-\nonumber \\
 &-& \int d^3\sigma_1\, \Big((1 + n(\tau, {\vec \sigma}_1))\,
 \Big[{{8\pi\, G}\over {c^3}}\, {{\delta\, {\cal M}(\tau, {\vec
 \sigma}_1)}\over {\delta\, \theta^i(\tau, \vec \sigma)}}{|}_{\theta^i=0} -
 {1\over 2}\, {{\delta\, {\cal S}(\tau, {\vec \sigma}_1)}\over
 {\delta\, \theta^i(\tau, \vec \sigma)}}{|}_{\theta^i=0} \Big] -\nonumber \\
 &-& {1\over 2}\, n(\tau, {\vec \sigma}_1)\, {{\delta\, {\cal T}(\tau,
 {\vec \sigma}_1)}\over {\delta\, \theta^i(\tau, \vec \sigma)}}{|}_{\theta^i=0}
 \Big)\, \Big],
 \end{eqnarray*}

\bigskip

\begin{eqnarray*}
   &&\int d^3\sigma_1\, \Big(1 + n(\tau ,{\vec \sigma}_1)\Big)\,
 {{\delta\, {\cal M}(\tau ,{\vec \sigma}_1)}\over
 {\delta\, \theta^i(\tau ,\vec \sigma)}}{|}_{\theta^i=0} =\nonumber \\
 &&{}\nonumber \\
 &=&{1\over 2}\, \sum_i\, \delta^3(\vec \sigma, {\vec
 \eta}_i(\tau))\, \eta_i\, \Big((1 + n)\nonumber \\
 &&{{{\tilde \phi}^{-2/3}\, \sum_{rsa}\, Q_a^{-2}\, ( V_{(i)ra}\,
 \delta_{sa} + \delta_{ra}\, V_{(i)sa})\,  \Big(
 \kappa_{ir}(\tau ) - {{Q_i}\over c}\, A_{\perp\, r}\Big)\,
 \Big(\kappa_{is}(\tau ) - {{Q_i}\over c}\,
 A_{\perp\, s}\Big)}\over {\sqrt{m_i^2\, c^2 +
 {\tilde \phi}^{-2/3}\, \sum_{a}\, Q_a^{-2}\,  \Big(
 \kappa_{ia}(\tau ) - {{Q_i}\over c}\, A_{\perp\, a}\Big)^2\,
 }}} \Big)(\tau ,\vec \sigma) +\nonumber \\
 &&{}\nonumber \\
  &+& \Big(1 + n(\tau ,\vec \sigma)\Big)\,  \Big({\tilde \phi}^{-1/3}\,
 {1\over {c}}\, \Big[\, \sum_{ars}\, Q_a^2\,
  V_{(i)ra}\,  \delta_{sa}
  \pi^r_{\perp}\, \pi^s_{\perp} +  \sum_{abr}\, Q_a^{-2}\, Q_b^{-2}\,
  V_{(i)ra}\, F_{rb}\, F_{ab} -\nonumber \\
  &-&  {1\over 2}\, \sum_{arsn}\, Q_a^2\,  \Big(V_{(i)ra}\,
  \delta_{sa} + V_{(i)sa}\,  \delta_{ra}\Big)\,
  \Big(2\, \pi^r_{\perp} - \sum_m\, \delta^{rm}\, \sum_i\, Q_i\, \eta_i\, \partial_m\,
  c(\vec \sigma , {\vec \eta}_i(\tau))\Big)\nonumber \\
 &&\delta^{sn}\, \sum_j\, Q_j\, \eta_j\, \partial_n\,
  c(\vec \sigma , {\vec \eta}_j(\tau))
  \Big]\Big)(\tau ,\vec \sigma),
 \end{eqnarray*}

\begin{eqnarray*}
    &&\int d^3\sigma_1\, [1 + n(\tau ,{\vec \sigma}_1)]\,
   {{\delta\, {\cal S}(\tau ,{\vec \sigma}_1
  )}\over {\delta\, \theta^i(\tau ,\vec \sigma )}}\,
  {|}_{\theta^i = 0} =\nonumber \\
  &=&- 2\,\Big[{\tilde \phi}^{1/3}\, \sum_{ra}\, Q_a^{-2}\, V_{(i)ra}\, \Big(
 \partial_a\, n\, \partial_r\, q -
 \sum_{\bar b}\, \gamma_{\bar ba}\, \partial_r\, R_{\bar b}) +
 \partial_r\, n\,  \partial_a\, q - \sum_{\bar b}\,
 \gamma_{\bar br}\, \partial_a\, R_{\bar b}) -\nonumber \\
 &-& (1 + n)\, \Big[2\, (2\, \partial_a\, q\,
  \partial_r\, q -  \partial_a\,
 \partial_r\, q) + \sum_{\bar b}\, (\gamma_{\bar ba} +
 \gamma_{\bar br})\, \partial_a\, \partial_r\, R_{\bar b}
 +\nonumber \\
 &+& 2\, \sum_{\bar b}\, (\gamma_{\bar ba}\, \partial_a\,
 q\, \partial_r\, R_{\bar b} + \gamma_{\bar br}\,
 \partial_r\, q\, \partial_a\, R_{\bar b}) -\nonumber \\
 &-& \sum_{\bar b\bar c}\, (2\, \gamma_{\bar br}\, \gamma_{\bar ca}
 + \delta_{\bar b\bar c})\, \partial_a\, R_{\bar b}\,
 \partial_r\, R_{\bar c} \Big]\Big) \Big](\tau ,\vec \sigma),
 \end{eqnarray*}

 \begin{eqnarray*}
    &&\int d^3\sigma_1\, n(\tau, {\vec \sigma}_1)\,
   {{\delta\, {\cal T}(\tau ,{\vec \sigma}_1)}\over
 {\delta\, \theta^i(\tau ,\vec \sigma )}}
 {|}_{\theta^i = 0} =\nonumber \\
  &&{}\nonumber \\
  &&{}\nonumber \\
 &=& - 2\, \Big[{\tilde \phi}^{1/3}\, \sum_{ra}\, Q_a^{-2}\, V_{(i)ra}\,
 \Big( \partial_r\, \partial_a\, n - 3\, (\partial_r\, n\,
 \partial_a\, q + \partial_a\, n\,
 \partial_r\, q) \Big)\Big](\tau ,\vec \sigma),
 \end{eqnarray*}

\begin{eqnarray*}
 F_i(\tau, \vec \sigma)  &=&{{8\pi\, G}\over {c^3}}\,
   \int d^3\sigma_1\, \sum_a\, {\bar n}_{(a)}(\tau ,{\vec \sigma}_1)\,
   {{\delta\, {\cal H}_{(a)}(\tau ,{\vec \sigma}_1
  )}\over {\delta\, \theta^i(\tau ,\vec \sigma )}}\,
  {|}_{\theta^i = 0} -\nonumber \\
 &-& {3\over 2}\, \Big((1 + n)\, \tilde \phi\,
 \sum_{m \not= n}\, {{\sigma_{(m)(n)}}\over {Q_m\, Q_n^{-1} - Q_n\, Q_m^{-1}}}
  \sum_{c \not= d}\, \sum_{tk}\, \epsilon_{mnt}\, \epsilon_{ikt}\,
 \epsilon_{kcd}\, Q_c\, Q_d^{-1}\, \sigma_{(c)(d)} \Big)(\tau ,\vec \sigma)
  =\nonumber \\
   &=& - \Big[{\tilde \phi}^{-1/3}\, \sum_a\, \Big(\sum_r\,
 \partial_r\, {\bar n}_{(a)}\, \Big[{{8\pi\, G}\over {c^3}}\,
 Q_a^{-1}\, \epsilon_{ira}\, \sum_{\bar b}\,
 (\gamma_{\bar ba}\,  - \gamma_{\bar br})\, \Pi_{\bar b} -\nonumber \\
 &-&  \tilde \phi\, \Big(\sum_{b\not= a}\,
 Q_b^{-1}\, \epsilon_{irb}\, \sigma_{(a)(b)} + Q_a^{-1}\, \sum_{b\not= r}\,
 Q_b\, Q_r^{-1}\, \epsilon_{iab}\, \sigma_{(r)(b)} \Big)\Big] +\nonumber \\
 &+&  \tilde \phi\, \partial_a\, {\bar
 n}_{(a)}\, Q_a^{-1}\, \sum_{b\not= c}\, \epsilon_{icb}\, Q_b\, Q_c^{-1}\,
 \sigma_{(b)(c)} \Big)\Big](\tau ,\vec \sigma) +\nonumber \\
 &&{}\nonumber \\
 &+& \Big[{\tilde \phi}^{-1/3}\, \sum_a\, {\bar n}_{(a)}\,
 \Big({{8\pi\, G}\over {c^3}}\,
 Q_a^{-1}\, \sum_b\, \epsilon_{iba}\, \Big[({{c^3}\over {12\pi\, G}}\,
 \tilde \phi\, \partial_b\, {}^3K
 - \sum_{\bar b}\, \partial_b\, R_{\bar b}\, \Pi_{\bar b}) +\nonumber \\
 &+& \sum_{\bar b}\, \gamma_{\bar bb}\, \partial_b\, \Pi_{\bar b} +
  \sum_{\bar b}\, (\gamma_{\bar ba} - \gamma_{\bar bb})\,
 (2\, \partial_b\, q +
  \partial_b\, \Gamma_a^{(1)})\, \Pi_{\bar b}
  +  {\cal M}_b\Big] +
  \end{eqnarray*}

\bea
 &+&  \tilde \phi\, \Big[Q_a^{-1}\, \sum_{b,
 c \not= b}\, Q_b\, Q_c^{-1}\, \Big(\epsilon_{iab}\, \partial_c\, \sigma_{(b)(c)}
 - \epsilon_{icb}\, \partial_a\, \sigma_{(b)(c)} + 4\,
 (\epsilon_{iab}\, \partial_c\, q -\nonumber \\
 &-& \epsilon_{icb}\, \partial_a\, q)\, \sigma_{(b)(c)}
 - (\epsilon_{iab}\, \partial_c\,  - \epsilon_{icb}\, \partial_a)\,
 (\Gamma_a^{(1)} - \Gamma_b^{(1)} + \Gamma_c^{(1)})\,
 \sigma_{(b)(c)}\Big)  -\nonumber \\
 &-& \sum_{r\, b \not= a}\, Q_b^{-1}\, \sigma_{(a)(b)}\, \epsilon_{irb}\,
 (2\,  \partial_r\, q +
 \partial_r\, \Gamma_a^{(1)} )\Big]\Big)\, \Big](\tau ,\vec \sigma) .\nonumber \\
 &&{}
 \label{2.10}
 \eea

\subsection{The Shift Functions}

From the $\tau$-preservation of the gauge fixings $\theta^i(\tau
,\vec \sigma) \approx 0$, see Eqs.(6.10) of I, we get the following
equations for the shift functions

\bea
 &&\Big(Q_b^{-1}\, \partial_b\, {\bar n}_{(a)} + Q_a^{-1}\, \partial_a\,
 {\bar n}_{(b)} - \Big[Q_b^{-1}\,  \Big(2\, \partial_b\,
 q + \partial_b\, \Gamma_a^{(1)}\Big)\,
 {\bar n}_{(a)} +\nonumber \\
 &+& Q_a^{-1}\,  \Big(2\,  \partial_a\, q +
 \partial_a\, \Gamma_b^{(1)}\Big)\,
 {\bar n}_{(b)}\Big]\Big)(\tau ,\vec \sigma) \approx\nonumber \\
 &\approx&  2\, \Big[{\tilde \phi}^{1/3}\, (1 + n)\,
 \sigma_{(a)(b)}{|}_{a \not= b}\Big](\tau ,\vec \sigma),
 \qquad a \not= b.
 \label{2.11}
 \eea

\subsection{The Instantaneous 3-Space and the Lapse Functions}

The preservation in $\tau$ of the gauge fixing constraint
${}^3K(\tau, \vec \sigma) \approx F(\tau, \vec \sigma)$ given in
Eqs. (\ref{2.1}) gives  Eq.(6.12) of paper I. The restriction of
this equation to our family of gauges (by using Eqs. (C22), (C10)
and (C17) of I and the super-Hamiltonian constraint) gives the
following equation for the determination of the lapse function (it
is the Raychaudhuri equation)

\begin{eqnarray*}
 &&\Big(\sum_a\, Q_a^{-1}\, \Big[\partial_a^2\, n + {1\over 2}\, \Big(2\,
 \partial_a\, q - {7\over 2}\,
 \partial_a\, \Gamma_a^{(1)}\Big)\,
 \partial_a\, n \Big] -\nonumber \\
 &-& {{4\pi\, G}\over {c^3}}\, (1 + n)\, \Big[{\tilde \phi}^{-1/3}\,
 {\cal M} - {1\over 2}\, {\tilde \phi}^{-1/6}\,
 \int d^3\sigma_1\, \Big(1 + n(\tau ,{\vec \sigma}_1)\Big)\,
 {{\delta\, {\cal M}(\tau ,{\vec \sigma}_1)}\over
 {\delta\, \phi(\tau ,\vec \sigma)}} \Big] -\nonumber \\
 &-&(1 + n)\, \Big[{1\over 3}\, {\tilde \phi}^{2/3}\, ({}^3K)^2
  + ({{8\pi\, G}\over {c^3}})^2\,
 {\tilde \phi}^{-4/3}\, \sum_{\bar a}\, \Pi^2_{\bar a} +
 {\tilde \phi}^{2/3}\, \sum_{ab, a \not= b}\, \sigma^2_{(a)(b)}
 \Big] -\nonumber \\
 &-&  {\tilde \phi}^{2/3}\, \Big[ -
\partial_{\tau}\, {}^3K + {\tilde \phi}^{-1/3}\,
\sum_a\, {\bar n}_{(a)}\, Q_a^{-1}\, \partial_a\, {}^3K
 \Big] \Big)(\tau, \vec \sigma) = 0,
 \end{eqnarray*}

\bea
   &&\int d^3\sigma_1\, \Big(1 + n(\tau ,{\vec \sigma}_1)\Big)\,
 {{\delta\, {\cal M}(\tau ,{\vec \sigma}_1)}\over
 {\delta\, \phi(\tau ,\vec \sigma)}} =\nonumber \\
 &&{}\nonumber \\
 &=& - 2\, \sum_i\, \delta^3(\vec \sigma, {\vec
 \eta}_i(\tau))\, \eta_i\, \Big((1 + n)\,
 {{{\tilde \phi}^{-5/6}\, \sum_{a}\, Q_a^{-2}\,
 \Big(\kappa_{ia}(\tau ) - {{Q_i}\over c}\,
  A_{\perp\, a}\Big)^2\, }\over {\sqrt{m_i^2\, c^2 +
 {\tilde \phi}^{-2/3}\, \sum_{a}\, Q_a^{-2}\,  \Big(
 \kappa_{ia}(\tau ) - {{Q_i}\over c}\, A_{\perp\, a}\Big)^2\,
 }}} \Big)(\tau ,\vec \sigma) -\nonumber \\
 &&{}\nonumber \\
  &-& \Big(1 + n(\tau ,\vec \sigma)\Big)\, \Big({\tilde \phi}^{-1/2}\,
 \Big[{1\over {c}}\, \sum_{ars}\, Q_a^2\,
  \delta_{ra}\, \delta_{sa}\, \pi^r_{\perp}\, \pi^s_{\perp} +
   {1\over {2c}}\, \sum_{ab}\, Q_a^{-2}\, Q_b^{-2}\, F_{ab}\, F_{ab}
   -\nonumber \\
 &-&  {1\over {c}}\, \sum_{arsn}\, Q_a^2\, \delta_{ra}\, \delta_{sa}\,
  \Big(2\, \pi^r_{\perp} - \sum_m\, \delta^{rm}\, \sum_i\, Q_i\, \eta_i\, \partial_m\,
  c(\vec \sigma , {\vec \eta}_i(\tau))\Big)\nonumber \\
 &&\delta^{sn}\, \sum_j\, Q_j\, \eta_j\, \partial_n\,
  c(\vec \sigma , {\vec \eta}_j(\tau))
 \Big]\Big)(\tau ,\vec \sigma).
 \label{2.12}
 \eea

\bigskip

We can also consider gauges in which in Eq.(\ref{2.1}) we have
${}^3K(\tau, \vec \sigma) \approx F(\tau, \vec \sigma, \vec \sigma -
{\vec \eta}_1(\tau), ..., \vec \sigma - {\vec \eta}_N(\tau))$. In
this case we have $\partial_{\tau}\, {}^3K = \partial_{\tau}\,
F{|}_{{\vec \eta}_i(\tau)} + \sum_i\, {{\partial\, F}\over
{\partial\, {\vec \eta}_i}}\, {\dot {\vec \eta}}_i(\tau)$ with
${\dot {\vec \eta}}_i(\tau)$ given by the first set of Hamilton
equations for the particles.

\subsection{The Dirac Multipliers}

Once the lapse and shift functions are known, the Dirac multipliers
appearing in the Dirac Hamiltonian, given in Eq.(3.48) of paper I,
are determined by the following equations

\bea
 \partial_{\tau}\, n(\tau ,\vec \sigma ) &\cir& \lambda_n(\tau
 ,\vec \sigma ),\nonumber \\
 \partial_{\tau}\, {\bar n}_{(a)}(\tau ,\vec \sigma ) &\cir&
 \lambda_{{\bar n}_{(a)}}(\tau ,\vec \sigma ).
 \label{2.13}
 \eea

\subsection{The Weak ADM Energy}

The weak ADM energy, given in Eqs. (6.3) and (C8) of I,  becomes
(${\cal S}$ is given in Eq.(\ref{2.5}))

  \bea
 {\hat E}_{ADM}{|}_{\theta^i = 0} &=& c\, \int d^3\sigma\,
 \Big[ {\cal M} {|}_{\theta^i = 0} -
  {{c^3}\over {16\pi\, G}}\, {\cal S}{|}_{\theta^i = 0} +
  \nonumber \\
  &+& {{4\pi\, G}\over {c^3}}\, {\tilde \phi}^{-1}\, \sum_{\bar a}\,
  \Pi^2_{\bar a} + {{c^3}\over {16\pi\, G}}\, \tilde \phi\, \sum_{a
  \not= b}\, \sigma^2_{(a)(b)} - {{c^3}\over {24\pi\, G}}\, \tilde
  \phi\, ({}^3K)^2 \Big](\tau ,\vec \sigma ),\nonumber \\
 &&{}
 \label{2.14}
 \eea

\noindent where  Eqs.(\ref{2.4}) and (\ref{2.5}) have to be used. As
noted in paper I the gauge momentum proportional to the inertial
York time ${}^3K$ (existing due to the Lorentz signature of
space-time) gives rise to a negative gauge kinetic term, without
analogue in ordinary gauge theories (electro-magnetism and
Yang-Mills theory).

\subsection{The Equations of Motion for the Tidal Variables}

By using Eqs.(6.13), (C9), (C15), (C23) and (C7) of I, we get the
expression of the momenta $\Pi_{\bar a}$ and then the second order
equations of motion for the tidal variables $R_{\bar a}$. For the
tidal momenta we get

 \bea
  \Pi_{\bar a}(\tau, \vec \sigma) &=& - {{c^3}\over {8\pi\, G}}\,
  \Big[\tilde \phi\, \sum_a\, \gamma_{\bar aa}\, \sigma_{(a)(a)}\Big](\tau
  ,\vec \sigma) \cir\nonumber \\
 &\cir& {{c^3}\over {8\pi\, G}}\, {{\tilde \phi(\tau ,\vec \sigma)}\over
 {1 + n(\tau, \vec \sigma)}}\, \Big[\partial_{\tau}\, R_{\bar a} +
 {\tilde \phi}^{-1/3}\, \sum_a\,  Q_a^{-1}\nonumber \\
 && \Big(\Big[\gamma_{\bar aa}\,(2\, \partial_a\,
 q + \partial_a\, \Gamma_a^{(1)})\, -
 \partial_a\, R_{\bar a}\Big]\, {\bar n}_{(a)} - \gamma_{\bar aa}\,
 \partial_a\, {\bar n}_{(a)} \Big)\Big](\tau ,\vec \sigma),\nonumber \\
 &&{}
 \label{2.15}
 \eea

As a consequence, for the tidal variables $R_{\bar a}$ we get
(Eq.(\ref{2.15}) is used to eliminate the dependence upon $\Pi_{\bar
a}$ of the equation of paper I)

\begin{eqnarray*}
 \partial^2_{\tau}\, R_{\bar a}(\tau ,\vec \sigma ) \cir\,
  &\approx&  \Big[{\tilde \phi}^{-1/3}\, \sum_a\,  Q_a^{-1}\,  {\bar
 n}_{(a)}\, \sum_{\bar b}\, (\gamma_{\bar aa}\, \gamma_{\bar ba}
 - \delta_{\bar a\bar b})\, \partial_a\,
 \partial_{\tau}\, R_{\bar b} +\nonumber \\
 &+&{\tilde \phi}^{-1/3}\, \sum_a\, Q_a^{-1}\, \Big[\Big(\gamma_{\bar aa}\,
 (2\,  \partial_a q +
 \partial_a\, \Gamma_a^{(1)}) -  \partial_a\,
 R_{\bar a} \Big)\, {\bar n}_{(a)} -\nonumber \\
 &-& \gamma_{\bar aa}\,  \partial_a\,
 {\bar n}_{(a)}\Big]\, \partial_{\tau}\, \Gamma_a^{(1)} -\nonumber \\
 &-& {\tilde \phi}^{-1}\, \Big[\partial_{\tau}\, R_{\bar a} + {2\over 3}\,
 {\tilde \phi}^{-1/3}\, \sum_a\, Q_a^{-1}\, \Big(\Big[\gamma_{\bar aa}\,
 (-   \partial_a\, q + \partial_a\, \Gamma_a^{(1)}) -\nonumber \\
 &-& \partial_a\, R_{\bar a}\Big]\, {\bar n}_{(a)} - \gamma_{\bar aa}\,
  \partial_a\, {\bar n}_{(a)}\Big)\Big]\, \partial_{\tau}\, \tilde
 \phi -\end{eqnarray*}

\begin{eqnarray*}
 &-& {1\over 3}\, {\tilde \phi}^{-4/3}\, \sum_a\, \gamma_{\bar aa}\,
 Q_a^{-1}\,  {\bar n}_{(a)}\, \partial_a\, \partial_{\tau}\,
 \tilde \phi +\nonumber \\
 &+&\Big[ \partial_{\tau}\, R_{\bar a} + {\tilde \phi}^{-1/3}\, \sum_a\,
 Q_a^{-1}\, \Big(\Big[\gamma_{\bar aa}\,  (2\, \partial_a\, q +
 \partial_a\, \Gamma_b^{(1)}) -\nonumber \\
 &-&  \partial_a\, R_{\bar a} \Big]\, {\bar n}_{(a)} -
 \gamma_{\bar aa}\,  \partial_a\, {\bar n}_{(a)}\Big)\Big]\,
 {{\partial_{\tau}\, n}\over {1 + n}} -\nonumber \\
 &-& {\tilde \phi}^{-1/3}\, \sum_a\, Q_a^{-1}\, \Big(\Big[\gamma_{\bar aa}\,
 (2\,  \partial_a\, q +
 \partial_a\, \Gamma_a^{(1)}) -  \partial_a\,
 R_{\bar a}\Big]\, \partial_{\tau}\, {\bar n}_{(a)} -\nonumber \\
 &-&\gamma_{\bar aa}\,  \partial_a\, \partial_{\tau}\, {\bar n}_{(a)}
  \Big)\Big](\tau ,\vec \sigma ) +
 \end{eqnarray*}

 \begin{eqnarray*}
  &+& {1\over 2}\, \Big({\tilde \phi}^{-1}\, (1 + n)\Big)(\tau ,\vec \sigma)\, \int
 d^3\sigma_1\, \Big[(1 + n)(\tau, {\vec \sigma}_1)\,
 {{\delta\, {\cal S}(\tau ,{\vec \sigma}_1)}\over
 {\delta\, R_{\bar a}(\tau, \vec \sigma)}}{|}_{\theta^i = 0}
  + n(\tau ,{\vec \sigma}_1)\,
 {{\delta\, {\cal T}(\tau ,{\vec \sigma}_1)}\over
 {\delta\, R_{\bar a}(\tau, \vec \sigma)}}{|}_{\theta^i = 0}\Big] +\nonumber \\
 &&{}\nonumber \\
  &+& \Big({\tilde \phi}^{-1}\, (1 + n)\Big)(\tau ,\vec \sigma)\,
 \Big({{{\tilde \phi}^{2/3}}\over {1 + n}}\, \sum_a\, Q_a^{-1}\,
 \Big[ \Big(\partial_a\, {\bar n}_{(a)} + {\bar n}_{(a)}\, (4\, \partial_a\, q
 - \partial_a\, \Gamma_a^{(1)} - {{\partial_a\, n}\over {1 + n}})\Big)\,
 \Big(\partial_{\tau}\, R_{\bar a} +\nonumber \\
 &+& {\tilde \phi}^{-1/3}\, \sum_c\,  Q_c^{-1}\,
 \Big[\Big(\gamma_{\bar ac}\,(2\, \partial_c\,
 q + \partial_c\, \Gamma_c^{(1)})\, -
 \partial_c\, R_{\bar a}\Big)\, {\bar n}_{(c)} - \gamma_{\bar ac}\,
 \partial_c\, {\bar n}_{(c)} \Big]\Big) +\nonumber \\
 &+& {\bar n}_{(a)}\, \partial_a\, \Big(\partial_{\tau}\, R_{\bar a} +
 {\tilde \phi}^{-1/3}\, \sum_c\,  Q_c^{-1}\,
  \Big[\Big(\gamma_{\bar ac}\,(2\, \partial_c\,
 q + \partial_c\, \Gamma_c^{(1)})\, -
 \partial_c\, R_{\bar a}\Big)\, {\bar n}_{(c)} - \gamma_{\bar ac}\,
 \partial_c\, {\bar n}_{(c)} \Big]\Big) \Big] +\nonumber \\
 &+& {\tilde \phi}^{2/3}\, \sum_{ab, a\not=
 b}\, Q_b^{-1}\, (\gamma_{\bar aa} - \gamma_{\bar ab})\,
 \Big[\partial_b\, {\bar n}_{(a)} - (2\,  \partial_b\, q
 + \partial_b\, \Gamma_a^{(1)})\,
 {\bar n}_{(a)}\Big]\, \sigma_{(a)(b)}\Big)(\tau, \vec \sigma) -
 \nonumber \\
 &&{}\nonumber \\
  &-& {{8\pi\, G}\over {c^3}}\, \Big({\tilde \phi}^{-1}\, (1 +
 n)\Big)(\tau ,\vec \sigma)\, \int
 d^3\sigma_1\, (1 + n)(\tau, {\vec \sigma}_1)\,
 {{\delta\, {\cal M}(\tau ,{\vec \sigma}_1)}\over
 {\delta\, R_{\bar a}(\tau, \vec \sigma)}}{|}_{\theta^i = 0},
  \end{eqnarray*}

\bigskip

\begin{eqnarray*}
   &&\int d^3\sigma_1\,\, [1 + n(\tau ,{\vec \sigma}_1)]\,
   {{\delta\, {\cal S}(\tau ,{\vec \sigma}_1)}\over
 {\delta\, R_{\bar a}(\tau ,\vec \sigma )}} {|}_{\theta^i = 0} =\nonumber \\
  &&{}\nonumber \\
  &&{}\nonumber \\
 &=&2\, \Big({\tilde \phi}^{1/3}\, \sum_a\, Q_a^{-2}\, \Big[\partial_a\, n\, \Big(2\,
 \gamma_{\bar aa}\,  \partial_a\, q - \sum_{\bar b}\,
 (2\, \gamma_{\bar aa}\, \gamma_{\bar ba} - \delta_{\bar a\bar b})\,
 \partial_a\, R_{\bar b}\Big) -\nonumber \\
 &-& (1 + n)\, \Big(2\, \gamma_{\bar aa}\, \Big(-  \partial_a^2\,
 q + 2\, (\partial_a\, q)^2\Big) +\nonumber \\
 &+&\sum_{\bar b}\, (2\, \gamma_{\bar aa}\,
 \gamma_{\bar ba} - \delta_{\bar a\bar b})\, (\partial_a^2\,
 R_{\bar b} + 2\,  \partial_a\, q\, \partial_a\,
 R_{\bar b} )  +\nonumber \\
 &+& \sum_{\bar b\bar c}\, \Big(2\, \gamma_{\bar ba}\,
 (\delta_{\bar a\bar c} - \gamma_{\bar aa}\, \gamma_{\bar ca}) -
 \gamma_{\bar aa}\, \delta_{\bar b\bar c}\Big)\, \partial_a\, R_{\bar b}\,
 \partial_a\, R_{\bar c} \Big) \Big]\Big)(\tau ,\vec \sigma ),
 \end{eqnarray*}

  \begin{eqnarray*}
   \int d^3\sigma_1\, n(\tau, {\vec \sigma}_1)\,
   {{\delta\, {\cal T}(\tau ,{\vec \sigma}_1)}\over
 {\delta\, R_{\bar a}(\tau ,\vec \sigma )}}
 {|}_{\theta^i = 0}
 &=& 2\, \Big[{\tilde \phi}^{1/3}\, \sum_a\, \gamma_{\bar aa}\, Q_a^{-2}\,
 \Big( \partial^2_a\, n - 6\,  \partial_a\,
 q\, \partial_a\, n \Big)\Big](\tau ,\vec \sigma),
 \end{eqnarray*}

\bea
   &&\int d^3\sigma_1\, \Big(1 + n(\tau ,{\vec \sigma}_1)\Big)\,
 {{\delta\,  {\cal M}(\tau ,{\vec \sigma}_1)}\over
 {\delta\, R_{\bar a}(\tau ,\vec \sigma)}} =\nonumber \\
 &&{}\nonumber \\
 &=& - \sum_i\, \delta^3(\vec \sigma, {\vec
 \eta}_i(\tau))\, \eta_i\, \Big((1 + n)\nonumber \\
 && {{{\tilde \phi}^{-2/3}\, \sum_{a}\, \gamma_{\bar aa}\, Q_a^{-2}\,
 \Big(\kappa_{ia}(\tau ) - {{Q_i}\over c}\,
 A_{\perp\, a}\Big)^2\, }\over {\sqrt{m_i^2\, c^2 +
 {\tilde \phi}^{-2/3}\, \sum_{a}\, Q_a^{-2}\,  \Big(
 \kappa_{ia}(\tau ) - {{Q_i}\over c}\, A_{\perp\, a}\Big)^2\,
 }}} \Big)(\tau ,\vec \sigma) +\nonumber \\
 &&{}\nonumber \\
 &+& \Big(1 + n(\tau ,\vec \sigma)\Big)\,  \Big({\tilde \phi}^{-1/3}\,
 \Big[{1\over {c}}\, \sum_{ars}\, \gamma_{\bar aa}\, Q_a^2\,
  \delta_{ra}\, \delta_{sa}\, \pi^r_{\perp}\, \pi^s_{\perp}  -
   {1\over {2c}}\, \sum_{ab}\, (\gamma_{\bar aa} + \gamma_{\bar ab})\,
   Q_a^{-2}\, Q_b^{-2}\, F_{ab}\, F_{ab} -\nonumber \\
  &-&   {1\over {c}}\, \sum_{arsn}\, \gamma_{\bar aa}\,
  Q_a^2\, \delta_{ra}\, \delta_{sa}\,
  \Big(2\, \pi^r_{\perp} - \sum_m\, \delta^{rm}\, \sum_i\, Q_i\, \eta_i\, \partial_m\,
  c(\vec \sigma , {\vec \eta}_i(\tau))\Big)\nonumber \\
 &&\delta^{sn}\,  \sum_j\, Q_j\, \eta_j\, \partial_n\,
  c(\vec \sigma , {\vec \eta}_j(\tau))
 \Big]\Big)(\tau ,\vec \sigma).\nonumber \\
 &&{}
 \label{2.16}
 \eea

\noindent The three integrals at the end of Eq.(\ref{2.16}) were
given in Eqs.(C28), (C29) and (C27) of paper I. The expression of
the last integral in Eqs.(\ref{2.16}) has been obtained by using the
super-momentum constraints (\ref{2.7}).
\medskip

To get the final form of the second order  equations for $R_{\bar
a}$ we have to use: a) Eq.(\ref{2.8}) for $\partial_{\tau}\, \tilde
\phi$; b) the Hamilton equations (\ref{2.13}) for the Dirac
multipliers in the 3-orthogonal gauges with $\partial_{\tau}\, n$
and $\partial_{\tau}\, {\bar n}_{(r)}$ determined by the solution of
Eqs. (\ref{2.12}) and (\ref{2.11}); c) Eq.(\ref{2.15}) for
$\Pi_{\bar a}$.

\subsection{The Equations of Motion for the Particles}

By using Eqs.(6.14) and (6.15) of paper I, the first half of the
Hamilton equations for the particles implies

 \begin{eqnarray*}
  \eta_i\, {\dot \eta}^r_i(\tau ) &\cir& \eta_i\,
  \Big({{{\tilde \phi}^{-2/3}\, (1 + n)\,  Q_r^{-2}\,
  \Big(\kappa_{ir}(\tau ) - {{Q_i}\over c}\, A_{\perp\, r}\Big)\,
 }\over {\sqrt{m_i^2\, c^2 + {\tilde \phi}^{-2/3}\, \sum_c\, Q^{-2}_c\,
 \Big( \kappa_{ic}(\tau ) - {{Q_i}\over c}\, A_{\perp\, c}\Big)^2\,
 }}} -\nonumber \\
 &-& \phi^{-2}\, Q_r^{-1}\,  {\bar n}_{(r)}
 \Big)(\tau ,{\vec \eta}_i(\tau )),
 \end{eqnarray*}

\bea
 &&\Downarrow\nonumber \\
 &&{}\nonumber \\
 && \kappa_{ir}(\tau ) =
 {{Q_i}\over c}\, A_{\perp\,r}(\tau ,{\vec \eta}_i(\tau ))
 + m_i\, c\,\,\ \Big[ {\tilde \phi}^{2/3}\,  Q_r^2\,
  \Big(\dot{\eta}^r_i(\tau ) + {\tilde \phi}^{-1/3}\,
 Q_r^{-1}\,  {\bar n}_{(r)}\Big)\,\, \Big(\Big(1 + n\Big)^2 -
 \nonumber \\
 &-& {\tilde \phi}^{2/3}\, \sum_c\,
 Q_c^2\,  \Big(\dot{\eta}^c_i(\tau ) + {\tilde \phi}^{-1/3}\,
  Q_c^{-1}\,  {\bar n}_{(c)} \Big)^2\, \Big)^{-1/2}\Big]
 (\tau ,{\vec \eta}_i(\tau )).
 \label{2.17}
 \eea

\noindent so that the second half of the Hamilton equations becomes

 \begin{eqnarray*}
 \eta_i\, {d\over {d\tau}}\,\,&& \Big(
  m_i\, c\,\,\ \Big({{ {\tilde \phi}^{2/3}\,  Q_r^2\,
 \Big(\dot{\eta}^r_i(\tau ) + {\tilde \phi}^{-1/3}\,
 Q_r^{-1}\,  {\bar n}_{(r)}\Big)}\over { \sqrt{\Big(1 + n\Big)^2 -
  {\tilde \phi}^{2/3}\, \sum_{c}\,
 Q_c^2\, \Big(\dot{\eta}^c_i(\tau ) + {\tilde \phi}^{-1/3}\,
 Q_c^{-1}\,  {\bar n}_{(c)} \Big)^2}}}
  \Big)(\tau, {\vec \eta}_i(\tau))\,\, \cir\nonumber \\
 &&{}\nonumber \\
 &\cir&\,\, \Big(- {{\partial}\over
 {\partial\, \eta_i^r}}\,  {\cal W}
 + {{\eta_i\, Q_i}\over c}\, ({\dot \eta}^s_i(\tau)\, {{\partial\, A_{\perp\, s}}\over
 {\partial\, \eta_i^r}} - {{d\, A_{\perp\, r}}\over {d\tau}}) +
 \eta_i\, {\check F}_{ir} \Big)(\tau ,{\vec
 \eta}_i(\tau)), \nonumber \\
 &&{}\nonumber \\
 &&{}\nonumber \\
 {\cal W} &=& \int d^3\sigma\, \Big[ (1 + n)\, {\cal W}_{(n)} +
 {\tilde \phi}^{-1/3}\, \sum_a\,
 Q_a^{-1}\, {\bar n}_{(a)}\,  {\cal W}_a\Big](\tau, \vec \sigma),\nonumber \\
  {\cal W}_{(n)}(\tau ,\vec \sigma) &=& - {1\over
  {2c}}\,\Big[{\tilde \phi}^{-1/3}\, \sum_a\, Q_a^2\,
 \left(2\, \pi_\perp^a - \delta^{am}\, \sum_i\, Q_i\, \eta_i\,
 {{\partial\, c(\vec \sigma, {\vec \eta}_i(\tau))}\over {\partial\, \sigma^m}}
 \right)\nonumber \\
 && \delta^{an}\, \sum_j\, Q_j\, \eta_j\, {{\partial\, c(\vec \sigma,
 {\vec \eta}_j(\tau))}\over {\partial\, \sigma^n}}
 \Big](\tau ,\vec \sigma),\nonumber \\
 &&{}\nonumber \\
 {\cal W}_r(\tau ,\vec \sigma) &=&  - {1\over c}\, \sum_s\,
 F_{rs}(\tau ,\vec \sigma)\, \delta^{sn}\, \sum_i\, Q_i\, \eta_i\,
 {{\partial\, c(\vec \sigma, {\vec \eta}_i(\tau))}\over {\partial\,
 \sigma^n}},\end{eqnarray*}

 \bea
 {\check F}_{ir} &=& m_i\, c\, \Big[\Big(1 + n\Big)^2 -
  {\tilde \phi}^{2/3}\, \sum_c\,
 Q_c^2\,  \Big(\dot{\eta}^c_i(\tau ) + {\tilde \phi}^{-1/3}\,
 Q_c^{-1}\,  {\bar n}_{(c)} \Big)^2\Big]^{-1/2}\nonumber \\
 &&\Big[- (1 + n)\, {{\partial\, n}\over {\partial\, \eta_i^r}} +
 {\tilde \phi}^{1/3}\, \sum_{a}\,  Q_a\, \Big( {{\partial\,
 {\bar n}_{(a)}}\over {\partial\, \eta_i^r}} -\nonumber \\
 &-&(2\,  \partial_r\, q +
 \sum_{\bar a}\, \gamma_{\bar aa}\, \partial_r\, R_{\bar a})\,
 {\bar n}_{(a)}\Big)\, \Big({\dot \eta}_i^a(\tau) + {\tilde \phi}^{-1/3}\,  Q_a^{-1}\,
 {\bar n}_{(a)}\Big) +\nonumber \\
 &+&{\tilde \phi}^{2/3}\, \sum_{a}\, Q_a^2\,
 (2\,  \partial_r\, q +
 \sum_{\bar a}\, \gamma_{\bar aa}\, \partial_r\, R_{\bar a})\,
 \Big({\dot \eta}_i^a(\tau) + {\tilde \phi}^{-1/3}\,  Q_a^{-1}\,
 {\bar n}_{(a)}\Big)^2 \,\, \Big].\nonumber \\
 &&{}
 \label{2.18}
 \eea

Here ${\cal W}$ is the non-inertial Coulomb potential, ${\check
F}_{ir}$ are inertial relativistic forces and the other terms
correspond to the non-inertial Lorentz force \cite{4}.

\subsection{The Equations of Motion for the Transverse
Electro-Magnetic Field}

Finally, from Eqs.(6.16) of I the Hamilton equations  for the
transverse electro-magnetic fields in the radiation gauge become
($P^{rs}_{\perp}(\vec \sigma) = \delta^{rs} - \sum_{uv}\,
\delta^{ru}\, \delta^{sv}\, {{\partial_u\, \partial_v}\over
{\Delta}}$)

\begin{eqnarray*}
 \partial_{\tau}\, A_{\perp\, r}(\tau ,\vec \sigma ) &\cir&
  \sum_{nua}\, \delta_{rn}\, P^{nu}_{\perp}(\vec \sigma)\, \Big[
 {\tilde \phi}^{-1/3}\, (1 + n)\, Q_a^2\, \delta_{ua}\,  \Big(\pi^a_{\perp} -
 \sum_m\, \delta^{am}\, \sum_i\, Q_i\, \eta_i\, {{\partial\,
 c(\vec \sigma, {\vec \eta}_i(\tau))}\over {\partial\, \sigma^m}}
 \Big) +\nonumber \\
 &+&  {\tilde \phi}^{-1/3}\, Q_a^{-1}\,  {\bar n}_{(a)}\, F_{au}
 \Big](\tau ,\vec \sigma),\nonumber \\
 &&{}\nonumber \\
 \partial_{\tau}\, \pi^r_{\perp}(\tau ,\vec \sigma) &\cir&
  \sum_{m}\, P^{rm}_{\perp}(\vec \sigma)\,  \Big(\sum_a\,
 \delta_{ma}\, \sum_i\, \eta_i\, Q_i\, \delta^3(\vec \sigma, {\vec
 \eta}_i(\tau))\, \nonumber \\
 &&\Big[{ { {\tilde \phi}^{-2/3}\, (1 + n)\, Q_a^{-2}\,
   \kappa_{ia}(\tau)}\over
 {\sqrt{m_i^2\, c^2 + {\tilde \phi}^{-2/3}\, \sum_b\, Q_b^{-2}\,
  \Big(\kappa_{ib}(\tau ) - {{Q_i}\over c}\,
 A_{\perp\, b}\Big)^2\, }}} -\nonumber \\
  &-& {\tilde \phi}^{-1/3}\, Q_a^{-1}\,
  {\bar n}_{(a)}\Big](\tau ,{\vec \eta}_i(\tau)) -
 \end{eqnarray*}

  \bea
 &-&\Big[2\, {\tilde \phi}^{-1/3}\, (1 + n)\, \sum_{ab}\, Q_a^{-2}\, Q_b^{-2}\,
 \delta_{ma}\, \Big(\partial_b\, F_{ab} - \Big[2\,
 \partial_b\, q + 2\, \partial_b\,
 (\Gamma_a^{(1)} + \Gamma_b^{(1)})\Big]\, F_{ab} \Big) +\nonumber \\
 &+&2\, {\tilde \phi}^{-1/3}\, \sum_{ab}\, Q_a^{-2}\, Q_b^{-2}\,
 \delta_{ma}\, \partial_b\, n\, F_{ab} -\nonumber \\
 &-& {\tilde \phi}^{-1/3}\, \sum_a\, {\bar n}_{(a)}\, Q_a^{-1}\,
  \Big(\partial_a\, \pi_{\perp}^m - \Big[2\, \partial_a\, q +
 \partial_a\, \Gamma_a^{(1)}\Big]\, \pi^m_{\perp} +\nonumber \\
 &+& \delta_{ma}\, \sum_n\,  \Big[ 2\, \partial_n\, q +
 \partial_n\, \Gamma_a^{(1)}\Big]\, \pi^n_{\perp} +\nonumber \\
 &+& \sum_i\, \eta_i\, Q_i\, \Big[  \Big(2\, \partial_a\, q +
 \partial_a\, \Gamma_a^{(1)}\Big)\, {{\partial\, c(\vec \sigma,
 {\vec \eta}_i(\tau)))}\over {\partial\, \sigma^m}} -
 {{\partial^2\, c(\vec \sigma, {\vec \eta}_i(\tau)))}
 \over {\partial\, \sigma^m\, \partial\, \sigma^a}}
 -\nonumber \\
 &-& \delta_{ma}\, \sum_n\, \Big( \Big[2\, \partial_n\, q +
 \partial_n\, \Gamma_a^{(1)}\Big]\, {{\partial\, c(\vec \sigma,
 {\vec \eta}_i(\tau)))}\over {\partial\, \sigma^n}} -
 {{\partial^2\, c(\vec \sigma, {\vec \eta}_i(\tau)))}
 \over {\partial\, \sigma^n\, \partial\, \sigma^n}} \Big)
 \Big] \Big) +\nonumber \\
 &+& {\tilde \phi}^{-1/3}\, \sum_a\, Q_a^{-1}\, \sum_i\, \eta_i\, Q_i\,
 \Big(\partial_a\, {\bar n}_{(a)}\,
  {{\partial\, c(\vec \sigma, {\vec \eta}_i(\tau))}\over {\partial\, \sigma^m}}
 - \delta_{ma}\, \sum_n\, \partial_n\, {\bar n}_{(a)}\,
 {{\partial\, c(\vec \sigma, {\vec \eta}_i(\tau))}\over {\partial\, \sigma^n}}
 \Big)\,\, \Big](\tau ,\vec \sigma)\, \Big).\nonumber \\
 &&{}
 \label{2.19}
 \eea

\subsection{The Weak ADM Poincare' Generators}

While the weak ADM energy ${\hat P}^{\tau}_{ADM} = {1\over c}\,
{\hat E}_{ADM}$ is given in Eq.(\ref{2.14}), Eqs.(2.22) and (3.47)
of paper I give the following expressions in the 3-orthogonal gauges
for the other weak ADM Poincare' generators (the last term in the
boosts was added in Ref.\cite{18})

\begin{eqnarray*}
 {\hat P}^r_{ADM} &=&
 \int d^3\sigma \, \Big[{}^3g^{rs}\, {\cal M}_{s}
 - 2\, {}^3\Gamma^{r}_{su}(\tau ,\vec \sigma )\, {}^3\Pi^{su}
 \Big](\tau ,\vec \sigma )\,
 =\nonumber \\
 &&{}\nonumber \\
 &=& 2\int d^3\sigma\,\Big\{
 \,\tilde{\phi}^{-2/3}\sum_{\bar{b}}\,Q_r^{-2} \Big(\, 2\,
 \gamma_{\bar{b}r}\, \partial_r\, q +
 \sum_{\bar{a}}(\gamma_{\bar{b}r}\gamma_{\bar{a}r}-
 {1\over 2}\, \delta_{\bar{a}\bar{b}})\,\partial_rR_{\bar{a}}\Big)
 \,\Pi_{\bar{b}} -\nonumber\\
 &&\nonumber\\
 &-& {{c^3}\over {12\pi\, G}}\, \tilde{\phi}^{1/3}\,Q_r^{-2} \Big(\, 4\, \partial_r\, q +
 \partial_r\, \Gamma_r^{(1)}\Big)\, {}^3K +\nonumber\\
 &&\nonumber\\
 &+&\frac{c^3}{8\pi G}\tilde{\phi}^{1/3}\sum_{d}\,Q_r^{-1}Q_d^{-1}
 \Big(\,2\, \partial_d\, q +
 \partial_d\, \Gamma_r^{(1)}\Big)
 \,\sigma_{(r)(d)} + \frac{1}{2}\tilde{\phi}^{-2/3}\,Q_r^{-2}\,{\cal
 M}_r\, \Big\}\, \approx 0,
 \end{eqnarray*}

\begin{eqnarray*}
 {\hat J}^{rs}_{ADM} &=&
  \int d^3\sigma\,  \Big[- 2 (\sigma^{r}\, {}^3\Gamma^{s} _{uv} -
 \sigma^{s}\, {}^3\Gamma^{r}_{uv})\, {}^3\Pi^{uv} +
 (\sigma^{r}\, {}^3g^{su} - \sigma^{s}\, {}^3g^{ru})\, {\cal
 M}_{u}\Big] (\tau ,\vec \sigma ) =\nonumber \\
 &&{}\nonumber \\
 &=& 2\, \int d^3\sigma\,\Big\{\,
 \sigma^r\,\Big[ \,\tilde{\phi}^{-2/3}\sum_{\bar{b}}\,Q_s^{-2}
 \Big(\, 2\, \gamma_{\bar{b}s}\, \partial_s\, q +
 \sum_{\bar{a}}(\gamma_{\bar{b}s}\gamma_{\bar{a}s}-(1/2)\delta_{\bar{a}\bar{b}})\,\partial_sR_{\bar{a}}\Big)
 \,\Pi_{\bar{b}} -\nonumber\\
 &&\nonumber\\
 &-&{{c^3}\over {12\pi\, G}}\, \tilde{\phi}^{1/3}\,Q_s^{-2}\,
 \Big(\,4\, \partial_s\, q +
 \partial_s\, \Gamma_s^{(1)}\Big)\, {}^3K +\nonumber\\
 &&\nonumber\\
 &+&\frac{c^3}{8\pi G}\tilde{\phi}^{1/3}\sum_{d}\,Q_s^{-1}Q_d^{-1}
 \Big(\,2\, \partial_d\, q +
 \partial_d\, \Gamma_s^{(1)}\Big)
 \,\sigma_{(s)(d)} + \frac{1}{2}\tilde{\phi}^{-2/3}\,Q_s^{-2}\,{\cal
 M}_s\, \Big] -\nonumber\\
 &&\nonumber\\
 &-&\sigma^s\,\Big[ \,\tilde{\phi}^{-2/3}\sum_{\bar{b}}\,Q_r^{-2}
 \Big(\,2\, \gamma_{\bar{b}r}\, \partial_r\, q +
 \sum_{\bar{a}}(\gamma_{\bar{b}r}\gamma_{\bar{a}r}-(1/2)\delta_{\bar{a}\bar{b}})\,\partial_rR_{\bar{a}}\Big)
 \,\Pi_{\bar{b}} -\nonumber\\
 &&\nonumber\\
 &-&{{c^3}\over {12\pi\, G}}\, \tilde{\phi}^{1/3}\,Q_r^{-2}\, \Big(4\, \partial_r\, q +
 \partial_r\, \Gamma_r^{(1)}\Big) \, {}^3K +\nonumber\\
 &&\nonumber\\
 &+&\frac{c^3}{8\pi G}\tilde{\phi}^{1/3}\sum_{d}\,Q_r^{-1}Q_d^{-1}
 \Big(\,2\, \partial_d\, q + \partial_d\, \Gamma_r^{(1)}\Big)
 \,\sigma_{(r)(d)} + \frac{1}{2}\tilde{\phi}^{-2/3}\,Q_r^{-2}\,{\cal
 M}_r\, \Big]\, \Big\},
 \end{eqnarray*}

\begin{eqnarray*}
 {\hat J}^{\tau r}_{ADM} &=& - {\hat J}^{r\tau}_{ADM} =
 \int d^3\sigma\,   \Big( \sigma^{r}\, \Big[ {{c^3}\over {16\pi\,
 G}}\, \sqrt{\gamma}\,\,  {}^3g^{ns}\, ({}^3\Gamma^{u}_{nv}\,
 {}^3\Gamma^{v}_{su} - {}^3\Gamma^{u}_{ns}\, {}^3\Gamma^{v}_{vu}) -
 \nonumber \\
 &-& {{8\pi\, G}\over {c^3\, \sqrt{\gamma}}}\, {}^3G_{nsuv}\,
 {}^3\Pi^{ns}\, {}^3\Pi^{uv} - {\cal M}\Big] +\nonumber \\
 &+& {{c^3}\over {16\pi\, G}}\, \delta^{r}_{u}\, ({}^3g_{vs} - \delta_{vs})\,
 \partial_{n}\, \Big[ \sqrt{\gamma}\, ({}^3g^{ns} \,
 {}^3g^{uv} - {}^3g^{nu}\, {}^3g^{sv})\Big] \Big) (\tau ,\vec \sigma
 ) \, =\end{eqnarray*}

\bea
 &=&  \int d^3\sigma\,\Big\{
 \sigma^r\,\Big[ \frac{c^3}{16\pi G} \,{\cal S}-{\cal M}-\frac{4\pi
 G}{c^3}\tilde{\phi}^{-1}
 \sum_{\bar{b}}\,\Pi_{\bar{b}}^2 -\nonumber\\
 &&\nonumber\\
 &-&\tilde{\phi}\, \frac{c^3}{16\pi G}\,
 \Big(\sum_{a \not= b}\,\sigma^2_{(a)(b)} - \frac{2
 }{3}\, ({}^3K)^2 \Big)\,\Big] -\nonumber\\
 &&\nonumber\\
 &-&\frac{c^3}{16\pi G} \tilde{\phi}^{-1/3}\, Q_r^{-2}\,
 \sum_s\, ({\tilde \phi}^{2/3} - Q_s^{-2})\, \Big[\delta_{rs}\,
 \partial_s\, (\Gamma_r^{(1)} + \Gamma_s^{(1)} + q) -
 \nonumber \\
 &-& \partial_r\, (\Gamma_r^{(1)} + \Gamma_s^{(1)} + q)
 \Big]\,\Big\}(\tau, \vec \sigma) \approx 0.\nonumber \\
 &&{}
 \label{2.20}
 \eea

As discussed in Section IIE of paper I, ${\hat J}^{\tau r}_{ADM}
\approx 0$ are the gauge-fixings for the rest-frame constraints
${\hat P}^r_{ADM} \approx 0$ eliminating the internal 3-center of
mass in the 3-spaces $\Sigma_{\tau}$, which are non-inertial rest
frames of the 3-universe.

\subsection{Dimensions}

In checking the validity of the previous formulas it is useful to
remember the dimensions of the relevant quantities:\medskip

$[\tau = ct ] = [x^{\mu}] = [\vec \sigma ] = [{\vec \eta}_i] = [l]$,
$[{\vec \kappa}_i] = [m_ic] = [P^{\mu}] = [E/c] = [m\, l\, t^{-1}]$,
$[S] = [\hbar ] = [J^{AB}] = [m\, l^2\, t^{-1}]$, $[T^{AB}] = [{\cal
M}] = [{\cal M}_r] = [{\cal H}] = [{\cal H}_{(a)}] = [m\, l^{-2}\,
t^{-1}]$, $[{}^4g] = [{}^3g] = [n] = [n_{(a)}] = [{}^3e_{(a)r}] =
[{\dot {\vec \eta}}_i] = [\theta_i] = [\tilde \phi] = [0]$,
$[{}^3\pi^r_{(a)}] = [{}^3{\tilde \Pi}^{rs}] = [\Pi_{\bar a}] =
[\pi_{\tilde \phi}] = [m\, l^{-1}\, t^{-1}]$, $[{}^3\omega_{r(a)}] =
[{}^3K_{rs}] = [{}^3K] = [\sigma_{(a)(b)}] = [l^{-1}]$, $[{}^3R] =
[{}^3\Omega_{rs(a)}] = [l^{-2}]$, $[Q_i] = [m^{1/2}\, l^{3/2}\,
t^{-1}]$, $[A_{\perp\, r}] = [Q\, l^{-1}] = [m^{1/2}\, l^{1/2}\,
t^{-1}]$, $[{{Q_i}\over c}\, A_{\perp r}] = [m\, l\, t^{-1}]$,
$[\pi^r_{\perp}] = [E^r_{\perp}] = [B_r] = [l^{-1}\, A_{\perp\, r}]
= [m^{1/2}\, l^{-1/2}\, t^{-1}]$, $[G = 6.7\, 10^{-8}\, cm^3\,
s^{-2}\, g^{-1}] = [m^{-1}\, l^3\, t^{-2}]$, $[G/c^3 = 2.5\,
10^{-39}\, sec/g] = [m^{-1}\, t] $, $[G/c^2 =\, 7.421\, 10^{-29}\,
cm/g] = [m^{-1}\, l] $.

\vfill\eject

\section{The Weak Field Approximation and the Linearization}

The standard decomposition used for the weak field approximation in
the harmonic gauges  is

\bea
 {}^4g_{\mu\nu} &=& {}^4\eta_{\mu\nu} + h_{\mu\nu},\qquad
  |h_{\mu\nu}|, |\partial_{\alpha}\, h_{\mu\nu}|, |\partial_{\alpha}\,
 \partial_{\beta}\, h_{\mu\nu}| << 1,\nonumber \\
 &&{}
 \label{3.1}
 \eea

\noindent where ${}^4\eta_{\mu\nu}$ is the flat metric in an
inertial frame of the background Minkowski space-time. This is
equivalent to take a 3+1 splitting of our space-time with an
inertial foliation, having Euclidean instantaneous 3-spaces, against
the equivalence principle and against the fact (explicitly shown in
paper I) that each solution of Einstein's equations has an
associated dynamically selected preferred 3+1 splitting.

\bigskip

In this Section we shall define a linearization of Hamilton-Dirac
equations in the (non-harmonic) 3-orthogonal Schwinger time gauges
(\ref{2.1}) using as background the asymptotic Minkowski 4-metric
existing in our asymptotically Minkowskian space-times. Actually we
look for the following decomposition to be done by using radar
4-coordinates adapted to an admissible (see Ref.\cite{4}) 3+1
splitting of space-time

\bea
 {}^4g_{AB}(\tau, \sigma^r) &=& {}^4g_{(1)AB}(\tau, \sigma^r) +
 O(\zeta^2)\, \rightarrow {}^4\eta_{AB(asym)}\,\, at\,  spatial\, infinity,\nonumber \\
 &&{}\nonumber \\
 &&{}^4g_{(1)AB}(\tau, \sigma^r) = {}^4\eta_{AB (asym)} + {}^4h_{(1)AB}(\tau,
 \sigma^r),\nonumber \\
 &&{}^4h_{(1)AB}(\tau, \sigma^r) = O(\zeta) \rightarrow 0\,\, at\,
   spatial\, infinity,
 \label{3.2}
 \eea

\noindent where $\zeta << 1$ is a small a-dimensional parameter, the
small perturbation ${}^4h_{(1)AB}$ has no intrinsic meaning in the
bulk and ${}^3g_{(1)rs}(\tau, \sigma^r) = - \sgn\,
{}^4g_{(1)rs}(\tau, \sigma^r)$ is the positive-definite 3-metric on
the instantaneous 3-space $\Sigma_{\tau}$. In our case the
instantaneous 3-spaces will deviated from flat Euclidean 3-spaces by
curvature effects of order $O(\zeta)$, in accord with the
equivalence principle.

\bigskip

We must make an assumption on the variables $R_{\bar a}$,
$\partial_{\tau}\, R_{\bar a}$, $\partial^2_{\tau}\, R_{\bar a}$,
$\phi$, $n$, ${\bar n}_{(a)}$, $\sigma_{(a)(b)}{|}_{a \not= b}$,
$\Pi_{\bar a}$ (or $\sigma_{(a)(a)}$), $\pi_{\tilde \phi} =
{{c^3}\over {12\pi\, G}}\, {}^3K$, ${\dot \eta}_i^r$, $\kappa_{ir}$,
$A_{\perp\, r}$, $\pi^r_{\perp}$ such that the coupled equations
(\ref{2.8}) and (\ref{2.9}) (contracted Bianchi identities),
(\ref{2.11}) (shift determination), (\ref{2.12}) (lapse
determination with free ${}^3K$), (\ref{2.15}) (expression of
$\Pi_{\bar a}$), (\ref{2.16}) (equations for $\partial_{\tau}^2\,
R_{\bar a}$), (\ref{2.17}) (expression of $\kappa_{ir}$),
(\ref{2.18}) (equations for $\eta_i^r$), (\ref{2.19}) (Hamilton
equations for $A_{\perp\, r}$ and $\pi^r_{\perp}$),  all have the
two members consistent (of the same order).
\medskip

Let us remember \cite{13} that to avoid coordinate singularities we
must always have $N(\tau, \vec \sigma) = 1 + n(\tau, \vec \sigma)
> 0$ (3-spaces at different times do not intersect each other),
$\sgn\, {}^4g_{\tau\tau}(\tau, \vec \sigma) > 0$ (no rotating disk
pathology) and ${}^3g_{rs}(\tau, \vec \sigma)$ with three distinct
positive eigenvalues.

\subsection{A Consistent Hamiltonian Linearization}

Let us see which assumptions are needed to get Eq.(\ref{3.2}) in the
3-orthogonal Schwinger time gauges (\ref{2.1}).\bigskip

The first assumption is that on each instantaneous 3-space
$\Sigma_{\tau}$ we have the following limitation of the
a-dimensional configurational tidal variables $R_{\bar a}$ in the
York canonical basis

\bea
 &&| R_{\bar a}(\tau ,\vec \sigma ) = R_{(1)\bar a}(\tau,
 \vec \sigma) |  = O(\zeta) << 1,\nonumber \\
 &&{}\nonumber \\
 &&|\partial_u\, R_{\bar a}(\tau ,\vec \sigma )| \sim {1\over L}
 O(\zeta),\qquad |\partial_u\, \partial_v\, R_{\bar a}(\tau ,\vec
 \sigma )| \sim {1\over {L^2}} O(\zeta),\nonumber \\
 && |\partial_{\tau}\, R_{\bar a}| = {1\over L}\, O(\zeta),\qquad
 |\partial^2_{\tau}\, R_{\bar a}| = {1\over {L^2}}\, O(\zeta),\qquad
 |\partial_{\tau}\, \partial_u\, R_{\bar a}| = {1\over {L^2}}\, O(\zeta),
 \nonumber \\
 &&{}\nonumber \\
 &&\Rightarrow\,\, Q_a(\tau, \vec \sigma) = e^{\sum_{\bar a}\,
 \gamma_{\bar aa}\, R_{\bar a}(\tau, \vec \sigma)} = 1 +
 \Gamma^{(1)}_a(\tau, \vec \sigma) + O(\zeta^2),\nonumber \\
 &&\qquad \Gamma_a^{(1)} = \sum_{\bar a}\, \gamma_{\bar aa}\,
 R_{\bar a} = O(\zeta),\qquad \sum_a\, \Gamma^{(1)}_a = 0,\qquad R_{\bar a} =
 \sum_a\, \gamma_{\bar aa}\, \Gamma_a^{(1)},
 \label{3.3}
 \eea

\noindent where $L$ is a {\it big enough characteristic length
interpretable as the reduced wavelength $\lambda / 2\pi$ of the
resulting GW's}. Therefore the tidal variables $R_{\bar a}$ are
slowly varying over the length $L$ and times $L/c$. This also
implies that the Riemann tensor ${}^4R_{ABCD}$, the Ricci tensor
${}^4R_{AB}$ and the scalar 4-curvature ${}^4R$ behave as ${1\over
{L^2}}\, O(\zeta)$. Also the intrinsic 3-curvature scalar of the
instantaneous 3-spaces $\Sigma_{\tau}$, given in Eqs.(\ref{2.5}), is
of order ${1\over {L^2}}\, O(\zeta)$. To simplify the notation we
use $R_{\bar a}$ for $R_{(1)\bar a}$ in the rest of the
paper.\medskip

As a consequence of the behavior of the Riemann tensor, the mean
radius of curvature  ${}^4{\cal R}$ of space-time is of order
${}^4{\cal R}^{-2} \approx {1\over {L^2}}\, O(\zeta)$. Therefore we
get that the {\it requirements of the weak field approximation are
satisfied}:\medskip

i) ${\cal A} = O(\zeta)$, if ${\cal A} \sim R_{\bar a}$ is the
amplitude of the GW;

ii) $\Big({L\over {{\cal R}}}\Big)^2 = O(\zeta)$, namely $L \approx
{{\lambda}\over {2\pi}} << {}^4{\cal R}$.

\bigskip

As a first attempt let us put $\phi = {\tilde \phi}^{1/6} = 1 +
\phi_{(o)} + \phi_{(1)} + O(\zeta^2)$, $n = n_{(o)} + n_{(1)} +
O(\zeta^2)$ and ${\bar n}_{(a)} = {\bar n}_{(o)(a)} + {\bar
n}_{(1)(a)} + O(\zeta^2)$, with $\phi_{(o)}, n_{(o)}, {\bar
n}_{(o)(a)} = O(1)$, $\phi_{(1)}, n_{(1)}, {\bar n}_{(1)(a)} \sim
O(\zeta)$, and with similar expansions for the other variables like
$\sigma_{(a)(b)}$, ${}^3K$, .... However, this implies \footnote{For
$|\phi_{(1)}| < 1$ we have $\phi^n = {\tilde \phi}^{n/6} = (1 +
\phi_{(o)})^n + n\, (1 + \phi_{(o)})^{n - 1}\, \phi_{(1)}\,
\mapsto_{\phi_{(o)} = 0}\, 1 + n\, \phi_{(1)} + O(\zeta^2)$,
$\phi^{-n} = {\tilde \phi}^{- n/6} = (1 + \phi_{(o)})^{-n} - n\, (1
+ \phi_{(o)})^{- n - 1}\, \phi_{(1)}\, \mapsto_{\phi_{(o)} = 0}\, 1
- n\, \phi_{(1)} + O(\zeta^2)$, $\partial_r\, q  = \phi^{-1}\,
\partial_r\, \phi = {1\over 6}\, {\tilde \phi}^{-1}\, \partial_r\,
\tilde \phi =  (1 + \phi_{(o)})^{-1}\, \partial_r\, \phi_{(o)} + (1
+ \phi_{(o)})^{-2}\, \Big[(1 + \phi_{(o)})\, \partial_r\, \phi_{(1)}
- \phi_{(1)}\, \partial_r\, \phi_{(o)}\Big]\, \mapsto_{\phi_{(o)} =
0}\, \partial_r\, \phi_{(1)} + O(\zeta^2)$.}

\bea
 - \sgn\, {}^4g_{rs} &=& {}^3g_{rs} = \phi^4\, \Big(1 + 2\, \sum_{\bar
 a}\, \gamma_{\bar ar}\, R_{\bar a}\Big)\, \delta_{rs} + O(\zeta^2)
 =\nonumber \\
 &=&\Big((1 + \phi_{(o)})^4 + 4\, (1 + \phi_{(o)})^3\, \phi_{(1)} + 2\,
 (1 + \phi_{(o)})^4\, \sum_{\bar a}\, \gamma_{\bar ar}\, R_{\bar
 a}\Big)\, \delta_{rs} + O(\zeta^2) =\nonumber \\
 &=&  (1 + \phi_{(o)})^4\, \delta_{rs} + O(\zeta),\nonumber \\
 &&{}\nonumber \\
 \sgn\, {}^4g_{\tau\tau} &=& (1 + n)^2 - \sum_a\, {\bar n}^2_{(a)}
  = (1 + n_{(o)})^2 - \sum_a\, {\bar n}^2_{(o)(a)}
 + O(\zeta),\nonumber \\
 &&{}\nonumber \\
 - \sgn\, {}^4g_{\tau r} &=& \phi^2\, Q_r\, {\bar n}_{(r)} = \Big((1 +
 \phi_{(o)})^2  + 2\, (1 + \phi_{(o)})\, \phi_{(1)} +\nonumber \\
 &+& (1 + \phi_{(o)})^2\, \sum_{\bar a}\,
 \gamma_{\bar ar}\, R_{\bar a}\Big)\, {\bar n}_{(r)} + O(\zeta^2)
 = (1 + \phi_{(o)})^2\, {\bar n}_{(o)(r)} + O(\zeta),
 \label{3.4}
 \eea

\bigskip

\noindent and the equations for $\phi_{(o)}$, $n_{(o)}$, ${\bar
n}_{(o)(a)}$ turn out to be  not linear.

\bigskip

Therefore  we must assume $\phi_{(o)} =  n_{(o)} = {\bar n}_{(o)(a)}
= 0$. In this way Eq.(\ref{3.2}) can be implemented in the following
way \footnote{ The "trace reversed" perturbation is ${}^4{\bar
h}_{(1)AB} = {}^4h_{(1)AB} - {1\over 2}\, {}^4\eta_{AB}\, \sgn\,
h_{(1)}$, ${\bar h}_{(1)} = - h_{(1)}$, ${}^4{\bar h}_{(1)\tau\tau}
= \sgn\, (6\, \phi_{(1)} - n_{(1)})$, ${}^4{\bar h}_{(1)\tau r} = -
\sgn\, {\bar n}_{(1)(r)}$, ${}^4{\bar h}_{(1)rs} = - \sgn\, [2\,
(\Gamma_r^{(1)} + 5\, \phi_{(1)}) - n_{(1)}]\, \delta_{rs}$.}

\bea
 \phi &=& 1 + \phi_{(1)} + O(\zeta^2),\qquad \tilde \phi = 1 +
 6\, \phi_{(1)} + O(\zeta^2),\nonumber \\
 N &=& 1 + n = 1 + n_{(1)} +
 O(\zeta^2),\qquad {\bar n}_{(a)} = {\bar n}_{(1)(a)} + O(\zeta^2),
 \nonumber \\
 &&{}\nonumber \\
 &&\Downarrow\qquad {}^4g_{(1)AB} = {}^4\eta_{AB(asym)} + {}^4h_{(1)AB}, \nonumber \\
 &&{}\nonumber \\
 {}^4h_{(1)\tau\tau} &=& 2\, \sgn\, n_{(1)} = O(\zeta),\nonumber \\
 {}^4h_{(1)\tau r} &=& - \sgn\, {\bar n}_{(1)(r)} = O(\zeta),\nonumber \\
 {}^4h_{(1)rs} &=& - 2\, \sgn\,(\Gamma_r^{(1)} + 2\, \phi_{(1)})\,
 \delta_{rs} = O(\zeta),\qquad \Gamma_r^{(1)} =
 \sum_{\bar a}\, \gamma_{\bar ar}\, R_{\bar a},\quad \sum_r\,
 \Gamma_r^{(1)} = 0.,\nonumber \\
 &&{}\nonumber \\
 &&\quad h_{(1)} = \sgn\, {}^4\eta^{AB}\,
 {}^4h_{(1)AB} = 2\, (n_{(1)} - 6\, \phi_{(1)}) = O(\zeta),
 \label{3.5}
 \eea

\noindent while the triads and cotriads become ${}^3{\bar
e}^r_{(1)(a)} = \delta^r_a\, (1 - \Gamma^{(1)}_r - 2\, \phi_{(1)}) +
O(\zeta^2)$ and ${}^3{\bar e}_{(1)(a)r} = \delta_{ra}\, (1 +
\Gamma^{(1)}_r + 2\, \phi_{(1)}) + O(\zeta^2)$, respectively.
Therefore we have ${}^4g_{\tau\tau} = \sgn\, [1 + 2\, n_{(1)}] +
O(\zeta^2)$, ${}^4g_{\tau r} = - \sgn\, {\bar n}_{(1)(r)} +
O(\zeta^2)$, ${}^4g_{rs} = - \sgn\, {}^3g_{rs} = - \sgn\,
\delta_{rs}\, [1 + 2\, (\Gamma_r^{(1)} + 2\, \phi_{(1)})] +
O(\zeta^2)$, $\tilde \phi = \phi^6 = \sqrt{det\, {}^3g_{rs}} = 1 +
6\, \phi_{(1)} + O(\zeta^2)$.

\bigskip

With these assumptions Eq.(\ref{2.3}) implies

 \bea
 {{8\pi\, G}\over {c^3}}\, \Pi_{\bar a}(\tau, \vec \sigma)\, &=&
  {{8\pi\, G}\over {c^3}}\, \Pi_{(1) \bar a}(\tau, \vec \sigma) =
 {1\over L}\, O(\zeta) \cir
 \Big[\partial_{\tau}\, R_{\bar a} - \sum_a\, \gamma_{\bar aa}\,
 \partial_a\, {\bar n}_{(1)(a)}\Big](\tau, \vec \sigma) + {1\over L}\,
 O(\zeta^2),\nonumber \\
 &&{}\nonumber \\
 &&\sigma_{(a)(a)} = \sigma_{(1)(a)(a)} = - {{8\pi\, G}\over
 {c^3}}\, \sum_{\bar a}\, \gamma_{\bar aa}\, \Pi_{(1) \bar a} +
 {1\over L}\, O(\zeta^2).
 \label{3.6}
 \eea

\noindent Let us remark that everywhere $\Pi_{(1) \bar a}$ appears
in the combination ${G\over {c^3}}\, \Pi_{(1) \bar a} = {1\over L}\,
O(\zeta)$, which behaves like $\partial_{\tau}\, R_{\bar a}$, i.e.
it varies slowly over $L$.\medskip

Finally the super-momentum constraints (\ref{2.7}), Eqs.(\ref{2.3})
and dimensional arguments require

\begin{eqnarray*}
 &&\sigma_{(a)(b)}{|}_{a\not= b} = \sigma_{(1)(a)(b)}{|}_{a\not= b}
 = {1\over L}\, O(\zeta),\nonumber \\
 &&\Rightarrow\quad {{8\pi\, G}\over {c^3}}\, \pi_i^{(\theta)} =
 {1\over L}\, O(\zeta^2) = \sum_{a\not= b}\, (\Gamma^{(1)}_a -
 \Gamma^{(1)}_b)\, \epsilon_{iab}\, \sigma_{(1)(a)(b)} + {1\over
 L}\, O(\zeta^3),
  \end{eqnarray*}

 \bea
 &&{}^3K = {{12\pi\, G}\over {c^3}}\, \pi_{\tilde \phi} =
 {}^3K_{(1)} = {{12\pi\, G}\over {c^3}}\, \pi_{(1) \tilde \phi} =
 {1\over L}\, O(\zeta),\nonumber \\
 &&{}\nonumber \\
 &&\Downarrow\nonumber \\
 &&{}\nonumber \\
 {}^3K_{rs} &=& {}^3K_{(1)rs} = {1\over L}\, O(\zeta) =\nonumber \\
 &=& (1 - \delta_{rs})\,
 \sigma_{(1)(r)(s)} + \delta_{rs}\, \Big[{1\over 3}\, {}^3K_{(1)} -
 \partial_{\tau}\, \Gamma_r^{(1)} + \sum_a\, (\delta_{ra} - {1\over 3})\,
 \partial_a\, {\bar n}_{(1)(a)}\Big] + {1\over L}\, O(\zeta^2).\nonumber \\
 &&{}
 \label{3.7}
 \eea

\noindent These equations imply that once we have found a solution
$\sigma_{(1)(a)(b)}{|}_{a \not b}(\tau, \vec \sigma)$ of the
linearization of the super-momentum constraints (\ref{2.7}), then we
can put $\pi^{(\theta)}_i(\tau, \vec \sigma) \approx 0$ in the York
canonical basis, which becomes adapted to 13 of the 14 constraints
after the linearization.

\medskip

Let us remark that the triad momenta and the standard ADM momenta,
given after Eq.(2.22) of paper I, have the following weak field
limit: $\pi_{(a)}^r = \delta_{ra}\, (\pi_{(1)\tilde \phi} +
\sum_{\bar b}\, \gamma_{\bar ba}\, \Pi_{(1)\bar b}) - {{c^3}\over
{8\pi\, G}}\, (1 - \delta_{rs})\, \sigma_{(1)(r)(s)} + {1\over L}\,
O(\zeta^2)$, ${}^3\Pi^{rs} = {1\over 4}\, ({}^3{\bar e}^r_{(a)}\,
{}^3{\bar \pi}^s_{(a)} + {}^3{\bar e}^s_{(a)}\, {}^3{\bar
\pi}^r_{(a)}) = {1\over 4}\,({}^3\pi^r_{(s)} + \pi^s_{(r)}) +
{1\over L}\, O(\zeta^2) = - {1\over 2}\, \delta_{rs}\,
(\pi_{(1)\tilde \phi} + \sum_{\bar b}\, \gamma_{\bar br}\,
\Pi_{(1)\bar b}) + {{c^3}\over {16\pi\, G}}\, (1 - \delta_{rs})\,
\sigma_{(1)(r)(s)} + {1\over L}\, O(\zeta^2)$.

\bigskip

Let us now consider our matter, i.e. positive-energy scalar
particles and the transverse electro-magnetic field in the radiation
gauge.

\bigskip

For the particles we have $\eta^r_i = O(1)$ and ${\dot \eta}^r_i =
O(1)$ (since $\tau = c\, t$, in the non-relativistic limit we have $
{\dot {\vec \eta}}_i = {\vec v}_i/c = O(1)\, \rightarrow_{c
\rightarrow \infty}\, 0$).

\bigskip

However, without further restrictions on the masses, the momenta and
the electro-magnetic field  Eqs.(\ref{2.4})  would imply ${\cal M} =
{\cal M}_{(o)} + {\cal M}_{(1)} + {{mc}\over {L^3}}\, O(\zeta^2)$,
with ${\cal M}_{(o)} = O(1)$, ${\cal M}_{(1)} = {{mc}\over {L^3}}\,
O(\zeta)$, and ${\cal M}_r = {\cal M}_{(o)\, r} = {{mc}\over
{L^3}}\, O(1)$ (${\cal M}$ and ${\cal M}_r$ are densities; $m$ is a
typical particle mass). But then Eqs.(\ref{2.6}) and (\ref{2.7}) for
the super-Hamiltonian and super-momentum constraints would not be
consistent. For instance Eq.(\ref{2.6}), whose unknown is $\phi = 1
+ \phi_{(1)} + O(\zeta^2)$, would be ${\cal M}_{(o)}(\tau, \vec
\sigma) + F_{(1)}[\phi_{(1)}, {\cal M}_{(1)}, ...](\tau, \vec
\sigma) + {{mc}\over {L^3}}\, O(\zeta^2) \approx 0$ with $F_{(1)} =
{{mc}\over {L^3}}\, O(\zeta)$.

\bigskip

To get a consistent approximation we must introduce a {\it
ultraviolet cutoff} $M$ on the masses and momenta of the particles
and on the electro-magnetic field so that

\bea
 {\cal M}(\tau, \vec \sigma) &=& {\cal
 M}^{(UV)}_{(1)}(\tau, \vec \sigma) + {\cal R}_{(2)}(\tau, \vec \sigma),\nonumber \\
 &&{}\nonumber \\
 && m_i = M\, O(\zeta),\qquad
 \int d^3\sigma\, {\cal M}_{(1)}^{(UV)}(\tau, \vec
 \sigma) = Mc\, O(\zeta),\qquad \int d^3\sigma\, {\cal R}_{(2)}(\tau, \vec \sigma)
 = Mc\, O(\zeta^2),\nonumber \\
 &&{}\nonumber \\
 &&{}\nonumber \\
 {\cal M}_r(\tau, \vec \sigma) &=&
 {\cal M}_{(1)r}(\tau, \vec \sigma),\qquad
 \int d^3\sigma\, {\cal M}^{(UV)}_{(1)r}(\tau, \vec \sigma) =
 Mc\, O(\zeta).
 \label{3.8}
 \eea

\medskip

Here $M$ is a finite mass defining the ultraviolet cutoff: $M\, c^2$
gives an estimate of the weak ADM energy of the 3-universe contained
in the instantaneous 3-spaces $\Sigma_{\tau}$, because it can be
assumed to be of the order of the mass Casimir of the asymptotic ADM
Poincare' group. The associated length scale is the gravitational
radius $R_M = 2M\, {G\over {c^2}} \approx 10^{-29}\, M$ \footnote{
The Earth mass $M_{Earth} = 5.98\, 10^{28}\, g$ gives rise to a
gravitational radius $R_{Earth} = {{2\, G\, M_{Earth}}\over {c^2}} =
0.888\, cm$. By comparison the Compton wavelength of an electron is
${{\hbar}\over {m_e\, c}} = 3.861592\, 10^{-11} cm$, the classical
electron radius is ${{e^2}\over {m_e\, c^2}} = 2.81794\, 10^{-13}\,
cm$ and the Planck length is $L_P = \sqrt{{{\hbar\, G}\over {c^3}}}
= 1.616\, 10^{-33} cm$.}.
\medskip

Therefore the description of particles in our approximation will be
reliable only if their masses and momenta are less of $Mc\,
O(\zeta)$ and at distances $r$ from the particles satisfying $r >
R_M$ (that is at each instant we must enclose each particle in a
sphere of radius $R_M$ and our approximation is not valid inside
these spheres). This will be clear in the next Section, where we
will obtain an equation like $\triangle\, \phi_{(1)}(\tau, \vec
\sigma) = {1\over {L^2}}\, O(\zeta) \sim {G\over {c^3}}\, {\check
{\cal M}}_{(1)}(\tau, \vec \sigma) + ...$ implying $\phi_{(1)}(\tau,
\vec \sigma) = O(\zeta) \sim {G\over {c^3}}\, {{m_i\, c}\over {|\vec
\sigma - {\vec \eta}_i(\tau)|}} + ... \sim {{R_M}\over {|\vec \sigma
- {\vec \eta}_i(\tau)|}}\, O(\zeta)$ namely $|\vec \sigma - {\vec
\eta}_i(\tau)| >> R_M$. Therefore our results in the weak field
approximation  can be trusted till a distance $d >> R_M$ from the
particles.

\medskip

From Eq.(\ref{2.4}) we get for the mass density and mass current
density

\begin{eqnarray*}
 {\cal M}^{(UV)}(\tau, \vec \sigma) &=&
 {\cal M}^{(UV)}_{(1)}(\tau, \vec \sigma) +
 {\cal M}^{(UV)}_{(2)}(\tau, \vec \sigma)
 + {\cal R}_{(3)}(\tau, \vec \sigma),
 \qquad \int d^3\sigma\, {\cal R}(\tau, \vec \sigma) =
 Mc\, O(\zeta^3)\nonumber \\
 &&{}\nonumber \\
 &&{}\nonumber \\
 {\cal M}^{(UV)}_{(1)}(\tau, \vec \sigma) &=&
 \sum_i\, \delta^3(\vec \sigma
 ,{\vec \eta}_i(\tau ))\, \eta_i\, \sqrt{m^2_i\, c^2 +
 \Big({\vec \kappa}_i(\tau ) - {{Q_i}\over c}\, {\vec A}_{\perp}\Big)^2}
 (\tau ,\vec \sigma ) +\nonumber \\
 &+& {1\over {2c}}\, \Big( \Big[\sum_a\,
 \Big((\pi^a_{\perp})^2 - \Big(2\, \pi^a_{\perp} - \sum_i\, Q_i\, \eta_i\,
 {{\partial\, c(\vec \sigma, {\vec \eta}_i(\tau))}\over {\partial\, \sigma^a}}
 \Big)\nonumber \\
 && \sum_j\, Q_j\, \eta_j\, {{\partial\, c(\vec \sigma, {\vec \eta}_j(\tau))}
 \over {\partial\, \sigma^a}} \Big) + {1\over 2}\, \sum_{ab}\, F^2_{ab}
 \Big]\Big)(\tau ,\vec \sigma),
 \end{eqnarray*}

 \bea
 {\cal M}^{(UV)}_{(2)}(\tau, \vec \sigma) &=&
 \sum_i\, \delta^3(\vec \sigma
 ,{\vec \eta}_i(\tau ))\, \eta_i\, \Big(\nonumber \\
 &&{{- 2\, \phi_{(1)}\, \Big({\vec \kappa}_i(\tau ) - {{Q_i}\over c}\,
 {\vec A}_{\perp}\Big)^2 -  \sum_a\, \Gamma_a^{(1)}\,
 \Big(\kappa_{ia}(\tau )
 - {{Q_i}\over c}\, A_{\perp a}\Big)^2}\over {\sqrt{m^2_i\, c^2 +
  \Big({\vec \kappa}_i(\tau ) -
 {{Q_i}\over c}\, {\vec A}_{\perp}\Big)^2}}}\Big)(\tau ,\vec \sigma ) -\nonumber \\
 &-& {1\over {2c}}\, \Big( 2\, \sum_a\,
 \Big[ \phi_{(1)} - \Gamma^{(1)}_a\Big]\,
 \Big[(\pi^a_{\perp})^2 - \Big(2\, \pi_{\perp}^a - \sum_i\, Q_i\, \eta_i\,
 {{\partial\, c(\vec \sigma, {\vec \eta}_i(\tau))}\over {\partial\, \sigma^a}}
 \Big)\nonumber \\
 && \sum_j\, Q_j\, \eta_j\, {{\partial\, c(\vec \sigma, {\vec \eta}_j(\tau))}
 \over {\partial\, \sigma^a}} \Big] +  \sum_{ab}\,
 \Big[  \phi_{(1)} + \Gamma^{(1)}_a
 + \Gamma^{(1)}_b\Big]\, F^2_{ab} \Big)(\tau ,\vec \sigma), \nonumber \\
 &&{}\nonumber \\
 &&\int d^3\sigma\, {\cal M}^{(UV)}_{(1)}(\tau, \vec
 \sigma) = Mc\, O(\zeta), \qquad \int d^3\sigma\, {\cal M}^{(UV)}_{(2)}(\tau, \vec
 \sigma) = Mc\, O(\zeta^2),\nonumber \\
 &&{}
 \label{3.9}
 \eea

 \medskip

 \bea
  {\cal M}_r(\tau ,\vec \sigma )&=&
   {\cal M}^{(UV)}_{(1)r}(\tau ,\vec \sigma )
  = \sum_{i=1}^N\, \eta_i\,
 \Big(\kappa_{ir}(\tau ) - {{Q_i}\over c}\,
 A_{\perp r}(\tau ,\vec \sigma)\Big)\, \delta^3(\vec
 \sigma ,{\vec \eta}_i(\tau )) -\nonumber \\
 &-& {1\over c}\, \sum_s\, F_{rs}(\tau ,\vec \sigma)\, \Big(
  \pi_\perp^s(\tau,  \vec{\sigma}) - \sum_n\, \delta^{sn}\, \sum_i\, Q_i\, \eta_i\,
 {{\partial\, c(\vec \sigma, {\vec \eta}_i(\tau))}\over {\partial\,
 \sigma^n}}\Big),\nonumber \\
 &&{}\nonumber \\
 &&\qquad \int d^3\sigma\, {\cal M}^{(UV)}_{(1)r}(\tau ,\vec \sigma )
 = Mc\, O(\zeta).
 \label{3.10}
 \eea

\bigskip

Therefore for the particles and the transverse electro-magnetic
field the validity of the weak field approximation requires

\bea
 &&{\vec \eta}_i(\tau) = O(1),\qquad {{{\vec \kappa}_i(\tau)}\over {m_i c}} =
 O(1),\qquad {{{\vec \kappa}_i(\tau)}\over {Mc}} =
 O(\zeta),\qquad {{m_i}\over M} \leq  O(\zeta),\nonumber \\
 &&{}\nonumber \\
 &&A_{\perp r}(\tau, \vec \sigma), \pi^r_{\perp}(\tau, \vec \sigma)
 = O(1),\qquad with\quad {{Q_i}\over c}\, A_{\perp r}(\tau, {\vec
 \eta}_i(\tau)) = Mc\, O(\zeta),\nonumber \\
 &&{}\nonumber \\
 &&\int d^3\sigma\,[{1\over c}\, \pi^r_{\perp}(\tau, \vec \sigma)]^2,\,\,
 {1\over c}\, F^2_{rs}(\tau, \vec \sigma),\,\, [{1\over c}\, F_{rs}\,
 \pi^s_{\perp}](\tau, \vec \sigma) = Mc\, O(\zeta).
 \label{3.11}
 \eea

\medskip

Moreover the boundary conditions at spatial infinity and the local
intensities for the transverse electro-magnetic field must be such
that the integral conditions in Eqs.(\ref{3.8}) hold. The last line
of Eqs.(\ref{3.11}) agrees with Eq.(\ref{3.8}) only if the radiation
part of the transverse electro-magnetic field is concentrated in
small volumes $V = V_o\, O(\zeta)$ with a sufficiently rapid decay
outside them. The restriction ${{Q_i}\over c}\, A_{\perp r}(\tau,
{\vec \eta}_i(\tau)) = Mc\, O(\zeta)$, dictated by Eq.(\ref{2.17}),
implies a bound on the value of the electric charges of the
Lienard-Wiechert transverse potential evaluated in Ref.\cite{14} in
special relativity: since this potential has the form ${{Q_i}\over
c}\, A_{\perp r\, LW}(\tau, \vec \sigma) = \sum_{j \not= i}\,
{{Q_i\, Q_j}\over {c\, 4\pi\, |\vec \sigma - {\vec
\eta}_j(\tau)|}}\, F_{jr}$ with $F_{jr} = O(1)$, it turns out that
for distances $|{\vec \eta}_i(\tau) - {\vec \eta}_j(\tau)|
> R_M$ the restriction implies that the product $e_i\, e_j$ of the
electric charges semiclassically simulated by the Grassmann
variables $Q_i\, Q_j$ must satisfy $e_i\, e_j < R_M\, M\, c^2$.

\medskip
Our results will be equivalent to a re-summation of the PN
expansions valid for small rest masses  still having relativistic
velocities (${{{\check {\vec \kappa}}^2_i}\over {m_i^2\, c^2}} =
O(1)$, ${{{\vec v}_i}\over c}= O(1)$).\bigskip

Let us remark that in this way the energy-momentum tensor $T^{AB}$
(its expression derives from Eq.(3.11) of paper I after having
expressed the metric components in the York canonical basis, after
having expressed the electro-magnetic field in the radiation gauge
as said after Eq.(3.35) of paper I and after having done the weak
field approximation) has the following behavior

\begin{eqnarray*}
  T^{\tau\tau} &=& {\cal M}_{(1)}^{(UV)} + {\cal R}_{(2)}^{\tau\tau},\nonumber \\
 T^{\tau r} &=& {\cal M}^{(UV)}_{(1)\, r} + {\cal R}_{(2)}^{\tau r},\nonumber \\
 T^{rs} &=& \sum_i\, \delta^3(\vec \sigma, {\vec \eta}_i)\, \eta_i\,
 {{(\kappa_{ir} - {{Q_i}\over c}\, A_{\perp\, r})\, (
 \kappa_{is} - {{Q_i}\over c}\, A_{\perp\, s})}\over {\sqrt{m_i^2\,
 c^2 + \sum_a\, (\kappa_{ia} - {{Q_i}\over c}\, A_{\perp\,
 a})^2}}} +\nonumber \\
 &+&{1\over c}\, \Big[- \Big(\pi^r_{\perp} - \sum_i\, \eta_i\, Q_i\,
 {{\partial\, c(\vec \sigma, {\vec \eta}_i(\tau))}\over {\partial\, \sigma^r}}\Big)\,
 \Big(\pi^s_{\perp} - \sum_j\, \eta_j\, Q_j\, {{\partial\, c(\vec \sigma, {\vec \eta}_j(\tau))}
 \over {\partial\, \sigma^s}}\Big) +\nonumber \\
 &+& {1\over 2}\, \delta^{rs}\,
 \Big(\sum_a (\pi^a_{\perp} - \sum_i\, \eta_i\, Q_i\, {{\partial\,
 c(\vec \sigma, {\vec \eta}_i(\tau))}\over {\partial\, \sigma^a}})^2 -
 {1\over 2}\, \sum_{ab}\, F^2_{ab}\Big) + \sum_a\,
 F_{ra}\, F_{sa}\Big] + {\cal R}_{(2)}^{rs} =\nonumber \\
 &=& T^{rs}_{(1)} + {\cal R}_{(2)}^{rs},
 \end{eqnarray*}

\begin{eqnarray*}
 &&\sum_r\, T^{rr}_{(1)} = \sum_i\, \delta^3(\vec \sigma - {\vec
 \eta}_i(\tau))\, \eta_i\, {{\sum_r\, \Big(\kappa_{ir} -
 {{Q_i}\over c}\, A_{\perp r}\Big)^2}\over {\sqrt{m_i^2\,
 c^2 + \sum_a\, (\kappa_{ia} - {{Q_i}\over c}\, A_{\perp\,
 a})^2}}} +\nonumber \\
 &+& {1\over {2c}}\, \Big[\sum_a\, (\pi^a_{\perp})^2 +
 {1\over 2}\, \sum_{ab}\, F^2_{ab} -\nonumber \\
 &-& \sum_{ai}\, \eta_i\, Q_i\, \Big(2\, \pi^a_{\perp} - \sum_{j \not= i}\,
 \eta_j\, Q_j\, \partial_a\, c(\vec \sigma, {\vec \eta}_j(\tau))\Big)\,
 \partial_a\, c(\vec \sigma, {\vec \eta}_i(\tau))\Big],
 \end{eqnarray*}

 \begin{eqnarray*}
 &&\sum_r\, \gamma_{\bar ar}\, T^{rr}_{(1)} = \sum_i\, \delta^3(\vec \sigma - {\vec
 \eta}_i(\tau))\, \eta_i\, {{\sum_r\, \gamma_{\bar ar}\, \Big(\kappa_{ir} -
 {{Q_i}\over c}\, A_{\perp r}\Big)^2}\over {\sqrt{m_i^2\,
 c^2 + \sum_a\, (\kappa_{ia} - {{Q_i}\over c}\, A_{\perp\,
 a})^2}}} -\nonumber \\
 &-& {1\over c}\,
 \sum_r\, \gamma_{\bar ar}\, \Big[(\pi^r_{\perp})^2 - \sum_a\, F^2_{ra}
 -\nonumber \\
 &-& \sum_i\, \eta_i\, Q_i\, \Big(2\, \pi^r_{\perp} - \sum_{j \not= i}\,
 \eta_j\, Q_j\, \partial_r\, c(\vec \sigma, {\vec \eta}_j(\tau))\Big)\,
 \partial_r\, c(\vec \sigma, {\vec \eta}_i(\tau))\Big],
 \end{eqnarray*}

 \bea
 && {\cal M}_{(2)}^{(UV)} = - \sum_a\, (\Gamma_a^{(1)} + 2\, \phi_{(1)})\,
 T^{aa}_{(1)} +\nonumber \\
 &+& {1\over c}\, \Big[\sum_a\, 3\, \Gamma_a^{(1)}\,
 \Big(\pi^a_{\perp} - \sum_i\, \eta_i\, Q_i\, {{\partial\,
 c(\vec \sigma, {\vec \eta}_i(\tau))}\over {\partial\, \sigma^a}}\Big)^2 -\nonumber \\
 &-& \sum_{ab}\, (\phi_{(1)} + {3\over 2}\, \Gamma_a^{(1)})\, F^2_{ab} \Big].
 \label{3.12}
 \eea

\medskip

Since, as said in Subsection IIE of paper I, we have $\nabla_A\,
T^{AB}(\tau, \vec \sigma) \cir 0$ from the Bianchi identities and
since ${}^4g_{AB} = {}^4\eta_{AB(asym)} + O(\zeta)$, we must have
$\partial_A\, T_{(1)}^{AB}(\tau, \vec \sigma) \cir 0 +
\partial_A\, {\cal R}^{AB}_{(2)}$. At the lowest order this implies

 \bea
  &&\partial_{\tau}\, {\cal M}^{(UV)}_{(1)} + \partial_r\,
 {\cal M}_{(1)r}^{(UV)} = 0 + \partial_A\, {\cal R}^{A\tau}_{(2)},\nonumber \\
 &&\partial_{\tau}\, {\cal M}^{(UV)}_{(1)r} + \partial_s\, T^{rs}_{(1)} =
 0 + \partial_A\, {\cal R}^{Ar}_{(2)},
 \label{3.13}
 \eea

\noindent as in inertial frames in Minkowski space-time. The
equation $\partial_A\, T_{(1)}^{AB}(\tau, \vec \sigma)\, \cir\, 0 +
\partial_A\, {\cal R}^{AB}_{(2)}$ implies $\partial_A\,
\Big(T_{(1)}^{AB}(\tau, \vec \sigma)\, \sigma^C - T^{AC}_{(1)}(\tau,
\vec \sigma)\, \sigma^B\Big)\, \cir\, 0 + \partial_A\, {\cal
R}^{ABC}_{(2)}$ (angular momentum conservation).

\bigskip

Finally let us consider Einstein's equations in radar 4-coordinates,
i.e. ${}^4R_{AB} - {1\over 2}\, {}^4g_{AB}\, {}^4R\, \cir\, {{8\pi\,
G}\over {c^3}}\, T_{AB}$. Since we have ${}^4R_{AB} = {1\over
{L^2}}\, O(\zeta)$ from Eqs.(\ref{3.3}) and ${{8\pi\, G}\over
{c^3}}\, \int d^3\sigma\, T_{(1)AB}(\tau, \vec \sigma) \approx R_M\,
O(\zeta)$ from Eqs. (\ref{3.8}) and (\ref{3.12}), we get from
Einstein's equations the following local estimate of the order of
$T_{(1)AB}(\tau, \vec \sigma)$

\beq
 T_{(1)AB}(\tau, \vec \sigma) \approx {{Mc}\over {R_M\, L^2}}\, O(\zeta).
 \label{3.14}
 \eeq

Therefore the support $V = V_o\, O(\zeta)$ of the radiation part of
the electro-magnetic field, defined after Eq.(\ref{3.11}), must have
$V_o \sim R_M\, L^2$ with $L \geq R_M$.

\bigskip

In conclusion, since the weak field linearized solution can be
trusted only at distances $d >> R_M$ from the particles, the GW's
described by our linearization must have a wavelength satisfying
$\lambda \approx L > d >> R_M$ (with the weak field approximation we
have $\lambda << {}^4{\cal R}$ without the slow motion assumption).
\medskip

If all the particles are contained in a compact set of radius $l_c$
(the source), the frequency $\nu = {c\over {\lambda}}$ of the
emitted GW's will be of the order  of the typical frequency
$\omega_s$ of the motion inside the source, where the typical
velocities are of the order $v \approx \omega_s\, l_c$. As a
consequence we get $\nu = {c\over {\lambda}} \approx \omega_s
\approx v/l_c$ or $\lambda \approx {c\over v}\, l_c >> R_M$, so that
we get ${v\over c} \approx {{l_c}\over {\lambda}} << {{l_c}\over
{R_M}}$ and $l_c >> R_M$ if ${v\over c} = O(1)$.
\medskip

If the velocities of the particles become non-relativistic, i.e. in
the slow motion regime with $v << c$ (for binary systems with total
mass m and held together by weak gravitational forces we have also
${v\over c} \approx \sqrt{{{R_m}\over {l_c}}} << 1$), we have
$\lambda >> l_c$ and we can have $l_c \approx R_M$.

\subsection{The Linearization Interpreted as the First Term of a
Hamiltonian Post-Minkowskian Expansion in the Non-Harmonic
3-Orthogonal Gauges}

In our class of asymptotically flat space-times near Minkowski
space-time the above linearization can be interpreted as the first
term in a HPM expansion with a UV cutoff on the matter. This is due
to the fact that most of the canonical variables in the York basis
parametrize deviations from Minkowski space-time with the Cartesian
4-coordinates of an inertial frame, which vanish if $G \rightarrow
0$ (for $G = 0$ we get $R_{\bar a} = n = {\bar n}_{(a)} =  \Pi_{\bar
a} = 0$, $\phi = 1$ in the Minkowski rest-frame instant form, where
we have also ${}^3K = \sigma_{(a)(b)} = 0$).\medskip

As a consequence, by using Eqs.(\ref{2.3}) we can write the
following HPM expansions (here we do not use $R_{\bar a}$ to mean
$R_{(1)\bar a}$ as is done in the weak field approximation)

\bea
 R_{\bar a} &=& \sum_{n=1}^{\infty}\, G^n\, R_{[n]\, \bar
 a},\qquad R_{(1)\bar a} = G\, R_{[1]\bar a},\nonumber \\
 n &=& \sum_{n=1}^{\infty}\, G^n\, n_{[n]},\qquad
 n_{(1)} = G\, n_{[1]},\nonumber \\
 {\bar n}_{(a)} &=& \sum_{n=1}^{\infty}\, G^n\, {\bar n}_{[n]\,
 (a)},\qquad {\bar n}_{(1)(a)} = G\, {\bar n}_{[1](a)},\nonumber \\
 \phi &=& e^q = {\tilde \phi}^{1/6} = 1 + \sum_{n=1}^{\infty}\, G^n\,
 \phi_{[n]},\qquad \phi_{(1)} = G\, \phi_{[1]},\nonumber \\
 {}^3K &=& {{12\pi}\over {c^3}}\, G\, \pi_{\tilde \phi} =
 \sum_{n=1}^{\infty}\, G^n\, {}^3K_{[n]},\qquad {}^3K_{(1)} =
 G\, {}^3K_{[1]}, \nonumber \\
 &&\qquad \pi_{\tilde \phi} = {{c^3}\over {12\pi}}\,
 \Big({}^3K_{[1]} + \sum_{n=2}^{\infty}\, G^{n-1}\, {}^3K_{[n]}\Big),\nonumber \\
 \sigma_{(a)(b)}{|}_{a \not= b} &=&  {{8\pi}\over {c^3}}\,
 \phi^{-6}\, \sum_i\, {{\epsilon_{abi}\, G\,
 \pi^{(\theta)}_i}\over {Q_a\, Q_b^{-1} - Q_b\, Q_a^{-1}}}
 = \sum_{n=1}^{\infty}\, G^n\, \sigma_{[n]\, (a)(b)}{|}_{a \not=
 b}, \qquad \sigma_{(1)\, (a)(b)}{|}_{a \not= b} = G\,
 \sigma_{[1]\, (a)(b)}{|}_{a \not= b}.\nonumber \\
 &&{}
 \label{3.15}
 \eea

\medskip

Moreover Eq.(\ref{3.6}) implies $G\, \Pi_{\bar a} =
\sum_{n=1}^{\infty}\, G^n\, \Pi_{[n]\, \bar a}$, so that from
Eq.(\ref{2.3}) we get  $\sigma_{(a)(a)} - {{8\pi}\over {c^3}}\,
\phi^{-6}\, \sum_{\bar a}\, \gamma_{\bar aa}\, G\, \Pi_{\bar a} =
\sum_{n=1}^{\infty}\, G^n\, \sigma_{[n](a)(a)}$. Finally, since we
have $Q_a\, Q_b^{-1} - Q_b\, Q_a^{-1} \rightarrow_{G \rightarrow
0}\, 2\, \sum_{\bar a}\, (\gamma_{\bar aa} - \gamma_{\bar ab})\, G\,
R_{[1]\bar a}$, from $G\, \sigma_{[1](a)(b)}{|}_{a \not= b} =
{{4\pi}\over {c^3}}\, \sum_i\, {{\epsilon_{abi}\,
\pi^{(\theta)}_i}\over {\sum_{\bar a}\, (\gamma_{\bar aa} -
\gamma_{\bar ab})\, R_{[1]\bar a}}}$ we also get $\pi_i^{(\theta)} =
 \sum_{n=1}^{\infty}\, G^n\, \pi_{[n]i}^{(\theta)}$.

\bigskip

The study of HPM at the second order will be done in a future paper.

\subsection{The HPM Linearization of the Gauge Fixings for Harmonic
Gauges}

In Eqs. (5.3) and (5.4) of paper I we expressed the Hamiltonian
version of  the gauge-fixing constraints $\chi^{\tau}(\tau, \vec
\sigma) \approx 0$ and $\chi^r(\tau, \vec \sigma) \approx 0$
selecting the family of 4-harmonic gauges in the York canonical
basis.\medskip

If we eliminate the gauge fixing $\theta^i(\tau, \vec \sigma)
\approx 0$ identifying the family of 3-orthogonal gauges and we
consider the angles (O(3) canonical coordinates of first kind)
$\theta^i(\tau, \vec \sigma)$ as small quantities
$\theta^i_{(1)}(\tau, \vec \sigma) = O(\zeta)$ (so that we have
$V_{sa}(\theta^i) = \delta_{sa} + \epsilon_{sai}\, \theta^i_{(1)} +
O(\zeta^2)$), we can extend our HPM linearization to arbitrary
gauges. With some calculatins it can be checked that the first half
of Eqs. (\ref{3.7}), regarding the super-momentum constraints, are
still valid with $\theta^i = \theta^i_{(1)} + O(\zeta^2) \not= 0$.
\medskip

Then the linearization of the harmonic gauge fixing
$\chi^{\tau}(\tau, \vec \sigma) \approx 0$  (see the second half of
Eqs. (5.3) of paper I) becomes

\beq
 \partial_{\tau}\, n_{(1)}(\tau, \vec \sigma) \approx - \Big(\sum_r\, \partial_r\,
 {\bar n}_{(1)(r)} + {}^3K_{(1)}\Big)(\tau, \vec
 \sigma).
 \label{3.16}
 \eeq

Instead the linearization of the harmonic gauge fixings
$\chi^r(\tau, \vec \sigma) \approx 0$   of Eqs. (5.4) of paper I
becomes

\beq
 \partial_{\tau}\, {\bar n}_{(1)(r)}(\tau, \vec \sigma) \approx -
 \Big(\partial_r\, n_{(1)} + 2\, \partial_r\, (\phi_{(1)} -
 \Gamma_r^{(1)})\Big)(\tau, \vec \sigma).
 \label{3.17}
 \eeq

Both the equations do not depend on $\theta_{(1)}(\tau, \vec
\sigma)$. As said in paper I all the gauge fixings for the gauge
variables $\theta_{(1)}(\tau, \vec \sigma)$ and ${}^3K_{(1)}(\tau,
\vec \sigma)$ compatible with these equations identify 4-harmonic
gauges.\medskip

The previous two equations imply the following wave equations for
the lapse and shift functions

\bea
 &&\Box\, n_{(1)}(\tau, \vec \sigma) \approx \Big(2\, \sum_r\, \partial_r^2\,
 (\phi_{(1)} - \Gamma_r^{(1)})\Big)(\tau, \vec
 \sigma),\nonumber \\
 &&\Box\, {\bar n}_{(1)(r)}(\tau, \vec \sigma) -
 \partial_r\, \sum_{s \not= r}\, {\bar n}_{(1)(s)}(\tau, \vec \sigma) \approx
 \nonumber \\
 &&\approx\,  \Big(2 \,
 \partial_r\, \partial_{\tau}\, (\Gamma_r^{(1)} - \phi_{(1)})
 + \partial_r\, {}^3K_{(1)}\Big)(\tau, \vec \sigma).
 \label{3.18}
 \eea

The solution of these hyperbolic equations requires initial data at
$\tau \rightarrow - \infty$. Instead in the family of 3-orthogonal
gauges we have elliptic equations on a fixed 3-space $\Sigma_{\tau}$
requiring data only on it.\medskip

If we denote $n_{(1)}^{(HH)}$ and ${\bar n}^{(HH)}_{(1)(r)}$ the
retarded solutions of Eqs.(\ref{3.18}), the 4-metric
${}^4g^{(HH)}_{(1)AB}(\bar \tau, {\vec {\bar \sigma}})$ in harmonic
radar 4-coordinates $(\bar \tau, {\vec {\bar \sigma}})$ will have
the form ${}^4g^{(HH)}_{(1)\tau\tau} = \sgn\, \Big(1 + 2\,
n^{(HH)}_{(1)}\Big)$, ${}^4g^{(HH)}_{(1)\tau r} = - \sgn\, {\bar
n}^{(HH)}_{(1)(r)}$, ${}^4g^{(HH)}_{(1)rs} = - \sgn\,
\Big(\delta_{rs} + A^{(HH)}_{(1)rs}\Big)$ with $A^{(HH)}_{(1)rs}$
depending on which harmonic gauge one chooses. The connection to the
4-metric ${}^4g_{(1)AB}(\tau, \vec \sigma)$ in the family of
3-orthogonal gauges is by means of a 4-coordinate transformation
$\tau = \bar \tau + a_{(1)}(\bar \tau, {\vec {\bar \sigma}})$,
$\sigma^r = {\bar \sigma}^r + b^r_{(1)}(\bar \tau, {\vec {\bar
\sigma}})$ implying ${}^4g^{(HH)}_{(1)AB}(\bar \tau, {\vec {\bar
\sigma}}) = {{\partial\, \sigma^C}\over {\partial\, {\bar
\sigma}^A}}\, {{\partial\, \sigma^D}\over {\partial\, {\bar
\sigma}^B}}\, {}^4g_{(1)CD}(\tau, \vec \sigma)$. As a consequence
one gets

\bea
 &&{{\partial\, a_{(1)}(\bar \tau, {\vec {\bar \sigma}})}\over
 {\partial\, \bar \tau}} = \Big(n_{(1)}^{(HH)} - n_{(1)}\Big)(\bar \tau,
 {\vec {\bar \sigma}}),\nonumber \\
 &&{{\partial\, b^r_{(1)}(\bar \tau, {\vec {\bar \sigma}})}\over
 {\partial\, \bar \tau}} = \Big({\bar n}_{(1)(r)}^{(HH)} -
 {\bar n}_{(1)(r)}\Big)(\bar \tau, {\vec {\bar \sigma}}),\nonumber \\
 &&{{\partial\, b^r_{(1)}(\bar \tau, {\vec {\bar \sigma}})}\over
 {\partial\, {\bar \sigma}^s}} + {{\partial\, b^s_{(1)}(\bar \tau,
 {\vec {\bar \sigma}})}\over {\partial\, {\bar \sigma}^r}} =
 \Big(2\, (\Gamma_r^{(1)} + 2\, \phi_{(1)})\, \delta_{rs} -
 A^{(HH)}_{(1)rs}\Big)(\bar \tau, {\vec {\bar \sigma}}).\nonumber \\
 &&{}
 \label{3.19}
 \eea

\vfill\eject

\section{The Solution of the Linearized Equations for $\tilde \phi$, $1 + n$,
$\pi_i^{(\theta)}$, ${\bar n}_{(a)}$ and the linearized ADM
Generators}

In this Section we find  the linearization of the super-Hamiltonian
and super-momentum constraints and then of the equations determining
the lapse and shift functions of our family of 3-orthogonal
Schwinger time gauges. The solutions of these linearized equations
allow to express ${\tilde \phi}_{(1)}$, $1 + n_{(1)}$,
$\sigma_{(1)(a)(b)}{|}_{a \not= b}$ (i.e. $\pi_i^{(\theta)}$),
${\bar n}_{(1)(a)}$ in terms of matter and of the tidal variables
$\Gamma^{(1)}_r = \sum_{\bar a}\, \gamma_{\bar ar}\, R_{\bar a}$. In
the next Section we will see that the linearized equations for the
tidal variables depend only on matter, so that at the end the
previous solutions will depend only on the matter. While the
equations of elliptic type solved in this Section will determine the
instantaneous inertial dependence on the matter of the 4-metric
(like the Coulomb potential in the radiation gauge in the case of
the electro-magnetic field), the wave equations for the tidal
variables in the next Section will determine the retarded dependence
on matter of the 4-metric. We will also see that at this order the
contracted Bianchi identities are identically satisfied. Then we
will evaluate the asymptotic ADM Poincare' generators till the
second order.

\subsection{The Super-Hamiltonian Constraint and the Lapse Function}

Let us first consider the determination of $\phi = {\tilde
\phi}^{1/6} = 1 + \phi_{(1)} + O(\zeta^2)$ by using the
super-Hamiltonian constraint (\ref{2.6}) and of the lapse function
by means of Eq. (\ref{2.12}).

\subsubsection{The Equation for $\phi_{(1)}$}

Since the linearized Laplace-Beltrami operator and the scalar
3-curvature in the family of 3-orthogonal gauges, see
Eq.(\ref{2.5}), have the expressions ($\triangle = \sum_a\,
\partial_a^2$ is the flat asymptotic Laplacian)\medskip

 \bea
 {\hat \triangle}{|}_{\theta^i = 0} &=& \triangle + O(\zeta),
 \qquad {\hat \triangle}{|}_{\theta^i = 0}\, \phi = \triangle\,
 \phi_{(1)} + O(\zeta^2), \nonumber \\
 &&{}\nonumber \\
 {}^3\hat R{|}_{\theta^i=0}\, &=&\, 2\, \sum_a\, \partial^2_a\,
 \Gamma_a^{(1)} + O(\zeta^2),
 \label{4.1}
 \eea

\noindent the super-Hamiltonian constraint given in Eq.(\ref{2.6})
gives rise to the following linearized elliptic equation for
$\phi_{(1)}$
\medskip

\beq
 \triangle\, \phi_{(1)}(\tau, \vec \sigma)\, \approx\, - {{2\pi\,
 G}\over {c^3}}\, {\cal M}^{(UV)}_{(1)}(\tau, \vec \sigma) +
 {1\over 4}\, \sum_a\, \partial^2_a\, \Gamma_a^{(1)}(\tau, \vec
 \sigma) + O(\zeta^2).
 \label{4.2}
 \eeq

\subsubsection{The Equation for the Lapse Function $1 + n_{(1)}$}

In the family of 3-orthogonal gauges Eqs.(\ref{3.7}) imply ${}^3K =
{}^3K_{(1)} = {1\over L}\, O(\zeta) \approx F_{(1)}$ with
$F_{(1)}(\tau, \vec \sigma)$ arbitrary numerical function of the
same order.\medskip

To find the linearization of Eq.(\ref{2.12}) for the lapse function
we need the following result

\bea
 &&\int d^3\sigma_1\, \Big(1 + n(\tau ,{\vec \sigma}_1)\Big)\,
 {{\delta\, {\cal M}(\tau ,{\vec \sigma}_1)}\over
 {\delta\, \phi(\tau ,\vec \sigma)}}\, =\nonumber \\
 &&{}\nonumber \\
 &&= - 2\, \sum_i\,
 \delta^3(\vec \sigma, {\vec \eta}_i(\tau))\, \eta_i\,
 \Big({{\sum_a\, \Big(\kappa_{ia}(\tau) - {{Q_i}\over c}\, A_{\perp\,
 a}\Big)^2}\over {\sqrt{m_i^2\, c^2 + \sum_a\, \Big(
 \kappa_{ia}(\tau) - {{Q_i}\over c}\, A_{\perp\,
 a}\Big)^2}}}\Big)(\tau, \vec \sigma) -\nonumber \\
 &&- {1\over c}\, \Big[\sum_a\, (\pi^a_{\perp})^2 + {1\over 2}\,
 \sum_{ab}\, F^2_{ab} - \sum_a\, \Big(2\, \pi^a_{\perp} - \sum_{k
 \not= j}\, \eta_k\, Q_k\,
 \partial_a\, c(\vec \sigma, {\vec \eta}_k(\tau))\Big)
 \nonumber \\
 &&\sum_j\, \eta_j\, Q_j\, \partial_a\, c(\vec \sigma, {\vec
 \eta}_j(\tau))\Big](\tau, \vec \sigma) - 2\, \sum_a\, {\cal
 R}^{aa}_{(2)} =\nonumber \\
 &&=\, - 2\, \sum_a\, T_{(1)}^{aa} - 2\, \sum_a\, {\cal
 R}^{aa}_{(2)},\qquad from\,\, Eq.(\ref{3.12}),
 \label{4.3}
 \eea

\medskip
\noindent where the following approximation was used\medskip

\bea
 &&{1\over {\sqrt{m_i^2\, c^2 + {\tilde \phi}^{-2/3}\, \sum_a\, Q_a^{-2}\,
 \Big(\kappa_{ia}(\tau) - {{Q_i}\over c}\, A_{\perp\,
 a}\Big)^2}}}(\tau, {\vec \eta}_i(\tau)) = \nonumber \\
 &&= \Big({1\over {\sqrt{m_i^2\, c^2 + \sum_a\, \Big(
 \kappa_{ia}(\tau) - {{Q_i}\over c}\, A_{\perp\, a}\Big)^2}}}\,
 \Big[1 +\nonumber \\
 &&+ {{\sum_a\, (\Gamma_a^{(1)} + 2\, \phi_{(1)})\,
 \Big(\kappa_{ia}(\tau) - {{Q_i}\over c}\, A_{\perp\,
 a}\Big)^2}\over {m_i^2\, c^2 + \sum_a\, \Big(
 \kappa_{ia}(\tau) - {{Q_i}\over c}\,
 A_{\perp\, a}\Big)^2}}\Big]\Big)(\tau, {\vec \eta}_i(\tau))
 + {1\over {Mc}}\, O(\zeta).
 \label{4.4}
 \eea

\bigskip
As a consequence, by using  Eqs. (\ref{4.3}) and (\ref{3.12}),
Eq.(\ref{2.12}) becomes the following linearized elliptic equation
for the lapse function
\medskip

\bea
 \triangle\, n_{(1)}(\tau, \vec \sigma)\, &\cir& - \partial_{\tau}\,
 {}^3K_{(1)}(\tau, \vec \sigma) + {1\over 2}\, \sum_a\, \partial_a^2\,
 (\Gamma_a^{(1)} - 4\, \phi_{(1)})(\tau, \vec \sigma) +
  {{4\pi\, G}\over {c^3}}\, \sum_a\, T_{(1)}^{aa}(\tau, \vec
 \sigma) =\nonumber \\
 &{\buildrel {(\ref{4.2})}\over =}& - \partial_{\tau}\,
 {}^3K_{(1)}(\tau, \vec \sigma) + {{4\pi\, G}\over {c^3}}\,
 \Big({\cal M}^{(UV)}_{(1)} + \sum_a\, T^{aa}_{(1)}\Big)(\tau, \vec
 \sigma).
 \label{4.5}
 \eea

\subsubsection{The Solutions for $\phi_{(1)}$ and $n_{(1)}$}

The solutions, vanishing at spatial infinity, of Eqs. (\ref{4.2})
and (\ref{4.5}) for $\phi_{(1)}$ and $n_{(1)}$ are (${\cal
M}_{(1)}^{(UV)}$ is given in Eq.(\ref{3.9}))\medskip

\bea
 \phi_{(1)}(\tau, \vec \sigma) \, &\cir& \Big[- {{2\pi\, G}\over
 {c^3}}\, {1\over {\triangle}}\, {\cal M}^{(UV)}_{(1)} +
 {1\over 4}\, \sum_c\, {{\partial_c^2}\over {\triangle}}\,
 \Gamma_c^{(1)}\Big](\tau, \vec \sigma) \cir\nonumber \\
 &&{}\nonumber \\
 &\cir& {G\over {2c^3}}\, \Big[\sum_i\, \eta_i\, {{\sqrt{m_i^2\, c^2 +
 \sum_c\, \Big(\kappa_{ic}(\tau) - {{Q_i}\over c}\,
 A_{\perp\, c}(\tau, {\vec \eta}_i(\tau))\Big)^2}}\over
 {|\vec \sigma - {\vec \eta}_i(\tau)| }} +\nonumber \\
 &+& {1\over {2\, c}}\, \int {{d^3\sigma_1}\over {
 |\vec \sigma - {\vec \sigma}_1| }} \Big( \Big[\sum_a\,
 \Big((\pi^a_{\perp})^2 - \Big(2\, \pi_{\perp}^a - \sum_i\, Q_i\, \eta_i\,
 {{\partial\, c(\vec \sigma, {\vec \eta}_i(\tau))}\over {\partial\, \sigma^a}}
 \Big)\nonumber \\
 && \sum_j\, Q_j\, \eta_j\, {{\partial\, c(\vec \sigma, {\vec \eta}_j(\tau))}
 \over {\partial\, \sigma^a}} \Big) + {1\over 2}\, \sum_{ab}\, F^2_{ab}
 \Big]\Big)(\tau ,{\vec \sigma}_1)\Big] -\nonumber \\
 &-& {1\over {16\pi}}\, \int d^3\sigma_1\, {{\sum_a\, \partial_{1a}^2\,
 \Gamma_a^{(1)}(\tau, {\vec \sigma}_1)}\over {|\vec \sigma - {\vec \sigma}_1|
 }},\qquad \partial_r\, q = \partial_r\, \phi_{(1)} + O(\zeta^2), \nonumber \\
 &&{}\nonumber \\
 {}^4g_{(1)rs}(\tau, \vec \sigma)&\cir& - \sgn\, \delta_{rs}\, [1 + 2\,
 (\Gamma_r^{(1)} + 2\, \phi_{(1)})(\tau, \vec \sigma)],
 \label{4.6}
 \eea

\bigskip

\bea
 n_{(1)}(\tau, \vec \sigma)\, &\cir& \Big[{{4\pi\, G}\over {c^3}}\,
 {1\over {\triangle}}\, \Big({\cal M}^{(UV)}_{(1)} +
 \sum_a\, T_{(1)}^{aa}\Big) - {1\over {\triangle}}\, \partial_{\tau}\,
 {}^3K_{(1)} \Big](\tau, \vec \sigma)\, \cir\nonumber \\
 &&{}\nonumber \\
 &\cir& - {G\over {c^3}}\, \Big[\sum_i\, \eta_i\, {{\sqrt{m_i^2\, c^2 +
 \sum_c\, \Big(\kappa_{ic}(\tau) - {{Q_i}\over c}\,
 A_{\perp\, c}(\tau, {\vec \eta}_i(\tau))\Big)^2}}\over {
 |\vec \sigma - {\vec \eta}_i(\tau)|}}\nonumber \\
 &&\Big(1 + {{\sum_c\, \Big(\kappa_{ic}(\tau) - {{Q_i}\over c}\,
 A_{\perp\, c}(\tau, {\vec \eta}_i(\tau))\Big)^2}\over {m_i^2\, c^2 +
 \sum_c\, \Big(\kappa_{ic}(\tau) - {{Q_i}\over c}\,
 A_{\perp\, c}(\tau, {\vec \eta}_i(\tau))\Big)^2}}
 \Big) +\nonumber \\
 &+& {1\over  c}\, \int {{d^3\sigma_1}\over {
 |\vec \sigma - {\vec \sigma}_1| }} \Big[\sum_a\,
 \Big((\pi^a_{\perp})^2 - \Big(2\, \pi_{\perp}^a - \sum_i\, Q_i\, \eta_i\,
 {{\partial\, c(\vec \sigma, {\vec \eta}_i(\tau))}\over {\partial\, \sigma^a}}
 \Big)\nonumber \\
 && \sum_j\, Q_j\, \eta_j\, {{\partial\, c(\vec \sigma, {\vec \eta}_j(\tau))}
 \over {\partial\, \sigma^a}} \Big) + {1\over 2}\, \sum_{ab}\, F^2_{ab}
 \Big](\tau ,{\vec \sigma}_1)\Big] +\nonumber \\
 &+& \int d^3\sigma_1\, {{ \partial_{\tau}\, {}^3K(\tau, {\vec \sigma}_1)}\over
 {4\pi\, |\vec \sigma - {\vec \sigma}_1|}},
 \nonumber \\
 &&{}\nonumber \\
 {}^4g_{(1)\tau\tau}(\tau, \vec \sigma) &\cir& \sgn\, [1 + 2\, n_{(1)}(\tau, \vec \sigma)].
 \label{4.7}
 \eea
\medskip

While $\phi_{(1)}$ depends upon the tidal variables, the lapse
function of these 3-orthogonal gauges is independent from them.
However, while the volume 3-element $\phi_{(1)}$ of the 3-space is
independent from the inertial gauge variable ${}^3K_{(1)}$, the
lapse function, connecting nearby instantaneous 3-space with
different local York times, depends upon $\partial_{\tau}\,
{}^3{\cal K}_{(1)} = \partial_{\tau}\,{1\over {\triangle}}\,
 {}^3K_{(1)}$. At this order the spatial 4-metric
${}^4g_{(1)rs}$ depends upon $2\, \phi_{(1)} + \Gamma_r^{(1)}$: the
first term describes the instantaneous inertial part of the
gravitational field in the 3-orthogonal gauges, while the tidal term
the retarded one. Instead the component ${}^4g_{(1)\tau\tau}$,
relevant for local proper time, depends only upon the instantaneous
inertial part of the gravitational field and upon the inertial gauge
variable $\partial_{\tau}\, {}^3{\cal K}_{(1)}$.

\subsection{The Super-Momentum Constraints and the Shift Functions}

Let us now consider the determination of $\sigma_{(1)(a)(b)}{|}_{a
\not= b}$, by means of the super-momentum constraints (\ref{2.7})
and the determination of the shift functions from  Eqs.
(\ref{2.11}). As said after Eq.(\ref{3.7}), such a solution
$\sigma_{(1)(a)(b)}{|}_{a \not= b}$ implies $\pi_{i}^{(\theta)} =
{{c^3}\over {8\pi\, G}}\, \sum_{a \not= b}\, \epsilon_{iab}\,
(\Gamma^{(1)}_a - \Gamma^{(1)}_b)\, \sigma_{(1)(a)(b)} \approx 0$ in
the York canonical basis of the linearized theory.

\subsubsection{The Equations for $\sigma_{(1)(a)(b)}{|}_{a \not=
b}$}

Eqs.(\ref{2.7}) for $\sigma_{(1)(a)(b)}{|}_{a \not= b}$ takes the
following linearized form (${\cal M}_{(1)a}^{(UV)}$ is given in
Eq.(\ref{3.10}))\medskip

\bea
 \sum_{b \not= a}\, \partial_b\, \sigma_{(1)(a)(b)}(\tau, \vec
 \sigma)\, &\approx& {2\over 3}\, \partial_a\, {}^3K_{(1)}(\tau, \vec
 \sigma) + {{8\pi\, G}\over {c^3}}\, \Big({\cal M}_{(1)a}^{(UV)}
 + \sum_{\bar a}\, \gamma_{\bar aa}\, \partial_a\, \Pi_{\bar a}\Big)(\tau, \vec
 \sigma) =\nonumber \\
 &&{}\nonumber \\
 &=& \Big(\partial_{\tau}\, \partial_a\, \Gamma_a^{(1)} +
 {1\over 3}\, \partial_a\, (\sum_c\, \partial_c\, {\bar n}_{(1)(c)})
 - \partial_a^2\, {\bar n}_{(1)(a)} +\nonumber \\
 &+& {2\over 3}\, \partial_a\, {}^3K_{(1)} + {{8\pi\, G}\over {c^3}}\,
 {\cal M}_{(1)a}^{(UV)}\Big)(\tau, \vec \sigma),
 \nonumber \\
 &&{}
 \label{4.8}
 \eea

\noindent where we have used Eq.(\ref{3.6}) for $\Pi_{\bar a}$ and
$\sum_{\bar a}\, \gamma_{\bar aa}\, \gamma_{\bar ab} = \delta_{ab} -
{1\over 3}$ (see before Eq.(\ref{2.3})).

\subsubsection{The Equations for the Shift Functions ${\bar
n}_{(1)(r)}$}

Eqs.(\ref{2.11}) with $a \not= b$ for the shift functions gives rise
to the following linearized equations
\medskip

 \beq
 \Big(\partial_b\, {\bar n}_{(1)(a)}(\tau, \vec \sigma) + \partial_a\,
 {\bar n}_{(1)(b)}(\tau, \vec \sigma)\Big){|}_{a \not= b} \, \cir\, 2\,
 \sigma_{(1)(a)(b)}{|}_{a \not= b}(\tau, \vec \sigma),\qquad
 a \not= b.
 \label{4.9}
 \eeq

\subsubsection{The Solutions for ${\bar n}_{(1)(a)}$ and
$\sigma_{(1)(a)(b)}{|}_{a \not= b}$}

By applying the operator $\partial_b$ to Eqs.(\ref{4.9}) and by
summing over $b$, we get\medskip

\beq
 \sum_{b \not= a}\, \Big[\partial_b^2\, {\bar n}_{(1)(a)} +
 \partial_a\, (\partial_b\, {\bar n}_{(1)(b)})\Big](\tau,
 \vec \sigma)\, \cir\, 2\, \sum_{b \not= a}\,
 \partial_b\, \sigma_{(1)(a)(b)}(\tau, \vec \sigma).
 \label{4.10}
 \eeq

\bigskip

By putting  Eqs.(\ref{4.8}) for $\sigma_{(1)(a)(b)}{|}_{a \not= b}$
into Eq.(\ref{4.10}), we get an equation containing only the shift
functions\medskip

\bea
 &&\sum_{b \not= a}\, \Big[\partial_b^2\, {\bar n}_{(1)(a)} +
 \partial_a\, (\partial_b\, {\bar n}_{(1)(b)})\Big](\tau, \vec \sigma)\,
 \cir\nonumber \\
 &&\cir\, \Big[{4\over 3}\, \partial_a\,{}^3K_{(1)} + {{16\pi\, G}\over {c^3}}\,
 {\cal M}^{(UV)}_{(1)a} - 2\, \partial_a^2\, {\bar n}_{(1)(a)}
 +\nonumber \\
 &&+{2\over 3}\, \partial_a\, (\sum_c\, \partial_c\, {\bar n}_{(1)(c)}) +
 2\, \partial_{\tau}\, \partial_a\, \Gamma_a^{(1)}\Big](\tau, \vec
 \sigma).
 \label{4.11}
 \eea
\medskip

Eqs.(\ref{4.11}) can be rewritten in the form (no more containing
the condition $b \not= a$)\medskip

\bea
 &&\Big[\triangle\, {\bar n}_{(1)(a)} + {1\over 3}\, \partial_a\,
 (\sum_b\, \partial_b\, {\bar n}_{(1)(b)})\Big](\tau, \vec \sigma)\,
 \cir \nonumber \\
 &&\cir \Big[{4\over 3}\, \partial_a\, {}^3K_{(1)} + 2\, \partial_{\tau}\,
 \partial_a\, \Gamma_a^{(1)} + {{16\pi\, G}\over {c^3}}\,
 {\cal M}_{(1)a}^{(UV)}\Big](\tau, \vec \sigma).
 \label{4.12}
 \eea
\medskip

If we apply $\sum_a\, \partial_a$ to Eqs.(\ref{4.12}) we get\medskip

\bea
 &&{4\over 3}\, \triangle\, \Big[\sum_a\, \partial_a\, {\bar
 n}_{(1)(a)}\Big](\tau, \vec \sigma)\, \cir\, \Big[{4\over 3}\,
 \triangle\, {}^3K_{(1)} +\nonumber \\
 &&+ {{16\pi\, G}\over {c^3}}\, \sum_a\,
 \partial_a\, {\cal M}_{(1)a}^{(UV)} + 2\, \sum_a\,
 \partial_{\tau}\, \partial_a^2\, \Gamma_a^{(1)}\Big](\tau, \vec
 \sigma),
 \label{4.13}
 \eea
\medskip

\noindent and this equation implies\medskip

\bea
 &&\sum_a\, \partial_a\, {\bar n}_{(1)(a)}(\tau, \vec \sigma) \, \cir\,
 \Big[{}^3K_{(1)} + {{12\pi\, G}\over {c^3}}\, {1\over {\triangle}}\,
 \sum_a\, \partial_a\, {\cal M}_{(1)a}^{(UV)} + {3\over 2}\,
 {1\over {\triangle}}\, \sum_a\, \partial_{\tau}\,
 \partial_a^2\, \Gamma_a^{(1)}\Big](\tau, \vec \sigma).
 \nonumber \\
 &&{}
 \label{4.14}
 \eea
\bigskip

As a consequence the final linearized equation (of elliptic type)
for the shift functions is\medskip

\bea
 \triangle\, {\bar n}_{(1)(a)}(\tau, \vec \sigma)\, &\cir&
 \Big[\partial_a\, {}^3K_{(1)} + {{4\pi\, G}\over {c^3}}\, \Big(4\,
 {\cal M}_{(1)a}^{(UV)} - {{\partial_a}\over {\triangle}}\,\, \sum_c\,
 \partial_c\, {\cal M}_{(1)c}^{(UV)}\Big) +\nonumber \\
 &+& {1\over 2}\, \Big(4\, \partial_{\tau}\, (\partial_a\, \Gamma_a^{(1)}) -
 {{\partial_a}\over {\triangle}}\, \sum_c\,
 \partial_{\tau}\, (\partial_c^2\, \Gamma_c^{(1)})\Big)\Big](\tau,
 \vec \sigma),
 \label{4.15}
 \eea
\medskip

\noindent whose solution, vanishing at spatial infinity and
depending upon the tidal variables, is \medskip

\bea
 {\bar n}_{(1)(a)}(\tau, \vec \sigma) \, &\cir& \Big[{{\partial_a}\over
 {\triangle}}\, {}^3K_{(1)} + {{4\pi\, G}\over {c^3}}\, {1\over
 {\triangle}}\, \Big(4\, {\cal M}_{(1)a}^{(UV)} -
 {{\partial_a}\over {\triangle}}\,\, \sum_c\,
 \partial_c\, {\cal M}_{(1)c}^{(UV)}\Big) +\nonumber \\
 &+& {1\over 2}\, \partial_{\tau}\, {{\partial_a}\over {\triangle}}\,
 \Big(4\, \Gamma_a^{(1)} -
 \sum_c\,  {{\partial_c^2}\over {\triangle}}\, \Gamma_c^{(1)}\Big)
 \Big](\tau, \vec \sigma) \cir\nonumber \\
 &&{}\nonumber \\
 &\cir& - {{4\, G}\over {c^3}}\, \sum_i\, \eta_i\, \Big[{{
 \kappa_{ia}(\tau) - {{Q_i}\over c}\, A_{\perp a}(\tau,
 {\vec \eta}_i(\tau))}\over {|\vec \sigma - {\vec \eta}_i(\tau)|}}
 -\nonumber \\
 &-& \sum_c\, \Big(\kappa_{ic}(\tau) - {{Q_i}\over c}\,
 A_{\perp c}(\tau, {\vec \eta}_i(\tau))\Big)\, \int {{d^3\sigma_1}\over
 {4\pi\, |\vec \sigma - {\vec \sigma}_1|\, |{\vec \sigma}_1
 - {\vec \eta}_i(\tau)|^3}}\nonumber \\
 &&\Big(\delta_{ac} - 3\, {{(\sigma_1^a
 - \eta_i^a(\tau))\, (\sigma_1^c - \eta_i^c(\tau))}\over
 {|{\vec \sigma}_1 - {\vec \eta}_i(\tau)|^2}}\Big)\Big]
 -\nonumber \\
 &-& \int {{d^3\sigma_1}\over {4\pi\, |\vec \sigma - {\vec \sigma}_1|}}\,
 \partial_{1a}\, \Big[\Big({}^3K_{(1)} + 2\, \partial_{\tau}\,
 \Gamma_a^{(1)}\Big)(\tau, {\vec \sigma}_1) +\nonumber \\
 &+& \int d^3\sigma_2\, {{\sum_c\, \partial_{\tau}\, \partial_{2c}^2\,
 \Gamma_c^{(1)}(\tau, {\vec \sigma}_2)}\over {8\pi\, |{\vec \sigma}_1 -
 {\vec \sigma}_2|}}\Big] +\nonumber \\
 &+& {G\over {c^4}}\, \int {{d^3\sigma_1}\over {|\vec \sigma - {\vec \sigma}_1|}}
 \, \sum_s\nonumber \\
 &&\Big[4\, F_{as}(\tau ,{\vec \sigma}_1)\, \Big(
  \pi_\perp^s(\tau,  {\vec \sigma}_1) - \sum_n\, \delta^{sn}\, \sum_i\, Q_i\, \eta_i\,
 {{\partial\, c({\vec \sigma}_1, {\vec \eta}_i(\tau))}\over {\partial\,
 \sigma_1^n}}\Big) + \nonumber \\
 &+& \sum_c\, \partial_{1a}\, \partial_{1c}\, \int {{d^3\sigma_2}\over
 {4\pi\, |{\vec \sigma}_1 - {\vec \sigma}_2|}}\nonumber \\
 &&F_{cs}(\tau ,{\vec \sigma}_2)\, \Big(
  \pi_\perp^s(\tau,  {\vec \sigma}_2) - \sum_n\, \delta^{sn}\, \sum_i\, Q_i\, \eta_i\,
 {{\partial\, c({\vec \sigma}_2, {\vec \eta}_i(\tau))}\over {\partial\,
 \sigma_2^n}}\Big) \Big],\nonumber \\
 &&{}\nonumber \\
 {}^4g_{(1)\tau r}(\tau, \vec \sigma) &\cir& - \sgn\, {\bar
 n}_{(1)(r)}(\tau, \vec \sigma).
 \label{4.16}
 \eea
\bigskip

Then the functions $\sigma_{(1)(a)(b)}{|}_{a \not= b}$ have the the
following expression\medskip

\begin{eqnarray*}
 \sigma_{(1)(a)(b)}{|}_{a \not= b}\, &\cir& {1\over 2}\,
 \Big(\partial_a\, {\bar n}_{(1)(b)} + \partial_b\, {\bar
 n}_{(1)(a)}\Big)(\tau, \vec \sigma)\, \cir\nonumber \\
 &&{}\nonumber \\
 &\cir& {{\partial_a\, \partial_b}\over {\triangle}}\, {}^3K_{(1)} + {{8\pi\,
 G}\over {c^3}}\, \Big[{1\over {\triangle}}\, \Big(\partial_a\,
 {\cal M}^{(UV)}_{(1)b} + \partial_b\, {\cal M}^{(UV)}_{(1)a}\Big)
  - {1\over 2}\, {{\partial_a\, \partial_b}\over {\triangle}}\, \sum_c\,
  {{\partial_c}\over {\triangle}}\, {\cal M}^{(UV)}_{(1)c}\Big] +\nonumber \\
 &+& \partial_{\tau}\, {{\partial_a\, \partial_b}\over {\triangle}}\,
 \Big(\Gamma_a^{(1)} + \Gamma_b^{(1)} - {1\over 2}\, \sum_c\,
 {{\partial_c^2}\over {\triangle}}\, \Gamma_c^{(1)}\Big) \cir
 \end{eqnarray*}

\begin{eqnarray*}
 &\cir&- {1\over 2}\, \sum_d\, (\delta_{ad}\, \partial_b + \delta_{bd}\,
 \partial_a)\quad
 \Big( {{4\, G}\over {c^3}}\, \sum_i\, \eta_i\, \Big[{{
 \kappa_{id}(\tau) - {{Q_i}\over c}\, A_{\perp d}(\tau,
 {\vec \eta}_i(\tau))}\over {|\vec \sigma - {\vec \eta}_i(\tau)|}}
 -\nonumber \\
 &-& \sum_c\, \Big(\kappa_{ic}(\tau) - {{Q_i}\over c}\,
 A_{\perp c}(\tau, {\vec \eta}_i(\tau))\Big)\, \int {{d^3\sigma_1}\over
 {4\pi\, |\vec \sigma - {\vec \sigma}_1|\, |{\vec \sigma}_1
 - {\vec \eta}_i(\tau)|^3}}\nonumber \\
 && \Big(\delta_{dc} - 3\, {{(\sigma_1^d
 - \eta_i^d(\tau))\, (\sigma_1^c - \eta_i^c(\tau))}\over
 {|{\vec \sigma}_1 - {\vec \eta}_i(\tau)|^2}}\Big)\Big]
 +\nonumber \\
 &+& \int {{d^3\sigma_1}\over {4\pi\, |\vec \sigma - {\vec \sigma}_1|}}\,
 \partial_{1d}\, \Big[\Big({}^3K_{(1)} + 2\, \partial_{\tau}\,
 \Gamma_d^{(1)}\Big)(\tau, {\vec \sigma}_1) +
  \int d^3\sigma_2\, {{\sum_c\, \partial_{\tau}\, \partial_{2c}^2\,
 \Gamma_c^{(1)}(\tau, {\vec \sigma}_2)}\over {8\pi\, |{\vec \sigma}_1 -
 {\vec \sigma}_2|}}\Big]\,\, \Big) +
 \end{eqnarray*}

\bea
 &+& {G\over {2\, c^4}}\, \Big(\partial_a\, \int {{d^3\sigma_1}\over
 {|\vec \sigma - {\vec \sigma}_1|}}\, \sum_s\,
 \Big[4\, F_{bs}(\tau ,{\vec \sigma}_1)\, \Big(
  \pi_\perp^s(\tau,  {\vec \sigma}_1) - \sum_n\, \delta^{sn}\, \sum_i\, Q_i\, \eta_i\,
 {{\partial\, c({\vec \sigma}_1, {\vec \eta}_i(\tau))}\over {\partial\,
 \sigma_1^n}}\Big) +\nonumber \\
 &+& \sum_c\, \partial_{1b}\, \partial_{1c}\, \int {{d^3\sigma_2}\over
 {4\pi\, |{\vec \sigma}_1 - {\vec \sigma}_2|}}\,
 F_{cs}(\tau ,{\vec \sigma}_2)\, \Big(
  \pi_\perp^s(\tau,  {\vec \sigma}_2) - \sum_n\, \delta^{sn}\, \sum_i\, Q_i\, \eta_i\,
 {{\partial\, c({\vec \sigma}_2, {\vec \eta}_i(\tau))}\over {\partial\,
 \sigma_2^n}}\Big) \Big] +\nonumber \\
 &+& \partial_b\, \int {{d^3\sigma_1}\over
 {|\vec \sigma - {\vec \sigma}_1|}}\, \sum_s\,
 \Big[4\, F_{as}(\tau ,{\vec \sigma}_1)\, \Big(
  \pi_\perp^s(\tau,  {\vec \sigma}_1) - \sum_n\, \delta^{sn}\, \sum_i\, Q_i\, \eta_i\,
 {{\partial\, c({\vec \sigma}_1, {\vec \eta}_i(\tau))}\over {\partial\,
 \sigma_1^n}}\Big) +\nonumber \\
 &+& \sum_c\, \partial_{1a}\, \partial_{1c}\, \int {{d^3\sigma_2}\over
 {4\pi\, |{\vec \sigma}_1 - {\vec \sigma}_2|}}\,
 F_{cs}(\tau ,{\vec \sigma}_2)\, \Big(
  \pi_\perp^s(\tau,  {\vec \sigma}_2) - \sum_n\, \delta^{sn}\, \sum_i\, Q_i\, \eta_i\,
 {{\partial\, c({\vec \sigma}_2, {\vec \eta}_i(\tau))}\over {\partial\,
 \sigma_2^n}}\Big) \Big] \Big).\nonumber \\
 &&{}
 \label{4.17}
 \eea

Both the functions ${\bar n}_{(1)(a)}$ and $\sigma_{(1)(a)(b)}{|}_{a
\not= b}$ depend upon matter, upon the tidal variables and upon the
spatial gradients of the inertial gauge function ${}^3{\cal K}_{(1)}
= {1\over {\triangle}}\, {}^3K_{(1)}$.
\medskip

Due to Eq.(\ref{4.16}) we have that gravito-magnetism (described by
${}^4g_{(1)\tau r}$) in the 3-orthogonal gauges depends on both the
instantaneous inertial and retarded parts of the gravitational field
and upon the gauge variable $\partial_r\, {}^3{\cal
K}_{(1)}$.\medskip

The integral appearing in the shift function has the following
expression $\int {{d^3\sigma_1}\over {4\pi\, |\vec \sigma - {\vec
\sigma}_1|\, |{\vec \sigma}_1 - {\vec
\eta}_i(\tau)|^3}}\,\Big(\delta^{ac} - 3\, {{(\sigma_1^a -
\eta_i^a(\tau))\, (\sigma_1^c - \eta_i^c(\tau))}\over {|{\vec
\sigma}_1 - {\vec \eta}_i(\tau)|^2}}\Big) = - {1\over 2}\, {1\over
{|\vec \sigma - {\vec \eta}_i(\tau)|}}\, \Big(\delta^{ac} -
{{(\sigma^a - \eta_i^a(\tau))\, (\sigma^c - \eta_i^c(\tau))}\over
{|\vec \sigma - {\vec \eta}_i(\tau)|^2}}\Big)$, so that the
contribution to gravito-magnetism coming from the mass current
density ${\cal M}_{(1)(r)}^{(UV)}(\tau, \vec \sigma)$ has the final
form $- {{2\, G}\over {c^3}}\, \sum_i\, {{\eta_i}\over {|\vec \sigma
- {\vec \eta}_i(\tau)|}}\, \Big(\kappa_{ia}(\tau) - {{Q_i}\over c}\,
A_{\perp a}(\tau, {\vec \eta}_i(\tau)) + {{(\sigma^a -
\eta^a_i(\tau))\, \Big[{\vec \kappa}_i(\tau) - {{Q_i}\over c}\,
{\vec A}_{\perp}(\tau, {\vec \eta}_i(\tau))\Big] \cdot (\vec \sigma
- {\vec \eta}_i(\tau))}\over {|\vec \sigma - {\vec
\eta}_i(\tau)|^2}}\Big)$.

\medskip

From Eqs.(\ref{3.6}) and (\ref{4.16}) we get that the tidal momenta
$\Pi_{\bar a}$ have the following expression

\bea
 {{8\pi\, G}\over {c^3}}\, \Pi_{\bar a}(\tau, \vec \sigma) &\cir&
 \partial_{\tau}\, R_{\bar a}(\tau, \vec \sigma) - \sum_a\,
 \gamma_{\bar aa}\, \Big[\partial_{\tau}\, {{\partial_a^2}\over
 {2\, \triangle}}\, (4\, \Gamma_a^{(1)} - {1\over {\triangle}}\,
 \sum_c\, \partial_c^2\, \Gamma_c^{(1)}) +\nonumber \\
 &+& {{4\pi\, G}\over {c^3}}\, {1\over {\triangle}}\, (4\, \partial_a\,
 {\cal M}^{(UV)}_{(1)a} - {{\partial_a^2}\over {\triangle}}\, \sum_c\,
 \partial_c\, {\cal M}^{(UV)}_{(1)c}) + {{\partial_a^2}\over {\triangle}}\,
 {}^3K_{(1)}\Big] =\nonumber \\
 &&{}\nonumber \\
 &=& \Big(\sum_{\bar b}\, M_{\bar a\bar b}\, \partial_{\tau} \, R_{\bar
 b} - \sum_a\, \gamma_{\bar aa}\, \Big[{{4\pi\, G}\over {c^3}}\,
 {1\over {\triangle}}\, (4\, \partial_a\,
 {\cal M}^{(UV)}_{(1)a} - {{\partial_a^2}\over {\triangle}}\, \sum_c\,
 \partial_c\, {\cal M}^{(UV)}_{(1)c}) +\nonumber \\
 &+&  {{\partial_a^2}\over {\triangle}}\,
 {}^3K_{(1)} \Big]\Big)(\tau, \vec \sigma),\nonumber \\
 &&{}\nonumber \\
 && M_{\bar a\bar b} = \delta_{\bar a\bar b} - \sum_a\, \gamma_{\bar
 aa}\, {{\partial_a^2}\over {\triangle}}\, \Big(2\, \gamma_{\bar ba} -
 {1\over 2}\, \sum_b\, \gamma_{\bar bb}\, {{\partial_b^2}\over
 {\triangle}}\Big),
 \label{4.18}
 \eea

\noindent where we introduced the operator $M_{\bar a\bar b}$, which
will be shown to be connected with the selection of the
traceless-transverse (TT) part of the 3-metric on $\Sigma_{\tau}$ in
Section VI. The relation between tidal momenta and tidal velocities
depends on the inertial gauge variable $\sum_a\, \gamma_{\bar aa}\,
\partial_a^2\, {}^3{\cal K}$.

\subsection{The Contracted Bianchi Identities}

The contracted Bianchi identity (\ref{2.8}) for $\partial_{\tau}\,
\phi_{(1)}$ has the following linearized form
\medskip

\beq
 \partial_{\tau}\, \phi_{(1)}(\tau, \vec \sigma)\, = 6\,
 \partial_{\tau}\, q_{(1)}(\tau, \vec \sigma)\, \cir\, {1\over
 6}\, \Big(\sum_a\, \partial_a\, {\bar n}_{(1)(a)} - {}^3K_{(1)}\Big)(\tau,
 \vec \sigma) + O(\zeta^2).
 \label{4.19}
 \eeq

\bigskip

By using the solutions (\ref{4.6}) and (\ref{4.16}) for $\phi_{(1)}$
and ${\bar n}_{(1)(a)}$, this equation is identically satisfied at
the lower order  due to the conservation $\partial_A\,
T_{(1)}^{A\tau} = 0$ at the lowest order, see Eq.(\ref{3.13}).

\bigskip

The Bianchi identities (\ref{2.9}) become\medskip

\beq
 \partial_{\tau}\, \pi^{(\theta)}_i \cir 0 + O(\zeta^2),
 \label{4.20}
 \eeq

\noindent consistently with Eq. (\ref{3.7}) , which gives
$\pi_i^{(\theta)}(\tau, \vec \sigma) = 0 + O(\zeta^2)$ in our
3-orthogonal gauges. \medskip

It can also be checked with a lengthy calculation that also the
contracted Bianchi identities (\ref{2.10}) for
$\sigma_{(1)(a)(b)}{|}_{a \not= b}$ are satisfied at this order.
This check requires the use of Eqs. (\ref{4.8}), (\ref{3.6}),
(\ref{4.9}), (\ref{4.17}), (\ref{4.6}), (\ref{4.7}), (\ref{3.13})
(i.e. of the already found solutions of the linearized Hamilton
equations) to get an expression which vanishes if we use
Eqs.(\ref{6.11}), which are a byproduct of the linearized second
order equations (\ref{6.4}) for the tidal variables as shown in
Section VI.\medskip

As a consequence possible problems with the generalized
gravitational Gribov problem, identified in Ref.\cite{5}, will
appear at higher orders.

\bigskip

Therefore the contracted Bianchi identities are identically
satisfied at the lowest order.

\subsection{The ADM Poincare' Generators}

The weak ADM Poincare' generators in the family of 3-orthogonal
gauges are given in Eqs. (\ref{2.14}) and (\ref{2.20}), with the
mass and momentum densities of Eqs.(\ref{2.4}). They are the
analogue of the internal Poincare' generators of Minkowski inertial
rest-frame instant form \cite{4}.\medskip

By using Eq.(\ref{3.6}) for the tidal momenta and Eqs. (\ref{2.4}),
(\ref{2.5}) and (\ref{2.14}), after a long but straightforward
calculation (including various integrations by parts) we get the
following form of the weak ADM energy at the second order
\footnote{While the terms $\sum_a\, \gamma_{\bar aa}\,
\partial_a\, {\bar n}_{(1)(a)}$ coming from the tidal momenta $\Pi_{\bar a}$
are {\it gravito-magnetic} potentials of the 3-orthogonal gauges
(they depend on the shift functions), the term $\sum_{a \not= b}\,
\sigma^2_{(1)(a)(b)}$, determined by the super-momentum constraints,
is a 3-coordinate dependent potential like ${\cal S}_{(2)}$ (coming
from the Gamma-Gamma term of the 3-curvature).}\medskip

\begin{eqnarray*}
 {1\over c}\, {\hat E}_{ADM}\, &=&\, \int d^3\sigma\, \Big[
 {\cal M}^{(UV)}_{(1)} + {\cal M}^{(UV)}_{(2)} +\nonumber \\
 &+&{{c^3}\over {16\pi\, G}}\, \Big(\sum_{\bar a}\, (\partial_{\tau}\,
 R_{\bar a} - \sum_a\, \gamma_{\bar aa}\, \partial_a\, {\bar n}_{(1)(a)})^2 +
 \sum_{a \not= b}\, \sigma^2_{(1)(a)(b)} -\nonumber \\
 &-& {2\over 3}\, ({}^3K_{(1)})^2 - {\cal S}_{(2)}{|}_{\theta^i = 0}
 \Big](\tau, \vec \sigma) + O(\zeta^3) =\nonumber \\
 &&{}\nonumber \\
 &=& M_{(1)}\, c + {1\over c}\, {\hat E}_{ADM (2)}  +
  Mc\, O(\zeta^3),\nonumber \\
 &&{}\nonumber \\
 &&{\cal S}_{(2)}{|}_{\theta^i = 0} = \sum_a\, \Big[8\,
 (\partial_a\, \phi_{(1)})^2 - 4\, \sum_{\bar a}\, \gamma_{\bar
 aa}\, \partial_a\, \phi_{(1)}\, \partial_a\, R_{\bar a}
 +\nonumber \\
 &&+ \sum_{\bar a\bar b}\, (2\, \gamma_{\bar aa}\, \gamma_{\bar ba}
 - \delta_{\bar a\bar b})\, \partial_a\, R_{\bar a}\, \partial_a\,
 R_{\bar b}\Big] + O(\zeta^3),
  \end{eqnarray*}

\begin{eqnarray*}
  M_{(1)}\, c &=&  \int d^3\sigma\, {\cal
 M}^{(UV)}_{(1)}(\tau, \vec \sigma) = \sum_i\, \eta_i\,  \sqrt{m^2_i\, c^2 +
 \Big({\vec \kappa}_i(\tau ) - {{Q_i}\over c}\,
 {\vec A}_{\perp}(\tau, {\vec \eta}_i(\tau))\Big)^2}
  +\nonumber \\
 &+& {1\over {2c}}\, \int d^3\sigma\, \Big( \Big[\sum_a\,
 \Big((\pi^a_{\perp})^2 - \Big(2\, \pi^a - \sum_i\, Q_i\, \eta_i\,
 {{c(\vec \sigma, {\vec \eta}_i(\tau))}\over {\partial\, \sigma^a}}
 \Big)\nonumber \\
 && \sum_j\, Q_j\, \eta_j\, {{c(\vec \sigma, {\vec \eta}_j(\tau))}
 \over {\partial\, \sigma^a}} \Big) + {1\over 2}\, \sum_{ab}\, F^2_{ab}
 \Big]\Big)(\tau ,\vec \sigma),
 \end{eqnarray*}

 \bea
 {1\over c}\, {\hat E}_{ADM (2)} &=& \int d^3\sigma\, \Big({\cal
 M}^{(UV)}_{(2)} +   \Big(\sum_a\, \partial_a\,
 {\cal M}^{(UV)}_{(1)a}\Big)\, {1\over {\triangle}}\, {}^3K_{(1)}
  +\nonumber \\
 &+& {{8\pi\, G}\over {c^3}}\, \Big[{1\over 4}\, {\cal M}^{(UV)}_{(1)}\,
 {1\over {\triangle}}\, {\cal M}^{(UV)}_{(1)} - \sum_a\,
 {\cal M}^{(UV)}_{(1)a}\, {1\over {\triangle}}\, {\cal M}^{(UV)}_{(1)a}
 - {1\over 4}\, \Big(\sum_a\, {{\partial_a}\over {\triangle}}\,
 {\cal M}^{(UV)}_{(1)a}\Big)^2\Big] +\nonumber \\
 &+& {{c^3}\over {16\pi G}}\, \sum_{\bar a\bar b}\,
  \Big[\partial_{\tau}\, R_{\bar a}\, M_{\bar a\bar b}\,
 \partial_{\tau}\, R_{\bar b} + \sum_a\, \partial_a\,
 R_{\bar a}\, M_{\bar a\bar b}\, \partial_a\, R_{\bar b}\Big] \,
 \Big)(\tau, \vec \sigma).
 \label{4.21}
 \eea

\noindent Eqs. (\ref{4.6}), (\ref{4.16}) and (\ref{4.17}) for
$\phi_{(1)}$, ${\bar n}_{(1)(a)}$ and $\sigma_{(1)(a)(b)}{|}_{a
\not= b}$ have been used to get ${\hat E}_{ADM(2)}$. Since, as we
will see in Section VII, the solution of the Hamilton equation for
the tidal variables $R_{\bar a}$ is proportional to ${G\over
{c^3}}$, we see that all the terms not in the first line of ${\hat
E}_{(2)ADM}$ are of order ${G\over {c^3}}$.

\medskip

Eq.(\ref{3.13}) implies $\partial_{\tau}\, M_{(1)}c = 0 + {{Mc}\over
L}\, O(\zeta^2)$.

\medskip

In ${\hat E}_{ADM(2)}$ we can see a kinetic term for the tidal
variables $R_{\bar a}$, with the operator $M_{\bar a\bar b}$
appearing in Eq.(\ref{4.18}) (the connected operator ${\tilde
M}_{ab}$ appearing in the boosts (\ref{4.24}) is defined in
Eq.(\ref{6.4}) of Section VI). Moreover, there is a potential for
the tidal variables and bilinear terms in the matter. Finally, there
is a term depending on the inertial gauge variable (non-local York
time) ${}^3{\cal K} = {1\over {\triangle}}\, {}^3K_{(1)}$ coupled to
the divergence of the matter current density. However at this order
the {\it negative definite quadratic term in the local York time
${}^3K_{(1)} = \triangle\, {}^3{\cal K}_{(1)}$ appearing} in
Eq.(\ref{2.14}) disappears due to the elimination of the tidal
momenta with Eq.(\ref{4.18}), the use of the solution of the
constraints and of some of the Hamilton equations plus integrations
by parts. As we will see in Eqs. (\ref{4.23}) and (\ref{4.24}) in
the six Lorentz generators  there is a dependence both on the local
and non-local York time at the second order.

\bigskip

For the other weak Poincare' generators the weak field approximation
of Eqs.(\ref{2.20}), by using Eqs.(\ref{4.18}) and by making
integrations by parts, gives (all the terms in $p^r_{(2)}$,
$j^{rs}_{(2)}$ and all the terms except the first in $j^{\tau
r}_{(2)}$ are of order ${G\over {c^3}}$ on the solution for the
tidal variables given in Section VII; the last term in the boost
generators is a surface term which can be dropped with our boundary
conditions)
\medskip

\begin{eqnarray*}
 {\hat P}^{r}_{ADM}\, &=&
  p^r_{(1)} + p^r_{(2)} + Mc\, O(\zeta^3) \approx 0,\nonumber \\
 &&{}\nonumber \\
 &&{}\nonumber \\
  p^r_{(1)} &=& \int d^3\sigma\, {\cal
 M}^{(UV)}_{(1)r}(\tau, \vec \sigma) = \sum_i\, \eta_i\,
 \Big(\kappa_{ir}(\tau ) - {{Q_i}\over c}\,
 A_{\perp r}(\tau ,{\vec \eta}_i(\tau))\Big) -\nonumber \\
 &-& {1\over c}\, \int d^3\sigma\, \sum_s\, F_{rs}(\tau ,\vec \sigma)\, \Big(
  \pi_\perp^s(\tau,  \vec{\sigma}) - \sum_n\, \delta^{sn}\, \sum_i\, Q_i\, \eta_i\,
 {{\partial\, c(\vec \sigma, {\vec \eta}_i(\tau))}\over {\partial\,
 \sigma^n}}\Big),
 \end{eqnarray*}

\bea
 p^r_{(2)} &=& - \int d^3\sigma\, \Big( {{c^3}\over {8\pi\, G}}\,
 \sum_{\bar a\bar b}\, \partial_r\, R_{\bar a}\, M_{\bar a\bar b}\,
 \partial_{\tau}\, R_{\bar b} +\nonumber \\
 &+& \sum_a\, {\cal M}_{(1)a}^{(UV)}\, {{\partial_r\,
 \partial_a}\over {\triangle}}\, \Big(2\, \Gamma_a^{(1)} -
 {1\over 2}\, \sum_c\, {{\partial_c^2}\over {\triangle}}\,
 \Gamma_c^{(1)}\Big) -\nonumber \\
 &-& {\cal M}_{(1)}^{(UV)}\,
 {{\partial_r}\over {\triangle}}\, {}^3K_{(1)}
 \Big)(\tau, \vec \sigma),
 \label{4.22}
 \eea

\bigskip

\begin{eqnarray*}
 {\hat J}^{rs}_{ADM} &=&
 j^{rs}_{(1)} + j^{rs}_{(2)} + Mc\, O(\zeta^3),\nonumber \\
  &&{}\nonumber \\
  &&{}\nonumber \\
  &&j^{rs}_{(1)} =  \int d^3\sigma\,
  \Big(\sigma^r\, {\cal M}^{(UV)}_{(1)s} - \sigma^s\,
  {\cal M}^{(UV)}_{(1)r}\Big)(\tau, \vec \sigma)
  =\nonumber \\
  &=&  \sum_i\, \eta_i\, \Big[\eta^r_i(\tau)\,\Big(\kappa_{is}(\tau )
  - {{Q_i}\over c}\, A_{\perp s}(\tau ,\vec \sigma)\Big) -
  \eta^s_i(\tau)\,  \Big(\kappa_{ir}(\tau ) - {{Q_i}\over c}\,
 A_{\perp r}(\tau ,\vec \sigma)\Big)  \Big] -\nonumber \\
 &-& {1\over c}\, \int d^3\sigma\, \Big[\sigma^r\,
 \sum_u\, F_{su}(\tau ,\vec \sigma)\, \Big(
  \pi_\perp^u(\tau,  \vec{\sigma}) - \sum_n\, \delta^{un}\, \sum_i\, Q_i\, \eta_i\,
 {{\partial\, c(\vec \sigma, {\vec \eta}_i(\tau))}\over {\partial\,
 \sigma^n}}\Big)  -\nonumber \\
 &-& \sigma^s\, \sum_u\, F_{ru}(\tau ,\vec \sigma)\, \Big(
  \pi_\perp^u(\tau,  \vec{\sigma}) - \sum_n\, \delta^{un}\, \sum_i\, Q_i\, \eta_i\,
 {{\partial\, c(\vec \sigma, {\vec \eta}_i(\tau))}\over {\partial\,
 \sigma^n}}\Big) \Big],
 \end{eqnarray*}

  \bea
 j^{rs}_{(2)} &=&   \int d^3\sigma\, \Big( {\cal
 M}_{(1)}^{(UV)}\, (\sigma^r\, \partial_s - \sigma^s\, \partial_r)\,
 {1\over {\triangle}}\, {}^3K_{(1)} -\nonumber \\
 &-&2\, \sum_u\, {\cal M}_{(1)u}^{(UV)}\, (\sigma^r\, \partial_s - \sigma^s\, \partial_r)\,
 {{\partial_u}\over {\triangle}}\, \Gamma_u^{(1)} - {1\over 2}\,
 \sum_c\, ({\cal M}_{(1)r}^{(UV)}\, {{\partial_s}\over {\triangle}} - {\cal M}_{(1)s}^{(UV)}\,
 {{\partial_r}\over {\triangle}})\, {{\partial_c^2}\over {\triangle}}\, \Gamma_c^{(1)}
 +\nonumber \\
 &+& {1\over 2}\, \sum_{uv}\, {\cal M}_{(1)u}^{(UV)}\, (\sigma^r\,
 \partial_s - \sigma^s\, \partial_r)\, {{\partial_u}\over
 {\triangle}}\, {{\partial_v^2}\over {\triangle}}\, \Gamma_v^{(1)} +
 2\, \Big({\cal M}_{(1)r}^{(UV)}\, {{\partial_s}\over {\triangle}}\,
 \Gamma_s^{(1)} - {\cal M}_{(1)s}^{(UV)}\, {{\partial_r}\over
 {\triangle}}\, \Gamma_r^{(1)}\Big) -\nonumber \\
 &-& {{c^3}\over {8\pi\, G}}\, \Big[\sum_{\bar a\bar b}\, (\sigma^r\, \partial_s
 - \sigma^s\, \partial_r)\, R_{\bar a}\, M_{\bar a\bar b}\,
 \partial_{\tau}\, R_{\bar b} + 2\, {}^3K_{(1)}\, \partial_r\,
 \partial_s\, (\Gamma^{(1)}_s - \Gamma_r^{(1)})  +\nonumber \\
 &+& 2\, \partial_{\tau}\, (\Gamma_r^{(1)} +  \Gamma_s^{(1)} -
 {1\over 2}\, \sum_c\, {{\partial_c^2}\over {\triangle}}\, \Gamma_c^{(1)})\,
 {{\partial_r\, \partial_s}\over {\triangle}}\, (\Gamma_s^{(1)} - \Gamma_r^{(1)})
 \Big]\Big)(\tau, \vec \sigma),\nonumber \\
 &&{}
 \label{4.23}
 \eea

\bigskip

\begin{eqnarray*}
 {\hat J}^{\tau r}_{ADM}\, &=& - {\hat J}^{r\tau}_{ADM} =
 j^{\tau r}_{(1)} + j^{\tau r}_{(2)}   + Mc\, O(\zeta^3) \approx 0,
 \end{eqnarray*}

 \begin{eqnarray*}
  &&j^{\tau r}_{(1)} = - j^{r\tau}_{(1)} = - \int d^3\sigma\,
  \sigma^r\, {\cal M}^{(UV)}_{(1)}(\tau, \vec \sigma) =
  - \sum_i\, \eta_i\, \eta^r_i\,  \sqrt{m^2_i\, c^2 +
 \Big({\vec \kappa}_i(\tau ) - {{Q_i}\over c}\,
 {\vec A}_{\perp}(\tau, {\vec \eta}_i(\tau))\Big)^2} +\nonumber \\
 &+& {1\over {2c}}\, \int d^3\sigma\, \sigma^r\, \Big( \Big[\sum_a\,
 \Big((\pi^a_{\perp})^2 - \Big(2\, \pi^a - \sum_i\, Q_i\, \eta_i\,
 {{c(\vec \sigma, {\vec \eta}_i(\tau))}\over {\partial\, \sigma^a}}
 \Big)\nonumber \\
 && \sum_j\, Q_j\, \eta_j\, {{c(\vec \sigma, {\vec \eta}_j(\tau))}
 \over {\partial\, \sigma^a}} \Big) + {1\over 2}\, \sum_{ab}\, F^2_{ab}
 \Big]\Big)(\tau ,\vec \sigma),
 \end{eqnarray*}

\bea
 j^{\tau r}_{(2)}&=& - \int d^3\sigma\,\sigma^r\,{\cal M}^{(UV)}_{(2)}
 (\tau ,\vec \sigma)+\\
 &-&\int d^3\sigma\,\sigma^r\,\Big\{ \frac{c^3}{16\pi
 G}\sum_{a,b}\,(\partial_a\Gamma_b^{(1)})^2 -\frac{c^3}{8\pi
 G}\sum_a\,\partial_a\Gamma_a^{(1)}\,\partial_a\Big(
 \Gamma_a^{(1)}-\frac{1}{2}\,\sum_c\,\frac{\partial_c^2\Gamma_c^{(1)}}{\Delta}\Big)-\\
 &-&\frac{c^3}{32\pi
 G}\sum_a\partial_a\Big(\sum_c\,\frac{\partial_c^2\Gamma_c^{(1)}}{\Delta}\Big)
 \partial_a\Big(\sum_d\,\frac{\partial_d^2\Gamma_d^{(1)}}{\Delta}\Big)+\\
 &+&\frac{1}{2}\sum_a\,\frac{\partial_a{\cal
 M}_{(1)}^{UV}}{\Delta}\,\partial_a\Big(\Gamma_a^{(1)}-
 \frac{1}{2}\,\sum_c\,\frac{\partial_c^2\Gamma_c^{(1)}}{\Delta}\Big)
 +\frac{32\pi G}{c^3}\sum_a\,\Big(\frac{\partial_a{\cal M}_{(1)}^{UV}}{\Delta}\Big)^2+\\
 &+&\frac{c^3}{16\pi
 G}\sum_{a,b}\,\Big(\widetilde{M}_{ab}\,\partial_\tau\Gamma^{(1)}_b\Big)^2
 +\frac{c^3}{16\pi G}\sum_{a\neq b}\,\Big[
 \frac{\partial_a\partial_b\partial_\tau}{\Delta}\Big(
 \Gamma^{(1)}_a+\Gamma^{(1)}_b-\frac{1}{2}\,\sum_c\,\frac{\partial_c^2\Gamma_c^{(1)}}{\Delta}
 \Big)\Big]^2-\\
 &-&2\sum_{a,b}\,\Big(\widetilde{M}_{ab}\,\partial_\tau\Gamma^{(1)}_b\Big)\frac{\partial_a{\cal
 M}_{(1)a}^{UV}}{\Delta}+ 2\sum_{a\neq b}\,
 \frac{\partial_a\partial_b\partial_\tau}{\Delta}\Big(
 \Gamma^{(1)}_a+\Gamma^{(1)}_b-\frac{1}{2}\,\sum_c\,\frac{\partial_c^2\Gamma_c^{(1)}}{\Delta}
 \Big)
 \frac{\partial_a{\cal M}_{(1)a}^{UV}}{\Delta}-\\
 &-&\frac{c^3}{8\pi
 G}\sum_{a,b}\,\Big(\widetilde{M}_{ab}\,\partial_\tau\Gamma^{(1)}_b\Big)
 \frac{\partial_a^2}{\Delta}\Big( {}^3K_{(1)}-\frac{4\pi
 G}{c^3}\sum_c\frac{\partial_c{\cal M}_{(1)c}^{UV}}{\Delta}\Big)
 +\\
 &+&\frac{c^3}{8\pi G}\sum_{a\neq b}\,
 \frac{\partial_a\partial_b\partial_\tau}{\Delta}\Big(
 \Gamma^{(1)}_a+\Gamma^{(1)}_b-\frac{1}{2}\,\sum_c\,\frac{\partial_c^2\Gamma_c^{(1)}}{\Delta}
 \Big) \frac{\partial_a\partial_b}{\Delta}\Big(
 {}^3K_{(1)}-\frac{4\pi G}{c^3}\sum_c\frac{\partial_c{\cal M}_{(1)c}^{UV}}{\Delta}\Big)+\\
 &+&\frac{c^3}{16\pi
 G}\sum_{a,b}\,\Big[\frac{\partial_a\partial_b}{\Delta}\Big(
 {}^3K_{(1)}-\frac{4\pi G}{c^3}\sum_c\frac{\partial_c{\cal M}_{(1)c}^{UV}}{\Delta}\Big)\Big]^2+\\
 &+&2\sum_{a,b}\,\frac{\partial_a{\cal
 M}_{(1)b}^{UV}}{\Delta}\frac{\partial_a\partial_b}{\Delta}\Big(
 {}^3K_{(1)}-\frac{4\pi G}{c^3}\sum_c\frac{\partial_c{\cal M}_{(1)c}^{UV}}{\Delta}\Big)-\\
 &-&\frac{c^3}{48\pi G}\Big( {}^3K_{(1)}+\frac{12\pi
 G}{c^3}\sum_c\frac{\partial_c{\cal M}_{(1)c}^{UV}}{\Delta}\Big)^2
 -\frac{c^3}{24\pi G}\Big({}^3K_{(1)}\Big)^2+\\
 &+&\frac{8\pi G}{c^3}\sum_{a,b}\,\Big[
 \Big(\frac{\partial_a{\cal M}_{(1)b}^{UV}}{\Delta}\Big)^2+
 \frac{\partial_a{\cal M}_{(1)b}^{UV}}{\Delta}\frac{\partial_b\,
 {\cal M}_{(1)a}^{UV}}{\Delta}\Big]\,\,\Big\}(\tau ,\vec \sigma) +\\
 &+&\int d^3\sigma\,\Big[ -\frac{3}{2}\,\frac{{\cal
 M}_{(1)}^{UV}}{\Delta}\,\partial_r\Gamma^{(1)}_r+\frac{3c^3}{16\pi
 G}\partial_r\Gamma^{(1)}_r\Big(\sum_c\,\frac{\partial_c^2\Gamma_c^{(1)}}{\Delta}
 \Big)\Big](\tau ,\vec \sigma) +\\
 &-&\int d^3\sigma\,\partial_r\,\Big\{ \frac{c^3}{16\pi
 G}\Big[2\Big(\Gamma^{(1)}_r\Big)^2-\sum_s\,\Big(\Gamma^{(1)}_s\Big)^2-
 \frac{1}{2}\Big(\sum_c\,\frac{\partial_c^2\Gamma_c^{(1)}}{\Delta}
 \Big)^2\Big]-\frac{2\pi G}{c^3}\Big(\frac{{\cal
 M}_{(1)}}{\Delta}\Big)^2\,\,\Big\}(\tau ,\vec \sigma). \nonumber \\
 &&{}
 \label{4.24}
 \eea

\bigskip

We see that at the lowest order we get an unfaithful realization of
the ten {\it internal} Poincare' generators $M_{(1)}\, c$,
$p^r_{(1)} \approx 0 + O(\zeta^2)$, $j^{rs}_{(1)}$, $j^{\tau
r}_{(1)} \approx 0 + O(\zeta^2)$ for the same matter in a rest-frame
instant form approximately valid [modulo corrections at
$O(\zeta^2)$] in an abstract Minkowski space-time with the Wigner
instantaneous 3-spaces coinciding with the asymptotic inertial
3-spaces at spatial infinity of our space-times. They are the {\it
internal} Poincare' generators of the system positive -energy
charged scalar particles plus transverse electro-magnetic field in
the radiation gauge studied in Refs.\cite{11,12}.\medskip

\medskip
Like in Minkowski space-time the internal 3-center of mass $\vec
\eta(\tau)$ of the isolated system {\it 3-universe} (i.e.
"gravitational field plus particles plus electro-magnetic field" on
the instantaneous 3-space $\Sigma_{\tau}$) and its conjugate
momentum are eliminated by the constraints $p^r_{(1)} \approx 0 +
O(\zeta^2)$ and $j^{\tau r}_{(1)} {\buildrel {def}\over =}\, -
M_{(1)}\, c\, \eta^r(\tau) \approx 0 + O(\zeta^2)$. They imply $
\eta^r(\tau) = - {1\over {M_{(1)}\, c}}\, \int d^3\sigma\,
\sigma^r\, {\cal M}^{(UV)}_{(1)}(\tau, \vec \sigma) \approx 0$, i.e.
the 3-center of mass of the 3-universe is in the origin of the
3-coordinates: the world-line  of the time-like observer is the
Fokker-Pryce 4-center of inertia $Y^{\mu}(\tau) = z^{\mu}(\tau, \vec
0)$ \cite{4,11,12}, to be interpreted as a version in the bulk of an
asymptotic inertial observer.

\bigskip

With the methods used in Refs.\cite{4,11,12} it could be shown that
in our asymptotically Minkowskian space-times there is a decoupled
(Newton-Wigner) canonical 4-center of mass ${\tilde
x}^{\mu}_{com}(\tau) = z^{\mu}(\tau, {\tilde {\vec \eta}}(\tau))$ of
the instantaneous 3-universes $\Sigma_{\tau}$, carrying a
pole-dipole structure (mass ${1\over c}\, {\hat E}_{ADM}$ and rest
spin ${\hat J}^{rs}_{ADM}$) and non-covariant with respect to an
{\it external} asymptotic ADM Poincare' group with the same
generators as in special relativity \cite{4,11,12}.

\subsection{The Effective Hamiltonian in the 3-Orthogonal Gauge with
a Given ${}^3K_{(1)}(\tau, \vec \sigma)$}

The restriction of the Hamilton equations of paper I to our family
of 3-orthogonal gauges, given in Section II, can also be generated
by using an effective Hamiltonian determined by the gauge-fixing
procedure.\medskip

Since our gauge fixings are $\theta^i(\tau, \vec \sigma) \approx 0$
and $\pi_{\tilde \phi}(\tau, \vec \sigma) - {{c^3}\over {12\pi\,
G}}\, F(\tau, \vec \sigma) \approx 0$ with ${}^3K(\tau, \vec \sigma)
\approx F(\tau, \vec \sigma)$ an arbitrary numerical function, we
must take into account the explicit $\tau$-dependence of this
function which implies that the effective Hamiltonian is no more
${1\over c}\, {\hat E}_{ADM}$ with the weak ADM energy given in
Eq.(\ref{4.21}).

\medskip

To find the effective Hamiltonian we should perform the following
time-dependent canonical transformation: $\pi_{\tilde \phi}(\tau,
\vec \sigma)\, \mapsto\, \pi^{'}_{\tilde \phi}(\tau, \vec \sigma) =
\pi_{\tilde \phi}(\tau, \vec \sigma) - {{c^3}\over {12\pi\, G}}\,
F(\tau, \vec \sigma)$, ${\tilde \phi}^{'}(\tau, \vec \sigma)\, =
\tilde \phi(\tau, \vec \sigma)$, with all the other canonical
variables fixed. Due to the explicit time-dependence the new
effective Dirac Hamiltonian is $H_D^{'} = H_D + {{c^3}\over {12\pi\,
G}}\, \int d^3\sigma\, {{\partial\, F(\tau, \vec \sigma)}\over
{\partial\, \tau}} \tilde \phi(\tau, \vec \sigma)$. In the new
canonical basis the gauge fixings are $\theta^i(\tau, \vec \sigma)
\approx 0$, $\pi^{'}_{\tilde \phi}(\tau, \vec \sigma) \approx 0$ and
have no explicit $\tau$-dependence.\medskip

However to go to the reduced phase space we have to find the Dirac
brackets and this would require the explicit solution of the
super-Hamiltonian and super-momentum constraints. This can be done
in the linearized theory.
\bigskip

From Eq.(\ref{4.18}) we get (see Section VI for the inverse of the
operator $M_{\bar a\bar b}$)

\beq
 \partial_{\tau}\, R_{\bar a} = \sum_{\bar b}\, M^{-1}_{\bar a\bar
 b}\, \Big({{8\pi\, G}\over {c^3}}\, \Pi_{\bar b} + \sum_a\, \gamma_{\bar ba}\,
 \Big[{{4\pi\, G}\over {c^3}}\,
 {1\over {\triangle}}\, (4\, \partial_a\,
 {\cal M}^{(UV)}_{(1)a} - {{\partial_a^2}\over {\triangle}}\, \sum_c\,
 \partial_c\, {\cal M}^{(UV)}_{(1)c}) + {{\partial_a^2}\over {\triangle}}\,
 {}^3K_{(1)} \Big] \Big),
 \label{4.25}
 \eeq

\noindent and this expression can be put in ${\hat E}_{ADM(2)}$ of
Eq. (\ref{4.21}). As a consequence, the effective Hamiltonian of the
linearized theory in the 3-orthogonal gauges is (use Eq.(\ref{4.6})
for $\tilde \phi = 1 + 6\, \phi_{(1)} + O(\zeta^2)$)

\bea
 H_{eff} &=& M_{(1)}\, c + {1\over c}\, {\hat E}_{ADM(2)} +
 \nonumber \\
 &+&{{c^3}\over {12\pi\, G}}\, \int d^3\sigma\, \partial_{\tau}\,
 {}^3K_{(1)}(\tau, \vec \sigma)\, \Big[1 + {6\over {\triangle}}\,
 \Big(- {{2\pi\, G}\over {c^3}}\, {\cal M}_{(1)}^{(UV)} + {1\over 4}\,
 \sum_b\, \partial_b^2\, \Gamma_b^{(1)}\Big)(\tau, \vec \sigma)\Big]
 =\nonumber \\
 &=&\int d^3\sigma\, \Big({\cal M}_{(1)}^{(UV)} + {\cal
 M}^{(UV)}_{(2)} + \nonumber \\
 &+& {{8\pi\, G}\over {c^3}}\, \Big[{1\over 4}\, {\cal M}^{(UV)}_{(1)}\,
 {1\over {\triangle}}\, {\cal M}^{(UV)}_{(1)} - \sum_a\,
 {\cal M}^{(UV)}_{(1)a}\, {1\over {\triangle}}\, {\cal M}^{(UV)}_{(1)a}
 - {1\over 4}\, \Big(\sum_a\, {{\partial_a}\over {\triangle}}\,
 {\cal M}^{(UV)}_{(1)a}\Big)^2\Big] +\nonumber \\
 &+&  \Big(\sum_a\, {{\partial_a}\over {\triangle}}\,
 {\cal M}^{(UV)}_{(1)a}\Big)\, {}^3K_{(1)}
 + {{c^3}\over {16\pi G}}\, \sum_{\bar a\bar b}\,
  \sum_a\, \partial_a\, R_{\bar a}\, M_{\bar a\bar b}\,
  \partial_a\, R_{\bar b}  +\nonumber \\
 &+& {{c^3}\over {16\pi G}}\, \sum_{\bar a\bar b}\,
 \Big({{8\pi\, G}\over {c^3}}\, \Pi_{\bar a} + \sum_a\, \gamma_{\bar
 aa}\, \Big[{{4\pi\, G}\over {c^3}}\,
 {1\over {\triangle}}\, (4\, \partial_a\,
 {\cal M}^{(UV)}_{(1)a} - {{\partial_a^2}\over {\triangle}}\, \sum_c\,
 \partial_c\, {\cal M}^{(UV)}_{(1)c}) + {{\partial_a^2}\over {\triangle}}\,
 {}^3K_{(1)} \Big] \Big)\nonumber \\
 && M^{-1}_{\bar a\bar b}\,
 \Big({{8\pi\, G}\over {c^3}}\, \Pi_{\bar b} + \sum_b\, \gamma_{\bar bb}\,
 \Big[{{4\pi\, G}\over {c^3}}\,
 {1\over {\triangle}}\, (4\, \partial_b\,
 {\cal M}^{(UV)}_{(1)b} - {{\partial_b^2}\over {\triangle}}\, \sum_d\,
 \partial_d\, {\cal M}^{(UV)}_{(1)d}) + {{\partial_b^2}\over {\triangle}}\,
 {}^3K_{(1)} \Big] \Big) +\nonumber \\
 &+&{{c^3}\over {12\pi\, G}}\,  \partial_{\tau}\,
 {}^3K_{(1)}\, \Big[1 + {6\over {\triangle}}\,
 \Big(- {{2\pi\, G}\over {c^3}}\, {\cal M}_{(1)}^{(UV)} + {1\over 4}\,
 \sum_b\, \partial_b^2\, \Gamma_b^{(1)}\Big)(\tau, \vec
 \sigma)\Big]\Big)(\tau, \vec \sigma).\nonumber \\
 &&{}
 \label{4.26}
 \eea

As expected we get ${d\over {d\tau}}\, \Big(M_{(1)}\, c^2 + {\hat
E}_{ADM(2)}\Big) = {{\partial}\over {\partial \tau}}\,
\Big(M_{(1)}\, c^2 + {\hat E}_{ADM(2)}\Big) + \{ M_{(1)}\, c^2 +
{\hat E}_{ADM(2)}, H_{eff} \} = 0$ (${{\partial}\over {\partial
\tau}}$ acts only on ${}^3K_{(1)}(\tau, \vec \sigma)$). One could
check that the Hamilton equations in the 3-orthogonal gauges are
generated by this effective Hamiltonian

\medskip

In conclusion, when ${}^3K_{(1)}(\tau, \vec \sigma) \approx
F_{(1)}(\tau, \vec \sigma)$ is $\tau$-dependent the effective
Hamiltonian is not the energy and there are additional inertial
effects like it happens in non-inertial frames in Minkowski
space-time \cite{4}.

\vfill\eject

\section{The Linearized Equations for the Particles and the
Electro-Magnetic Field}

\subsection{The Equations of Motion of the Particles}

Let us now study the weak field approximation of the Hamilton
equations (\ref{2.18}) for the particles. Since in this
approximation we have
\medskip

\bea
 &&{1\over {\sqrt{(1 + n)^2 - \phi^4\, \sum_c\, Q_c^2\, ({\dot
 \eta}^c_i(\tau) + \phi^{-2}\, Q_c^{-1}\, {\bar n}_{(c)})^2}}}(\tau,
 {\vec \eta}_i(\tau))\, =\nonumber \\
 &&{}\nonumber \\
 &&=\,  {1\over {\sqrt{1 - {\dot {\vec \eta}}^2_i(\tau) }}}\,
 \Big[1 - {{n_{(1)} - \sum_c\,
 {\dot \eta}_i^c(\tau)\, [{\bar n}_{(1)(c)} + (\Gamma_c^{(1)} + 2\,
 \phi_{(1)})\, {\dot \eta}^c_i(\tau)]}\over {1 - {\dot {\vec \eta}}^2_i(\tau)
 }}\Big](\tau ,{\vec \eta}_i(\tau)) + O(\zeta^2),\nonumber \\
 &&{}
 \label{5.1}
 \eea
\bigskip

\noindent we get that the particle momenta (\ref{2.17}) have the
following expression in terms of the particle velocities\medskip

\begin{eqnarray*}
 {{\kappa_{ir}(\tau)}\over {m_i\, c}}\, &\cir& {{Q_i}\over {m_i\,c^2}}\, A_{\perp\,
 r}(\tau, {\vec \eta}_i(\tau)) + {1\over {\sqrt{1 -
 {\dot {\vec \eta}}^2_i(\tau)   }}}\, \Big[{\dot
 \eta}^r_i(\tau)\, \Big(1 + 2\, (\Gamma_r^{(1)} + 2\, \phi_{(1)}) -
 \nonumber \\
 &-& {{n_{(1)} - \sum_c\, {\dot \eta}_i^c(\tau)\, [{\bar n}_{(1)(c)} +
 (\Gamma_c^{(1)} + 2\, \phi_{(1)})\, {\dot \eta}^c_i(\tau)]}\over {1
 - {\dot {\vec \eta}}^2_i(\tau)}} \Big) + {\bar
 n}_{(1)(r)}\Big](\tau ,{\vec \eta}_i(\tau)) =\nonumber \\
 &=& {{ {\dot \eta}^r_i(\tau)}\over {\sqrt{1 -
 {\dot {\vec \eta}}^2_i(\tau)   }}} + {M\over {m_i}}\, O(\zeta) +
 {M\over {m_i}}\, O(\zeta^2),\nonumber \\
 &&{}\nonumber \\
 &&or
 \end{eqnarray*}

\begin{eqnarray*}
 {{\kappa_{ir}(\tau)}\over {m_i\, c}}\, &\cir&
 {{Q_i}\over {m_i\,c^2}}\, A_{\perp\,
 r}(\tau, {\vec \eta}_i(\tau)) +\nonumber \\
 &+& {1\over {\sqrt{1 -
 {\dot {\vec \eta}}^2_i(\tau)}}}\, \Big[{\dot
 \eta}^r_i(\tau)\, \Big(1 - {{8\pi\, G}\over {c^3}}\, {1\over
 {\triangle}}\, {\cal M}^{(UV)}_{(1)} +2\, \Gamma^{(1)}_r +
  \sum_b\, {{\partial_b^2}\over {\triangle}}\, \Gamma_b^{(1)}
 -\nonumber \\
 &-& {1\over {1 - {\dot {\vec \eta}}^2_i(\tau)}}\, \Big[
 {{4\pi\, G}\over {c^3}}\, {1\over {\triangle}}\, ({\cal M}_{(1)}^{(UV)}
 + \sum_b\, T_{(1)}^{bb})  -\nonumber \\
 &-& \sum_c\, {\dot \eta}^c_i(\tau)\, \Big({{4\pi\, G}\over {c^3}}\,
 {1\over {\triangle}}\, (4\, {\cal M}^{(UV)}_{(1)c} - {{\partial_c}\over
 {\triangle}}\, \sum_b\, \partial_b\, {\cal M}^{(UV)}_{(1)b}) +
 \nonumber \\
 &+& {1\over 2}\, \partial_{\tau}\, {{\partial_c}\over {\triangle}}\,
 (4\, \Gamma_c^{(1)} - \sum_b\, {{\partial_b^2}\over {\triangle}}\,
 \Gamma_b^{(1)})\Big) +
 \end{eqnarray*}

\bea
 &+& \sum_c ({\dot \eta}^c_i(\tau))^2\, \Big(- {{4\pi\, G}\over
 {c^3}}\, {1\over {\triangle}}\, {\cal M}_{(1)}^{(UV)} +
 \Gamma_c^{(1)} + {1\over 2}\, \sum_b\, {{\partial_b^2}\over {\triangle}}\,
 \Gamma_b^{(1)} \Big)\, \Big] \Big) +\nonumber \\
 &+& {{4\pi\, G}\over {c^3}}\, {1\over {\triangle}}\, (4\,
 {\cal M}^{(UV)}_{(1)r} - {{\partial_r}\over {\triangle}}\,
 \sum_b\, \partial_b\, {\cal M}^{(UV)}_{(1)b}) +
 {1\over 2}\, \partial_{\tau}\, {{\partial_r}\over
 {\triangle}}\, (4\, \Gamma_r^{(1)} -
 \sum_b\, {{\partial_b^2}\over {\triangle}}\, \Gamma_b^{(1)}) +
 \nonumber \\
 &+& {1\over {\triangle}}\, \Big( \partial_r\, {}^3K_{(1)}
 + {{ {\dot \eta}^r_i(\tau)}\over {1 - {\dot {\vec \eta}}^2_i(\tau)}}
 \, \partial_{\tau}\, {}^3K_{(1)} \Big)\Big](\tau ,{\vec \eta}_i(\tau)),\nonumber \\
 &&{}\nonumber \\
 \label{5.2}
 \eea

\noindent where Eqs. (\ref{4.6}), (\ref{4.7}) and (\ref{4.16}) were
used to find the final expression. This result is implied by
Eqs.(\ref{3.11}): the ultraviolet cutoff implies a small deviation
from the expression in the free case.
\medskip

In the last expression the mass density ${\cal M}_{(1)}^{(UV)}$, the
mass current density ${\cal M}_{(1)r}^{(UV)}$ and the tidal
variables $\Gamma_a^{(1)}$, which will be shown to depend on the
stress tensor $T_{(1)}^{rs}$ in Sections VI and VII, have to be
evaluated by using the lowest order expression $\kappa_{ir}(\tau) -
{{Q_i}\over c}\, A_{\perp r}(\tau, {\vec \eta}_i(\tau))\, \mapsto
{{m_i\, c\, {\dot \eta}^r_i(\tau)}\over {\sqrt{1 - {\dot {\vec
\eta}}_i^2(\tau)}}}$ (so that $\sqrt{m_i^2c^2 + \Big({\vec
\kappa}_i(\tau) - {{Q_i}\over c}\, {\vec A}_{\perp}(\tau, {\vec
\eta}_i(\tau))\Big)^2} \mapsto {{m_i\, c}\over {\sqrt{1 - {\dot
{\vec \eta}}_i(\tau)}}}$).

\medskip

The particle momenta depend on the $\tau$- and spatial-derivatives
of the inertial gauge variable ${}^3{\cal K}_{(1)} = {1\over
{\triangle}}\, {}^3K_{(1)}$. If the York time ${}^3K_{(1)}$ would
depend also on the particle positions, we should make the following
replacement $\partial_{\tau}\, {}^3K_{(1)} \mapsto
\partial_{\tau}\, {}^3K_{(1)}{|}_{{\vec \eta}_i} + \sum_b\,
{\dot \eta}^b_i(\tau)\, \partial_b\, {}^3K_{(1)}$.

\bigskip

As a consequence, we can get the following second order form of the
equations of motion (\ref{2.17}) for the particles, implied by the
Hamilton equations (we use the solutions (\ref{4.6}), (\ref{4.7}),
(\ref{4.16}) for $\phi_{(1)}$, $n_{(1)}$, ${\bar n}_{(1)(r)}$),
\medskip

\begin{eqnarray*}
 &&\eta_i\,  {d\over {d\tau}}\, \Big[{{ {\dot
 \eta}^r_i(\tau)}\over {\sqrt{1 - {\dot {\vec \eta}}^2_i(\tau)}}}\,
 \Big(1 + 2\, (\Gamma_r^{(1)} + 2\,
 \phi_{(1)}) -\nonumber \\
 &&- {{n_{(1)} - \sum_c\, {\dot \eta}_i^c(\tau)\, [{\bar
 n}_{(1)(c)} + (\Gamma_c^{(1)} + 2\, \phi_{(1)})\, {\dot
 \eta}^c_i(\tau)]}\over {1 - {\dot {\vec \eta}}^2_i(\tau)}}
 \Big) +\nonumber \\
 &&+ {{{\bar n}_{(1)(r)}}\over {\sqrt{1 -
 {\dot {\vec \eta}}^2_i(\tau)}}}\,\Big](\tau ,{\vec
 \eta}_i(\tau))\, \cir
 \end{eqnarray*}

\begin{eqnarray*}
 &\cir& \eta_i\,  {d\over {d\tau}}\, \Big(
 {{{\dot \eta}^r_i(\tau)}\over {\sqrt{1 - {\dot {\vec
 \eta}}_i^2(\tau)}}}\, \Big[1 + 2\, \Gamma_r^{(1)} -
 {{8\pi\, G}\over {c^3}}\, {1\over {\triangle}}\, {\cal
 M}_{(1)}^{(UV)} + \sum_c\, {{\partial_c^2}\over {\triangle}}\,
 \Gamma_c^{(1)} -\nonumber \\
 &-& {1\over {1 - {\dot {\vec \eta}}_i^2(\tau)}}\, \Big(
 {{4\pi\, G}\over {c^3}}\, {1\over {\triangle}}\,
 ({\cal M}_{(1)}^{(UV)} + \sum_a\, T^{aa}_{(1)}) -
 {{\partial_{\tau}}\over {\triangle}}\, {}^3K_{(1)} -\nonumber \\
 &-&\sum_d\, {\dot \eta}^d_i(\tau)\, \Big[{{\partial_d}\over
 {\triangle}}\, {}^3K_{(1)} + {{4\pi\, G}\over {c^3}}\, {1\over
 {\triangle}}\, (4\, {\cal M}_{(1)d}^{(UV)} - {{\partial_d}\over {\triangle}}\,
 \sum_c\, \partial_c\, {\cal M}_{(1)c}^{(UV)}) +\nonumber \\
 &+&{1\over 2}\, {{\partial_d}\over {\triangle}}\, \partial_{\tau}\, (4\, \Gamma_d^{(1)}
 - \sum_c\, {{\partial_c^2}\over {\triangle}}\, \Gamma_c^{(1)})
 + {\dot \eta}^d_i(\tau)\, \Big(\Gamma_d^{(1)} - {{6\pi\, G}\over
 {c^3}}\, {1\over {\triangle}}\, {\cal M}_{(1)}^{(UV)} + {1\over
 2}\, \sum_c\, {{\partial_c^2}\over {\triangle}}\, \Gamma_c^{(1)}
 \Big)\Big]\Big)\Big] +\nonumber \\
 &+& {1\over {\sqrt{1 - {\dot {\vec \eta}}^2_i(\tau)}}}\,
 \Big[{{\partial_r}\over {\triangle}}\, {}^3K_{(1)} +
 {{4\pi\, G}\over {c^3}}\, {1\over {\triangle}}\, (4\,
 {\cal M}_{(1)r}^{(UV)} - {{\partial_r}\over {\triangle}}\,
 \sum_c\, \partial_c\, {\cal M}_{(1)c}^{(UV)}) +\nonumber \\
 &+& {1\over 2}\, {{\partial_r}\over {\triangle}}\, \partial_{\tau}\, (4\, \Gamma_r^{(1)}
 - \sum_c\, {{\partial_c^2}\over {\triangle}}\, \Gamma_c^{(1)})
  \Big](\tau, {\vec \eta}_i(\tau)) \cir\nonumber \\
 &&{}\nonumber \\
 &&\cir\, {{\eta_i\, Q_i}\over {m_i\, c^2}}\, F_{rs}(\tau ,{\vec
 \eta}_i(\tau))\, {\dot\eta}^s_i(\tau) -
 {{\eta_i}\over {m_i\, c}}\, {{\partial\, {\cal W}[\tau
 ,{\vec \eta}_i(\tau)]}\over {\partial\, \eta^r_i}} +
 {{\eta_i}\over {m_i\, c}}\, {\check F}_{ir}(\tau ,{\vec \eta}_i(\tau)),
 \end{eqnarray*}

\begin{eqnarray*}
 {\cal W}[\tau,{\vec \eta}_i(\tau)] &=& \int d^3\sigma\,
 \sum_a \Big[\Big(1 + n_{(1)} + 2\, (\Gamma_a^{(1)} -
 \phi_{(1)})\Big)\, {\cal W}_{(n)a} +\nonumber \\
 &+& \Big(1 - (\Gamma_a^{(1)} + 2\,
 \phi_{(1)})\Big)\, {\cal W}_a\Big](\tau, \vec \sigma) =\nonumber \\
 &=&\int d^3\sigma\, \sum_a \Big[\Big( 1+ {{4\pi\, G}\over {c^3}}\,
 {1\over {\triangle}}\, (2\, {\cal M}_{(1)}^{(UV)} + \sum_b\, T_{(1)}^{bb})
 + 2\, \Gamma_a^{(1)} -\nonumber \\
 &-& {1\over 2}\, \sum_b\,
 {{\partial_b^2}\over {\triangle}}\, \Gamma_b^{(1)} - {1\over {\triangle}}\,
 \partial_{\tau}\, {}^3K_{(1)}\Big) \, {\cal W}_{(n)a} +\nonumber \\
 &+& \Big(1 + {{4\pi\, G}\over {c^3}}\, {1\over {\triangle}}\, {\cal
 M}_{(1)}^{(UV)} - \Gamma_a^{(1)} - {1\over 2}\,
 \sum_b\, {{\partial_b^2}\over {\triangle}}\, \Gamma_b^{(1)}\Big)\, {\cal W}_a
 \Big](\tau, \vec \sigma),\nonumber \\
 &&{}\nonumber \\
 &&{}\nonumber \\
 &&{\cal W}_{(n)a}(\tau ,\vec \sigma) = - {1\over {2 c}}\,
 \Big(2\, \pi^a_{\perp}(\tau ,\vec \sigma) - \sum_{j\not=
 i,k}\, \eta_j\, Q_j\, \partial_a\, c(\vec \sigma, {\vec
 \eta}_j(\tau))\Big)\, \sum_{k\not= i}\, \eta_k\, Q_k\,
 \partial_a\, c(\vec \sigma, {\vec \eta}_k(\tau)),\nonumber \\
 &&{\cal W}_a(\tau ,\vec \sigma) = - {1\over c}\, \sum_s\,
 F_{as}(\tau ,\vec \sigma)\, \sum_{k\not= i}\, \eta_k\, Q_k\,
 \partial_s\, c(\vec \sigma, {\vec \eta}_k(\tau)),
 \end{eqnarray*}

\bea
 {\check F}_{ir}(\tau,{\vec \eta}_i(\tau)) &=& {{m_i\, c}\over {\sqrt{1 -
 {\dot {\vec \eta}}^2_i(\tau)}}}\, \Big[\sum_a\, {\dot
 \eta}^a_i(\tau)\, \Big({{\partial\, {\bar n}_{(1)(a)}}\over
 {\partial\, \eta^r_i}} +
  {{\partial\, (\Gamma_a^{(1)} + 2\, \phi_{(1)})}\over
 {\partial\, \eta_i^r}}\, {\dot
 \eta}^a_i(\tau)\Big) -\nonumber \\
 &-& {{\partial\, n_{(1)}}\over {\partial\,
 \eta^r_i}}\Big](\tau,{\vec \eta}_i(\tau)) =\nonumber \\
 &=&{{m_i\, c}\over {\sqrt{1 - {\dot {\vec \eta}}^2_i(\tau)
 }}}\,  {{\partial}\over {\partial
 \eta^r_i}}\, \Big[\sum_a\, {\dot \eta}^a_i(\tau)\, \Big({{4\pi\, G}\over
 {c^3}}\, {1\over {\triangle}}\, (4\, {\cal M}^{(UV)}_{(1)a} -
 {{\partial_a}\over {\triangle}}\, \sum_c\, \partial_c\, {\cal M}^{(UV)}_{(1)c})
 +\nonumber \\
 &+& {1\over 2}\, \partial_{\tau}\, {{\partial_a}\over
 {\triangle}}\, (4\, \Gamma_a^{(1)} -
 \sum_c\, {{\partial_c^2}\over {\triangle}}\, \Gamma_c^{(1)}) + {{\partial_a}\over
 {\triangle}}\, {}^3K_{(1)} +\nonumber \\
 &+& {\dot \eta}^a_i(\tau)\, \Big[- {{4\pi\, G}\over {c^3}}\,
 {1\over {\triangle}}\, {\cal M}^{(UV)}_{(1)} + \Gamma_a^{(1)} +
 {1\over 2}\, \sum_c\, {{\partial_c^2}\over {\triangle}}\, \Gamma_c^{(1)}
 \Big]\Big) -\nonumber \\
 &-&{{4\pi\, G}\over {c^3}}\, {1\over {\triangle}}\, ({\cal M}_{(1)}^{(UV)}
 + \sum_a\, T_{(1)}^{aa}) - {1\over {\triangle}}\, \partial_{\tau}\,
 {}^3K_{(1)}\Big](\tau,{\vec \eta}_i(\tau)),\nonumber \\
 &&{}
 \label{5.3}
 \eea

\noindent where ${\cal W}$ is the non-inertial Coulomb potential
term and ${\check F}_{ir}$ are generalized inertial forces (now
function of the inertial and tidal components of the gravitational
field). As a consequence, the deviation from free motions is at the
first order, consistently with the weak field approximation.\medskip

Let us remark that in absence of the electro-magnetic field the
final form of the equations of motion of particle $i$ does not
depend upon the mass $m_i$ like it happens for test particles
following geodesics. Therefore the masses $m_i$ are playing both the
role of inertial and gravitational mass of the particles: their
equality implies that the final form of the equations for particle
$i$ only depends on the masses $m_{j \not= i}$ present in the mass
density, in the mass current density and in the tidal variables.
\medskip

These equations of motion for dynamical (not test) scalar particles
with a definite sign of the energy (implied by the Grassmann charges
$\eta_i$ as shown in paper I) are not thought  as the point limit of
small extended objects as in Ref.\cite{16} and do not contain terms
corresponding to a gravitational self-force \cite{16,17} because
$\eta_i^2\, m_i^2 = 0$ (only terms $\eta_i\, \eta_j\, m_i\, m_j$
with $i \not= j$ appear). As shown in Section VII with this
description we can still get the energy balance when GW's are
emitted.

\bigskip

The Newtonian limit of the equations of motion for the particle in
absence of the electro-magnetic field will be studied in the third
paper \cite{b}.

\subsection{The Equations of Motion of the Transverse
Electro-Magnetic Field}

The weak field limit of Eqs.(\ref{2.19}) is\medskip

\begin{eqnarray*}
 \partial_{\tau}\, A_{\perp r}(\tau ,\vec \sigma) &\cir& \sum_{na}\,
 \delta_{rn}\, P^{na}_{\perp}(\vec \sigma)\, \Big([1 + n_{(1)} - 2\,
 (\Gamma_a^{(1)} - \phi_{(1)})]\, \pi^a_{\perp} +\nonumber \\
 &+& \sum_b\, {\bar n}_{(1)(b)}\, F_{ba} \Big)(\tau, \vec \sigma)
 = \sum_n\, \delta_{rs}\, \pi^s_{\perp}(\tau, \vec \sigma) + O(\zeta),
 \end{eqnarray*}

 \begin{eqnarray*}
 \partial_{\tau}\, \pi^r_{\perp}(\tau, \vec \sigma) &\cir& \sum_a\,
 P^{ra}_{\perp}(\vec \sigma)\, \Big(\sum_i\, \eta_i\, Q_i\,
 \delta^3(\vec \sigma, {\vec \eta}_i(\tau))\, \Big[{{\kappa_{ia}(\tau)}
 \over {\sqrt{m_i^2c^2 + \sum_b\, \Big(\kappa_{ib}(\tau) -
 {{Q_i}\over c}\, A_{\perp b}\Big)^2}}}\, \Big( 1 +\nonumber \\
 &+& n_{(1)} - 2\, (\Gamma_a^{(1)} + 2\, \phi_{(1)}) + {{\sum_b\,
 (\Gamma_b^{(1)} + 2\, \phi_{(1)})\, \Big(\kappa_{ib}(\tau) -
 {{Q_i}\over c}\, A_{\perp b}\Big)^2}\over {m_i^2c^2 + \sum_b\,
 \Big(\kappa_{ib}(\tau) - {{Q_i}\over c}\, A_{\perp b}\Big)^2}}
 \Big) - {\bar n}_{(1)(a)} \Big](\tau, {\vec \eta}_i(\tau)) -
 \end{eqnarray*}

\bea
 &-& \Big[ \sum_b\, \Big([1 + n_{(1)} - 2\, (\Gamma_a^{(1)} + \Gamma_b^{(1)}
 + \phi_{(1)})]\, \partial_b\, F_{ab} -\nonumber \\
 &-& [\partial_b\, \phi_{(1)} - \partial_b\, n_{(1)} + \partial_b\, (\Gamma^{(1)}_{a} +
 \Gamma^{(1)}_{b})]\, F_{ab}\Big)  - \sum_b\, {\bar
 n}_{(1)(b)}\, \partial_b\, \pi^a_{\perp} +\nonumber \\
 &+& \sum_i\, \eta_i\, Q_i\, \sum_b\, \Big(\partial_b\, {\bar
 n}_{(1)(b)}\, {{\partial\, c(\vec \sigma, {\vec \eta}_i(\tau))}\over
 {\partial\, \sigma^a}} - \partial_b\, {\bar n}_{(1)(a)}\,
 {{\partial\, c(\vec \sigma, {\vec \eta}_i(\tau))}\over {\partial\, \sigma^b}}
 \Big)\Big](\tau, \vec \sigma)\Big).
 \label{5.4}
 \eea

The explicit expression of these equations in terms of the
instantaneous inertial and retarded gravitational quantities can be
obtained by using Eqs. (\ref{4.6}), (\ref{4.7}) and (\ref{4.16}). At
the lowest order and using Eq.(\ref{5.2}) these equations imply the
special relativistic wave equation $\Box\, A_{\perp r}(\tau, \vec
\sigma)\, \cir\, \sum_{nsm}\, \delta_{rn}\, P^{ns}_{\perp}(\vec
\sigma)\, \delta_{sm}\, \sum_i\, \eta_i\, Q_i\, {\dot
\eta}^m_i(\tau)\, \delta^3(\vec \sigma, {\vec \eta}_i(\tau))$, whose
Lienard-Wiechert solution was found and put in Hamiltonian form in
Ref.\cite{14}.
\medskip

The first of Eqs.(\ref{5.4}) can be inverted by iteration and the
resulting form of the transverse electro-magnetic momenta is

\bea
 \pi^r_{\perp}(\tau, \vec \sigma) &\cir& \sum_a\, P_{\perp}^{ra}(\vec \sigma)\,
 \Big([1 - n_{(1)} + 2\, (\Gamma_a^{(1)} - \phi_{(1)})]\,
 \partial_{\tau}\, A_{\perp\, a} -\nonumber \\
 &-& \sum_b\, {\bar n}_{(1)(b)}\, F_{ba}\Big)(\tau, \vec \sigma) + O(\zeta^2).
 \label{5.5}
 \eea

If we put Eq.(\ref{5.5}) into the second of Eqs.(\ref{5.4}) we get a
wave equation for $A_{\perp\, r}(\tau, \vec \sigma)$ containing
extra terms in $\partial_{\tau}\, A_{\perp\, a}$ and $\partial_b\,
\partial_{\tau}\, A_{\perp\, a}$, which will be studied elsewhere.
By using the solutions $\phi_{(1)}$, $n_{(1)}$, ${\bar n}_{(1)(a)}$,
we see that this equation has as sources both  the matter and the
tidal variables. Since in the next two Sections we will show that
the tidal variables $\Gamma_r^{(1)}$ are determined by the stress
tensor $T_{(1)}^{rs}$ in a retarded way, the final equations for the
transverse electro-magnetic field depend on the matter both in an
(action-at-a-distance) instantaneous way and in a retarded way. If
one could find a Lienard-Wiechert-type solution of these equations,
the final form of the particle equations (\ref{5.3}) would be of
coupled integro-differential equations instead of differential
equations like it happens in absence of the gravitational field.

\vfill\eject

\section{The Linearized Second Order Equations for the Tidal
Variables $R_{\bar a}$ in the 3-Orthogonal Gauges}

In Section IV we solved the equations of elliptic type for
$\phi_{(1)}$, $n_{(1)}$, ${\bar n}_{(1)(r)}$ and
$\sigma_{(1)(a)(b)}{|}_{a \not= b}$. The solutions depend on the
tidal variables $R_{\bar a}$.\medskip

We must now study the linearization of the second order equations
(\ref{2.16}) for the tidal variables by using the solutions of
Section IV. We will see that also in these non-harmonic 3-orthogonal
gauges we get wave equations, but they will also contain the
information that the final 3-metric is traceless and transverse
(TT).

\subsection{The Linearization of Eqs.(\ref{2.16}).}

For the three integrals appearing in Eqs.(\ref{2.16}) we get the
following linearization:\medskip

a) the last integral in Eq.(\ref{2.16}))  becomes\medskip

\bea
 &&\int d^3\sigma_1\, \Big(1 + n(\tau ,{\vec \sigma}_1)\Big)\,
 {{\delta\, {\cal M}(\tau ,{\vec \sigma}_1)}\over
 {\delta\, R_{\bar a}(\tau ,\vec \sigma)}}\, =\nonumber \\
 &&{}\nonumber \\
 &&=\, - \sum_i\,
 \delta^3(\vec \sigma, {\vec \eta}_i(\tau))\, \eta_i\,
 \Big({{\sum_a\, \gamma_{\bar aa}\, \Big(\kappa_{ia}(\tau)
 - {{Q_i}\over c}\, A_{\perp\, a}\Big)^2}\over {\sqrt{m_i^2\, c^2
 + \sum_a\, \Big(\kappa_{ia}(\tau) - {{Q_i}\over c}\, A_{\perp\,
 a}\Big)^2}}}\Big)(\tau, \vec \sigma) +\nonumber \\
 &&+ {1\over c}\, \Big[\sum_a\,
 \gamma_{\bar aa}\, (\pi^a_{\perp})^2 - \sum_{ab}\,
 \gamma_{\bar aa}\, F^2_{ab} - \sum_a\,
 \gamma_{\bar aa}\Big(2\, \pi^a_{\perp} - \sum_{k \not= j}\, \eta_k\,
 Q_k\, \partial_a\, c(\vec \sigma, {\vec \eta}_k(\tau))\Big)
 \nonumber \\
 &&\sum_j \, \eta_j\, Q_j\, \partial_a\, c(\vec \sigma, {\vec
 \eta}_j(\tau))\Big](\tau, \vec \sigma) + O(mc\, \zeta^2)
 =\nonumber \\
 &=& -  \sum_a\, \gamma_{\bar aa}\, T^{aa}_{(1)} + O(\zeta^2),
 \qquad from\,\,  Eq.(\ref{3.12}).
 \label{6.1}
 \eea

\bigskip

b) the first integral in Eq.(\ref{2.16})  becomes\medskip

\bea
 &&\int d^3\sigma_1\,\, [1 + n(\tau ,{\vec \sigma}_1)]\,
   {{\delta\, {\cal S}(\tau ,{\vec \sigma}_1)}\over
 {\delta\, R_{\bar a}(\tau ,\vec \sigma )}} {|}_{\theta^i = 0}\,
 =\nonumber \\
 &&=\, 4\, \sum_a\, \gamma_{\bar aa}\, \partial^2_a\, \phi_{(1)}(\tau,
 \vec \sigma) - 2\, \sum_{a\bar b}\, (2\, \gamma_{\bar aa}\,
 \gamma_{\bar ba} - \delta_{\bar a\bar b})\, \partial^2_a\, R_{\bar
 b}(\tau, \vec \sigma) + O(\zeta^2).
 \label{6.2}
 \eea

\bigskip

c) the second integral in Eq.(\ref{2.16})  becomes\medskip

\beq
 \int d^3\sigma_1\, n(\tau, {\vec \sigma}_1)\,
   {{\delta\, {\cal T}(\tau ,{\vec \sigma}_1)}\over
 {\delta\, R_{\bar a}(\tau ,\vec \sigma )}}{|}_{\theta^i = 0}\,
 =\, 2\, \sum_a\, \gamma_{\bar aa}\, \partial^2_a\, n_{(1)}(\tau
 ,\vec \sigma) + O(\zeta^2),
 \label{6.3}
 \eeq

\bigskip

\noindent As a consequence,  the linearization of the second order
equations (\ref{2.16}) for the tidal variables $R_{\bar a}$
is\medskip

\begin{eqnarray*}
 \partial_{\tau}^2\, R_{\bar a}(\tau, \vec \sigma)\, &\cir& \triangle\,
 R_{\bar a}(\tau, \vec \sigma) + \sum_a\, \gamma_{\bar aa}\, \Big[
 \partial_{\tau}\, \partial_a\, {\bar n}_{(1)(a)} +\nonumber \\
 &+& \partial_a^2\, n_{(1)} + 2\, \partial_a^2\, \phi_{(1)} - 2\,
 \partial_a^2\, \Gamma_a^{(1)} + {{8\pi\, G}\over {c^3}}\,
 T_{(1)}^{aa} \Big](\tau, \vec \sigma),\nonumber \\
 &&{}\nonumber \\
 &&\Downarrow\qquad by\, using\, Eqs.(\ref{4.6}),\, (\ref{4.7}),\,
 (\ref{4.16})
 \end{eqnarray*}

 \bea
 \Box\, \sum_{\bar b}\, M_{\bar a\bar b}\,
 R_{\bar b}\,\,  &\cir&\,\,\, E_{\bar a},\nonumber \\
 &&{}\nonumber \\
 &&\quad M_{\bar a\bar b} = \delta_{\bar a\bar b} - \sum_a\, \gamma_{\bar
 aa}\, {{\partial_a^2}\over {\triangle}}\, \Big(2\, \gamma_{\bar
 ba} - {1\over 2}\, \sum_b\, \gamma_{\bar bb}\, {{\partial_b^2}\over
 {\triangle}}\Big),\nonumber \\
 &&\quad E_{\bar a} = {{4\pi\, G}\over {c^3}}\, \sum_a\, \gamma_{\bar
 aa}\, \Big[{{\partial_{\tau}\, \partial_a}\over {\triangle}}\,
 \Big(4\, {\check {\cal M}}^{(UV)}_{(1)a}
 - {{\partial_a}\over {\triangle}}\, \sum_c\, \partial_c\,
 {\check {\cal M}}^{(UV)}_{(1)c}\Big) +\nonumber \\
 &&\qquad + 2\, T_{(1)}^{aa}
 + {{\partial_a^2}\over {\triangle}}\, \sum_b\, T_{(1)}^{bb}\Big],
 \nonumber \\
 &&{}\nonumber \\
 &&\Downarrow\nonumber \\
 &&{}\nonumber \\
  \Box\, \sum_b\, {\tilde M}_{ab}\, \Gamma_b^{(1)}\,\,
 &\cir&\,\,\, \sum_{\bar a}\, \gamma_{\bar aa}\,
 E_{\bar a},\nonumber \\
 &&{}\nonumber \\
 &&\quad {\tilde M}_{ab} = \sum_{\bar a\bar b}\, \gamma_{\bar aa}\,
 \gamma_{\bar bb}\, M_{\bar a\bar b} = \delta_{ab}\, \Big(1
 - 2\, {{\partial_a^2}\over {\triangle}}\Big) + {1\over 2}\,
 \Big(1 + {{\partial_a^2}\over {\triangle}}\Big)\,
 {{\partial_b^2}\over {\triangle}},\nonumber \\
 &&\qquad \sum_a\, {\tilde M}_{ab} = 0,\qquad M_{\bar a\bar b} = \sum_{ab}\, \gamma_{\bar aa}\,
 \gamma_{\bar bb}\, {\tilde M}_{ab},
 \label{6.4}
 \eea

\noindent where we used Eqs.(\ref{4.6}), (\ref{4.7}), (\ref{4.16})
and $\sum_{\bar a}\, \gamma_{\bar aa}\, \gamma_{\bar ab} =
\delta_{ab} - {1\over 3}$, $\sum_a\, \gamma_{\bar aa}\, \gamma_{\bar
ba} = \delta_{\bar a\bar b}$, $\sum_a\, \gamma_{\bar aa} = 0$.

\medskip

Therefore we get the massless wave equation (with the flat
d'Alambertian $\Box$ associated to the asymptotic Minkowski metric)
not for the tidal variable $R_{\bar a}$ but for the quantity
$\sum_{\bar b}\, M_{\bar a\bar b}\, R_{\bar b}$, where $M_{\bar
a\bar b}$ is the spatial operator already found in Eq.(\ref{4.18}).

\subsection{The Meaning of the Operators $M_{\bar a\bar b}$ and
${\tilde M}_{ab}$}

Let us show that the operators $M_{\bar a\bar b}$ and ${\tilde
M}_{ab}$ are present to select the TT part of the spatial metric
${}^4g_{(1)rs} = - \sgn\, {}^3g_{(1)rs} = - \sgn\, \delta_{rs} +
{}^4h_{(1)rs} = - \sgn\, \delta_{rs}\, \Big(1 + 2(\Gamma_r^{(1)} +
2\, \phi_{(1)})\Big)$, namely that they define the polarization
pattern of the gravitational waves in these non-harmonic
3-orthogonal gauges.
\medskip

In Ref.\cite{19} it is shown that in every gauge we can make the
following decomposition of $h_{(1)rs}(\tau, \vec \sigma)$

\beq
 {}^4h_{(1)rs} = {}^4h^{TT}_{(1)rs} + {1\over 3}\, \delta_{rs}\,
 H_{(1)} + {1\over 2}\, (\partial_r\, \epsilon_{(1)s} +
 \partial_s\, \epsilon_{(1)r}) + (\partial_r\, \partial_s - {1\over
 3}\, \delta_{rs}\, \triangle)\, \lambda_{(1)},
 \label{6.5}
 \eeq\medskip

\noindent with $\sum_r\, \partial_r\, \epsilon_{(1)r} = 0$ and
${}^4h_{(1)rs}^{TT}$ traceless and transverse, i.e. $\sum_r\,
{}^4h^{TT}_{(1)rr} = 0$, $\sum_r\, \partial_r\, {}^4h^{TT}_{(1)rs} =
0$. The functions $H_{(1)}$, $\lambda_{(1)}$ and $\epsilon_{(1)r}$
have the following expression

\begin{eqnarray*}
 H_{(1)} &=& {}^4h_{(1)} = \sum_{rs}\, \delta^{rs}\, {}^4h_{(1)rs} =
 \sum_r\, {}^4h_{(1)rr} = - 12\, \sgn\, \phi_{(1)},\nonumber \\
 \lambda_{(1)} &=& {3\over 2}\, {1\over {\triangle^2}}\, \sum_{uv}\,
 \Big(\partial_u\, \partial_v\, - {1\over 3}\, \delta_{uv}\,
 \triangle\Big)\, {}^4h_{(1)uv} = - 3\, \sgn\,  \sum_u\,
 {{\partial_u^2}\over {\triangle^2}}\, \Gamma_u^{(1)},\nonumber \\
 \epsilon_{(1)r} &=& {2\over {\triangle}}\, \Big(\sum_u\,
 \partial_u\, {}^4h_{(1)ur} - {{\partial_r}\over {\triangle}}\,
 \sum_{uv}\, \partial_u\, \partial_v\, {}^4h_{(1)uv}\Big) =\nonumber \\
 &=& - 4\, \sgn\, {{\partial_r}\over {\triangle}}\, \Big(
 \Gamma_r^{(1)} - \sum_u\, {{\partial_u^2}\over {\triangle}}\,
 \Gamma_u^{(1)} \Big),
 \end{eqnarray*}

\bea
 &&\Downarrow\nonumber \\
 &&{}\nonumber \\
 {}^4h_{(1)rs} &=& {}^4h^{TT}_{(1)rs} - \sgn\, \Big[ \Big(4\, \phi_{(1)}
 - \sum_u\, {{\partial_u^2}\over {\triangle}}\, \Gamma_u^{(1)}\Big)\,
 \delta_{rs} +\nonumber \\
 &+& 2\, {{\partial_r\, \partial_s}\over {\triangle}}\,
 \Big(\Gamma_r^{(1)} + \Gamma_s^{(1)} - {1\over 2}\, \sum_u\,
 {{\partial_u^2}\over {\triangle}}\, \Gamma_u^{(1)}\Big).
 \label{6.6}
 \eea

\noindent In the last line of this equation we have given the form
of the 3-metric implied by Eq.(\ref{6.5}).

\bigskip

As a consequence the TT part of the spatial metric is independent
from $\phi_{(1)}$ and has the expression\medskip

\bea
 {}^4h^{TT}_{(1)rs} &=& {}^4h_{(1)rs} + {1\over 2}\, \delta_{rs}\,
 \Big(\sum_{uv}\, {{\partial_u\, \partial_v}\over {\triangle}}\,
 {}^4h_{(1)uv} - \sum_u\, {}^4h_{(1)uu}\Big) +\nonumber \\
 &+& {1\over 2}\,
 {{\partial_r\, \partial_s}\over {\triangle}}\, \Big(\sum_{uv}\,
 {{\partial_u\, \partial_v}\over {\triangle}}\, {}^4h_{(1)uv} +
 \sum_u\, {}^4h_{(1)uu}\Big) - \sum_u\, {{\partial_u}\over
 {\triangle}}\, \Big(\partial_r\, {}^4h_{(1)us} + \partial_s\,
 {}^4h_{(1)ur}\Big) =\nonumber \\
 &=& - \sgn\, \Big[\Big(2\, \Gamma_r^{(1)} +
 \sum_u\, {{\partial_u^2}\over {\triangle}}\, \Gamma_u^{(1)}\Big)\,
 \delta_{rs} - 2\, {{\partial_r\, \partial_s}\over {\triangle}}\,
 (\Gamma_r^{(1)} + \Gamma_s^{(1)}) + {{\partial_r\,
 \partial_s}\over {\triangle}}\, \sum_u\, {{\partial_u^2}\over
 {\triangle}}\, \Gamma_u^{(1)}\Big] =\nonumber \\
 &=&\sum_{uv}\, {\cal P}_{rsuv}\, {}^4h_{(1)uv},\nonumber \\
 &&{}\nonumber \\
 {\cal P}_{rsuv} &=& {1\over 2}\, (\delta_{ru}\, \delta_{sv} +
 \delta_{rv}\, \delta_{su}) - {1\over 2}\, \Big(\delta_{rs} -
 {{\partial_r\, \partial_s}\over {\triangle}}\Big)\, \delta_{uv}
 + {1\over 2}\, \Big(\delta_{rs} + {{\partial_r\, \partial_s}\over
 {\triangle}}\Big)\, {{\partial_u\, \partial_v}\over {\triangle}}
 -\nonumber \\
 &-& {1\over 2}\, \Big[{{\partial_u}\over {\triangle}}\,
 (\delta_{rv}\, \partial_s + \delta_{sv}\, \partial_r)
 + {{\partial_v}\over {\triangle}}\, (\delta_{ru}\, \partial_s
 + \delta_{su}\, \partial_r)\Big],
 \label{6.7}
 \eea

\noindent where ${\cal P}_{rsuv}$ is the projector extracting the TT
part of the spatial metric. It satisfies $\sum_{uv}\, {\cal
P}_{rsuv}\, {\cal P}_{uvmn} = {\cal P}_{rsmn} = {\cal P}_{srmn} =
{\cal P}_{rsnm}$, $\sum_r\, \partial_r\, {\cal P}_{rsuv} = \sum_u\,
\partial_u\, {\cal P}_{rsuv} = 0$, $\sum_r\, {\cal P}_{rruv} =
\sum_u\, {\cal P}_{rsuu} = 0$.\medskip

With plane waves, i.e. ${}^4h_{(1)rs}(\tau, \vec \sigma) =
A_{rs}(\tau)\, e^{i\, \vec n \cdot \vec \sigma} + cc$, we recover
the standard projector ${\cal P}_{rsuv} \mapsto \Lambda_{rsuv} =
P_{ru}\, P_{sv} - {1\over 2}\, P_{rs}\, P_{uv}$ with $P_{rs} =
\delta_{rs} - n_r\, n_s$ ($\vec n$ is a unit vector orthogonal to
the wave-front).

\bigskip

The diagonal elements ${}^4h^{TT}_{(1)aa}$ of the TT 3-metric
contain the operator ${\tilde M}_{ab}$ of Eq.(\ref{6.4}) since we
have\medskip

\bea
 {}^4h^{TT}_{(1)aa} &=& - 2\, \sgn\, \Big[\Big(1 - 2\,
 {{\partial_a^2}\over {\triangle}}\Big)\, \Gamma_a^{(1)} + {1\over
 2}\, \Big(1 + {{\partial_a^2}\over {\triangle}}\Big)\, \sum_c\,
 {{\partial_c^2}\over {\triangle}}\, \Gamma_c^{(1)}\Big] =\nonumber \\
 &=& - 2\, \sgn\, \sum_c\, {\tilde M}_{ac}\,
 \Gamma_c^{(1)},\nonumber \\
 &&{}\nonumber \\
 \Rightarrow&& \Box\, {}^4h^{TT}_{(1)aa} \cir - 2\, \sgn\,
 \sum_{\bar a}\, \gamma_{\bar aa}\, E_{\bar a}.
 \label{6.8}
 \eea
 \bigskip

If we apply the decomposition (\ref{6.5}) to the stress tensor
$T^{rs}_{(1)}$ of Eqs.(\ref{3.12}), we get

\bea
 T_{(1)}^{ rs} &=& {1\over 3}\,  {\tilde H}_{(1)}\, \delta_{rs} +
 T^{(TT)}_{(1) rs} + {1\over 2}\, (\partial_r\, {\tilde \epsilon}_{(1) s} +
 \partial_s\, {\tilde \epsilon}_{(1) r}) + (\partial_r\, \partial_s -
 {1\over 3}\, \delta_{rs}\, \triangle)\, {\tilde \lambda}_{(1)},\nonumber \\
 &&{}\nonumber \\
 &&\sum_r\, \partial_r\, {\tilde \epsilon}_{(1) r} = \sum_r\,
 \partial_r\, T_{(1)}^{(TT) rs} = \sum_r\, T_{(1)}^{(TT) rr} =
 0,\nonumber \\
 &&{}\nonumber \\
 {\tilde H}_{(1)} &=&  \sum_r\, T^{rr}_{(1)},\nonumber \\
 {\tilde \lambda}_{(1)} &=& {3\over 2}\, {1\over {\triangle^2}}\, \sum_{rs}\,
 (\partial_r\, \partial_s - {1\over 3}\, \delta_{rs}\, \triangle )\,
 T^{rs}_{(1)},\nonumber \\
 {\tilde \epsilon}_{(1) r} &=& 2\, \sum_u\, (\partial_u\, T^{ru}_{(1)} -
 {1\over 3}\, \partial_r\, T^{uu}_{(1)}) - 2\, {{\partial_r}\over
 {\triangle}}\, \sum_{uv}\, (\partial_u\, \partial_v - {1\over 3}\,
 \delta_{uv}\, \triangle)\, T^{uv}_{(1)},\nonumber \\
 &&{}\nonumber \\
  T^{(TT)ab}_{(1)} &=& \sum_{cd}\, {\cal P}_{abcd}\, T_{(1)}^{cd},
 \nonumber \\
 T^{(TT)aa}_{(1)} &=& \sum_{cd}\, \Big[\delta_{ac}\,
 \delta_{ad} - 2\, \delta_{ad}\, {{\partial_a\, \partial_c}\over
 {\triangle}} + {1\over 2}\, (1 + {{\partial_a^2}\over
 {\triangle}})\, {{\partial_c\, \partial_d}\over {\triangle}} -
 {1\over 2}\, \delta_{cd}\, (1 - {{\partial_a^2}\over
 {\triangle}})\Big]\, T^{cd}_{(1)},\nonumber \\
 &&{}
 \label{6.9}
 \eea

\noindent where $T^{(TT)ab}_{(1)}$ is the TT part of the stress
tensor.\bigskip

By using Eqs.(\ref{3.13}), i.e. $\partial_{\tau}\, {\cal
M}^{(UV)}_{(1)}\, \cir\,\, - \sum_c\,\partial_c\, {\cal M}^{(UV)}_c
+ \partial_A\, {\cal R}_{(2)}^{A\tau}$, $\partial_{\tau}\,  {\cal
M}^{(UV)}_a \cir - \sum_c\, \partial_c\, T_{(1)}^{ca} +
\partial_A\, {\cal R}_{(2)}^{Aa}$, so that $\partial_{\tau}^2\,
{\cal M}^{(UV)}_{(1)}\, \cir\,\, \sum_{cd}\, \partial_c\,
\partial_d\, T_{(1)}^{cd} + \partial_A\, (\partial_{\tau}\, {\cal R}_{(2)}^{A\tau}
+ \sum_c\, \partial_c\, {\cal R}_{(2)}^{Ac})$, we get the following
expression for the source term appearing in the second member of
Eq.(\ref{6.4})

\beq
 \sum_{\bar a}\, \gamma_{\bar aa}\, E_{\bar a}\, \cir\,\,
  {{8\pi\, G}\over {c^3}}\, T^{(TT)aa}_{(1)},\qquad
  E_{\bar a}\, \cir\, {{8\pi\, G}\over {c^3}}\, \sum_a\,
  \gamma_{\bar aa}\, T_{(1)}^{(TT) aa}.
 \label{6.10}
 \eeq
\medskip

Therefore we get that the TT part of the 3-metric satisfies the wave
equation

\bea
 \Box\, {}^4h^{TT}_{(1)aa} &\cir& - 2\, \sgn\, \sum_{\bar a}\,
 \gamma_{\bar aa}\, E_{\bar a}\, \cir\,\, - \sgn\, {{16\pi\, G}\over {c^3}}\,
 T^{(TT)aa}_{(1)},\nonumber \\
 &&{}\nonumber \\
 &&\Downarrow\nonumber \\
 &&{}\nonumber \\
 \Box\, {}^4h^{TT}_{(1)ab}\, &\cir& - \sgn\, {{16\pi\, G}\over {c^3}}\,
 T^{(TT)ab}_{(1)}.
 \label{6.11}
 \eea

\bigskip

The second line of Eqs.(\ref{6.11}) derives from the first line due
to the transversality property of ${}^4h^{TT}_{(1)rs}$ and
$T^{(TT)}_{(1) rs}$, which implies $\sum_{b \not= a}\, \partial_b\,
\Big(\Box\, {}^4h^{TT}_{(1)ba} + \sgn\, {{16\pi\, G}\over {c^3}}\,
T^{(TT)}_{(1) ba}\Big)\, =\, - \partial_a\, \Big(\Box\,
{}^4h^{TT}_{(1)aa} + \sgn\, {{16\pi\, G}\over {c^3}}\, T^{(TT)}_{(1)
aa}\Big)\, \cir\,\, 0$.

\medskip

Therefore, even if we are not in a harmonic gauge, the Hamilton
equations in the 3-orthogonal gauges imply {\it the massless wave
equation for the TT part of the spatial metric}.
\medskip

Once Eq.(\ref{6.11}) is solved for ${}^4h^{(TT)}_{(1)aa} = - 2\,
\sgn\, \sum_b\, {\tilde M}_{ab}\, \Gamma_b^{(1)}$, see
Eq.(\ref{6.8}), in terms of $T_{(1)}^{(TT) aa}$, we can find the
solution for $R_{\bar a} = \sum_a\, \gamma_{\bar aa}\,
\Gamma_a^{(1)}$ if we succeed to invert the operator ${\tilde
M}_{ab}$. Then, if we put this solution in the last line of Eqs.
(\ref{6.6}), we can find the spatial metric ${}^4h_{(1)rs}$ in the
3-orthogonal gauges.

\subsection{A Generalized TT Gauge}

At this stage the 4-metric of Eqs.(\ref{3.2}) and (\ref{3.5}) has
the following form

\begin{eqnarray*}
 {}^4g_{(1)AB} &=& {}^4\eta_{AB} + \sgn\, \left(
 \begin{array}{ccc}
 2\, n_{(1)} &{}& - {\bar n}_{(1)(r)} \\ &&\\
 - {\bar n}_{(1)(s)} &{}& - 2\, \Big(\Gamma_r^{(1)}
 + 2\, \phi_{(1)}\Big)\, \delta_{rs}
 \end{array} \right) + O(\zeta^2) =\nonumber \\
 &&{}\nonumber \\
  &=&{}^4\eta_{AB} + \sgn\, \left(
 \begin{array}{ccc}
 - 2\,\frac{\partial_\tau\,}{\Delta}\, {}^3K_{(1)} + \alpha(matter)
 &{}& - \frac{\partial_r}{\Delta}\, {}^3K_{(1)} + A_r(\Gamma_a^{(1)}) + \beta_r(matter)\\
 &&\\
  - \frac{\partial_s}{\Delta}\, {}^3K_{(1)} + A_s(\Gamma_a^{(1)}) + \beta_s(matter)&{}&
 \Big[B_r(\Gamma_a^{(1)}) + \gamma(matter)\Big]\, \delta_{rs}
 \end{array}
 \right) +\nonumber \\
 &+& O(\zeta^2),
 \end{eqnarray*}

\bea
 A_r(\Gamma_a^{(1)}) &=& - {1\over 2}\, \partial_{\tau}\,
 {{\partial_r}\over {\triangle}}\, \Big(4\, \Gamma_r^{(1)} -
 \sum_c\, {{\partial_c^2}\over {\triangle}}\, \Gamma_c^{(1)}\Big),
 \nonumber \\
 B_r(\Gamma_a^{(1)}) &=& - 2\, \Big(\Gamma_r^{(1)} + {1\over 2}\, \sum_c\,
 {{\partial_c^2}\over {\triangle}}\, \Gamma_c^{(1)}\Big),\nonumber \\
 \alpha(matter) &=& {{8\pi\, G}\over {c^3}}\, {1\over {\triangle}}\,
 \Big({\cal M}_{(1)}^{(UV)} + \sum_c\, T^{cc}_{(1)}\Big),\nonumber \\
 \beta_r(matter) &=& - {{4\pi\, G}\over {c^3}}\, {1\over
 {\triangle}}\, \Big(4\, {\cal M}_{(1)r}^{(UV)} - {{\partial_r}\over {\triangle}}\,
 \sum_c\, \partial_c\, {\cal M}_{(1)c}^{(UV)}\Big),\nonumber \\
 \gamma(matter) &=& {{8\pi\, G}\over {c^3}}\, {1\over
 {\triangle}}\, {\cal M}_{(1)}^{(UV)},
 \label{6.12}
 \eea

\noindent by using the expression of  $\phi_{(1)}$, $n_{(1)}$ and
${\bar n}_{(1)(r)}$ given in Eqs. (\ref{4.6}), (\ref{4.7}) and
(\ref{4.16}), respectively. We have explicitly shown the dependence
upon the inertial gauge variable ${}^3{\cal K}_{(1)} = {1\over
{\triangle}}\, {}^3K_{(1)}$, the instantaneous inertial effects
$\alpha(matter)$, $\beta_r(matter)$, $\gamma(matter)$ and the
retarded tidal effects $A_r(\Gamma_a^{(1)})$, $B_r(\Gamma_a^{(1)})$.
\bigskip

Let us consider the following coordinate transformation on the
3-space $\Sigma_{\tau}$ (endorsing it with $\tau$-dependent new
radar 3-coordinates)

\bea
 \bar \tau = \tau, && {\bar \sigma}^r = \sigma^r - \Psi_{(1)}^r(\tau, \vec
 \sigma),\nonumber \\
 &&{}\nonumber \\
 &&\Psi_{(1)}^r(\tau, \vec \sigma) = - {1\over 2}\, {{\partial_r}\over {\triangle}}\,
  \Big(4\, \Gamma_r^{(1)} - \sum_c\, {{\partial_c^2}\over {\triangle}}\,
 \Gamma_c^{(1)}\Big)(\tau, \vec \sigma) = O(\zeta),\nonumber \\
 &&{}
 \label{6.13}
 \eea

\noindent and the associated new 4-metric ${}^4{\bar
g}_{(1)AB}(\tau, {\vec {\bar \sigma}}) = {{\partial\, \sigma^C}\over
{\partial\, {\bar \sigma}^A}}\, {{\partial\, \sigma^D}\over
{\partial\, {\bar \sigma}^B}}\, {}^4g_{(1)CD}(\tau, \vec \sigma)$
replacing the one of Eq.(\ref{6.12}). The inverse transformation is
$\tau = \bar \tau$, $\sigma^r = {\bar \sigma}^r + \Psi_{(1)}(\tau,
{\vec {\bar \sigma}}) + O(\zeta^2)$.\medskip

Since we have ${{\partial\, {\bar \sigma}^r}\over {\partial\,
\sigma^s}} = \delta^r_s - {{\partial\, \Psi_{(1)}^r(\tau, \vec
\sigma)}\over {\partial\, \sigma^s}}$, ${{\partial\, {\bar
\sigma}^r}\over {\partial\, \tau}} = - {{\partial\,
\Psi_{(1)}^r(\tau, \vec \sigma)}\over {\partial\, \tau}}$, we get
${{\partial\, \sigma^r}\over {\partial\, {\bar \sigma}^s}} =
\delta^r_s + {{\partial\, \Psi_{(1)}^r(\tau, {\vec {\bar
\sigma}})}\over {\partial\, {\bar \sigma}^s}} + O(\zeta^2)$,
${{\partial\, \sigma^r}\over {\partial\, \bar \tau}} = {{\partial\,
\Psi_{(1)}^r(\tau, {\vec {\bar \sigma}})}\over {\partial\,  \tau}} +
O(\zeta^2)$. In particular, by using Eqs.(\ref{4.6}), (\ref{4.7})
and (\ref{4.16}), we get (since $\bar \tau = \tau$ and ${\vec {\bar
\sigma}}$, we always use $\tau$)

\begin{eqnarray*}
 {}^4{\bar g}_{(1)\tau\tau}(\tau, {\vec {\bar \sigma}}) &=&
 {}^4g_{(1)\tau\tau}(\tau, \vec \sigma) + O(\zeta^2) =
 \sgn\, \Big[1 + 2\, n_{(1)}\Big](\tau, {\vec {\bar \sigma}}) +
 O(\zeta^2) =\nonumber \\
 &=&  \sgn\, \Big[- 2\, {{\partial_{\tau}}\over {\triangle}}\,
 {}^3K_{(1)} + {{8\pi\, G}\over {c^3}}\, {1\over
 {\triangle}}\, \Big({\cal M}^{(UV)}_{(1)} + \sum_c\,
 T^{cc}_{(1)}\Big)\Big](\tau, {\vec {\bar \sigma}}) + O(\zeta^2),\nonumber \\
 {}^4{\bar g}_{(1)\tau r}(\tau, {\vec {\bar \sigma}}) &=&
 {}^4g_{(1)\tau r}(\tau, \vec \sigma) - \sgn\, {{\partial\,
 \Psi_{(1)}(\tau, {\vec {\bar \sigma}})^r}\over {\partial\, \tau}}
 + O(\zeta^2)=\nonumber \\
 &=& - \sgn\, \Big({\bar n}_{(1)(r)} + {{\partial\, \Psi_{(1)}^r}\over
 {\partial\, \tau}}\Big)(\tau, {\vec {\bar \sigma}}) +
 O(\zeta^2) =\nonumber \\
 &=& - \sgn\, \Big[{{\partial_r}\over {\triangle}}\,
 {}^3K_{(1)} + {{4\pi\, G}\over {c^3}}\, {1\over {\triangle}}\,
 \Big(4\, {\cal M}^{(UV)}_r - {{\partial_r}\over {\triangle}}\,
 \sum_c\, \partial_c\, {\cal M}_c^{(UV)}\Big)\Big](\tau,
 {\vec {\bar \sigma}}) + O(\zeta^2),
 \end{eqnarray*}

 \bea
 {}^4{\bar g}_{(1)rs}(\tau, {\vec {\bar \sigma}}) &=&
 {}^4g_{rs}(\tau, \vec \sigma) - \sgn\, \Big(\partial_r\, \Psi_{(1)}^s +
 \partial_s\, \Psi_{(1)}^r\Big)(\tau, {\vec {\bar \sigma}}) +
 O(\zeta^2) =\nonumber \\
 &=& - \sgn\, \Big(\delta_{rs}\, \Big[1 + 2\, (\Gamma^{(1)}_r +
 2\, \phi_{(1)})\Big](\tau, \vec \sigma)
 + \Big[\partial_r\, \Psi_{(1)}^s +
 \partial_s\, \Psi_{(1)}^r\Big](\tau, {\vec {\bar \sigma}})\Big)
  + O(\zeta^2) =\nonumber \\
 &=& {}^4\eta_{rs} + \Big[{}^4h^{TT}_{(1) rs}
 + \sgn\, {{8\pi\, G}\over {c^3}}\, {{\delta_{rs}}\over {\triangle}}\,
 {\cal M}^{(UV)}_{(1)}\Big](\tau, {\vec {\bar \sigma}})
  + O(\zeta^2).\nonumber \\
 &&{}
 \label{6.14}
 \eea

\noindent In the last line we used Eq.(\ref{4.6}) for $\phi_{(1)}$
to recover the TT 3-metric ${}^4h^{TT}_{(1) rs}$ of Eq.(\ref{6.7}).
As a consequence it turns out that the new 4-metric in the new radar
4-coordinates has the following expression

\bea
 {}^4{\bar g}_{AB} &=& {}^4\eta_{AB} +
 \sgn\, \left(
 \begin{array}{ccc}
 -2\,\frac{\partial_\tau\,}{\Delta}\, {}^3K_{(1)} +
 \alpha(matter)
 &{}&- \frac{\partial_r}{\Delta}\, {}^3K_{(1)} +
 \beta_r(matter)\\
 &&\\
 - \frac{\partial_s}{\Delta}\, {}^3K_{(1)} +
 \beta_s(matter)&{}&
 \sgn\, {}^4h^{TT}_{(1) rs} + \delta_{rs}\,  \gamma(matter)
 \end{array}
 \right) + O(\zeta^2).\nonumber \\
 &&{}
 \label{6.15}
 \eea

\noindent  Therefore the coordinate transformation (\ref{6.13})
leads to a {\it generalized TT-gauge} whose  3-metric is not
3-orthogonal due to the presence of the TT 3-metric. Also in absence
of matter it differs from the usual harmonic ones, whose
instantaneous 3-spaces are all Euclidean, for the non-spatial terms
depending upon the inertial gauge variable ${}^3{\cal K}_{(1)} =
{1\over {\triangle}}\, {}^3K_{(1)}(\tau, \vec \sigma)$ (the HPM form
of the gauge freedom in clock synchronization).
\medskip

If the matter sources have a compact support and if  the matter
terms ${1\over {\triangle}}\, {\cal M}_{(1)}^{(UV)}$ and ${1\over
{\triangle}}\, {\cal M}_{(1)r}^{(UV)}$ are negligible in the
radiation zone far away from the sources, then  Eq.(\ref{6.15})
gives a {\it spatial TT-gauge} with still the explicit dependence on
the inertial gauge variable ${}^3{\cal K}_{(1)}$ (non existing in
Newtonian gravity), which together with matter and the tidal
variables, determines the non-Euclidean nature of the instantaneous
3-spaces.

\subsection{Inversion of the Operator ${\tilde M}_{ab}$ and the
Resulting Form of the Second Order Equations for $R_{\bar a}$.}

Since Eq.(\ref{6.8}) gives $\sum_b\, {\tilde M}_{ab}\,
\Gamma_b^{(1)} = - {{\sgn}\over 2}\, {}^4h^{TT}_{(1)aa}$ with the
diagonal elements of the TT 3-metric satisfying the wave equation
(\ref{6.11}), i.e. $\Box\, {}^4h^{TT}_{(1)aa} \cir - \sgn\,
{{16\pi\, G}\over {c^3}}\, T^{(TT)aa}_{(1)}$, to find the tidal
variables $R_{\bar a} = \sum_a\, \gamma_{\bar aa}\, \Gamma_a^{(1)}$
associated to one solution of the wave equation we have to invert
the operator ${\tilde M}_{ab}$.

\medskip

If we introduce the functions

\bea
 g_a &=& - {{\sgn}\over 2}\, {}^4h^{TT}_{(1) aa} =  \sum_b\, {\tilde
 M}_{ab}\, \Gamma_b^{(1)} = \Gamma^{(1)}_a + \frac{1}{2}\,
 \left(1+\frac{\partial_a^2}{\Delta}\right)\sum_c\,\frac{\partial_c^2}{\Delta}\,
 \Gamma^{(1)}_c-2\frac{\partial_a^2}{\Delta}\,\Gamma^{(1)}_a =\nonumber \\
 &&{}\nonumber \\
 && \sum_a\, g_a = 0,\qquad \Box\, g_a\, \cir\, {{8\pi\, G}\over
 {c^3}}\, T_{(1)}^{(TT)aa},
 \label{6.16}
 \eea

\noindent we can write

\beq
 \Gamma^{(1)}_a = g_a - \partial_a\, \Psi^a_{(1)} - {1\over 3}\,
 \sum_c\, \partial_c\, \Psi^c_{(1)},
 \label{6.17}
 \eeq

\noindent where $\Psi^a_{(1)}$ is the function generating the
coordinate transformation (\ref{6.13}).\medskip

If we define the quantities

\bea
  z_{ab}{|}_{a \not= b} &=& {}^4h^{TT}_{(1) ab}{|}_{a \not= b}
 = - \frac{\partial_a\partial_b}{\Delta}\,
 \left(\Gamma^{(1)}_a + \Gamma^{(1)}_b - \frac{1}{2}\,
 \sum_c\, \frac{\partial_c}{\Delta}\,
 \Gamma^{(1)}_c\,\right) =\nonumber \\
 &=& 2\, \Big(\partial_a\, \Psi^b_{(1)} + \partial_b\,
 \Psi^a_{(1)}\Big){|}_{a \not= b},\qquad (a\neq b),\nonumber \\
 &&{}\nonumber \\
 &&{}\nonumber \\
 &&\sum_b\, \partial_b\, {}^4h^{TT}_{(1)ab} = 0,\quad \Rightarrow
 \quad \sum_{b \not= a}\, \partial_b\, z_{ba} = - \partial_a\, g_a,
 \nonumber \\
 &&{}\nonumber \\
 \Rightarrow && z_{ab}{|}_{a \not= b} = z_{ab}[g_c]{|}_{a \not= b}.
 \label{6.18}
 \eea

\noindent they turn out to be functionals of the $g_a$'s expressible
in terms of the functions $\Psi^a_{(1)}$ of Eq.(\ref{6.13}).
Therefore, also the functions $\Psi^a_{(1)}$ can be viewed as
functionals $\Psi^a_{(1)}[g_r]$ of the $g_a$'s.
\medskip

As a consequence, the TT 3-metric of Eq.(\ref{6.7}) can be written
in the form

\begin{eqnarray*}
 {}^4h^{TT}_{(1) ab} &=& - 2\, \sgn\, \Big[\Big(\Gamma_r^{(1)} +
 \sum_u\, {{\partial_u^2}\over {2\, \triangle}}\, \Gamma_u^{(1)}\Big)\,
 \delta_{rs} -  {{\partial_r\, \partial_s}\over {\triangle}}\,
 (\Gamma_r^{(1)} + \Gamma_s^{(1)}) + {{\partial_r\,
 \partial_s}\over {2\, \triangle}}\, \sum_u\, {{\partial_u^2}\over
 {\triangle}}\, \Gamma_u^{(1)}\Big] =
 \end{eqnarray*}

 \bea
&=& \left(
\begin{array}{ccc}
g_1&z_{12}[g]&z_{13}[g]\\
z_{12}[g]&g_2&z_{23}[g]\\
z_{13}[g]&z_{23}[g]&g_3
\end{array}
\right) .
 \label{6.19}
 \eea

\bigskip

The transversality property of the TT metric, shown in
Eq.(\ref{6.18}),  allows to express $z_{ab}[g_c]$ as the following
functional of $g_a$

\bea
 z_{23}[g_c](\tau, \vec \sigma) &=&\frac{1}{2}\,\int^{\sigma^2, \sigma^3}
  d{\tilde \sigma}^2\,d{\tilde \sigma}^3\,\left[
 -\partial_3^2\,g_3-\partial_2^2\,g_2+\partial_1^2\,g_1
 \right](\tau, \sigma^1, {\tilde \sigma}^2, {\tilde \sigma}^3),\nonumber\\
 z_{13}[g_c](\tau, \vec \sigma) &=&\frac{1}{2}\,\int^{\sigma^1, \sigma^3}
  d{\tilde \sigma^1}\,d{\tilde \sigma^3}\,\left[
 -\partial_1^2\,g_1-\partial_3^2\,g_3+\partial_2^2\,g_2
 \right](\tau, {\tilde \sigma}^1, \sigma^2, {\tilde \sigma}^3),\nonumber\\
 z_{12}[g_c](\tau, \vec \sigma) &=&\frac{1}{2}\,\int^{\sigma^1, \sigma^2}
 d{\tilde \sigma^1}\,d{\tilde \sigma^2}\,\left[
 - \partial_2^2\,g_2-\partial_1^2\,g_1+\partial_3^2\,g_3 \right](\tau,
 {\tilde \sigma}^1, {\tilde \sigma}^2, \sigma^3).\nonumber \\
 &&{}
 \label{6.20}
  \eea

By using $\sum_a\, g_a = 0$, Eq(\ref{6.20}) can be put in the
following form

\beq
 z_{ab}[g_c]{|}_{a \not= b}(\tau, \vec \sigma) = - \frac{1}{2}\,
 \int_{a \not= b}^{\sigma^a, \sigma^b}\,
 d{\tilde \sigma}^a\, d{\tilde \sigma}^b\left[
 (\Delta-\partial^2_a)\,g_b+(\Delta-\partial^2_b)\,g_a \right](\tau,
 {\tilde \sigma}^a, {\tilde \sigma}^b, \sigma^{c \not= a,b}).
 \label{6.21}
 \eeq

\bigskip

Due to Eqs.(\ref{6.18}), the functionals $\Psi^a_{(1)}[g_c]$ of the
$g_c$'s are related to the functionals given in Eqs.(\ref{6.21}) by
the equations

\beq
 \Big(\partial_a\, \Psi^b_{(1)}[g_c] + \partial_b\, \Psi^a_{(1)}[g_c]\Big){|}_{a \not= b}
 = z_{ab}[g_c]{|}_{a \not= b}.
 \label{6.22}
 \eeq

 \medskip

The solution $\Psi^a_{(1)}[g_c]$ of this equation is
($\epsilon^2_{aef}$ is symmetric in $e$ and $f$ and vanishes if
$a=e$ or $a=f$)

 \bea
 \Psi_{(1)}^1[g_c](\tau, \vec \sigma)&=&{1\over 2}\, \int^{\sigma^2, \sigma^3}\,
  d{\tilde \sigma}^2\,d{\tilde \sigma}^3\,\Big(
 \,\partial_3\,z_{12}[g_c]+\partial_2\,z_{13}[g_c]-\partial_1\,z_{23}[g_c]
 \Big)(\tau, \sigma^1, {\tilde \sigma}^2, {\tilde \sigma}^3),\nonumber\\
 \Psi_{(1)}^2[g_c](\tau, \vec \sigma)&=&{1\over 2}\, \int^{\sigma^1, \sigma^3}
 d{\tilde \sigma}^1\,d{\tilde \sigma^3}\,\Big(
 \,\partial_3\,z_{23}[g_c]+\partial_2\,z_{12}[g_c]-\partial_1\,z_{12}[g_c]
 \Big)(\tau, {\tilde \sigma}^1, \sigma^2, {\tilde \sigma}^3),\nonumber\\
 \Psi_{(1)}^3[g_c](\tau, \vec \sigma)&=&{1\over 2}\, \int^{\sigma^1, \sigma^2}\,
  d{\tilde \sigma^1}\,d{\tilde \sigma^2}\,\Big(
 \,\partial_1\,z_{23}[g_c]+\partial_3\,z_{12}[g_c]-\partial_2\,z_{13}[g_c]
 \Big)(\tau, {\tilde \sigma}^1, {\tilde \sigma}^2, \sigma^3),\nonumber \\
 &&{}\nonumber \\
 &&or\nonumber \\
 &&{}\nonumber \\
 \Psi_{(1)}^a[g_c](\tau, \vec \sigma) &=& {1\over 2}\, \sum_{ef}\, (\epsilon_{aef})^2\,
 \int^{\sigma^e, \sigma^f}\, d{\tilde \sigma}^e\,d{\tilde \sigma}^f\,\Big(
 \,\partial_e\,z_{fa}[g_c]+\partial_f\,z_{ea}[g_c]-\partial_a\,z_{ef}[g_c]
 \Big)(\tau, \sigma^a, {\tilde \sigma}^e, {\tilde \sigma}^f).  \nonumber \\
 &&{}
 \label{6.23}
 \eea

\medskip

By using Eq.(\ref{6.21}) and $\sum_a\, g_a = 0$ we get the following
form of Eqs.(\ref{6.23})

\bea
 \Psi_{(1)}^a[g_c](\tau, \vec \sigma) &=&- \sum_{ef}\frac{(\epsilon_{aef})^2}{4}\,
 \int^{\sigma^e, \sigma^f}\, d{\tilde \sigma}^e
 d{\tilde \sigma}^f\, \int^{{\tilde \sigma}^e, {\tilde \sigma}^f}\,
 d{\hat \sigma}^e d{\hat \sigma}^f\,
  \int^{\sigma^a} d{\tilde \sigma}^a\nonumber \\
  && \Big[(\Delta-\partial^2_e)\Delta\,g_f+(\Delta-\partial^2_f)\Delta\,g_e-
 \partial_e^2\,\partial_f^2\,(g_e+g_f)\Big](\tau, {\tilde \sigma}^a,
 {\hat \sigma}^e, {\hat \sigma}^f).
 \label{6.24}
 \eea

\bigskip

Then by using Eqs.(\ref{6.17}) we get

\bea
 \Gamma_a^{(1)} &=& g_a - \partial_a\, \Psi_{(1)}^a[g_b] -
  {1\over 3}\, \sum_c\, \partial_c\, \Psi_{(1)}^c[g_b] =\nonumber \\
  &&{}\nonumber \\
  &=& \sum_{bc}\, {\tilde M}^{-1}_{ab}\,  {\tilde M}_{bc}\, \Gamma_c^{(1)}
  = \sum_b\, {\tilde M}^{-1}_{ab}\, g_b,
  \label{6.25}
  \eea

\noindent and this is a definition of the inverse operator ${\tilde
M}^{-1}_{ab}$ by means of Eq.(\ref{6.24}).

\bigskip

If $g_c \, \cir\, \rho_c$ is a solution of the wave equation $\Box\,
g_c \, \cir\, {{8\pi\, G}\over {c^3}}\, T_{(1)}^{(TT)cc}$ (see
Eqs.(\ref{6.11}) and (\ref{6.16})), then the associated tidal
variables (our physical radiative degrees of freedom replacing the
TT 3-metric of the standard approach in harmonic gauges) are

\beq
 R_{\bar a}\, =\, \sum_a\, \gamma_{\bar aa}\, \Gamma_a^{(1)}\,
 \cir\, \sum_{ab}\, \gamma_{\bar aa}\, {\tilde M}^{-1}_{ab}\, \rho_b
 = \sum_a\, \gamma_{\bar aa}\, \Big(\rho_a - \partial_a\,
 \Psi_{(1)}^a[\rho_b] - {1\over 3}\, \sum_c\, \partial_c\,
 \Psi^c_{(1)}[\rho_b]\Big),
 \label{6.26}
 \eeq

\noindent with the functional $\Psi^a_{(1)}[\rho_b]$ of
Eqs.(\ref{6.23}). Eqs.(\ref{6.4}), (\ref{6.10}), (\ref{6.11}) imply
that these tidal variables are solutions of the wave equations

\bea
 \Box\, \sum_b\, {\tilde M}_{ab}\, \Gamma_b^{(1)} &\cir&
 {{8\pi\, G}\over {c^3}}\,  T_{(1)}^{(TT) aa},\nonumber \\
 &&{}\nonumber \\
 \Box\, \sum_{\bar b}\, M_{\bar a\bar b}\, R_{\bar b} &\cir&
 {{8\pi\, G}\over {c^3}}\, \sum_a\, \gamma_{\bar aa}\,
 T_{(1)}^{(TT)aa}.
 \label{6.27}
 \eea

\medskip

In the next Section we will study   the solutions for the tidal
variables, because they are the {\it HPM gravitational waves with
asymptotic background} in our family of 3-orthogonal gauges with
non-Euclidean 3-spaces $\Sigma_{\tau}$. \medskip

From Eqs.(\ref{2.3}), (\ref{3.6}), and by using $\sum_{\bar a}\,
\gamma_{\bar aa}\, \gamma_{\bar ab} = \delta_{ab} - {1\over 3}$, we
get that the extrinsic curvature tensor of our 3-spaces in our
family of 3-orthogonal gauges is the following first order quantity

\bea
 {}^3K_{(1)rs}\, =\, \sigma_{(1)(r)(s)}{|}_{r \not= s} +
 \delta_{rs}\, \Big({1\over 3}\, {}^3K_{(1)} - \partial_{\tau}\, \Gamma_r^{(1)}
 + \partial_r\, {\bar n}_{(1)(r)} - \sum_a\, \partial_a\, {\bar n}_{(1)(a)}\Big),
 \nonumber \\
 &&{}
 \label{6.28}
 \eea

\noindent with ${\bar n}_{(1)(r)}$ and $\sigma_{(1)(r)(s)}{|}_{r
\not= s}$ given in Eqs.(\ref{4.16}) and (\ref{4.17}), respectively
(they depend on ${}^3{\cal K}_{(1)} = {1\over {\triangle}}\,
{}^3K$). Therefore, our (dynamically determined) 3-spaces have a
first order deviation from Euclidean 3-spaces, embedded in the
asymptotically flat space-time, determined by both instantaneous
inertial matter effects and retarded tidal ones. Moreover the
inertial gauge variable ${}^3K_{(1)}$ (non existing in Newtonian
gravity) is still free. From Eqs.(\ref{2.5}) we see that the
intrinsic 3-curvature of these non-Euclidean 3-spaces is ${}^3{\hat
R}{|}_{\theta^i=0} = 2\, \sum_a\,
\partial_a^2\, \Gamma_a^{(1)}$: it is determined only by the tidal
variables, i.e. by the HPM gravitational waves propagating inside
these 3-spaces.
\medskip

If we use the coordinate system of Eqs. (\ref{6.13}) (\ref{6.15}) to
go in the generalized TT gauge, we can introduce the standard
polarization pattern of gravitational waves for ${}^4h^{TT}_{rs}$
(see Refs.\cite{19,20,21}) and then the inverse transformation
allows to rewrite it in our family of 3-orthogonal gauges.

\vfill\eject

\section{Post-Minkowskian Gravitational Waves with Asymptotic Background in the
3-Orthogonal Gauges}

In this Section we study the solutions of the linearized equations
for the tidal variables $R_{\bar a}(\tau, \vec \sigma)$ and the TT
3-metric ${}^4h^{TT}_{(1)rs}(\tau, \vec \sigma)$, namely the PM
gravitational waves with asymptotic background in the family of
3-orthogonal gauges. When needed we assume the validity of
multipolar expansion of the energy-momentum tensor, which is
reviewed in Appendix B.

\subsection{The Retarded Solution}

Since in Eqs.(\ref{6.27}) we have the flat wave operator $\Box =
\partial^2_{\tau} - \triangle$ associated with the asymptotic
Minkowski 4-metric, we give the solution as a retarded integral over
the past flat null cone attached to the point $(\tau, \vec \sigma)$
on the instantaneous 3-space $\Sigma_{\tau}$ at time $\tau$ by using
the retarded Green function $G(\tau, \vec \sigma; \tau^{'}, {\vec
\sigma}^{'}) = - \theta(\tau - \tau^{'})\, {{\delta(\tau - |\vec
\sigma - {\vec \sigma}^{'}| - \tau^{'}])}\over {4\pi\, |\vec \sigma
- {\vec \sigma}^{'}|}}$ ($\Box\, G(\tau, \vec \sigma; \tau^{'},
{\vec \sigma}^{'}) = \delta(\tau^{'} - \tau)\, \delta^3(\vec \sigma
- {\vec \sigma}^{'})$).
\bigskip

The retarded solution of Eqs.(\ref{6.27}) is

\bea
 R_{\bar a}(\tau, \vec \sigma) &=& \sum_a\, \gamma_{\bar aa}\,
 \Gamma_a^{(1)}(\tau, \vec \sigma)
 \cir \sum_{ab}\, \gamma_{\bar aa}\, {\tilde M}^{-1}_{ab}(\tau, \vec \sigma)\, \Big(
 F_b^{TT(hom)}(\tau, \vec \sigma) -\nonumber \\
 &-& {{2\, G}\over {c^3}}\, \int d^3\sigma_1\, {{T^{(TT)bb}_{(1)}(\tau
 - |\vec \sigma - {\vec \sigma}_1|; {\vec \sigma}_1)}\over {
 |\vec \sigma - {\vec \sigma}_1|}} \Big),\nonumber \\
 &&{}\nonumber \\
 &&\Gamma^{(1)}_a = \sum_{\bar a}\, \gamma_{\bar aa}\, R_{\bar a},
 \label{7.1}
 \eea

\noindent where $F_a^{(hom)}(\tau, \vec \sigma)$ is a homogeneous
solution of the flat wave operator ($\Box\, F_a^{(hom)} = 0$) and
the non-local operator ${\tilde M}^{-1}_{ab}$ is defined by means of
Eq.(\ref{6.26}).

\bigskip

The condition of {\it no-incoming radiation} is $F^{(hom)}_a(\tau,
\vec \sigma) = 0$: it uses the flat light-cone of the asymptotic
Minkowski metric at $\tau \rightarrow - \infty$. Therefore there are
only outgoing gravitational waves.\medskip

With our matter (point particles plus the electro-magnetic field) we
do not need to solve the Hamilton equations independently outside
and inside the matter sources with the subsequent matching of the
solutions like it is usually done with compact matter sources. Even
if $T_{(1)}^{(TT)aa}$ is assumed to have compact support, what is
relevant in Eqs.(\ref{7.1}) is the behavior far from the support of
the non-local quantity determined by ${\tilde M}^{-1}_{ab}$ applied
to the retarded integral.
\bigskip

Since we have $T^{TT}_{(1)}{}_{ab}(\tau,\vec{\sigma}) = {\cal
P}_{abcd}\, T^{cd}_{(1)}(\tau,\vec{\sigma})$, with the operator
${\cal P}_{abcd} = {\cal P}_{abcd}(\vec \sigma)$ defined in
Eq.(\ref{6.7}), the chosen solution (\ref{7.1}) has the following
form

\bea
 R_{\bar a}(\tau, \vec \sigma) &\cir& - {{2G}\over {c^3}}\,
 \sum_{ab}\, \gamma_{\bar aa}\, {\tilde M}^{-1}_{ab}(\tau, \vec \sigma)\, \int
 d\tau_1\, d^3\sigma_1\, \theta(\tau - \tau_1)\, {{\delta(\tau - |\vec
 \sigma - {\vec \sigma}_1| - \tau_1])}\over {|\vec \sigma
 -{\vec \sigma}_1|}}\nonumber \\
 && \sum_{uv}\, {\cal P}_{bbuv}({\vec \sigma}_1)\,
 T^{uv}_{(1)}(\tau_1, {\vec \sigma}_1) =\nonumber \\
 &=& - {{2G}\over {c^3}}\, \sum_{ab}\, \gamma_{\bar aa}\, {\tilde M}^{-1}_{ab}(\tau, \vec
 \sigma)\, \int {{d^3\sigma_1}\over {|\vec \sigma - {\vec \sigma}_1|}}\,
 \Big( \sum_{uv}\, {\cal P}_{bbuv}({\vec \sigma}_1)\,
 T_{(1)}^{uv}(\tau, {\vec \sigma}_1)
 \Big)_{\tau \rightarrow \tau - |\vec \sigma - {\vec \sigma}_1|}.
 \label{7.2}
 \eea

\bigskip

Analogously the solution for the TT 3-metric ${}^4h^{TT}_{(1)ab}$,
satisfying the wave equation (\ref{6.11}),  is

\beq
 {}^4h^{TT}_{(1)rs}(\tau, \vec \sigma) \cir \sgn\, - {4\, G\over {c^3}}\,
 \int d^3\sigma_1\, {{ \Big( \sum_{uv}\, {\cal P}_{rsuv}({\vec \sigma}_1)\,
 T_{(1)}^{uv}(\tau, {\vec \sigma}_1)
 \Big)_{\tau \rightarrow \tau - |\vec \sigma - {\vec \sigma}_1|} }\over
 {|\vec \sigma - {\vec \sigma}_1|}}.
 \label{7.3}
 \eeq

\medskip

Let us remark that, since we have $\Box\, {\cal P}_{abcd} = {\cal
P}_{abcd}\, \Box$, another solution of Eqs.(\ref{6.11}) is (in
general it is a solution differing from Eq.(\ref{7.3}) by a
homogeneous solution of the wave equation)

\beq
  {}^4{\tilde h}^{TT}_{(1)rs}(\tau,\vec{\sigma})\, =\, - \sgn\,
 \frac{4G}{c^3}\, \sum_{uv}\, {\cal P}_{rsuv}(\vec \sigma)\, \int d^3\sigma_1\,
 \frac{T^{uv}_{(1)}(\tau - \mid\vec{\sigma} - \vec{\sigma}_1\mid,
 \vec{\sigma_1})}{\mid\vec{\sigma} - \vec{\sigma}_1\mid}.
 \label{7.4}
 \eeq

\bigskip

To compare Eq.(\ref{7.4}) with Eq.(\ref{7.3}) we need the following
integral representation of the operator ${\cal P}_{rsuv}(\vec
\sigma)$ \footnote{We have $\triangle\, {1\over {4\pi\, |\vec
\sigma|}} = - \delta^3(\vec \sigma)$, $\partial_r\, |\vec \sigma -
{\vec \sigma}_1| =  {{\sigma^r - \sigma_1^r}\over {|\vec \sigma -
{\vec \sigma}_1|}}$, $\partial_r\, {1\over {|\vec \sigma - {\vec
\sigma}_1|}} = {{\sigma^r - \sigma_1^r}\over {|\vec \sigma - {\vec
\sigma}_1|^3}}$, $\partial_r\, \partial_s\, {1\over {|\vec \sigma -
{\vec \sigma}_1|}} = {{\delta^{rs}\, |\vec \sigma - {\vec
\sigma}_1|^2 - 3\, (\sigma^r - \sigma_1^r)\, (\sigma^s -
\sigma^s_1)}\over {|\vec \sigma - {\vec \sigma}_1|^5}}$,
$\partial_r\, f(\tau - |\vec \sigma - {\vec \sigma}_1|) = -
{{\sigma^r - \sigma_1^r}\over {|\vec \sigma - {\vec \sigma}_1|}}\,
\partial_{\tau}\, f(\tau - |\vec \sigma - {\vec \sigma}_1|)$.}

\bea
 f^{TT}_{rs}(\tau,\vec{\sigma}) &=& \sum_{uv}\,{\cal P}_{rsuv}(\vec \sigma)\,
 f^{uv}(\tau,\vec{\sigma})  {\buildrel {def}\over =} \int
 d^3\sigma_1\, \sum_{uv} d^{TT}_{rsuv}(\vec{\sigma} - \vec{\sigma}_1) \cdot\,
 f^{uv}(\tau,\vec{\sigma_1}), \nonumber \\
 &&{}\nonumber \\
 d^{TT}_{rsuv}(\vec \sigma - {\vec \sigma}_1)&=& {1\over 2}\, \Big(
 (\delta_{ru}\, \delta_{sv} + \delta_{rv}\, \delta_{su} - \delta_{rs}\,
 \delta_{uv})\, \delta^3(\vec \sigma - {\vec \sigma}_1) +\nonumber \\
 &+&\sum_{ab}\, \Big[\delta_{ua}\, (\delta_{rv}\, \delta_{sb} + \delta_{sv}\,
 \delta_{rb}) + \delta_{va}\, (\delta_{ru}\, \delta_{sb} + \delta_{su}\,
 \delta_{rb}) - \delta_{ra}\, \delta_{sb}\, \delta_{uv} \Big]\nonumber \\
 &&{{\delta^{ab}\, |\vec \sigma - {\vec \sigma}_1|^2 - 3\, (\sigma^a -
 \sigma^a_1)\, (\sigma^b - \sigma^b_1)}\over {4\pi\, |\vec \sigma -
 {\vec \sigma}_1|^5}} +\nonumber \\
 &+& \delta_{rs}\, \int d^3\sigma_2\, {{\delta^{rs}\, |\vec \sigma -
 {\vec \sigma}_2|^2 - 3\, (\sigma^r - \sigma^r_2)\, (\sigma^s -
 \sigma^s_2)}\over {4\pi\, |\vec \sigma - {\vec \sigma}_2|^5}}
 \nonumber \\
 &&{{\delta^{uv}\, |{\vec \sigma}_2 -
 {\vec \sigma}_1|^2 - 3\, (\sigma^u_2 - \sigma^u_1)\, (\sigma^v_2 -
 \sigma^v_1)}\over {4\pi\, |{\vec \sigma}_2 - {\vec \sigma}_1|^5}}\, \Big),
 \nonumber \\
 &&{}
 \label{7.5}
 \eea

\noindent where the integral kernel $d^{TT}_{abcd}$ is a function
only of the difference $\vec \sigma - {\vec \sigma}_1$.

\bigskip

As a consequence, the two retarded solutions (\ref{7.3}) and
(\ref{7.4})  for the TT 3-metric with the no-incoming radiation
condition take the form

 \beq
 {}^4h^{TT}_{(1)}{}_{rs}(\tau,\vec{\sigma}) =
 - \sgn\, \frac{4G}{c^3}\, \int d^3\sigma_1\, \int d^3\sigma_2\,
 \sum_{uv}\, d^{TT}_{rsuv}(\vec{\sigma}_1 - \vec{\sigma}_2)\,
 \frac{T^{uv}_{(1)}(\tau - \mid\vec{\sigma} - \vec{\sigma}_1\mid,
 \vec{\sigma}_2)}{\mid\vec{\sigma} - \vec{\sigma}_1\mid},
 \label{7.6}
 \eeq

\beq
 {}^4{\tilde h}^{TT}_{(1)}{}_{rs}(\tau,\vec{\sigma}) =
 - \sgn\, \frac{4G}{c^3}\, \int d^3\sigma_2\, \sum_{uv}\,
 d^{TT}_{rsuv}(\vec{\sigma} - \vec{\sigma}_2)\, \int d^3\sigma_1\,
 \frac{T^{uv}_{(1)}(\tau - \mid\vec{\sigma}_2 -
 \vec{\sigma}_1\mid,\vec{\sigma}_1)}{\mid\vec{\sigma}_2
 - \vec{\sigma}_1\mid},
 \label{7.7}
 \eeq

\noindent respectively. To make the comparison either an explicit
form of the energy-momentum tensor or its multipolar expansion is
needed.

\subsection{PM Gravitational Waves with Asymptotic Background:
Multipolar Expansion and the Quadrupole Emission Formula}

To look for the PM relativistic quadrupole formula we need the
multipolar expansion of the energy-momentum tensor in terms of
relativistic Dixon multipoles expressed in our rest-frame instant
form of dynamics. In Appendix B there is a review of such multipoles
based upon Ref.\cite{22}. See chapter 3 of Ref.\cite{20} for a
review of the standard multipole expansions used in the literature
(see Refs. \cite{23,24}).

\subsubsection{The multipolar expansion for the tidal variables $R_{\bar a}$ }

By using the multipolar expansion (\ref{a2}) centered on the center
of energy $w^{\mu}_E(\tau) = z^{\mu}(\tau, 0)$ ($\vec \eta(\tau) =
0$) and the operator ${\cal P}_{rsuv}$ defined in Eq.(\ref{6.7}),
from Eq.(\ref{7.2}) we get by making integrations by parts (assumed
valid)

\begin{eqnarray*}
 &&- {{2\, G}\over {c^3}}\, \int d^3\sigma_1\, {{T^{(TT)bb}_{(1)}(\tau
 - |\vec \sigma - {\vec \sigma}_1|; {\vec \sigma}_1)}\over {
 |\vec \sigma - {\vec \sigma}_1|}} =\nonumber \\
 &&{}\nonumber \\
 &&= - {{2\pi\, G}\over {c^3}}\, \sum_{n=0}^{\infty}\, {{(-)^n}\over {n!}}\,
 \sum_{uvr_1..r_n}\, \int d^3\sigma_1\, {{q^{r_1..r_n | uv}(\tau -
 |\vec \sigma - {\vec \sigma}_1|)}\over {|\vec \sigma - {\vec \sigma}_1|}}
 \nonumber \\
 &&\Big(\delta_{au}\, \delta_{av} - {1\over 2}\, (1 - {{\partial_{1a}^2}\over
 {\triangle_1}})\, \delta_{uv} + {1\over 2}\, (1 + {{\partial^2_{1a}}\over
 {\triangle_1}})\, {{\partial_{1u}\, \partial_{1v}}\over {\triangle_1}}
 -\nonumber \\
 &&- (\delta_{au}\, {{\partial_{1v}}\over {\triangle_1}} + \delta_{av}\,
 {{\partial_{1u}}\over {\triangle_1}})\, \partial_{1a} \Big)\, {{\partial^n\,
 \delta^3({\vec \sigma}_1)}\over {\partial\, \sigma^{r_1}_1\,
 ... \partial\, \sigma^{r_n}_1}} =
 \end{eqnarray*}

 \bea
 &=& - {{2\, G}\over {c^3}}\, \sum_{n=0}^{\infty}\, {{(-)^n}\over {n!}}\,
 \sum_{r_1..r_n}\, \Big[\Big(\partial_{1r_1}...\partial_{1r_n}\,
 {{q^{r_1..r_n | aa}(\tau - |\vec \sigma - {\vec \sigma}_1|)}\over
 {|\vec \sigma - {\vec \sigma}_1|}}\Big){|}_{{\vec \sigma}_1 = 0} -
 \nonumber \\
 &-& {1\over 2}\, \sum_u\, \Big(\partial_{1r_1}...\partial_{1r_n}\,
 {{q^{r_1..r_n | uu}(\tau - |\vec \sigma - {\vec \sigma}_1|)}\over
 {|\vec \sigma - {\vec \sigma}_1|}}\Big){|}_{{\vec \sigma}_1 = 0}
 +\nonumber \\
 &+& {1\over 2}\, \sum_{uv}\, \Big(\partial_{1r_1}...\partial_{1r_n}\,
 {{q^{r_1..r_n | uv}(\tau - |\vec \sigma - {\vec \sigma}_1|)}\over
 {|\vec \sigma - {\vec \sigma}_1|}}\Big){|}_{{\vec \sigma}_1 = 0}
 -\nonumber \\
 &-& {1\over 2}\, \sum_u\, \int {{d^3\sigma_2}\over {4\pi\, |{\vec \sigma}_2|}}\,
 \partial^2_{2a}\, \partial_{2r_1}...\partial_{2r_n}\,
 {{q^{r_1..r_n | uu}(\tau - |\vec \sigma - {\vec \sigma}_2|)}\over
 {|\vec \sigma - {\vec \sigma}_2|}} -\nonumber \\
 &-& {1\over 2}\, \sum_{uv}\, \int {{d^3\sigma_2}\over {4\pi\, |{\vec \sigma}_2|}}\,
 \partial^2_{2a}\, \partial_{2r_1}...\partial_{2r_n}\,
 {{q^{r_1..r_n | uv}(\tau - |\vec \sigma - {\vec \sigma}_2|)}\over
 {|\vec \sigma - {\vec \sigma}_2|}} -\nonumber \\
 &-&2\, \sum_u\, \int {{d^3\sigma_2}\over {4\pi\, |{\vec \sigma}_2|}}\,
  \int {{d^3\sigma_3}\over {4\pi\, |{\vec \sigma}_2 - {\vec \sigma}_3|}}\,
 \partial_{3a}\, \partial_{3u}\,  \partial_{3r_1}...\partial_{3r_n}\,
 {{q^{r_1..r_n | au}(\tau - |\vec \sigma - {\vec \sigma}_3|)}\over
 {|\vec \sigma - {\vec \sigma}_3|}} \Big].
 \label{7.8}
 \eea

The $n = 0$ term has the following expression

\begin{eqnarray*}
 && - {{2\, G}\over {c^3}}\, \Big[{{q^{| aa}(\tau - |\vec \sigma|) - {1\over 2}\,
 \sum_u\, q^{| uu}(\tau - |\vec \sigma|) + {1\over 2}\, \sum_{uv}\,
 q^{| uv}(\tau - |\vec \sigma|) }\over {|\vec \sigma|}} -\nonumber \\
 &-& {1\over 2}\, \int {{d^3\sigma_2}\over {4\pi\, |{\vec \sigma}_2|}}\,
 \partial^2_{2a}\, {{\sum_u\, q^{| uu}(\tau - |\vec \sigma -
 {\vec \sigma}_2|) + \sum_{uv}\, q^{| uv}(\tau - |\vec \sigma -
 {\vec \sigma}_2|) }\over {|\vec \sigma - {\vec \sigma}_2|}}
 -\nonumber \\
 &-& 2\, \sum_u\, \int {{d^3\sigma_2}\over {4\pi\, |{\vec \sigma}_2|}}\,
  \int {{d^3\sigma_3}\over {4\pi\, |{\vec \sigma}_2 - {\vec \sigma}_3|}}\,
 \partial_{3a}\, \partial_{3u}\, {{ q^{| uu}(\tau - |\vec \sigma -
 {\vec \sigma}_3|) }\over {|\vec \sigma - {\vec \sigma}_3|}} =
 \end{eqnarray*}

\bea
 &=& - {G\over {c^3}}\, \Big[{{\partial^2_{\tau}\, q^{aa| \tau\tau}(\tau
 - |\vec \sigma|) - {1\over 2}\, \sum_u\, \partial^2_{\tau}\,
 q^{uu | \tau\tau}(\tau - |\vec \sigma|) + {1\over 2}\, \sum_{uv}\,
 \partial^2_{\tau}\, q^{uv | \tau\tau}(\tau - |\vec \sigma|) }\over
 {|\vec \sigma|}} -\nonumber \\
 &-& {1\over 2}\, \int {{d^3\sigma_2}\over {4\pi\, |{\vec \sigma}_2|}}\,
 \partial^2_{2a}\, {{\sum_u\, \partial^2_{\tau}\, q^{uu | \tau\tau}(\tau - |\vec \sigma -
 {\vec \sigma}_2|) + \sum_{uv}\, \partial^2_{\tau}\, q^{uv | \tau\tau}(\tau - |\vec \sigma -
 {\vec \sigma}_2|) }\over {|\vec \sigma - {\vec \sigma}_2|}}
 -\nonumber \\
 &-& 2\, \sum_u\, \int {{d^3\sigma_2}\over {4\pi\, |{\vec \sigma}_2|}}\,
  \int {{d^3\sigma_3}\over {4\pi\, |{\vec \sigma}_2 - {\vec \sigma}_3|}}\,
 \partial_{3a}\, \partial_{3u}\, {{ \partial^2_{\tau}\, q^{uu | \tau\tau}(\tau
 - |\vec \sigma - {\vec \sigma}_3|) }\over {|\vec \sigma - {\vec \sigma}_3|}}
  =\nonumber \\
 &&{}\nonumber \\
 &=& - {G\over {c^3}}\, {{\sum_{uv}\, {\cal P}_{aauv}\, \partial^2_{\tau}\,
 q^{uv | \tau\tau}(\tau - |\vec \sigma|)}\over {|\vec \sigma|}} =
 - {G\over {c^3}}\, {{\partial^2_{\tau}\, q^{(TT) aa
 | \tau\tau}(\tau - |\vec \sigma|)}\over {|\vec \sigma|}},\nonumber \\
 &&{}\nonumber \\
 \Rightarrow&& R_{\bar a}(\tau, \vec \sigma) = -
 {G\over {c^3}}\, \sum_{ab}\, \gamma_{\bar aa}\, {\tilde
 M}^{-1}_{ab}\, {{\partial^2_{\tau}\, q^{(TT) aa
 | \tau\tau}(\tau - |\vec \sigma|)}\over {|\vec \sigma|}} +
 (higher\, multipoles).\nonumber \\
 &&{}
  \label{7.9}
  \eea

\noindent In the last expression we used Eq.(\ref{a13}) and
(\ref{a14}) and $\sum_r\, {\cal P}_{rruv} = 0$. The last line used
Eq.(\ref{6.26}).

\bigskip

The first term of Eq.(\ref{7.9}) gives the standard {\it quadrupole
emission} formula modified by the non-local operator ${\tilde
M}^{-1}_{ab}$ of Eq.(\ref{6.26}). The higher terms would give the
contribution of the mass octupole, momentum quadrupole and so on
(see Section 3.4 of Ref.\cite{20}).

\subsubsection{The multipolar expansion for the TT 3-metric  ${}^4h^{TT}_{(1)rs}$}

By putting in the solutions (\ref{7.6}) and (\ref{7.7}) the
multipolar expansion (\ref{a2}) with $\eta^r(\tau) = 0$ we get

 \bea
 {}^4h^{TT}_{(1)}{}_{rs}(\tau,\vec{\sigma})
 &=&- \sgn\, \frac{4G}{c^3}\,\sum_{n=0}^\infty\,\frac{(-)^n}{n!} \int
 d^3\sigma_1\, \sum_{r_1..r_nuv}\, \frac{q^{r_1...r_n\mid
 uv}(\tau - \mid\vec{\sigma} - \vec{\sigma}_1\mid)
 }{\mid\vec{\sigma} - \vec{\sigma}_1\mid}\nonumber \\
 &&\frac{\partial^n}{\partial\sigma^{r_1}_1...
 \partial\sigma^{r_n}_1}d^{TT}_{rsuv}(\vec{\sigma}_1).
 \label{7.10}
 \eea

\bea
 {}^4{\tilde h}^{TT}_{(1)}{}_{rs}(\tau,\vec{\sigma})
 &=&- \sgn\, \frac{4G}{c^3}\sum_{n=0}^\infty\,\frac{(-)^n}{n!} \int
 d^3\sigma_2\, \sum_{uv}\, d^{TT}_{rsuv}(\vec{\sigma} -
 \vec{\sigma}_2)\,
 \sum_{r_1..r_n}\, \frac{\partial^n}{\partial\sigma^{r_1}_2...\partial\sigma^{r_n}_2}
 \, \left( \frac{q^{r_1...r_n\mid uv}(\tau - \mid\vec{\sigma}_2\mid)
 }{\mid\vec{\sigma}_2\mid} \right) =\nonumber \\
 &=& {}^4h^{TT}_{(1)}{}_{rs}(\tau,\vec{\sigma}).
 \label{7.11}
 \eea

These two expressions can be shown to coincide with the change of
variable $\vec{\sigma}-\vec{\sigma}_1=\vec{\sigma}_2$
($d^3\sigma_1=d^3\sigma_2$) and by making suitable integrations by
parts (we assume that the integrations by parts can be done).
\medskip

We shall use Eq.(\ref{7.11}), for whose manipulation we need the
formula

\bea
 &&\frac{\partial^n}{\partial\sigma^{r_1}_2...\partial\sigma^{r_n}_2}
 \left( \frac{q^{r_1...r_n\mid uv}(\tau-\mid\vec{\sigma}_2\mid)
 }{\mid\vec{\sigma}_2\mid}
 \right)=\nonumber \\
 &&{}\nonumber \\
 &=&\sum_{k=0}^n\,\frac{n!}{k!(n-k)!}\,
 \frac{\partial^k}{\partial\sigma^{r_1}_2...\partial\sigma^{r_k}_2}
 \left(\frac{1}{\mid\vec{\sigma}_2\mid}\right) \,
 \frac{\partial^{n-k} q^{r_1...r_n\mid
 uv}(\tau-\mid\vec{\sigma}_2\mid)
 }{\partial\sigma^{r_{k+1}}_2...\partial\sigma^{r_n}_2}=\nonumber \\
 &&{}\nonumber \\
 &=&\sum_{k=0}^n\, \frac{n!}{k!(n-k)!}\,(-)^{n-k}
 \frac{\partial^k}{\partial\sigma^{r_1}_2...\partial\sigma^{r_k}_2}
 \left(\frac{1}{\mid\vec{\sigma}_2\mid}\right) \,
 \partial^{n-k}_\tau
 q^{r_1...r_n\mid uv}(\tau-\mid\vec{\sigma}_2\mid)
 \,n_{(2)\,r_{k+1}}...n_{(2)\,r_n},\nonumber \\
 &&{}\nonumber \\
 &&{}\nonumber \\
 &&n_r=\frac{\sigma^r}{\mid\vec{\sigma}\mid},\qquad
 n_{(2)\,r}=\frac{\sigma^r_2}{\mid\vec{\sigma}_2\mid}.
 \label{7.12}
 \eea

Therefore Eq.(\ref{7.11}) can be put in the form

\bea
 {}^4h^{TT}_{(1)}{}_{rs}(\tau,\vec{\sigma})
 &=&- \sgn\, \frac{4G}{c^3}\, \sum_{n=0}^\infty\,\sum_{k=0}^n\, \frac{(-)^{2n-k}}{k!(n-k)!}\,
 \int d^3\sigma_2\, \sum_{uv}\, d^{TT}_{rsuv}(\vec{\sigma}-\vec{\sigma}_2)\times\nonumber\\
 &&\nonumber\\
 &&\sum_{r_1..r_n}\,\frac{\partial^k}{\partial\sigma^{r_1}_2...\partial\sigma^{r_k}_2}
 \left(\frac{1}{\mid\vec{\sigma}_2\mid}\right) \,
 \partial^{n-k}_\tau
 q^{r_1...r_n\mid uv}(\tau-\mid\vec{\sigma}_2\mid)
 \,n_{(2)\,r_{k+1}}...n_{(2)\,r_n}.\nonumber \\
 &&{}
 \label{7.13}
 \eea

To study the behavior of Eq.(\ref{7.13}) at big distances, i.e. $r =
 | \vec \sigma | >> 1$, we use the following results shown in
Appendix B of Ref.\cite{25}

\bea
 && f^{uv}(\vec{\sigma})\mapsto\frac{A^{uv}}{r}\,\Rightarrow\,
 f^{TT}{}_{rs}(\vec{\sigma})=\int
 d^3\sigma_2\,\sum_{uv}\, d^{TT}_{rsuv}(\vec{\sigma} - \vec{\sigma}_2)\,f^{uv}(\vec{\sigma}_2)
 \mapsto\,\frac{B_{rs}}{r}+{\cal O}(1/r^3),\nonumber \\
 &&{}\nonumber \\
 && f^{uv}(\vec{\sigma})\mapsto\frac{A^{uv}}{r^2}\,\Rightarrow\,
 f^{TT}{}_{rs}(\vec{\sigma}) = \int d^3\sigma_2\, \sum_{uv}\, d^{TT}_{rsuv}(\vec{\sigma}
 - \vec{\sigma}_2)\,f^{uv}(\vec{\sigma}_2)\mapsto\,\frac{B_{rs}}{r^2}+{\cal O}(1/r^3),
 \nonumber \\
 &&{}\nonumber \\
 && f^{uv}(\vec{\sigma})\mapsto\frac{A^{uv}}{r^n}\,\Rightarrow\,
 f^{TT}{}_{rs}(\vec{\sigma}) = \int d^3\sigma_2\, \sum_{uv}\, d^{TT}_{rsuv}(\vec{\sigma}
 - \vec{\sigma}_2)\,f^{uv}(\vec{\sigma}_2)\mapsto {\cal O}(1/r^3),\qquad n\ge
 3.\nonumber \\
 &&{}
 \label{7.14}
 \eea

As a consequence at great distance only the terms with $k = 0$ give
the dominant contribution in Eq.(\ref{7.13})

 \bea
 {}^4h^{TT}_{(1)}{}_{rs}(\tau,\vec{\sigma})
 &=& - \sgn\, \frac{4G}{c^3}\, \sum_{n=0}^\infty\, \int
 d^3\sigma_2\, \sum_{uv}\, d^{TT}_{rsuv}(\vec{\sigma} - \vec{\sigma}_2)\nonumber \\
 &&\sum_{r_1..r_n}\, n_{(2)\,r_1}...n_{(2)\,r_n}
 \, \frac{\partial^n_\tau q^{r_1...r_n\mid
 uv}(\tau - \mid\vec{\sigma}_2\mid) }{\mid\vec{\sigma}_2\mid} +
 O(1/r^2).
 \label{7.15}
 \eea

As shown in Refs. \cite{19} (see also Ref. \cite{24}) we have the
following result

\bea
 f^{uv}(\vec{\sigma})&=& f^{uv}(\mid\vec{\sigma}\mid),\nonumber \\
 &&{}\nonumber \\
 &&\Downarrow\nonumber \\
 &&{}\nonumber \\
 f^{TT}{}_{rs}(\vec{\sigma}) &=& \int d^3\sigma_2\, \sum_{uv}\, d^{TT}_{rsuv}(\vec{\sigma}
 - \vec{\sigma}_2)\,f^{uv}(\mid\vec{\sigma}_2\mid) = \sum_{uv}\, \Lambda_{rsuv}(n)\,
 f^{uv}(\mid\vec{\sigma}\mid) + O(1/r^2),\nonumber \\
 &&{}
 \label{7.16}
 \eea

\noindent where $\Lambda_{abcd}(n)$ is the algebraic projector
(defined after Eq.(\ref{6.7}) for plane wave solutions)

\bea
 &&\Lambda_{abcd}(n) = (\delta_{ac} - n_a\,n_c)\, (\delta_{bd} - n_b\,n_c)
 - \frac{1}{2}\, (\delta_{ab} - n_a\,n_b)\, (\delta_{cd} - n_c\,n_d),
 \qquad n_r = \frac{\sigma^r}{\mid\vec{\sigma}\mid}.\nonumber \\
 &&{}
 \label{7.17}
 \eea

Therefore at great distances we can write

 \bea
 &&{}^4h^{TT}_{(1)}{}_{rs}(\tau,\vec{\sigma})
 = - \sgn\, \frac{4G}{c^3}\,\sum_{uv}\, \Lambda_{rsuv}(n)\,\sum_{n=0}^\infty\,
 \sum_{r_1..r_n}\, n_{r_1}...n_{r_n} \, \frac{\partial^n_\tau q^{r_1...r_n\mid
 uv}(\tau - \mid\vec{\sigma}\mid) }{\mid\vec{\sigma}\mid} +
 O(1/r^2).\nonumber \\
 &&{}
 \label{7.18}
 \eea

Eq.(\ref{7.18}) coincides with Eq.(3.34) of Ref.\cite{20}
\footnote{In Ref.\cite{20} it is shown that for small velocities
inside the source of gravitational waves the temporal derivatives of
the stress tensor multipoles are negligible giving terms of order
$O(v^2/c^2)$.} and the solution can be put in the form

 \bea
 {}^4h^{TT}_{(1)}{}_{rs}(\tau,\vec{\sigma})
 &=& - \sgn\, \frac{4G}{c^3}\, \sum_{uv}\, \Lambda_{rsuv}(n) \, \frac{ q^{\mid
 uv}(\tau - \mid\vec{\sigma}\mid) }{\mid\vec{\sigma}\mid} +  \nonumber \\
 &+&(higher\, multipoles) +  O(1/r^2).
  \label{7.19}
  \eea

Then Eq.(\ref{a13}), i.e. $q^{|uv} = {1\over 2}\,
\partial^2_{\tau}\, q^{uv|\tau\tau}$, leads to the standard quadrupole emission
formula. Again one could evaluate the contribution of higher
multipoles (mass octupole, momentum quadrupole,...) as in
Ref.\cite{20}.

\subsubsection{The Far Field of Time-Dependent Sources}

Let us look at the far field expression of our linearized 4-metric,
given in Eqs.(\ref{6.12}) in 3-orthogonal gauges.

\bigskip

By assuming matter sources with compact support and by using the
multipolar expansion of Appendix B and Eq.(\ref{7.9}), this equation
can be rewritten in the following form in the far wave zone ($r =
|\vec \sigma|$)

\bea
 {}^4g_{(1)\tau\tau} &=& \sgn\, \Big[1 - {{{\cal A}}\over r} -
 2\, {{\partial_{\tau}}\over {\triangle}}\, {}^3K_{(1)}\Big],
 \qquad {\cal A} = {{8\pi\, G}\over {c^3}}\, \Big(M_{(1)}c +
 \sum_i\, \eta_i\, {{{\vec \kappa}_i^2}\over {\sqrt{m_i^2c^2 +
 {\vec \kappa}_i^2}}}\Big),\nonumber \\
 &&{}\nonumber \\
 {}^4g_{(1)\tau r} &=& - \sgn\, \Big[{\cal N}_r + {{\partial_r}\over {\triangle}}\,
 {}^3K_{(1)}\Big],\nonumber \\
 &&\qquad {\cal N}_r = - {{4\pi\, G}\over {c^3}}\, {{(\vec \sigma \times
 {\vec j}_{(1)})^r}\over {r^3}} + [\partial_{\tau}\, and\, \partial_{\tau}^2\,
 (mass\, quadrupole) +\nonumber \\
 &&\qquad +(higher\, multipoles)],\nonumber \\
 &&{}\nonumber \\
 {}^4g_{(1)rs} &=& - \sgn\, \delta_{rs}\, \Big[1 + {{8\pi\, G}\over
 {c^3}}\, {{M_{(1)}c}\over r} + [\partial_{\tau}^2\, (mass\, quadrupole)
 + (higher\, multipoles)]\Big],\nonumber \\
 &&{}
 \label{7.20}
 \eea

\noindent with the two asymptotic Poincare' charges $M_{(1)}c$ and
${\vec j}_{(1)}$ given in Eqs.(\ref{4.21})  and Eq.(\ref{4.23})
respectively. Eqs.(\ref{a7}) and (\ref{a12}) have been used to get
the result for ${}^4g_{(1)\tau r}$. The last term in ${\cal A}$ has
been evaluated by omitting the electro-magnetic field and is
negligible in the non-relativistic limit. The shift function  ${\cal
N}_r$ gives the description of gravito-magnetism, Lense-Thirring
effect included (see for instance chapter 6 of Ref.\cite{2}), in the
non-harmonic 3-orthogonal gauges.
\medskip

We see that the results in the 3-orthogonal gauges are of the same
type as in the standard harmonic gauges as can be seen by comparing
Eq.(\ref{7.20}) with Eqs. (13.30) and (13.32)   of Ref. \cite{26}.
The only new terms are those involving the inertial gauge variable
${}^3{\cal K}_{(1)} = {1\over {\triangle}}\, {}^3K_{(1)}$.

\bigskip

Eqs.(\ref{7.20}) are compatible with the (direction-independent)
boundary conditions at spatial infinity for the 4-metric of our
class of asymptotically flat space-times, given after Eqs. (2.7) of
paper I (see also Eqs.(5.5) of Ref.\cite{5}).

\subsection{PM Gravitational Waves with Asymptotic Background:
the Energy Balance}

Having found the GW's of the HPM linearization of gravity, the next
step is to check the energy balance: the energy emitted by matter in
the form of GW's must be present in the gravitational field and
there should be a back-reaction on matter.\medskip

In the standard approach with compact sources there are various way
to evaluate the energy balance:\medskip

1) One can introduce the Landau-Lifschitz energy-momentum
pseudo-tensor $t^{\mu\nu}_{LL}$ of the gravitational field and use
the conservation law $\partial_{\mu}\, [- {}^4g\, (T^{\mu\nu} +
t_{LL}^{\mu\nu})] \cir 0$ to evaluate $dE/dt$ in the far-field zone
by using the quadrupolar approximation for GW's (see for instance
Ref.\cite{26}). This method is also used in the MPM + MPN approach
of Refs.\cite{23,27}, where the back-reaction (starting at the 2.5PN
order) of the GW's on the source can be taken into account till
3.5PN.\medskip

2) The same results are obtained with ADM Hamiltonian methods in
Ref. \cite{28} by taking a time-average of the work done by the
quadrupole radiation-reaction force appearing in the Hamilton
equations of the particles. Here it is emphasized the analogy with
electro-dynamics due to the appearance of the analogue of the Schott
term and it shown that there are no runaway solutions.\medskip

3) Instead in Ref.\cite{20} the coarse-graining method is used to
find an effective energy-momentum tensor for the gravitational field
at the second order, from which the increase in the energy of the
gravitational field due to the emission of GW is evaluated.\medskip

All these methods give the same result. The complications come from
the problems of regularization of the gravitational self-force
\cite{16,17} in the evaluation of the back reaction.\bigskip

Here we will show that we can recover the standard result without
having the gravitational self-force, due to the Grassmann
regularization ($\eta_i^2\, m_i^2 = 0$) of the gravitational
self-energies, by using the conservation of the ADM energy
(\ref{4.21}). The effect will result from interference terms
$\eta_i\, m_i\, \eta_j\, m_j$ with $i \not= j$ like it happens for
the Larmor formula of the electro-magnetic case if Grassmann-valued
electric charges are used to regularize the electro-magnetic
self-energies \cite{29}. For this calculation we consider only point
particles as matter ignoring the electro-magnetic field, because we
want to make a comparison with the treatments with compact sources
(the calculations with the electro-magnetic field should add the
assumption that such a field is localized in compact regions).
\bigskip

By making some integrations by parts (assumed valid with our
boundary conditions), the ADM energy (\ref{4.21}) has the form
(${\cal M}_{(2)}^{(UV)}$ is given in Eq.(\ref{3.12}))

\bea
  {1\over c}\, {\hat E}_{ADM} &=& \int d^3\sigma\,
  \Big({\cal M}^{(UV)}_{(1)} + {\cal M}^{(UV)}_{(2)} +
 {{2\pi\, G}\over {c^3}}\,  {\cal M}^{(UV)}_{(1)}\,
 {1\over {\triangle}}\, {\cal M}^{(UV)}_{(1)} -\nonumber \\
 &-&\sum_c\,  {\cal M}^{(UV)}_{(1)c}\,
 \Big[{{\partial_c}\over {\triangle}}\, {}^3K_{(1)} -
 {{8\pi\, G}\over {c^3}}\, {1\over {\triangle}}\, \Big(
 {\cal M}^{(UV)}_{(1)c} - {1\over 4}\, \sum_d\, {{\partial_c\,
 \partial_d}\over {\triangle}}\, {\cal M}^{(UV)}_{(1)d}\Big)\Big]
  +\nonumber \\
 &+& {{c^3}\over {16\pi G}}\, \sum_{\bar a\bar b}\,
  \Big[\partial_{\tau}\, R_{\bar a}\, M_{\bar a\bar b}\,
 \partial_{\tau}\, R_{\bar b} + \sum_a\, \partial_a\,
 R_{\bar a}\, M_{\bar a\bar b}\, \partial_a\, R_{\bar b}\Big] \,
 \Big)(\tau, \vec \sigma) + O(\zeta^3).\nonumber \\
 &&{}
 \label{7.21}
 \eea

Since we have:\hfill\break

1) $R_{\bar a} = \sum_a\, \gamma_{\bar aa}\, \Gamma_a^{(1)}$
($\sum_r\, \Gamma_r^{(1)} = 0$);\hfill\break

2) ${\tilde M}_{ab} = \sum_{\bar a\bar b}\, \gamma_{\bar aa}\,
\gamma_{\bar bb}\, M_{\bar a\bar b}$ from
Eq.(\ref{6.4});\hfill\break

3) $\sum_b\, {\tilde M}_{ab}\, \Gamma_b^{(1)} = - {{\sgn}\over 2}\,
{}^4h^{TT}_{(1)aa}$ from Eq.(\ref{6.8});\hfill\break

4) $\delta_{rs}\, \Gamma_r^{(1)} = - {{\sgn}\over 2}\,
\Big({}^4h_{(1)rs} - {1\over 3}\, \delta_{rs}\, \sum_v\,
{}^4h_{(1)vv}\Big)$ from ${}^4h_{(1)rs} = - 2\, \sgn\, \delta_{rs}\,
(\Gamma_r^{(1)} + 2\, \phi_{(1)})$ with $2\, \phi_{(1)} = -
{{\sgn}\over 6}\, \sum_v\, {}^4h_{(1)vv}$;\hfill\break

5) $\delta_{rs}\, \Gamma_r^{(1)} = - {{\sgn}\over 2}\,
\Big({}^4h_{(1)rs}^{TT} + {1\over 2}\, (\partial_r\, \epsilon_{(1)s}
+ \partial_s\, \epsilon_{(1)r}) + (\partial_r\, \partial_s - {1\over
3}\, \delta_{rs}\, \triangle)\, \lambda_{(1)}\Big)$ from
Eq.(\ref{6.5});\hfill\break

6) $\sum_r\, {}^4h^{TT}_{(1)rr} = \sum_r\,
\partial_r\, {}^4h^{TT}_{(1)rs} = 0$;\hfill\break

\noindent  the term bilinear in the gradients of $R_{\bar a}$ in
Eq.(\ref{7.21}) can be manipulated in such a way that, after an
integration by parts, the final form of Eq.(\ref{7.21}) becomes

\bea
   {1\over c}\, {\hat E}_{ADM} &=& \int d^3\sigma\,
  \Big({\cal M}^{(UV)}_{(1)} + {\cal M}^{(UV)}_{(2)} +
 {{2\pi\, G}\over {c^3}}\,  {\cal M}^{(UV)}_{(1)}\,
 {1\over {\triangle}}\, {\cal M}^{(UV)}_{(1)} -\nonumber \\
 &-&\sum_c\,  {\cal M}^{(UV)}_{(1)c}\,
 \Big[{{\partial_c}\over {\triangle}}\, {}^3K_{(1)} -
 {{8\pi\, G}\over {c^3}}\, {1\over {\triangle}}\, \Big(
 {\cal M}^{(UV)}_{(1)c} - {1\over 4}\, \sum_d\, {{\partial_c\,
 \partial_d}\over {\triangle}}\, {\cal M}^{(UV)}_{(1)d}\Big)\Big]
  +\nonumber \\
 &+& {{c^3}\over {64\pi G}}\, \sum_{rs}\,
  \Big[\Big(\partial_{\tau}\, {}^4h^{TT}_{(1)rs}\Big)^2
 + \sum_c \Big(\partial_c\, {}^4h^{TT}_{(1)rs}\Big)^2 \Big] \,
 \Big)(\tau, \vec \sigma) + O(\zeta^3) =\nonumber \\
 &{\buildrel {def}\over =}& {1\over c}\, \int d^3\sigma\,
 \rho_{E(1+2)}(\tau, \vec \sigma) + O(\zeta^3),
 \label{7.22}
 \eea

\noindent where in the last line we introduced the (coordinate
dependent) density $\rho_{E(1+2)}(\tau, \vec \sigma)$ of the ADM
energy ${\hat E}_{ADM}$ till the order $O(\zeta^2)$. This energy
density is the sum of three terms: a matter term
$\rho^{(matter)}_{E(1+2)}(\tau, \vec \sigma)$, a radiation term
$\rho^{(rad)}_{E(1+2)}(\tau, \vec \sigma)$ (the GW) and an
interaction term $\rho^{(int)}_{E(1+2)}(\tau, \vec \sigma)$ (the
interaction of GW's with matter: it controls both the emission and
the back-reaction). Therefore we have

\begin{eqnarray*}
 \rho_{E(1+2)}(\tau, \vec \sigma) &=& \rho^{(matter)}_{E(1+2)}(\tau, \vec \sigma)
 + \rho^{(rad)}_{E(1+2)}(\tau, \vec \sigma) +
 \rho^{(int)}_{E(1+2)}(\tau, \vec \sigma),
 \end{eqnarray*}

 \begin{eqnarray*}
 \rho^{(matter)}_{E(1+2)}(\tau, \vec \sigma) &=& {\cal
 M}_{(1)}(\tau, \vec \sigma)\, c = \sum_i\, \delta^3(\vec \sigma
 - {\vec \eta}_i(\tau))\, \eta_i\, c\, \sqrt{m_i^2\, c^2 +
 {\vec \kappa}_i^2(\tau)},
 \end{eqnarray*}

\begin{eqnarray*}
 \rho^{(rad)}_{E(1+2)}(\tau, \vec \sigma) &=& {{c^4}\over {64\pi\, G}}\,
 \sum_{rs}\, \Big[\Big(\partial_{\tau}\, {}^4h^{TT}_{(1)rs}\Big)^2
 + \sum_c \Big(\partial_c\, {}^4h^{TT}_{(1)rs}\Big)^2 \Big] (\tau, \vec
 \sigma),
 \end{eqnarray*}

\bea
 \rho^{(int)}_{E(1+2)}(\tau, \vec \sigma) &=& {\cal
 M}_{(2)}(\tau, \vec \sigma)\, c + \Big({{2\pi\, G}\over {c^2}}\,
 {\cal M}^{(UV)}_{(1)}\, {1\over {\triangle}}\, {\cal M}^{(UV)}_{(1)}
 -\nonumber \\
 &-&\sum_c\,  {\cal M}^{(UV)}_{(1)c}\,
 \Big[{{\partial_c}\over {\triangle}}\, {}^3K_{(1)} -
 {{8\pi\, G}\over {c^3}}\, {1\over {\triangle}}\, \Big(
 {\cal M}^{(UV)}_{(1)c} - {1\over 4}\, \sum_d\, {{\partial_c\,
 \partial_d}\over {\triangle}}\, {\cal M}^{(UV)}_{(1)d}\Big)\Big]
 \Big)(\tau, \vec \sigma)) =\nonumber \\
 &&{}\nonumber \\
 &=& \sum_i\, \delta^3(\vec \sigma - {\vec \eta}_i(\tau))\, \eta_i\,
 \Big[c\, \sum_b\, {{\kappa^2_{ib}(\tau)}\over {\sqrt{m_i^2\, c^2 +
 {\vec \kappa}_i^2(\tau)}}}\, \Big(\Gamma_b^{(1)} + {1\over 2}\,
 \sum_d\, {{\partial_d^2}\over {\triangle}}\, \Gamma_d^{(1)}\Big)(\tau,
 {\vec \eta}_i(\tau)) -\nonumber \\
 &-&c\, \sum_b\, {{\kappa^2_{ib}(\tau)}\over {\sqrt{m_i^2\, c^2 +
 {\vec \kappa}_i^2(\tau)}}}\, {G\over {4\pi}}\, \sum_{j \not= i}\,
 \eta_j\,\int d^3\sigma\, {{\sqrt{m_j^2\, c^2 +
 {\vec \kappa}_j^2(\tau)}}\over {|{\vec \eta}_i(\tau) - \vec \sigma|\,
 |\vec \sigma - {\vec \eta}_j(\tau)|}} -\nonumber \\
 &-& c\, \sum_b\, \kappa_{ib}(\tau)\, \Big({{\partial_b}\over
 {\triangle}}\, {}^3K_{(1)}\Big)(\tau, {\vec \eta}_i(\tau)) +
 {{2\, G}\over {c^2}}\, \sum_b\, \sum_{j \not= i}\, \eta_j\,
 {{\kappa_{ib}(\tau)\, \kappa_{jb}(\tau)}\over {|{\vec \eta}_i(\tau)
 - {\vec \eta}_j(\tau)|}} +\nonumber \\
 &+& {G\over {8\pi\, c^2}}\, \sum_b\, \kappa_{ib}(\tau)\, \sum_d\,
 \kappa_{id}(\tau)\, \sum_{j \not= i}\, \eta_j\, \int d^3\sigma\,
 {1\over {|{\vec \eta}_i(\tau) - \vec \sigma|}}\, \partial_b\, \partial_d\,
 {1\over {|\vec \sigma - {\vec \eta}_j(\tau)|}} -\nonumber \\
 &-& {G\over {2\, c^2}}\, \sum_{j \not= i}\, \eta_j\, {{\sqrt{m_i^2\, c^2 +
 {\vec \kappa}_i^2(\tau)}\, \sqrt{m_j^2\, c^2 +
 {\vec \kappa}_j^2(\tau)}}\over {{\vec \eta}_i(\tau)
 - {\vec \eta}_j(\tau)}} \Big].
 \label{7.23}
 \eea

\bigskip

Let us divide the 3-space $\Sigma_{\tau}$  in two regions by means
of a sphere $S$ of big radius $R >> l_c$: a) an inner region
$V_{(inner)}$ with a compact sub-region $V_c$ of linear dimension
$l_c$ containing all the particles (and the electro-magnetic field
if we would add it); b) an asymptotic far region $V_{(far)}$. Let
$n^r = \sigma^r/|\vec \sigma|$ be a unit 3-vector.\medskip

Since we have ${\hat E}_{ADM} = {\hat E}_{ADM}^{V_{(far)}} + {\hat
E}_{ADM}^{V_{(inner)}}$ and $\rho^{(matter)}_{E(1+2)}(\tau, \vec
\sigma){|}_{\vec \sigma \in V_{far}} = \rho^{(int)}_{E(1+2)}(\tau,
\vec \sigma){|}_{\vec \sigma \in V_{far}} = 0$, we get ${\hat
E}_{ADM}^{V_{(far)}} = \int_{V_{(far)}} \, d^3\sigma\,
\rho^{(rad)}_{E(1+2)}(\tau, \vec \sigma)$.\medskip

Since ${\hat E}_{ADM}$ is a constant, we have ${{d\, {\hat
E}_{ADM}}\over {d\tau}} = 0$ so that we get

\bea
 {{d\, {\hat E}_{ADM}^{V_{(inner)}}}\over {d\tau}} &=& - {{d\,
 {\hat E}_{ADM}^{V_{(far)}}}\over {d\tau}} = - \int_{V_{(far)}}\,
 d^3\sigma\, \partial_{\tau}\, \rho^{(rad)}_{E(1+2)}(\tau, \vec \sigma)
 =\nonumber \\
 &&{}\nonumber \\
 &=& - {{c^4}\over {32\pi\, G}}\, \int_{V_{(far)}}\,
 d^3\sigma\, \sum_{rs}\,  \Big[\partial_{\tau}\, {}^4h^{TT}_{(1)rs}\,
  \partial^2_{\tau}\, {}^4h^{TT}_{(1)rs}
 + \sum_c \partial_c\, {}^4h^{TT}_{(1)rs}\,
 \partial_{\tau}\, \partial_c\, {}^4h^{TT}_{(1)rs} \Big]
(\tau, \vec \sigma) =\nonumber \\
 &=& - {{c^4}\over {32\pi\, G}}\, \int_{V_{(far)}}\,
 d^3\sigma\, \sum_{rs}\, \Big[\sum_c\, \partial_c\,
 \Big(\partial_{\tau}\, {}^4h^{TT}_{(1)rs}\, \partial_c\,
 {}^4h^{TT}_{(1)rs}\Big) + \partial_{\tau}\, {}^4h^{TT}_{(1)rs}\,
 \Box\, {}^4h^{TT}_{(1)rs}\Big](\tau, \vec \sigma) \cir\nonumber \\
 &\cir& - {{c^4}\over {32\pi\, G}}\, \int_{V_{(far)}}\,
 d^3\sigma\, \sum_{rs}\, \sum_c\, \partial_c\,
 \Big(\partial_{\tau}\, {}^4h^{TT}_{(1)rs}\, \partial_c\,
 {}^4h^{TT}_{(1)rs}\Big)(\tau, \vec \sigma) + O(1/r^2),
 \label{7.24}
 \eea

\noindent where we have done an integration by parts and used
Eqs.(\ref{6.11}) (their second member is zero because the non-local
TT quantity $T_{(1)}^{(TT)rs}(\tau, \vec \sigma)$ is of order
$O(1/r^2)$ ($r = |\vec \sigma|$) in $V_{(far)}$).
\medskip

If in the far region we use the solution (\ref{7.19}), we have
$\partial_c\, {}^4h^{TT}_{(1)rs}(\tau, \vec \sigma) = - n^c\,
\partial_{\tau}\, {}^4h^{TT}_{(1)rs}(\tau, \vec \sigma) + O(1/r^2)$
and $\sum_c\, n^c\, \partial_c\, {}^4h^{TT}_{(1)rs}(\tau, \vec
\sigma) = - \partial_{\tau}\, {}^4h^{TT}_{(1)rs}(\tau, \vec \sigma)
+ O(1/r^2)$ because $r = |\vec \sigma| > R >> l_c$. Therefore we get
(the sphere $S$ is parametrized with the angles $\theta$ and
$\varphi$)

\bea
 {{d\, {\hat E}_{ADM}^{V_{(inner)}}}\over {d\tau}} &=& - {{d\,
 {\hat E}_{ADM}^{V_{(far)}}}\over {d\tau}} \cir\nonumber \\
 &\cir& - {{c^4}\over {32\pi\, G}}\, \int_{V_{(far)}}\,
 d^3\sigma\, \sum_c\, \partial_c\, \Big[n^c\, \sum_{rs}\,
 \Big(\partial_{\tau}\, {}^4h^{TT}_{(1)rs}\Big)^2 \Big](\tau, \vec \sigma)
 + O(1/r^2) =\nonumber \\
 &=&  {{c^4}\over {32\pi\, G}}\,  R^2\, \int_{S}\, d(cos\, \theta)\,
 d\varphi\, \sum_{rs}\, \Big(\partial_{\tau}\, {}^4h^{TT}_{(1)rs}\Big)^2(\tau,
 \vec \sigma) + O(1/r^2) =\nonumber \\
 &&{}\nonumber \\
 &{\buildrel {(\ref{7.19})}\over =}&  {G\over {2\pi\, c^2}}\,
 \int_{S}\, d(cos\, \theta)\, d\varphi\, \sum_{rsuv}\,
 \Lambda_{rsuv}(n)\, \partial_{\tau}\, q^{|rs}(\tau - R)\,
 \partial_{\tau}\, q^{|uv}(\tau - R) + O(1/r^2) =\nonumber \\
 &=&  {G\over {5\, c^2}}\, \sum_{rs}\, [\partial^3_{\tau}\,
 q^{rs|\tau\tau}(\tau - R)]\, [\partial^3_{\tau}\,
 q^{uv|\tau\tau}(\tau - R)] + O(1/r^2).
 \label{7.25}
 \eea

\noindent where in the last line we used Eq.(\ref{a13}) and
$\int_{S}\, d(cos\, \theta)\, d\varphi\, \Lambda_{rsuv}(n) =
{{2\pi}\over {15}}\, (11\, \delta_{ru}\, \delta_{sv} - 4\,
\delta_{rs}\, \delta_{uv} + \delta_{rv}\, \delta_{su})$ (see
Eq.(3.74) of Ref.\cite{20}). The change of sign is due to the fact
that the unit vector $n^c$ is minus the normal to the sphere
$S$.\medskip

Therefore Eq.(\ref{7.25}) reproduces the standard result for the
total radiated power also named the total gravitational luminosity
${\cal L}$ of the source (see for instance Eqs. (1.153) and (3.75)
of Ref. \cite{20}) \footnote{Since $\tau = c\, t$, Eq.(\ref{7.25})
can be rewritten as ${{d\, {\hat E}_{ADM}^{V_{(inner)}}}\over {dt}}
= {G\over {5\, c^7}}\, \Big({{\partial\, q^{rs|\tau\tau}}\over
{\partial\, t^3}}\Big)^2$. However, our mass quadrupole
$q^{rs|\tau\tau}$ is equal to $c\, Q^{rs}$, where $Q^{rs}$ is the
mass quadrupole of Ref.\cite{20}. As a consequence,  we have ${{d\,
{\hat E}_{ADM}^{V_{(inner)}}}\over {dt}} = {G\over {5\, c^5}}\,
\Big({{\partial\, Q^{rs}}\over {\partial\, t^3}}\Big)^2$ as in
Ref.\cite{20}.}.
\medskip

\bigskip

See Appendix C for the evaluation of the balance equations for the
3-momentum and the angular momentum by using the conservation of the
corresponding ADM generators (\ref{4.22}) and (\ref{4.23}).

\subsection{PM Gravitational Waves with Asymptotic Background:
Detection with the Geodetic Deviation Equation}

In Ref.\cite{20} there is a full discussion of the detectors of
gravitational waves and of the reference frames for the observers
looking for them. After discussing the observers using local
inertial frames (with Riemann normal coordinates) or freely falling
frames (with Fermi normal coordinates), there is a discussion of
observers using a TT frame (where particles at rest before the
arrival of the gravitational wave remain at rest after their
arrival) and of their use of the geodesic equation and of the
geodetic deviation equation. Then there is a discussion of the
proper detector frame used by experimentalists (in it locally one
uses an Euclidean Newtonian 3-space): now the solution of the
geodetic deviation equation, giving the coordinate displacement
$\vec \xi(t)$ of a test mass of mass m induced by a gravitational
wave (zero in the TT frame), can be put in the form $m\, {{d^2\,
\xi^i(t)}\over {dt^2}} = F^i$ with the Newton force $F^i = {m\over
2}\, {{d^2\, {}^4h^{TT}_{ij}}\over {dt^2}}\, \xi^j$. However, it is
possible to give a coordinate-independent description of the effect
of a gravitational wave on a test mass by using a notion of proper
distance between two nearby geodesics, see Refs.\cite{19,21}.
\medskip

All these approaches could be reproduced with our formalism. In this
Subsection we give first a discussion of geodesics in our
generalized TT gauge as our alternative to the TT frame and then we
discuss the proper distance and the geodesic deviation equation.

\subsubsection{The Geodetic Equation for Test Particles}

Let us consider the behavior of a test particle in presence of a
gravitational wave in the distant wave zone far away from the
sources. To this end let ignore the sources by putting ${\cal
M}_{(1)}^{(UV)} = 0$, ${\cal M}_{(1)a}^{(UV)} = 0$, $T_{(1)}^{ab}
=0$ (they are assumed to have compact support and to give vanishing
terms $E_{\bar a} = 0$ in Eqs.(\ref{6.4}) evaluated in the far zone)
and consider a gravitational wave solution of the wave equation $
\Box\,h^{TT}_{(1)aa}(\tau, \vec \sigma)\, \cir\, 0$, see
Eq.(\ref{6.8}).

\medskip

In this case the 4-metric is given in Eq.(\ref{6.12}) with all the
matter terms omitted (so that we have $\phi_{(1)}\, =\,
\frac{1}{4\,\Delta}\,\sum_c\,\frac{\partial_c^2}{\Delta}\,\Gamma^{(1)}_c$)
and with ${}^3{\cal K}_{(1)} = 0$ (here we assume that this
3-orthogonal gauge is relevant near the detector).

\bigskip

By using the embedding $z^{\mu}(\tau, \vec \sigma) =
\epsilon^{\mu}_A\, \sigma^A$ discussed in the Introduction, we
describe the world-line ${\tilde x}^{\mu}(s) = \epsilon^{\mu}_A\,
\sigma^A(s) = \epsilon^{\mu}_{\tau}\, \tau(s) + \epsilon^{\mu}_r\,
{\tilde \eta}^r(s)$ of the test particle as a time-like geodesic
with parameter $s$ ($ {{d^2\, \sigma^A(s)}\over {ds^2}} +
{}^4\Gamma^A_{BC}(\sigma^E(s))\, {{d \sigma^B(s)}\over {ds}}\, {{d
\sigma^C(s)}\over {ds}} = 0$). If we choose for s the proper time of
the test particle, we have $s = s(\tau) = \sqrt{\sgn\,
{}^4g_{(1)\tau\tau}(\tau(s), {\vec {\tilde \eta}}(s))}$ and ${\tilde
x}^{\mu}(s(\tau)) = x^{\mu}(\tau) = \epsilon^{\mu}_A\, \eta^A(\tau)
= \epsilon^{\mu}_{\tau}\, \tau + \epsilon^{\mu}_r\, \eta^r(\tau)$.
We have

\beq
 \frac{ds}{d\tau}=1+n_{(1)}(\tau,\eta^u(\tau)) +
 O(\zeta^2).
 \label{7.26}
 \eeq

In the weak field approximation the geodesic equation becomes the
following equation for $\vec \eta(\tau)$

\bea
 \frac{d^2\eta^r(\tau)}{d\tau^2}&=&-\left(
 {}^4\Gamma^r_{(1)\tau\tau}+2\,{}^4\Gamma^r_{(1)\tau
 u}\,\dot{\eta}^u+\,{}^4\Gamma^r_{(1)uv}\,\dot{\eta}^u\dot{\eta}^v
 \right)+\nonumber\\
 &&\nonumber\\
 &+& \left( {}^4\Gamma^\tau_{(1)\tau\tau}+2\,{}^4\Gamma^\tau_{(1)\tau
 u}\,\dot{\eta}^u+\,{}^4\Gamma^\tau_{(1)uv}\,\dot{\eta}^u\dot{\eta}^v
 \right)\,\dot{\eta}^r + \frac{1}{L}O(\zeta^2).
 \label{7.27}
 \eea

\medskip

Let us consider the following Christoffel symbol

\bea
 \sgn\, {}^4\Gamma^r_{(1)\tau\tau}&=&\frac{\partial\bar{n}_{(1)}}{\partial\sigma^r}
 +\frac{\partial\, n_{(1)(r)}}{\partial\tau} +
 \frac{1}{L}O(\zeta^2)=\nonumber\\
 &&\nonumber\\
 &=&\partial^2_\tau\, \left[
 \frac{2}{\Delta}\,\partial_a\Gamma_a^{(1)} - \frac{\partial_a}{2\,\Delta}\,
 \sum_c\,\frac{\partial_c^2}{\Delta}\,\Gamma_c^{(1)}
 \right] + \frac{1}{L}O(\zeta^2) =\nonumber \\
 &=& \partial^2_{\tau}\, \Psi^r_{(1)} + O(\zeta^2),
 \label{7.28}
 \eea

\noindent where we used the function $\Psi^r_{(1)}$ defined in
Eq.(\ref{6.13}).

As a consequence the geodetic equation can be written in the form

\bea
 \frac{d^2}{d\tau^2}\left[\eta^r(\tau)-\Psi^r_{(1)}(\tau,\eta^r(\tau))\right]&=&
 -2\left(\,{}^4\Gamma^r_{(1)\tau
 u}-\frac{\partial\Psi^r_{(1)}}{\partial\tau\partial\sigma^u}\right)\,\dot{\eta}^u-
 \left(\,{}^4\Gamma^r_{(1)uv}-\frac{\partial\Psi^r_{(1)}}{\partial\sigma^v
 \partial\sigma^u}\right)\,\dot{\eta}^u\dot{\eta}^v+\nonumber\\
 &&\nonumber\\
 &+& \left( {}^4\Gamma^\tau_{(1)\tau\tau}+2\,{}^4\Gamma^\tau_{(1)\tau
 u}\,\dot{\eta}^u+\,{}^4\Gamma^\tau_{(1)uv}\,\dot{\eta}^u\dot{\eta}^v
 \right)\,\dot{\eta}^r + \frac{1}{L}O(\zeta^2).
 \label{7.29}
 \eea

\medskip

A special set of geodesics solution of Eq.(\ref{7.29}) is {\it
implicitly} defined by the condition

\beq
 \eta^r(\tau)-\Psi^r_{(1)}(\tau,\eta^r(\tau))=\,constant.
  \label{7.30}
 \eeq

\medskip
In this case we have

\beq
 \dot{\eta}^r=\frac{\partial\Psi^r_{(1)}}{\partial\tau}+
 \frac{\partial\Psi^r_{(1)}}{\partial\sigma^s}\,\dot{\eta}^s,
 \label{7.31}
 \eeq

\noindent so that in the weak field approximation we get

\beq
 \dot{\eta}^r=\frac{\partial\Psi^r_{(1)}}{\partial\tau} +
 O(\zeta^2) = O(\zeta).
 \label{7.32}
 \eeq

As a consequence the right side of Eq.(\ref{7.29}) is zero modulo
terms of order $\frac{1}{L}\, O(\zeta^2)$:
$\frac{d^2}{d\tau^2}\left[\eta^r(\tau)-\Psi^r_{(1)}(\tau,\eta^r(\tau))\right]=0+\frac{1}{L}\,
O(\zeta^2)$.\medskip

Moreover in the weak field approximation the 4-velocity of the
geodesic (\ref{7.30}) turns out to be

\beq
 u^A = {{d \sigma^A(s)}\over {ds}} = {{d
 \eta^A(\tau)}\over {d\tau}}\, {{d\tau(s)}\over {ds}}
 =\left(\, u^{\tau} = 1-n_{(1)};\, u^r = \frac{\partial\Psi^r_{(1)}}{\partial\tau}\right).
 \label{7.33}
 \eeq

\bigskip

Let us remark that, since we have
$\dot{\eta}^r=\frac{\partial\Psi^r_{(1)}}{\partial\tau} + O(\zeta^2)
= O(\zeta)$ along these geodesics, then we get
$\frac{dA}{dt}=\frac{\partial A}{\partial\tau}+\frac{\partial
A}{\partial\sigma^s}\,\dot{\eta}^s=\frac{\partial A}{\partial\tau} +
\frac{1}{L}\, O(\zeta^2)$ for every quantity $A(\tau,\sigma^u)$ of
order $O(\zeta)$.

\bigskip

The solution (\ref{7.30}) selects a special family of geodesics
whose meaning can be clarified by remembering the coordinate
transformation (\ref{6.13}) leading to a generalized TT gauge, whose
associated 4-metric is given in Eq.(\ref{6.15}) with the matter
terms omitted.

\medskip

The new Christoffel symbols ${}^4\overline{\Gamma}^{\,A}{}_{BC} =
\frac{\partial\overline{\sigma}^{\,A}} {\partial\sigma^A}\,\left(
{}^4\Gamma^D{}_{EF} \frac{\partial\sigma^E}
{\partial\overline{\sigma}^{\,B}}\,
\frac{\partial\sigma^F}{\partial\overline{\sigma}^{\,C}}+
\frac{\partial^2\sigma^D}{\partial\overline{\sigma}^{\,B}
\partial\overline{\sigma}^{\,C}}
\right)$ imply

\beq
 \overline{\Gamma}^{\,r}_{(1)\tau\tau} = 0 + \frac{1}{L}\, O(\zeta^2).
 \label{7.34}
 \eeq

This equation is the sufficient and necessary condition to get the
result that the coordinate lines $\overline{\sigma}^{\,r}=constant$
be geodetic. This consequence of the solution (\ref{7.26}) is in
accord with the choice of constant spatial coordinates for the usual
TT gauge done in Ref. \cite{19}, pp. 13-16.

\subsubsection{Detection of Gravitational Waves}

As shown in Ref. \cite{21} the main observable for the detection of
gravitational waves is the {\it proper distance} between two nearby
geodesics.\medskip

The geodetic deviation ${\cal E}^A(\tau)$ is the infinitesimal
4-vector orthogonal to the 4-velocity $u^A(\tau)$ of the reference
geodesic: $ u_A(\tau)\, {\cal E}^A(\tau) = 0$. If $\eta^A(\tau) =
(\tau, \eta^r(\tau))$ is the reference geodesic and $\eta^A(\tau) +
{\cal E}^A(\tau)$ the nearby one, the proper distance between them
is the invariant

\beq
 D(\tau) = \sqrt{{}^4g_{AB}(\tau,\eta^u(\tau))\,{\cal E}^A(\tau)\,{\cal
 E}^B(\tau)}.
 \label{7.35}
 \eeq

\medskip

Following Ref.\cite{19} and consistently with the weak field
approximation we assume $D(\tau)<<L$, namely that the proper
distance is less of the wavelength of the gravitational wave to be
detected.
\medskip

Given the geodetic deviation equation for ${\cal E}^A(\tau)$

\beq
 (u^A(\tau) \,\nabla_A)\, (u^B(\tau)\, \nabla_B)\,{\cal
 E}^C(\tau) = - {}^4R^C{}_{EFG}(\eta^D(\tau))\, u^E(\tau)\,
 {\cal E}^F(\tau)\, u^G(\tau),
  \label{7.36}
 \eeq

\noindent we get the following equation for the proper distance (a
scalar quantity)

\bea
 (u^A(\tau)\,\nabla_A)^2\, D(\tau) &=& {{d^2}\over {ds^2}}\, \tilde
 D(s(\tau)) =\nonumber \\
 &=& \Big(1-2\,\bar{n}_{(1)}(\eta^E(\tau)\Big)\,
 \frac{d^2D(\tau)}{d\tau^2}- \frac{d\,\bar{n}_{(1)}(\eta^E(\tau))}{d\tau}\,
 \frac{d\, D(\tau)}{d\tau} + \frac{1}{L}\, O(\zeta^2) =\nonumber \\
 &&{}\nonumber \\
 &=&  - {}^4R_{ABCD}(\eta^E(\tau))\, u^B(\tau)\, u^D(\tau)\,
 \frac{{\cal E}^A(\tau)\, {\cal E}^C(\tau)}{D(\tau)}.\nonumber \\
 &&{}
  \label{7.37}
 \eea
\medskip

In the weak field approximation Eq.(\ref{7.37}) becomes

\bea
 &&\Big(1-2\,\bar{n}_{(1)}(\eta^E(\tau)\Big)\, \frac{d^2}{d\tau^2}D(\tau) -
 \frac{d\,\bar{n}_{(1)}(\eta^E(\tau))}{d\tau}\, \frac{d}{d\tau}D(\tau) =\nonumber \\
 &&{}\nonumber \\
 &&= -{}^4R_{(1)\tau r \tau s}(\tau,\eta^u(\tau))\, \frac{{\cal
 E}^r(\tau)\,{\cal E}^s(\tau)}{D(\tau)} + \frac{1}{L}\,
 O(\zeta^2),
 \label{7.38}
 \eea

\noindent with the following expression for the Riemann tensor
implied by Eq.(\ref{6.12})

\bea
 {}^4R_{(1)\tau s\tau r}&=&-
 \frac{\partial^2\, n_{(1)}}{\partial\sigma^s\partial\sigma^r}-
 \frac{1}{2}\frac{\partial}{\partial\tau}
 \left(\frac{\partial\bar{n}_{(1)(r)}}{\partial\sigma^s}+
 \frac{\partial\bar{n}_{(1)(r)}}{\partial\sigma^s}\right)
 +\delta_{rs}\,\frac{\partial^2}{\partial\tau^2}\left(\Gamma^{(1)}_a+2\phi_{(1)}\right)+
 \frac{1}{L^2}\, O(\zeta^2)=\nonumber\\
 &&\nonumber\\
 &=&-\frac{1}{2}\,\partial_\tau^2\,h^{TT}_{(1)rs}+\frac{1}{L^2}\,
 O(\zeta^2).
 \label{7.39}
 \eea
\medskip

By using the result $\frac{dA}{dt} = \frac{\partial A}{\partial\tau}
+ \frac{1}{L}\, O(\zeta^2)$, valid for every quantity
$A(\tau,\sigma^u)$ of order $O(\zeta)$, we get the following final
form of the equation for the proper distance

\bea
 &&\Big(1 - 2\, \bar{n}_{(1)}(\tau, \eta^u(\tau))\Big)\, \frac{d^2}{d\tau^2}D(\tau) -
 \frac{d\,\bar{n}_{(1)}(\eta^E(\tau))}{d\tau}\, \frac{d}{d\tau}D(\tau) =
 \nonumber \\
 &&{}\nonumber \\
 &&+ \frac{1}{2}\, \frac{d^2}{d\tau^2}\,h^{TT}_{rs}(\tau,\eta^u(\tau))\,\frac{{\cal
 E}^r(\tau)\,{\cal E}^s(\tau)}{D(\tau)} + \frac{1}{L}\,
 O(\zeta^2).
 \label{7.40}
 \eea

\bigskip

To study Eq.(\ref{7.40}) we follow the method of Ref.\cite{19}.

The structure of Eqs. (\ref{7.36}), (\ref{7.37}) or (\ref{7.40})
suggests that the geodetic deviation  ${\cal E}^A(\tau)$ may be
parametrized as the sum of a constant deviation ${\cal E}^A_o$ plus
a small corrective term determined by the weak gravitational field.
To this end let us introduce an arbitrary constant direction with
the unit constant 4-vector $n^A$, ${}^4g_{AB}\,n^A\,n^B = 1$, so
that the constant part of the deviation is given by ${\cal E}^A_o =
D_o\, n^A$ with $D_o = const.$. If $\delta{\cal E}^A(\tau)$, with
$\frac{\delta{\cal E}^A(\tau)}{D_o} = O(\zeta)$, is the small
corrective term, then the geodetic deviation is parametrized in the
following form

\beq
 {\cal E}^A(\tau) =\, D_o\, \left(\,n^A + \,\frac{\delta{\cal
 E}^A(\tau)}{D_o} +  O(\zeta^2)\,\right).
 \label{7.41}
 \eeq
\medskip

The induced parametrization of the proper distance is

\beq
 D(\tau)=D_o\,\left(1+\,\frac{\delta D(\tau)}{D_o}+\,\,
 O(\zeta^2)\,\right),
 \label{7.42}
 \eeq

\noindent with $\frac{\delta D(\tau)}{D_o} = {}^4g_{(1)AB}\, n^A\,
\frac{\delta{\cal E}^B(\tau)}{D_o} = O(\zeta)$.

\bigskip

With these parametrizations Eq. (\ref{7.40}) takes the form

\beq
 \frac{d^2}{d\tau^2}\frac{\delta D(\tau)}{D_o}=
 +\frac{1}{2}\,\frac{d^2}{d\tau^2}\,h^{TT}_{(1)rs}(\tau,
 \eta^u(\tau))\,n^r\,{n^s}+\frac{1}{L^2}\,
 O(\zeta^2).
 \label{7.43}
 \eeq

\noindent allowing to find the solution for $\frac{\delta
D(\tau)}{D_o}$ at the lowest order.
\medskip

As a consequence we get the following expression for the proper
distance

\beq
 D(\tau)    = D_o +
 + \frac{1}{2}\,h^{TT}_{(1)rs}(\tau,\eta^u(\tau))\,n^r\,{n^s} +
 O(\zeta^2).
 \label{7.44}
 \eeq

The choice of the direction $n^r$ allows to discuss the effects of
the polarization of the gravitational wave on the detectors as it is
done in Ref.\cite{19,21}.

\vfill\eject

\section{Conclusions}

We have defined a consistent HPM linearization of ADM tetrad gravity
in the York canonical basis in the family of non-harmonic
3-orthogonal Schwinger time gauges parametrized by the numerical
value ${}^3K_{(1)}(\tau, \vec \sigma)$ of the York time, the
inertial gauge variable describing the general relativistic remnant
of the freedom in clock synchronization. The non-Euclidean
instantaneous 3-spaces $\Sigma_{\tau}$ (a first order deformation of
the Euclidean 3-spaces of inertial Minkowski frames) are dynamically
determined except for for the value of the trace ${}^3K_{(1)}$ of
their extrinsic curvature tensor. The 4-metric has an asymptotic
Minkowski background at spatial infinity. PN expansions are avoided
by introducing a ultraviolet cutoff.\medskip

The PM solutions, ${\tilde \phi}_{(1)}(\tau, \vec \sigma) = 1 + 6\,
\phi_{(1)}(\tau, \vec \sigma)$ (the 3-volume element), $1 +
n_{(1)}(\tau, \vec \sigma)$ (the lapse function), ${\bar
n}_{(1)(r)}(\tau, \vec \sigma)$ (the shift functions),
$\sigma_{(1)(r)(s)}{|}_{r \not= s}(\tau, \vec \sigma)$ (the
non-diagonal elements of the shear of the Eulerian observers), of
the constraints and of the Hamilton equations implying the
preservation in $\tau$ of the gauge-fixings, have both an
instantaneous action-at-a-distance part depending on the
instantaneous value of the matter energy-momentum tensor and a part
depending on the tidal variables. At the PM level all previous
quantities do not depend on the York time ${}^3K_{(1)}(\tau, \vec
\sigma)$ but on the spatially non-local function ${}^3{\cal
K}_{(1)}(\tau, \vec \sigma) = {1\over {\triangle}}\,
{}^3K_{(1)}(\tau, \vec \sigma)$ (it can be named non-local York
time).\medskip

In these non-harmonic gauges two functions $\sum_{\bar b}\, M_{\bar
a\bar b}(\vec \sigma)\, R_{\bar a}(\tau, \vec \sigma)$, with the
operator $M_{\bar a\bar b}(\vec \sigma)$ containing only spatial
derivatives, of the tidal variables $R_{\bar a}(\tau, \vec \sigma)$
satisfy a wave equation with the flat asymptotic d'Alambertian. It
has been shown that the operator $M_{\bar a\bar b}(\vec \sigma)$
contains the information for determining the TT part of the 3-metric
on $\Sigma_{\tau}$. By using a no-incoming radiation condition with
respect to the flat asymptotic metric, we get a retarded solution
for the tidal variables in terms of the matter energy-momentum
tensor which describes PM TT GW's with asymptotic background
propagating at the velocity of light $c$ \footnote{See Ref.\cite{30}
for the problem of the velocity of propagation of the gravitational
field in general relativity and in bimetric theories.}. A multipolar
expansion of the energy-momentum tensor in terms of Dixon multipoles
allows to get the standard quadrupole emission formula as the
dominant part. Also the rate of variation of the energy is correct
due to the Grassmann regularization \footnote{In the
electro-magnetic case \cite{14} the Grassmann-valued electric
charges allow to find the effective potential (Coulomb plus Darwin)
corresponding to the one-photon exchange Feynman diagram. Therefore
the problem of electro-magnetic self-energies is pushed to the level
of loop diagrams and becomes part of the problem of renormalization
of QED. At the classical level the Grassmann regularization allows
to make sense of the classical equations of motion (not well defined
due to essential singularities at the charge location): the
replacement of the Grassmann-valued electric charges with the
standard electric charge is equivalent to a classical
renormalization of scalar charged particles. With gravity the same
mechanism is at work, except that we do not yet have an accepted
renormalizible theory of quantum gravity. Our procedure is a
classical regularization and in some sense we are identifying the
effective potential connected to the one graviton exchange.} of the
self-energies of the point particles (no gravitational self-energy)
due to the existence of the conserved ADM energy. Thereforall the
main properties of GW's are reproduced in our HPM approach. Only the
hereditary and memory tails (coming from the matching of MPM and MPN
solutions as said in Appendix A) are missing at this order: to study
them we have to go to higher orders in the HPM expansion (see
Section IIIB), which should correspond to the MPM solutin in the far
wave zone.
\medskip

All these results lead to a PM solution (modulo the choice of the
inertial gauge variable ${}^3{\cal K}_{(1)} = {1\over {\triangle}}\,
{}^3K_{(1)}$) for the gravitational field and identify a class of PM
Einstein space-times.

\bigskip

These results are {\it gauge dependent} because the York time
${}^3K_{(1)}$ is an inertial gauge variable. Therefore we have to
face the {\it gauge problem in general relativity}. The gauge
freedom of space-time 4-diffeomorphisms implies that a gauge choice
is equivalent to the choice of a set of 4-coordinates in the atlas
of the space-time 4-manifold.\medskip

The standard approach to the gauge problem is to try to describe
physical properties in terms of {\it gauge invariant} quantities,
i.e. in terms of 4-scalars. At the Hamiltonian level, where the
Hamiltonian gauge group (whose generators are the first-class
constraints) is equivalent to the 4-diffeomorphism group of
space-time only {\it on-shell} (i.e. on the solutions of Einstein
equations; see for instance Refs.\cite{15}), the Hamiltonian
gauge-invariant (off-shell) quantities are the {\it Dirac
observables} (DO), which have zero Poisson bracket with the
constraints. If we would know the solution of the super-Hamiltonian
and super-momentum constraints in the form $\hat \phi = \tilde \phi
- F[\theta^i, \pi_{\tilde \phi}, R_{\bar a}, \Pi_{\bar a}, matter]
\approx 0$, ${\hat \pi}^{(\theta)}_i = \pi^{(\theta)}_i -
G_i[\theta^i, \pi_{\tilde \phi}, R_{\bar a}, \Pi_{\bar a}, matter]
\approx 0$, we could look for a canonical transformation from the
York canonical basis to a Shanmugadhasan basis containing $n$,
${\bar n}_{(r)}$, ${\hat \theta}^i$, ${\hat \pi}^{(\theta)}_i
\approx 0$, $\hat \phi \approx 0$, ${\hat \pi}_{\hat \phi}$, ${\hat
R}_{\bar a}$, ${\hat \Pi}_{\bar a}$, $ DO_{(matter)}$. In this basis
${\hat \theta}^i$ and ${\hat \pi}_{\hat \phi}$, $n$, ${\bar
n}_{(r)}$,  would be the (primary and secondary) inertial gauge
variables and  ${\hat R}_{\bar a}$, ${\hat \Pi}_{\bar a}$, $
DO_{(matter)}$ the DO's (non-local quantities as functions of the
original variables). However no-one is able to find such a basis.
Also the most recent works of Refs.\cite{31} contain existence
proofs but no workable algorithm for finding the Dirac observables
of the gravitational field. Moreover it is not clear how many of
these DO are 4-scalars. Hopefully the 3-scalar tidal DO's ${\hat
R}_{\bar a}$, ${\hat \Pi}_{\bar a}$, may be replaced with two pairs
of 4-scalar DO's connected with the eigenvalues of the Weyl tensor
\cite{32} by means of a canonical transformation (this conjecture is
under investigation). For the transverse electro-magnetic field one
expects that the final DO's can be chosen as tetrad-dependent
4-scalars by using the Newman-Penrose formalism \cite{32}.

\bigskip

On the other side at the experimental level the description of
baryon matter  is intrinsically coordinate-dependent, namely is
connected with the conventions used by physicists, engineers and
astronomers for the modeling  of space-time.\medskip

A) The description of satellites around the Earth is done by means
of NASA coordinates \cite{33} either in ITRS (frame fixed on the
Earth surface) or in GCRS (frame centered on the Earth center) (see
Ref.\cite{34}).
\medskip

B) The description of planets and other objects in the Solar System
uses BCRS (a quasi-inertial Minkowski frame, if perturbations from
the Milky Way are ignored \footnote{Essentially it is defined as a
{\it quasi-inertial system}, {\it non-rotating} with respect to some
selected fixed stars, in Minkowski space-time with nearly-Euclidean
Newton 3-spaces. The qualification {\it quasi-inertial} is
introduced  to take into account general relativity, where inertial
frames exist only locally. It can also be considered as a PM
space-time with 3-spaces having a very small extrinsic curvature.}),
centered in the barycenter of the Solar System] and ephemerides (see
Ref.\cite{34}).
\medskip

C) In astronomy the positions of stars and galaxies are determined
from the data (luminosity, light spectrum, angles) on the sky as
living in a 4-dimensional nearly-Galilei space-time with the
celestial ICRS \cite{35} frame  considered as a "quasi-inertial
frame" (all galactic dynamics is Newtonian gravity), in accord with
the assumed validity of the cosmological and Copernican principles.
Namely one assumes a homogeneous and isotropic cosmological
Friedmann-Robertson - Walker solution of Einstein equations (the
standard $\Lambda$CDM cosmological model). In it the constant
intrinsic 3-curvature of instantaneous 3-spaces is  nearly zero as
implied by the CMB data\cite{36}, so that Euclidean 3-spaces (and
Newtonian gravity) can be used. However, to reconcile all the data
with this 4-dimensional reconstruction one must postulate the
existence of dark matter and dark energy as the dominant components
of the classical universe after the recombination 3-surface!

\medskip

As a consequence of the dependence on coordinates of the description
of matter, our proposal for solving the gauge problem in our
Hamiltonian framework with non-Euclidean 3-spaces is to choose a
gauge (i.e. a 4-coordinate system) in non-modified Einstein gravity
which is in agreement with the observational conventions in
astronomy. Since ICRS has diagonal 3-metric, our 3-orthogonal gauges
are a good choice. We are left with the inertial gauge variable
${}^3{\cal K}_{(1)} = {1\over {\triangle}}\, {}^3K_{(1)}$ (not
existing in Newtonian gravity). The suggestion is to try to fix
${}^3{\cal K}_{(1)}$ in such a way to eliminate dark matter as much
as possible, by reinterpreting it as a relativistic inertial effect
induced by the shift from Euclidean 3-spaces to non-Euclidean ones
(independently from cosmological assumptions). As a consequence,
ICRS should be reformulated not as a {\it quasi-inertial} reference
frame in Galilei space-time, but as a reference frame in a PM
space-time with ${}^3K_{(1)}$ deduced from the data connected to
dark matter: as a consequence {\it what is called dark matter would
be an indicator of the non-Euclidean nature of 3-spaces as
3-submanifolds of space-time (extrinsic curvature effect), whose
internal 3-curvature can be very small if it is induced by GW's}.
Then automatically BCRS would be its quasi-Minkowskian approximation
(quasi-inertial reference frame in Minkowski space-time) for the
Solar System. This point of view could also be useful for the ESA
GAIA mission (cartography of the Milky Way) \cite{37} and for the
possible anomalies inside the Solar System \cite{38}.

\bigskip

As a consequence in the third paper \cite{b} we will study the PN
expansion of our HPM solution for a system of point particles
without electro-magnetic field. There we will determine the explicit
dependence of the equations of motion of the particles, of the
proper time of time-like observers, of the time-like and null
geodesics, of the redshift of light and of the luminosity distance
upon the time- and spatial-gradients of the non-local York time
${}^3{\cal K}_{(1)} = {1\over {\triangle}}\, {}^3K_{(1)}$) in the PM
space-time. Then we will study  the slow motion limit making a PN
expansion at all ${n\over 2}PN$ orders with a detailed study of the
order 0.5PN.
\medskip

We will show that the main observational supports for the existence
of dark matter (rotation curves of galaxies, mass of galaxies from
the virial theorem and from gravitational lensing; see for instance
Refs.\cite{39}) can be translated in restrictions on the non-local
York time ${}^3{\cal K}_{(1)}(\tau, \vec \sigma)$ at the location of
galaxies. If we could find a global phenomenological parametrization
${}^3{\cal K}^{(phen)}_{(1)}(\tau, \vec \sigma)$ over all the
3-universe (in our PM space-time and with some average for the
$\tau$-dependence), we could get a phenomenological determination of
the York time ${}^3{\cal K}^{(phen)}_{(1)}(\tau, \vec \sigma) =
\triangle\, {}^3{\cal K}^{(phen)}_{(1)}(\tau, \vec \sigma)$ to be
used as an observational clock synchronization to be used to define
the 3-spaces of a PM ICRS.

\vfill\eject

\appendix

\section{The Standard Approach to Gravitational Waves by using
Einstein's Equations in the Family of 4-Harmonic Gauges}

The standard description of GW's (see for instance Ref.\cite{20}) is
done in the family of 4-harmonic gauges after a decomposition of the
4-metric ${}^4g_{\mu\nu}$ of asymptotically flat space-times in a
flat Minkowski background ${}^4\eta_{\mu\nu}$ plus a small
perturbation, ${}^4g_{\mu\nu} = {}^4\eta_{\mu\nu} + {}^4h_{\mu\nu}$,
$|{}^4h_{\mu\nu}| << 1$. The linearized Einstein equations in
harmonic gauges have the form $\Box\, {}^4{\bar h}_{\mu\nu} = -
\sgn\, {{16\pi\, G}\over {c^3}}\, T_{\mu\nu}$, $\partial^{\nu}\,
{}^4{\bar h}^{\mu\nu} = 0$ (${}^4{\bar h}_{\mu\nu} = {}^4h_{\mu\nu}
- {1\over 2}\, {}^4\eta_{\mu\nu}\, {}^4h$, ${}^4h =
{}^4\eta^{\mu\nu}\, {}^4h_{\mu\nu}$, $\sgn = \pm$ according to the
signature convention for the 4-metric), where $T_{\mu\nu}$ is the
matter energy-momentum tensor \footnote{Our $T_{\mu\nu}$ is ${1\over
c}\, T_{\mu\nu}$ of Ref.\cite{20}.} satisfying $\partial^{\nu}\,
T_{\mu\nu} = 0$ at this order.\medskip

It turns out that with an allowed coordinate transformation inside
the family of harmonic gauges, i.e. ${}^4{\bar h}_{\mu\nu} \mapsto
{}^4{\bar h}^{'}_{\mu\nu} = {}^4{\bar h}_{\mu\nu} -
(\partial_{\mu}\, \xi_{\nu} + \partial_{\nu}\, \xi_{\mu} -
{}^4\eta_{\mu\nu}\, \partial_{\rho}\, \xi^{\rho})$, one can identify
a transverse-traceless (TT) harmonic gauge in which the only remnant
of the gravitational field are the two polarizations of
GW's.\medskip

This weak field description is assumed valid in the wave zone far
away from the matter sources, which are assumed in slow motion
(Post-Newtonian (PN) approximation $v << c$) and with a small
self-gravity \footnote{See Ref.\cite{20} for a review of the problem
of self-gravity of extended compact objects, in particular binary
systems, and of the effacement principle of the internal structure
which becomes relevant only at the order 5PN.} (if $d$ is the linear
dimension of the source with mass $M$ and $R_M = {{2\, G\, M}\over
{c^2}}$ its Schwarzschild radius, one has ${{R_M}\over d} \approx
({v\over c})^2 = \epsilon << 1$, i.e. $d >> R_M$; $\epsilon^{n/2} =
({v\over c})^n$ is the ${n\over 2}\, PN$ order).\medskip

In this way one is led to the interpretation of a (spin 2) GW
propagating in the Euclidean 3-spaces of an inertial frame in
Minkowski space-time.\medskip

Since the equivalence principle forbids the existence of global
inertial frames in Einstein space-times, the next step is to replace
the above decomposition with one with respect to a curved background
${}^4{\bar g}_{\mu\nu}$, i.e. ${}^4g_{\mu\nu}= {}^4{\bar g}_{\mu\nu}
+ {}^4h_{\mu\nu}^{(\bar g)}$, such that the GW's are only ripples on
this background \footnote{As shown in chapter 1 of Ref.\cite{20}
this decomposition makes sense if: A) either ${}^4{\bar g}_{\mu\nu}$
has a scale of spatial variation $L_B$ and the wavelength of the GW
is $\lambda << L_B$ (and $|{}^4h_{\mu\nu}| << \lambda/L_B << 1$); B)
or ${}^4{\bar g}_{\mu\nu}$ contains only frequencies less than
$\nu_B$ and the frequency of the GW is $\nu >> \nu_B$ (this case is
the more relevant for detectors). Moreover, a detector of dimension
$L_D$ will react only to GW's with $\lambda >> L_D$. For GW's with
frequency $10^{-4} - 1\, Hz$ (to be detected by LISA) the
wave-length is of order $10^6 - 10^{10}\, cm$ ($\lambda\, \nu = c$).
If the frequency is $1 - 10^4\, Hz$ (to be detected by LIGO, VIRGO),
the wave-length is of order $10^{10} - 10^{14}\, cm$. }. With this
formalism and a coarse-grained description \footnote{Either a
spatial average on many wavelengths $\lambda$ of the GW or a
temporal average on several periods $1/\nu$ of the GW (the method
used in the detectors).} one can evaluate the energy loss associated
to the emission of GW's and take into account the back-reaction on
the background which is therefore modified.\medskip

However the most advanced description of GW's is done with the
Damour-Blanchet approach \cite{23} or with the equivalent DIRE
approach of Ref.\cite{27} (see chapter 5 of Ref.\cite{20} for an
overall review; see also Ref.\cite{19}). In these approaches one
uses the decomposition ${}^4g_{\mu\nu} = {}^4\eta_{\mu\nu} +
{}^4h_{\mu\nu}$ and writes a so-called relaxed form of Einstein
equations for the quantity ${}^4{\bf h}^{\mu\nu} = \sqrt{|{}^4g|}\,
{}^4g^{\mu\nu} - {}^4\eta^{\mu\nu}$, namely $\Box\, {}^4{\bf
h}^{\mu\nu} = \sgn\, {{16\pi\, G}\over {c^3}}\, \tau^{\mu\nu}$,
where the effective energy-momentum tensor is $\tau^{\mu\nu} =
|{}^4g|\, T^{\mu\nu} + {{c^3}\over {16\pi\, G}}\,
\Lambda^{\mu\nu}({}^4{\bf h})$ with $\Lambda^{\mu\nu}({}^4{\bf h}) =
{{16\pi\, G}\over {c^3}}\, |{}^4g|\, t^{\mu\nu}_{LL} +
\partial_{\alpha}\, {}^4{\bf h}^{\mu\beta}\, \partial_{\beta}\,
{}^4{\bf h}^{\nu\alpha} - {}^4{\bf h}^{\alpha\beta}\,
\partial_{\alpha}\, \partial_{\beta}\, {}^4{\bf h}^{\mu\nu}$.
$t^{\mu\nu}_{LL}$ is the Landau-Lifschitz energy-momentum
pseudo-tensor of the gravitational field, which satisfies the
ordinary conservation law $\partial_{\nu}\, [\sqrt{|{}^4g|}\,
(T^{\mu\nu} + t^{\mu\nu}_{LL})] = 0$ due to Einstein equations. The
ordinary equations of motion for the matter (i.e.
${}^4\nabla_{\nu}\, T^{\mu\nu} = 0$) are obtained by restricting the
solutions of the relaxed Einstein equations to the harmonic gauges
by requiring $\partial_{\nu}\, {}^4{\bf h}^{\mu\nu} = 0$. \medskip

Since the matter is supposed to have compact support of size $d$ and
to be in slow motion ($\sqrt{\epsilon} = {v\over c} \approx
\sqrt{{{R_M}\over d}} << 1$ and typically with a wavelength of GW's
satisfying $\lambda >> d$), the solution of the relaxed Einstein
equations is obtained in three steps:\hfill\break

A) In the far wave zone ($r >> \lambda$ and $d < r < \infty$), where
$T^{\mu\nu} = 0$, one makes a Post-Minkowskian (PM) expansion
${}^4{\bf h}^{\mu\nu} = \sum_{n=1}^{\infty}\, G^n\, {}^4{\bf
h}^{\mu\nu}_n$ and uses the restriction $\partial_{\nu}\, {}^4{\bf
h}^{\mu\nu} = 0$ to harmonic gauges. The iterative PM solution
(including a homogeneous solution of the wave equation) is expressed
in terms of {\it retarded} (symmetric trace free or STF) {\it
radiative multipoles} of the gravitational field. Only a finite
number of multipoles in ${}^4{\bf h}_1^{\mu\nu}$ are taken into
account to avoid the problem of the Green function of the wave
operator, which would require the knowledge of ${}^4{\bf
h}_1^{\mu\nu}$ also in the near region where the PM expansion does
not hold. With this regularization a multipolar PM (MPM) solution is
found with the property that by making a PN expansion one finds
${}^4{\bf h}_n^{00} = O(1/c^{2n})$, ${}^4{\bf h}^{oi}_n = O(1/c^{2n
+ 1})$, ${}^4{\bf h}^{ij}_n = O(1/c^{2n})$.\medskip

B) In the near zone ($r << \lambda$; the exterior near zone is $d <
r << \lambda$) one makes a PN expansion of ${}^4h_{\mu\nu}$ with
${}^4h_{00} = \sum_{n=1}^{\infty}\, {}^4h_{00}^{(2n)}$, ${}^4h_{0i}
= \sum_{n=1}^{\infty}\, {}^4h_{0i}^{(2n+1)}$, ${}^4h_{ij} =
\sum_{n=1}^{\infty}\, {}^4h_{ij}^{(2n)}$ with ${}^4h_{\mu\nu}^{(n)}
= O(({v\over c})^n)$ ($v/c \approx \sqrt{R_M/d}$) and of the
energy-momentum tensor $T^{00} = \sum_{n=0}^{\infty}\,
{}^{(2n)}T^{00}$, $T^{0i} = \sum_{n=1}^{\infty}\,
{}^{(2n+1)}T^{0i}$, $T^{ij} = \sum_{n=1}^{\infty}\,
{}^{(2n)}T^{ij}$. Since the Newtonian approximation corresponds to
${}^4h_{00}^{(2)}$, ${}^4h_{oi} = 0$, ${}^4h_{ij} = 0$, the 1PN
order contains ${}^4h_{00}^{(4)}$, ${}^4h_{0i}^{(3)}$,
${}^4h_{ij}^{(2)}$, the 2PN order contains ${}^4h_{00}^{(2)}$,
${}^4h_{0i}^{(3)}$, ${}^4h_{ij}^{(2)}$, and so on. Beyond some order
one finds divergences connected to the inversion of the Laplacian
operator which require the introduction of a regularization of the
Poisson integrals. This is due to the fact that it is not possible
to rebuild a retarded solution from its expansion for small
retardation without going outside the near zone (i.e. there are
divergencies for $r \rightarrow \infty$). To avoid these problems
one introduces a multipolar PN (MPN) expansion and uses it in the
relaxed Einstein equations with ${}^4{\bf h}^{\mu\nu} =
\sum_{n=2}^{\infty}\, {1\over {c^n}}\, {}^4h_{(n)}^{\mu\nu}$,
$\tau^{\mu\nu} = \sum_{n=-2}^{\infty}\, {1\over {c^n}}\,
\tau^{\mu\nu}_{(n)}$. By using a regularization prescription one
finds retarded solutions (regular at $r = 0$) in terms of {\it STF
matter multipoles}. Also a homogeneous solution of the wave equation
is needed: its regularity at $r = 0$ requires that it is a half
retarded minus half advanced solution. Finally to get the equations
of motion of matter and their PN expansion till order 3.5PN one has
to impose the harmonic gauge condition and to take into account the
back-reaction from the emission of GW's: this introduces the lacking
terms ${}^4h_{00}^{(2n+1)}$, ${}^4h_{0i}^{(2n)}$,
${}^4h_{ij}^{(2n+1)}$, and becomes relevant at the 2.5PN order
[$O(({v\over c})^5)$].
\medskip

C) Then one matches the MPM and the MPN solutions in the overlap of
the near and far zones: this allows to express the radiative
multipoles in terms of the matter ones. Now one can study the limit
at future null infinity ($r \rightarrow \infty$ with $u = r - t/c$
fixed) to test the nature of GW's. At higher orders hereditary terms
(tails starting from 1.5PN [$O(({v\over c})^3)$] and non-linear
(Christodoulou) memory starting from 3PN (see Ref.\cite{42} for a
review) appear, showing that GW's propagate not only on the flat
light-cone but also inside it (i.e. with all possible speeds $0 \leq
v \leq c$): there is an instantaneous wavefront followed by a tail
traveling at lower speed (it arrives later and then fades away) and
a persistent zero-frequency non-linear memory.

\medskip

Today there is control on the solution and on the matter equations
of motion till order 3.5PN [$O(({v\over c})^7)$] (for binaries see
the review in chapter 4 of Ref.\cite{20}) and well established
connections with numerical relativity (see the review in
Ref.\cite{43}) especially for the binary black hole problem (see the
review in Ref.\cite{44}).

\medskip

In Refs.\cite{28,45} there is a Hamiltonian approach starting from a
PN expansion of the ADM formalism in suitable non-harmonic gauges
(generalized isotropic ones with ${}^3K = 0$), which allow to
recover the previous results in harmonic gauges till the order
3PN.\medskip

However in this formulation  GW's propagate in the background
Euclidean 3-space implied by the decomposition ${}^4g_{\mu\nu} =
{}^4\eta_{\mu\nu} + {}^4h_{\mu\nu}$ and it is not clear how to
visualize them in the non-Euclidean instantaneous 3-spaces of the
global non-inertial frames implied by the equivalence principle.

\hfill\eject

\section{The Multipole Moments of the Matter Energy-Momentum
Tensor}

In Ref.\cite{22} there is a study of the relativistic Dixon
multipoles \cite{46} \footnote{As shown in this paper, strictly
speaking the multipolar expansion makes sense only if the
energy-momentum tensor is an analytic function of the 3-coordinates.
However see Ref.\cite{47} for the relaxation of this condition.} of
the energy-momentum tensor of relativistic matter systems (for
instance point particles plus the electro-magnetic field) in the
rest-frame instant form of dynamics in Minkowski space-time. Since
in HPM gravity in our asymptotically flat space-times and in its
non-inertial rest-frame developed in this paper  we have only small
deviations from such a scheme, we can apply this formalism (without
using the general relativistic Dixon multipoles \cite{48}) to the
energy-momentum tensor given in Eqs. (\ref{3.12}), which gives rise
to the ten internal Poincare' generators given in Eqs. (\ref{4.21})
- (\ref{4.24}) at lowest order $O(\zeta)$. We refer to chapter 3 of
Ref.\cite{20} for a review of the standard types of multipoles used
for the study of gravitational waves from compact sources.

\bigskip

Moreover at the lowest order we have $\partial_A\, T_{(1)}^{AB}\,
\cir\,\, 0$, $\partial_A\, (T^{AB}_{(1)}\, \sigma^c - T^{AC}_{(1)}\,
\sigma^B)\, \cir\,\, 0$, with $T_{(1)}^{\tau\tau} = {\cal
M}^{(UV)}_{(1)}$, $T_{(1)}^{\tau r} = {\cal M}^{(UV)}_{(1)r}$, see
after Eq.(\ref{3.12}).
\bigskip

Let $w^{\mu}(\tau) = z^{\mu}(\tau, \vec \eta(\tau))$ be the
time-like world-line of an arbitrary {\it center of motion} of the
particle system. On the instantaneous 3-space $\Sigma_{\tau}$ we can
define the following Dixon multipoles of the matter energy-momentum
tensor $T_{(1)}^{AB}(\tau, \vec \sigma)$ with respect to the point
$\vec \eta(\tau)$

\beq
 q^{r_1 ... r_n\,|\, AB}(\tau) = \int d^3\sigma\, \Big(\sigma^{r_1} -
 \eta^{r_1}(\tau)\Big) ... \Big(\sigma^{r_n} - \eta^{r_n}(\tau)\Big)\,
 T_{(1)}^{AB}(\tau, \vec \sigma),
 \label{a1}
 \eeq

\noindent and the following multipolar expansion of $T_{(1)}^{AB}$

\bea
 T_{(1)}^{AB}(\tau, \vec \sigma) &=& \sum_{n = 0}^{\infty}\, {{(-)^n}\over {n!}}\,
 \sum_{r_1...r_n}\, q^{r_1 ... r_n\,|\, AB}(\tau)\, {{\partial^n\,
 \delta^3(\vec \sigma, \vec \eta(\tau))}\over {\partial\, \sigma^{r_1}\,
 ... \partial\, \sigma^{r_n}}},\nonumber \\
 &&{}\nonumber \\
 T_{(1)}^{(TT)ab}(\tau, \vec \sigma) &=& \sum_{uv}\,
 \Big({\cal P}_{abuv}\, T_{(1)}^{uv}\Big)(\tau, \vec \sigma),
 \label{a2}
 \eea

\noindent where Eq.(\ref{6.9}) has to be used for the TT tensor.

 \medskip

To connect them to the standard Dixon multipoles of the space-time
energy-momentum tensor $T^{\mu\nu}(x = z(\tau, \vec \sigma)) =
\Big[z^{\mu}_A\, z^{\nu}_B\, T_{(1)}^{AB}\Big](\tau, \vec \sigma) +
O(\zeta^2)$ we must use the adapted embedding $z^{\mu}(\tau, \vec
\sigma)$ discussed in the Introduction.\medskip

By using Eqs.(\ref{4.21})-(\ref{4.24}), the relevant multipoles
are:\medskip

1a) the {\it mass monopole}

\beq
 q^{|\, \tau\tau} = \int d^3\sigma\, {\cal
 M}^{(UV)}_{(1)}(\tau, \vec \sigma) = M_{(1)}\, c;
 \label{a3}
 \eeq

\medskip
1b) the {\it mass dipole} (Eq.(\ref{4.24}) is used)

\bea
 q^{r\, |\, \tau\tau} &=& \int d^3\sigma\, (\sigma^r - \eta^r(\tau))\,
 {\cal M}^{(UV)}_{(1)}(\tau, \vec \sigma) = j^{\tau r}_{(1)}
 - M_{(1)}\, c\, \eta^r(\tau) \approx\nonumber \\
 &\approx& - M_{(1)}\, c\, \eta^r(\tau);
 \label{a4}
 \eea

\medskip
1c) the {\it mass quadrupole} (Eq.(\ref{4.24}) is used)

\bea
 q^{rs\, |\, \tau\tau} &=& \int d^3\sigma\, (\sigma^r - \eta^r(\tau))\,
 (\sigma^s - \eta^s(\tau))\, {\cal M}^{(UV)}_{(1)}(\tau,
 \vec \sigma) \approx \nonumber \\
 &\approx& \int d^3\sigma\, \sigma^r\,
 \sigma^s\, {\cal M}^{(UV)}_{(1)}(\tau, \vec \sigma) +
 M_{(1)}\, c\, \eta^r(\tau)\, \eta^s(\tau);
 \label{a5}
 \eea
\medskip

2a) the {\it momentum monopole} (Eq.(\ref{4.22}) is used)

\beq
 q^{|\, \tau r} = \int d^3\sigma\, {\cal
 M}^{(UV)}_{(1)r}(\tau, \vec \sigma) = p^r_{(1)} \approx 0;
 \label{a6}
 \eeq

\medskip

2b) the {\it momentum dipole}

\beq
 q^{r\, |\, \tau s} = \int d^3\sigma\, (\sigma^r - \eta^r(\tau))\,
 {\cal M}^{(UV)}_{(1)s}(\tau, \vec \sigma) \approx \int
 d^3\sigma\, \sigma^r\, {\cal M}^{(UV)}_{(1)s}(\tau, \vec
 \sigma),
 \label{a7}
 \eeq

\noindent whose antisymmetric part is $q^{r\, |\, \tau s} - q^{s\,
|\, \tau r} = - 2\, j^{rs}_{(1)}$ due to Eq.(\ref{4.23}) [as a
consequence we have $q^{r\, |\, \tau s} = {1\over 2}\, (q^{r\, |\,
\tau s} + q^{s\, |\, \tau r}) - j^{rs}_{(1)}$];
\medskip

3) the {\it stress tensor monopole}

 \beq
 q^{|\, rs} = \int d^3\sigma\, T^{rs}_{(1)}(\tau, \vec \sigma) = q^{|\, sr}.
 \label{a8}
 \eeq

\bigskip

If we choose as center of motion of the mass distribution the {\it
center of energy} $w_E^{\mu}(\tau)$, we must put equal to zero the
mass dipole and this implies $\eta^r(\tau) \approx 0$ (namely
$w_E^{\mu}(\tau) = z^{\mu}(\tau, 0)$ coincides with the origin of
3-coordinates)

\bea
 &&q^{r\, |\, \tau\tau} \approx 0,\qquad \Rightarrow\qquad \eta^r(\tau)
 \approx 0,\nonumber \\
 &&{}\nonumber \\
 \Rightarrow&& q^{rs|\tau\tau} = \int d^3\sigma\, \sigma^r\,
 \sigma^s\, {\cal M}^{(UV)}_{(1)}(\tau, \vec \sigma).
 \label{a9}
 \eea

\medskip

In this case the non-zero lowest multipoles are the mass monopole
$M_{(1)}\, c$, the mass quadrupole $q^{rs\, |\, \tau\tau}$, the
momentum dipole $q^{r\, |\, \tau s}$ and the stress tensor monopole
$q^{|\, rs}$.
\bigskip

The lowest order conservation law $\partial_A\, T_{(1)}^{AB}\,
\cir\, 0$ gives $\partial_{\tau}\, {\cal M}^{(UV)}_{(1)}\, \cir\, -
\sum_r\, \partial_r\, {\cal M}_{(1)r}^{(UV)}$ and $\partial_{\tau}\,
{\cal M}^{(UV)}_{(1)r} \, \cir\, - \sum_s\, \partial_s\,
T^{rs}_{(1)}$, see Eqs.(\ref{3.13}). By integrating over the whole
3-space with the matter density having compact support (or suitable
fall-off at spatial infinity) we get

\beq
 \partial_{\tau}\, M_{(1)} \, \cir\, 0 + O(\zeta^2),\qquad
 \partial_{\tau}\, p^r_{(1)} \, \cir\, 0 + O(\zeta^2).
 \label{a10}
 \eeq
 \medskip

The conservation law $\partial_A\, (T_{(1)}^{AB}\, \sigma^c -
T^{AC}_{(1)}\, \sigma^B) \, \cir\, 0$ implies

\beq
 \partial_{\tau}\, j^{rs}_{(1)}\, \cir 0 + O(\zeta^2),\qquad
  \partial_{\tau}\, j^{\tau r}_{(1)}\, \cir\, 0 + O(\zeta^2).
  \label{a11}
  \eeq
  \medskip

By using $\partial_A\, [T_{(1)}^{AB}\, \sigma^r\, \sigma^s] \,
\cir\, T_{(1)}^{rB}\, \sigma^s + T_{(1)}^{sB}\, \sigma^r$ we get
$\partial_{\tau}\, q^{rs\, |\, \tau\tau}\, \cir\, q^{r\, |\, \tau s}
+ q^{s\, |\, \tau r}$, so that we have

\beq
 q^{r\, |\, \tau s} = {1\over 2}\, \partial_{\tau}\, q^{rs\, |\,
 \tau\tau} - j^{rs}_{(1)}.
 \label{a12}
 \eeq
 \medskip

By using $\partial_A\, \partial_B\, (T_{(1)}^{AB}\, \sigma^C\,
\sigma^D)\, \cir\, 2\, T_{(1)}^{CD}$ we get

\beq
 2\, q^{|\, rs}\, \cir\, \partial_{\tau}^2\, q^{rs\, |\, \tau\tau}.
 \label{a13}
 \eeq
\bigskip

Therefore the relevant multipoles are expressible in terms of
$M_{(1)}$, $j^{rs}_{(1)}$, $\partial_{\tau}\, q^{rs\,|\, \tau\tau}$
and $\partial_{\tau}^2\, q^{rs\,|\, \tau\tau}$, where $q^{rs\,|\,
\tau\tau}$ is the mass quadrupole.

\bigskip

However in Eq.(\ref{6.11}) we have $T^{(TT)rs}_{(1)}$, whose stress
tensor monopole is $q^{(TT)|\, rs} = \int d^3\sigma\,
T^{(TT)rs}_{(1)}(\tau, \vec \sigma)$, and not $T^{rs}_{(1)}$. From
Eq.(\ref{6.9}) we have $T^{(TT)rs}_{(1)} = \sum_{uv}\, {\cal
P}_{rsuv}\, T^{uv}_{(1)}$. As a consequence we get $q^{(TT)|\, rs} =
\int d^3\sigma\, T^{(TT)rs}_{(1)}(\tau, \vec \sigma)\, =\, \int
d^3\sigma\, \sum_{uv}\, {\cal P}_{rsuv}\, T^{uv}_{(1)}(\tau, \vec
\sigma)$.
\bigskip

By using Eq.(\ref{6.9}) we get (the surface terms vanish with the
assumed support and boundary conditions)

\bea
 q^{|\, rs}\, &=& \int d^3\sigma\, T^{rs}_{(1)}(\tau, \vec
 \sigma)\, =\, \int d^3\sigma\, \Big[{1\over 3}\, {\tilde H}_{(1)}\,
 \delta^{rs} + T^{(TT)rs}_{(1)}\Big](\tau, \vec \sigma)\, =\nonumber \\
 &=& q^{(TT)|\, rs} + {1\over 3}\, \delta^{rs}\, \sum_u\, q^{|\,
 uu}, \nonumber \\
 &&{}\nonumber \\
 \Rightarrow&& q^{(TT)|\, rs} \, =\,  q^{|\, rs} - {1\over 3}\, \delta^{rs}\,
 \sum_u\,  q^{|\, uu}\, \cir\, {1\over 2}\, \partial^2_{\tau}\,
 (q^{rs\, |\, \tau\tau}  - {1\over 3}\, \delta^{rs}\, \sum_u\,
 q^{uu\, |\, \tau\tau}).\nonumber \\
 &&{}
 \label{a14}
 \eea

Therefore the monopole of the TT stress tensor $T_{(1)}^{(TT) rs}$
is connected to the second time derivative of the mass quadrupole as
usual.\bigskip

We also need higher multipoles $q^{r_1..r_n|rs} \approx \int
d^3\sigma\, \sigma^{r_1} ... \sigma^{r_n}\, T_{(1)}^{rs}(\tau, \vec
\sigma)$ and their TT analogues. For the dipole we get
\medskip

\beq
 \int d^3\sigma\, \partial_A\, \partial_B\, \Big(\sigma^{r_1}\,
 \sigma^r\, \sigma^s\, T_{(1)}^{AB}(\tau, \vec \sigma)\Big) =
 \partial^2_{\tau}\, q^{r_1rs| \tau\tau} = 2\, \Big(q^{r_1|rs} + q^{r| r_1s}
 + q^{s|r_1r}\Big).
 \label{a15}
 \eeq

We see that the second time derivative of the mass octupole is
connected to a combination of stress tensor dipoles. For $r=s=a$ we
have $\partial_{\tau}^2\, q^{r_1aa | \tau\tau} = 2\, (q^{r_1 | aa} +
2\, q^{a | r_1a})$. However $q^{r_1 | aa}$ cannot be expressed only
in terms of the mass octupole.

\bigskip

The same pattern holds  for the higher multipoles $q^{r_1..r_n|rs}$.
In particular with same methods used in Ref.\cite{20} we get the
following results for the stress tensor monopole $q^{|rs}$ and
dipole $q^{u|rs}$

\bea
 &&q^{|rs} = q^{|sr} = \partial_{\tau}\, q^{r|\tau s},\nonumber \\
 &&{}\nonumber \\
 &&q^{r|us} + q^{s|ru} = \partial_{\tau}\, q^{rs|\tau u},\nonumber \\
 &&{}\nonumber \\
 &&\partial_{\tau}\, q^{r|us} + \partial_{\tau}\, q^{s|ur} =
 \partial_{\tau}^2\, q^{rs|\tau u},\nonumber \\
 &&{}\nonumber \\
 &&2\, (\partial_{\tau}^{u|rs} + \partial_{\tau}\, q^{r|su} +
 \partial_{\tau}\, q^{s|ur}) = \partial_{\tau}^3\, q^{rsu|\tau\tau},
 \nonumber \\
 &&{}\nonumber \\
 &&\partial_{\tau}\, q^{u|rs} = {1\over 6}\, \partial_{\tau}^3\,
 q^{rsu|\tau\tau} + {1\over 3}\, (\partial_{\tau}^2\, q^{ur|\tau s} +
 \partial_{\tau}^2\, q^{us|\tau r} - 2\, \partial_{\tau}^2\, q^{rs|\tau
 u}).\nonumber \\
&&{}
 \label{a16}
 \eea

\noindent in terms of the momentum dipole $q^{r|\tau u}$ and
quadrupole $q^{rs|\tau u}$ and mass octupole $q^{rsu|\tau\tau}$.

\vfill\eject

\section{The Balance of Momentum and Angular Momentum for GW's}

Here we give the evaluation of the balance equations for the
3-momentum and the angular momentum by using the conservation of the
corresponding ADM generators (\ref{4.22}) and (\ref{4.23}).
\medskip

\subsection{The Balance of Momentum for the GW's}

By using Eqs.(\ref{6.4}) and (\ref{6.8}) the first term in
$p^r_{(2)ADM}$ in Eq.(\ref{4.22}) can be written in the form

\bea
 \int d^3\sigma\, && \Big(\sum_{\bar a\bar b}\, \partial_r\, R_{\bar a}\,
 M_{\bar a\bar b}\, \partial_{\tau}\, R_{\bar b}\Big)(\tau, \vec \sigma) =
 \int d^3\sigma\,  \Big(\sum_{ab}\, \partial_r\, \Gamma_a^{(1)}\,
 {\tilde M}_{ab}\, \Gamma_b^{(1)}\Big)(\tau, \vec \sigma) =
 \nonumber \\
 &=& \int d^3\sigma\,  \Big(\sum_{rs}\, \partial_r\, {}^4h^{TT}_{(1)rs}\,
 \partial_{\tau}\, {}^4h^{TT}_{(1)rs}\Big)(\tau, \vec \sigma).
 \label{b1}
 \eea

\medskip

Then, like in Eq.(\ref{7.23}) for the ADM energy, the ADM momentum
(\ref{4.22}) can be written in the form

\bea
 {\hat P}^r_{(1+2)ADM} &=& \int d^3\sigma\, \Big(\rho^{(matter)\, r}_{p(1+2)}
 + \rho^{(rad)\, r}_{p (1+2)} + \rho^{(int)\, r}_{p (1+2)}\Big)(\tau,
 \vec \sigma),\nonumber \\
 &&{}\nonumber \\
 \rho^{(matter)\, r}_{p(1+2)}(\tau, \vec \sigma) &=& {\cal
 M}_{(1)r}^{(UV)}(\tau, \vec \sigma) = \sum_i\, \delta^3(\vec \sigma
 - {\vec \eta}_i(\tau))\, \eta_i\, \kappa_{ir}(\tau),\nonumber \\
 \rho^{(rad)\, r}_{p(1+2)}(\tau, \vec \sigma) &=& {{c^3}\over {32\pi\, G}}\,
 \sum_{uv}\, \Big(\partial_r\, {}^4h^{TT}_{(1)uv}\, \partial_{\tau}\,
 {}^4h^{TT}_{(1)uv}\Big)(\tau, \vec \sigma),\nonumber \\
 \rho^{(int)\, r}_{p(1+2)}(\tau, \vec \sigma) &=& \Big(
 {\cal M}_{(1)r}^{(UV)}\, \Big(\sum_a\, {{\partial_a^2}\over
 {\triangle}}\, \Gamma_a^{(1)} - 2\, \Gamma_r^{(1)}\Big) +
 {1\over 2}\, \sum_{as}\, {\cal M}_{(1)s}^{(UV)}\, \partial_r\,
 \partial_s\, {{\partial_a^2}\over {\triangle}}\, \Gamma_a^{(1)}
 -\nonumber \\
 &-& {{8\pi\, G}\over {c^3}}\, {\cal M}_{(1)r}^{(UV)}\, {1\over
 {\triangle}}\, {\cal M}_{(1)}^{(UV)} - {\cal M}_{(1)}^{(UV)}\,
 {{\partial_r}\over {\triangle}}\, {}^3K_{(1)}
 \Big)(\tau, \vec \sigma).\nonumber \\
 &&{}
 \label{b2}
 \eea

\medskip

Like after Eq.(\ref{7.23}) let us divide the 3-space $\Sigma_{\tau}$
in two regions by means of a sphere $S$ of big radius $R >> l_c$: a)
an inner region $V_{(inner)}$ with a compact sub-region $V_c$ of
linear dimension $l_c$ containing all the particles (and the
electro-magnetic field if we would add it); b) an asymptotic far
region $V_{(far)}$. Let $n^r = \sigma^r/|\vec \sigma|$ be a unit
3-vector.\medskip

Since we have ${\hat P}^r_{ADM} = {\hat P}_{ADM}^{r\, V_{(far)}} +
{\hat P}_{ADM}^{r\, V_{(inner)}}$ and $\rho^{(matter)\,
r}_{p(1+2)}(\tau, \vec \sigma){|}_{\vec \sigma \in V_{far}} =
\rho^{(int)\, r}_{p(1+2)}(\tau, \vec \sigma){|}_{\vec \sigma \in
V_{far}} = 0$, we get ${\hat P}_{ADM}^{r\, V_{(far)}} =
\int_{V_{(far)}} \, d^3\sigma\, \rho^{(rad)\, r}_{p(1+2)}(\tau, \vec
\sigma)$.\medskip

Since ${\hat P}^r_{ADM} \approx 0$ is a constant, we have ${{d\,
{\hat P}^r_{ADM}}\over {d\tau}} = 0$ so that we get the following
result in place of  Eq.(\ref{7.24})

\bea
 {{d\, {\hat P}_{ADM}^{r\, V_{(inner)}}}\over {d\tau}} &=& - {{d\,
 {\hat P}_{ADM}^{r\, V_{(far)}}}\over {d\tau}} = - \int_{V_{(far)}}\,
 d^3\sigma\, \partial_{\tau}\, \rho^{r\, (rad)}_{p(1+2)}(\tau, \vec \sigma)
 =\nonumber \\
 &&{}\nonumber \\
 &=& - {{c^3}\over {32\pi\, G}}\, \int_{V_{(far)}}\,
 d^3\sigma\, \sum_{rs}\,  \Big[\partial_r\, {}^4h^{TT}_{(1)rs}\,
  \partial^2_{\tau}\, {}^4h^{TT}_{(1)rs}
 +  \partial_r\, \partial_{\tau}\, {}^4h^{TT}_{(1)rs}\,
 \partial_{\tau}\,  {}^4h^{TT}_{(1)rs} \Big](\tau, \vec \sigma).
 \nonumber \\
 &&{}
 \label{b3}
 \eea

By using the implication $\partial_{\tau}\, {}^4h^{TT}_{(1)rs} = -
n^c\, \partial_c\, {}^4h^{TT}_{(1)rs} + O(1/R^2)$ ($n^c =
\sigma^c/|\vec \sigma|$) of Eqs.(\ref{7.19}), Eqs (\ref{7.25}) are
replaced by the following expression

\bea
 {{d\, {\hat P}_{ADM}^{r\, V_{(inner)}}}\over {d\tau}} &=& - {{d\,
 {\hat P}_{ADM}^{r\, V_{(far)}}}\over {d\tau}} =\nonumber \\
 &=& - {{c^3}\over {32\pi\, G}}\, \int_{V_{(far)}}\,
 d^3\sigma\,  \sum_{rs}\, \sum_c\, \partial_c\,
 \Big(n^c\, \partial_{\tau}\, {}^4h^{TT}_{(1)rs}\, \partial_r\,
 {}^4h^{TT}_{(1)rs}\Big)(\tau, \vec \sigma) + O(1/R^2) =\nonumber \\
 &=&  {{c^3}\over {32\pi\, G}}\, R^2\, \int_S\, d(cos\, \theta)\, d\varphi
 \sum_{rs}\, \Big(\partial_{\tau}\, {}^4h^{TT}_{(1)rs}\, \partial_r\,
 {}^4h^{TT}_{(1)rs}\Big)(\tau, \vec \sigma) + O(1/R^2) =\nonumber \\
  &&{}\nonumber \\
 &{\buildrel {(\ref{7.19})}\over =}&  {G\over {2\pi\, c^3}}\,
 \int_{S}\, d(cos\, \theta)\, d\varphi\, \sum_{swuv}\,
 \Lambda_{swuv}(n)\, \partial_{\tau}\, q^{|sw}(\tau - R)\,
 \partial_r\, q^{|uv}(\tau - R) + O(1/r^2) =\nonumber \\
 &=&  {G\over {8\, c^2}}\, \int_{S}\, d(cos\, \theta)\, d\varphi\, \sum_{swuv}\,
 \Lambda_{swuv}(n)\,  [\partial^3_{\tau}\,
 q^{sw|\tau\tau}(\tau - R)]\, [\partial_r\, \partial^2_{\tau}\,
 q^{uv|\tau\tau}(\tau - R)] + O(1/r^2),\nonumber \\
 &&{}
 \label{b4}
 \eea

\noindent after having used $q^{|rs} = {1\over 2}\,
\partial_{\tau}^2\, q^{rs|\tau\tau}$, see Eq.(\ref{a13}).
This is the standard result for the momentum balance of GW's (see
for instance Eqs. (1.164) and (3.83) of Ref.\cite{20}).

\subsection{The Balance of Angular Momentum for GW's}

By using Eqs.(\ref{6.4}) and (\ref{6.8}) the last two lines of
$j^{rs}_{(2)ADM}$ in Eq.(\ref{4.23}) can be written in the form (see
Eq.(2.51) of Ref.\cite{20}, where the angular momentum is defined
with an overall minus signa with respect to us)

\bea
 {{c^3}\over {64\pi\, G}} &&
 \int d^3\sigma\,  \Big(
 \sum_{uv}\, \partial_{\tau}\, {}^4h^{TT}_{(1)uv}\, (\sigma^r\,
 \partial_s - \sigma^s\, \partial_r)\, {}^4h^{TT}_{(1)uv} -
 \nonumber \\
 &-& \sum_u\, ({}^4h^{TT}_{(1)ru}\, \partial_{\tau}\, {}^4h^{TT}_{(1)su}
 - {}^4h^{TT}_{(1)su}\, \partial_{\tau}\, {}^4h^{TT}_{(1)ru})
 \Big)(\tau, \vec \sigma).
 \label{b5}
 \eea

\medskip

Then, like in Eq.(\ref{7.23}) for the ADM energy, the ADM angular
momentum (\ref{4.23}) can be written in the form

\bea
 {\hat J}^{rs}_{(1+2)ADM} &=& \int d^3\sigma\, \Big(\rho^{(matter)\, rs}_{j(1+2)}
 + \rho^{(rad)\, rs}_{j (1+2)} + \rho^{(int)\, rs}_{j (1+2)}\Big)(\tau,
 \vec \sigma),\nonumber \\
 &&{}\nonumber \\
 \rho^{(matter)\, rs}_{j(1+2)}(\tau, \vec \sigma) &=& \sigma^r\, {\cal
 M}_{(1)s}^{(UV)}(\tau, \vec \sigma) - \sigma^s\, {\cal
 M}_{(1)r}^{(UV)}(\tau, \vec \sigma),\nonumber \\
 \rho^{(rad)\, rs}_{j(1+2)}(\tau, \vec \sigma) &=& {{c^3}\over {64\pi\, G}}\,
 \Big[\sum_{uv}\, \Big(\partial_{\tau}\, {}^4h^{TT}_{(1)uv}\, (\sigma^r\,
 \partial_s - \sigma^s\, \partial_r)\, {}^4h^{TT}_{(1)uv}\Big)
 -\nonumber \\
 &-& \sum_u\, \Big({}^4h^{TT}_{(1)ru}\, \partial_{\tau}\, {}^4h^{TT}_{(1)su}
 - {}^4h^{TT}_{(1)su}\, \partial_{\tau}\, {}^4h^{TT}_{(1)ru}\Big)
 \Big](\tau, \vec \sigma),\nonumber \\
 \rho^{(int)\, rs}_{j(1+2)}(\tau, \vec \sigma) &=& \Big(
 {1\over 2}\, {\cal
 M}_{(1)}^{(UV)}\, (\sigma^r\, \partial_s - \sigma^s\, \partial_r)\,
 {1\over {\triangle}}\, {}^3K_{(1)} -\nonumber \\
 &-& 4\, \Big[ \sigma^r\,
 {\cal M}_{(1)s}^{(UV)}\, (\Gamma_r^{(1)} - {{4\pi\, G}\over {c^3}}\, {1\over
 {\triangle}}\, {\cal M}_{(1)}^{(UV)} + {1\over 2}\, \sum_c\, {{\partial_c^2}
 \over {\triangle}}\, \Gamma_c^{(1)}) -\nonumber \\
 &-& \sigma^s\, {\cal M}_{(1)r}^{(UV)}\, (\Gamma_r^{(1)} - {{4\pi\, G}\over {c^3}}\, {1\over
 {\triangle}}\, {\cal M}_{(1)}^{(UV)} + {1\over 2}\, \sum_c\, {{\partial_c^2}
 \over {\triangle}}\, \Gamma_c^{(1)})\Big] +\nonumber \\
 &+& \sum_u\, {\cal M}_{(1)u}^{(UV)}\, (\sigma^r\, \partial_s - \sigma^s\, \partial_r)\,
 {{\partial_u}\over {\triangle}}\, \Gamma_u^{(1)} - {1\over 4}\,
 \sum_c\, ({\cal M}_{(1)r}^{(UV)}\, {{\partial_s}\over {\triangle}} - {\cal M}_{(1)s}^{(UV)}\,
 {{\partial_r}\over {\triangle}})\, {{\partial_c^2}\over {\triangle}}\, \Gamma_c^{(1)}
 -\nonumber \\
 &-& {1\over 4}\, \sum_{uv}\, {\cal M}_{(1)u}^{(UV)}\, (\sigma^r\,
 \partial_s - \sigma^s\, \partial_r)\, {{\partial_u}\over
 {\triangle}}\, {{\partial_v^2}\over {\triangle}}\, \Gamma_v^{(1)} +
 {\cal M}_{(1)r}^{(UV)}\, {{\partial_r}\over {\triangle}}\,
 \Gamma_s^{(1)} - {\cal M}_{(1)s}^{(UV)}\, {{\partial_s}\over
 {\triangle}}\, \Gamma_r^{(1)} \Big)(\tau, \vec \sigma).\nonumber \\
 &&{}
 \label{b6}
 \eea

\medskip

Like after Eq.(\ref{7.23}) let us divide the 3-space $\Sigma_{\tau}$
in two regions by means of a sphere $S$ of big radius $R >> l_c$: a)
an inner region $V_{(inner)}$ with a compact sub-region $V_c$ of
linear dimension $l_c$ containing all the particles (and the
electro-magnetic field if we would add it); b) an asymptotic far
region $V_{(far)}$. Let $n^r = \sigma^r/|\vec \sigma|$ be a unit
3-vector.\medskip

Since we have ${\hat J}^{rs}_{ADM} = {\hat J}_{ADM}^{rs\, V_{(far)}}
+ {\hat J}_{ADM}^{rs\, V_{(inner)}}$ and $\rho^{(matter)\,
rs}_{j(1+2)}(\tau, \vec \sigma){|}_{\vec \sigma \in V_{far}} =
\rho^{(int)\, rs}_{j(1+2)}(\tau, \vec \sigma){|}_{\vec \sigma \in
V_{far}} = 0$, we get ${\hat J}_{ADM}^{rs\, V_{(far)}} =
\int_{V_{(far)}} \, d^3\sigma\, \rho^{(rad)\, rs}_{j(1+2)}(\tau,
\vec \sigma)$.\medskip

Since ${\hat J}^{rs}_{ADM}$ is a constant, we have ${{d\, {\hat
J}^{rs}_{ADM}}\over {d\tau}} = 0$ so that we get the following
result in place of  Eq.(\ref{7.24}) (again by using Eqs.(\ref{7.19})
and $n^c = \sigma^c/|\vec \sigma|$)

\bea
 {{d\, {\hat J}_{ADM}^{rs\, V_{(inner)}}}\over {d\tau}} &=& - {{d\,
 {\hat J}_{ADM}^{rs\, V_{(far)}}}\over {d\tau}} = - \int_{V_{(far)}}\,
 d^3\sigma\, \partial_{\tau}\, \rho^{rs\, (rad)}_{j(1+2)}(\tau, \vec \sigma)
 =\nonumber \\
 &&{}\nonumber \\
 &=&  \int_{V_{(far)}}\, d^3\sigma\, \Big[n^c\, \partial_c\,  \rho^{rs\,
 (rad)}_{j(1+2)}\Big](\tau, \vec \sigma) + O(1/R^2) =\nonumber \\
 &=& \int_{V_{(far)}}\, d^3\sigma\, \partial_c\, \Big[n^c\,   \rho^{rs\,
 (rad)}_{j(1+2)}\Big](\tau, \vec \sigma) + O(1/R^2). \nonumber \\
 &&{}
 \label{b7}
 \eea
\medskip

As a consequence, we get

\bea
 {{d\, {\hat J}_{ADM}^{rs\, V_{(inner)}}}\over {d\tau}} &=& - {{d\,
 {\hat J}_{ADM}^{rs\, V_{(far)}}}\over {d\tau}} = \nonumber \\
 &=& {{c^3}\over {12\pi\, G}}\, R^2\, \int_S\, d(cos\, \theta)\, d\varphi
 \Big[ \sum_{uv}\, \Big(\partial_{\tau}\, {}^4h^{TT}_{(1)uv}\, (\sigma^r\,
 \partial_s - \sigma^s\, \partial_r)\, {}^4h^{TT}_{(1)uv}\Big)
 -\nonumber \\
 &-& \sum_u\, \Big({}^4h^{TT}_{(1)ru}\, \partial_{\tau}\, {}^4h^{TT}_{(1)su}
 - {}^4h^{TT}_{(1)su}\, \partial_{\tau}\, {}^4h^{TT}_{(1)ru}\Big)
 \Big](\tau, \vec \sigma) + O(1/R^2), \nonumber \\
 &&{}
 \label{b8}
 \eea

\noindent in accord with Eq.(2.61) of Ref.\cite{20} for the balance
of angular momentum.

\vfill\eject


\begin{thebibliography}{}


\bibitem{1}D.Alba and L.Lusanna, {\it The Einstein-Maxwell-Particle System
in the York Canonical Basis of ADM Tetrad Gravity: I) The Equations
of Motion in Arbitrary Schwinger Time Gauges.} (0907.4087).



\bibitem{2}I.Ciufolini and J.A.Wheeler, {\it Gravitation and
Inertia} (Princeton Univ.Press, Princeton, 1995).

\bibitem{3}J.Isenberg and J.E.Marsden, {\it The York Map is a Canonical
Transformation}, J.Geom.Phys. {\bf 1}, 85 (1984).\hfill\break
 J.W.York jr, {\it Gravitational Degrees of Freedom and the Initial
Value Problem}, Phys.Rev.Lett. {\bf 26}, 1656 (1971); {\it Role of
Conformal Three Geometry in the Dynamics of Gravitation},
Phys.Rev.Lett. {\bf 28}, 1082 (1972); {\it Kinematics and Dynamics
of General Relativity}, in {\it Sources of Gravitational Radiation},
Battelle-Seattle Workshop 1978, ed.L.L.Smarr (Cambridge Univ.Press,
Cambridge, 1979).\hfill\break
 A.Qadir and J.A.Wheeler, {\it York's Cosmic
Time Versus Proper Time}, in {\it {}From SU(3) to Gravity},
Y.Ne'eman's festschrift, eds. E.Gotsma and G.Tauber (Cambridge
Univ.Press, Cambridge, 1985).


\bibitem{4}D.Alba and L.Lusanna,
 {\it Charged Particles and the Electro-Magnetic Field in
Non-Inertial Frames: I. Admissible 3+1 Splittings of Minkowski
Spacetime and the Non-Inertial Rest Frames},  Int.J.Geom.Methods in
Physics {\bf 7}, 33 (2010) (0908.0213) and {\it II. Applications:
Rotating Frames, Sagnac Effect, Faraday Rotation, Wrap-up Effect
(0908.0215)},  Int.J.Geom.Methods in Physics, {\bf 7}, 185 (2010).


\bibitem{15} L.Lusanna and
M.Pauri, {\it General Covariance and the Objectivity of Space-Time
Point-Events}, talk at the Oxford Conference on Spacetime Theory
(2004) (gr-qc/0503069); {\it Explaining Leibniz equivalence as
difference of non-inertial Appearances: Dis-solution of the Hole
Argument and physical individuation of point-events}, History and
Philosophy of Modern Physics {\bf 37}, 692 (2006) (gr-qc/0604087);
{\it The Physical Role of Gravitational and Gauge Degrees of Freedom
in General Relativity. I: Dynamical Synchronization and Generalized
Inertial Effects; II: Dirac versus Bergmann Observables and the
Objectivity of Space-Time}, Gen.Rel.Grav. {\bf 38}, 187 and 229
(2006) (gr-qc/0403081 and 0407007); {\it Dynamical Emergence of
Instantaneous 3-Spaces in a Class of Models of General Relativity},
in {\it Relativity and the Dimensionality of the World}, ed.
V.Petkov (Springer Series Fundamental Theories of Physics, Berlin,
2007) (gr-qc/0611045).





\bibitem{5}D.Alba and L.Lusanna, {\it The York Map as a Shanmugadhasan
Canonical Transformationn in Tetrad Gravity and the Role of
Non-Inertial Frames in the Geometrical View of the Gravitational
Field}, Gen.Rel.Grav. {\bf 39}, 2149 (2007) (gr-qc/0604086,
v2).\hfill\break
 D.Alba and L.Lusanna, {\it The York Map as a Shanmugadhasan
Canonical Transformationn in Tetrad Gravity and the Role of
Non-Inertial Frames in the Geometrical View of the Gravitational
Field} (gr-qc/0604086, v1).



\bibitem{6}L.Lusanna, {\it The Rest-Frame Instant Form of Metric Gravity},
Gen.Rel.Grav. {\bf 33}, 1579 (2001)(gr-qc/0101048).

\bibitem{7}L.Lusanna and S.Russo, {\it A New Parametrization for Tetrad Gravity},
Gen.Rel.Grav. {\bf 34}, 189 (2002)(gr-qc/0102074).

\bibitem{8}R.De Pietri, L.Lusanna, L.Martucci and S.Russo, {\it Dirac's
Observables for the Rest-Frame Instant Form of Tetrad Gravity in a
Completely Fixed 3-Orthogonal Gauge}, Gen.Rel.Grav. {\bf 34}, 877
(2002) (gr-qc/0105084).

\bibitem{9}J.Agresti, R.De Pietri, L.Lusanna and L.Martucci, {\it
Hamiltonian Linearization of the Rest-Frame Instant Form of Tetrad
Gravity in a Completely Fixed 3-Orthogonal Gauge: a Radiation Gauge
for Background-Independent Gravitational Waves in a Post-Minkowskian
Einstein Spacetime}, Gen.Rel.Grav. {\bf 36}, 1055 (2004)
(gr-qc/0302084).

\bibitem{10}L. Lusanna and D. Nowak-Szczepaniak, {\it  The Rest-Frame Instant Form
 of Relativistic Perfect Fluids with Equation of State $\rho = \rho (\eta, s)$
 and of Nondissipative Elastic Materials.}, Int. J. Mod. Phys. {\bf A15}, 4943
 (2000).\hfill\break
 D.Alba and L.Lusanna, {\it Generalized Eulerian Coordinates for
Relativistic Fluids: Hamiltonian Rest-Frame Instant Form, Relative
Variables, Rotational Kinematics}  Int.J.Mod.Phys. {\bf A19}, 3025
(2004) (hep-th/020903).

\bibitem{11} D.Alba, H.W.Crater and L.Lusanna, {\it Towards Relativistic
Atom Physics. I. The Rest-Frame Instant Form of Dynamics and a
Canonical Transformation for a system of Charged Particles plus the
Electro-Magnetic Field}, Canad.J.Phys. {\bf 88}, 379 (2010)
(0806.2383).



\bibitem{12}D.Alba, H.W.Crater and L.Lusanna, {\it Towards Relativistic
Atom Physics. II. Collective and Relative  Relativistic Variables
for a System of Charged Particles plus the Electro-Magnetic Field},
Canad.J.Phys. {\bf 88}, 425 (2010) (0811.0715).

\bibitem{13} D.Alba, H.W.Crater and L.Lusanna, \textit{Hamiltonian
Relativistic Two-Body Problem: Center of Mass and Orbit
Reconstruction}, J.Phys. {\bf A40}, 9585 (2007) (gr-qc/0610200).

\bibitem{14} H.Crater and L.Lusanna, {\it The Rest-Frame Darwin Potential from
 the Lienard-Wiechert Solution in the Radiation Gauge},
 Ann.Phys.(N.Y.) {\bf 289}, 87 (2001)(hep-th/0001046).\hfill\break
 D.Alba, H.Crater and L.Lusanna, {\it The Semiclassical
 Relativistic Darwin Potential for Spinning Particles in the
 Rest-Frame Instant Form: Two-Body Bound States with Spin 1/2
 Constituents}, Int.J.Mod.Phys. {\bf A16}, 3365 (2001) (hep-th/0103109).





\bibitem{16}E.Poisson, {\it The Motion of Point Particles in Curved
Space-Time}, Living Rev.Rel. {\bf 7}, 6(2004) (gr-qc/0306052).

\bibitem{17}Y.Mino, M.Sasaki and T.Tanaka, {\it Gravitational Radiation
Reaction to a Particle Motion}, Phys.Rev. {\bf D55}, 3457 (1997)
(gr-qc/9606018).\hfill\break
 T.C.Quinn and R.M.Wald, {\it }, Phys.Rev.{\bf D56}, 3381 (1997)
 (gr-qc/9610053).\hfill\break
 S.Detweiler and B.F.Whiting, {\it Self-Force via a Green's Function
 Decomposition}, Phys.Rev. {\bf D67}, 024025 (2003) (gr-qc/0202086). \hfill\break
 S.Detweiler, {\it Perspective on Gravitational Self-Force
 Analyses}, Class.Quantum Grav. {\bf 22}, S681 (2005) (gr-qc/0501004).\hfill\break
 R.M.Wald, {\it Introduction to Gravitational Self-Force}, to appear in the
 Proceedings of {\it Mass and its Motion}, Orleans, France, 23-25 Jun 2008
 (0907.0412).\hfill\break
 S.E.Gralla and R.M.Wald, {\it Derivation of Gravitational
 Self-Force}, to appear in the Proceedings of {\it Mass and its Motion}, Orleans,
 France, 23-25 Jun 2008 (0907.0414).\hfill\break
 A.Pound, {\it A New Derivation of the Gravitational Self-Force},
 (2009) (0907.5197).\hfill\break
 L.Barack, {\it Gravitational Self-Force in Extreme Mass-Ratio
 Inspirals}, Class. Quantum Grav. {\bf 26}, 213001 (2009)
 (0908.1664).\hfill\break
 I.Vega, P.Diener, W.Tichy and S.Detweiler, {\it Self-Force with (3+1)
 Codes: A Primer for Numerical Relativists}, Phys.Rev. {\bf D80}, 084021
 (2009) (0908.2138).


\bibitem{b}D.Alba and L.Lusanna, {\it The Einstein-Maxwell-Particle
System in the York Canonical Basis of ADM Tetrad Gravity: III) The
Post-Minkowskian N-Body Problem, its Post-Newtonian Limit in
Non-Harmonic 3-Orthogonal Gauges and Dark Matter as an Inertial
Effect. } (arXiv: 1009.1794).


\bibitem{18}R.Beig and \'O Murchadha, {\it The Poincaré Group as the
Symmetry Group of Canonical general Relativity,}, Ann.Phys.(N.Y.)
{\bf 174}, 463 (1987).

\bibitem{19}E.E.Flanagan and S.A.Hughes, {\it The Basics of Gravitational Wave
Theory}, arXiv, gr-qc/0501041.


\bibitem{20}M.Maggiore, {\it Gravitational Waves} (Oxford Univ.
Press, Oxford, 2008).

\bibitem{21}B.F.Schutz, {\it A First Course in General Relativity}
(Cambridge Univ.Press, Cambridge, 1985).

\bibitem{22}D.Alba, L.Lusanna and M.Pauri, {\it Multipolar Expansions for Closed
and Open Systems of Relativistic Particles}, J.Math.Phys. {\bf 46},
062505 (2005).\hfill\break
 D.Alba, L.Lusanna and M.Pauri, {\it New Directions in
Non-Relativistic and Relativistic Rotational and Multipole
Kinematics for N-Body and Continuous Systems} (2005), in {\it Atomic
and Molecular Clusters: New Research}, ed.Y.L.Ping (Nova Science,
New York, 2006) (hep-th/0505005).


\bibitem{23}L.Blanchet, {\it Gravitational radiation from post-Newtonian
sources and inspiralling compact binaries},  Living Rev. Rel. {\bf
9}, 4 (2006); {\it Post-Newtonian Theory and the Two-Body Problem},
(0907.3596).\hfill\break
 T.Damour, {\it Gravitational Radiation and the Motion of Compact Bodies},
 in {\it Gravitational Radiation}, ed. N.Deruelle and
 T.Piran (North-Holland, Amsterdam, 1983), pp.59-144; {\it The Problem
 of Motion in Newtonian and Einsteinian Gravity }, in {\it Three
 Hundred Years of Gravitation}, ed. S.Hawking and W.Israel (Cambridge
 Univ.Pres, Cambridge, 1987), pp.128-198.


\bibitem{24} K.Thorne, {\it Gravitational Radiation}, in {\it Three
 Hundred Years of Gravitation}, ed. S.Hawking and W.Israel (Cambridge
 Univ.Pres, Cambridge, 1987), pp. 330-458; {\it The Theory of Gravitational
 Radiation: An Introductory Review}, in {\it Gravitational Radiation}, ed. N.Deruelle and
 T.Piran (North-Holland, Amsterdam, 1983), pp.1;  {\it Multipole Expansions
 of Gravitational Radiation}, Rev.Mod.Phys. {\bf 52}, 299 (1980).


\bibitem{25}R.Arnowitt, S.Deser and C.W.Misner, {\it Wave Zone in General Relativity}, Phys.Rev.
{\bf 121}, 1556 (1961).


\bibitem{26}H.Stephani, {\it General Relativity} (Cambridge Univ.
Press, Cambridge, second edition 1996).


\bibitem{27}M.E.Pati and C.M.Will, {\it Post-Newtonian
Gravitational Radiation and Equations of Motion via Direct
Integration of the Relaxed Einstein Equations: Foundations},
Phys.Rev. {\bf D62}, 124015; {\it II. Two-Body Equations of Motion
to Second Post-Newtonian Order and Radiation Reaction to 3.5
Post-Newtonian Order}, Phys.Rev. {\bf D65}, 104008 (2002); C.Will,
{\it III. Radiaction Reaction for Binary Systems with Spinning
Bodies}, Phys.Rev. {\bf D71}, 084027 (2005); H.Wang and C.Will, {\it
IV. Radiation Reaction for Binary Systems with Spin-Spin Coupling},
Phys.Rev. {\bf D75}, 064017 (2007); T.Mitchell and C.Will, {\it V.
Evidence for the Strong Equivalence Principle in Second
Post-Newtonian Order}, Phys.Rev. {\bf D75}, 124025 (2007).

\bibitem{28}G.Schaefer, {The Gravitational Quadrupole
Radiation-Reaction Force and the Canonical Formalism of ADM},
Ann.Phys. (N.Y.) {\bf 161}, 81 (1985); {\it The ADM Hamiltonian and
the Postlinear Approximation}, Gen.Rel.Grav. {\bf 18}, 255 (1985).
\hfill\break
 T.Ledvinka, G.Schaefer and J.Bicak, {\it Relativistic
Closed-Form Hamiltonian for Many-Body Gravitating Systems in the
Post-Minkowskian Approximation}, Phys.Rev.Lett. {\bf 100}, 251101
(2008) (0807.0214).


\bibitem{29}D.Alba and L.Lusanna, {\it The Lienard-Wiechert
Potential of Charged Scalar Particles and their Relation to Scalar
Electrodynamics in the Rest-Frame Instant Form}, Int.J.Mod.Phys.
{\bf 13}, 2791 (1998).

\bibitem{30}
S.M.Kopeikin and E.B.Fomalont, {\it Gravitomagnetism, Causality and
Aberration of gravity in the Gravitational Light-Ray Deflection
Experiments}, Gen.Rel.Grav. {\bf 39}, 1583 (2007) (gr-qc/0510077);
{\it Aberration and the Fundamental Speed of Gravity in the Jovian
Deflection Experiment}, Found.Physics {\bf 36}, 1244 (2006)
(astro-ph/0311063).\hfill\break
 S.M.Kopeikin and V.V.Makarov, {\it Gravitational Bending of Light
 by Planetary Multipoles and its Measurement with Microarcsecond
 Astronomical Interferometers}, Phys.Rev. {\bf D75}, 062002
 (2007) (astro-ph/0611358).\hfill\break
 S.M.Kopeikin and W.T.Ni, {\it Laser Ranging Delay in the Bi-Metric
 Theory of Gravity}, to appear in the proceedings of International
 Conference on Lasers, Clocks, and Drag-Free Key Technologies for
 Future Exploration in Space, Bremen, Germany, 30 May - 1 Jun 2005
    (gr-qc/0601071).\hfill\break
 S.Carlip, {\it  Model-Dependence of Shapiro Time Delay and the "Speed
 of Gravity/Speed of Light" Controversy}, Class.Quantum Grav. {\bf 21}, 3803 (2004)
 (gr-qc/0403060).\hfill\break
 C.Will, {\it Propagation Speed of Gravity and the Relativistic Time
 Delay}, Astrophys.J. {\bf 590}, 683 (2003) (astro-ph/0301145).

\bibitem{31}K.Giesel, J.Tambornine and T.Thieman, {\it  LTB Spacetimes
 in terms of Dirac Observables}, arXiv 0906.0569,\hfill\break
 T.Thiemann, {\it Reduced Phase Space Quantization and Dirac Observables},
 Class.Quantum Grav. {\bf 23}, 1163 (2006) (0411031).\hfill\break
 B.Dittrich, {\it Partial and Complete Observables for Hamiltonian
 Constrained Systems}, Class. Quantum Grav. {\bf 39}, 1891 (2007)
 (0411013); {\it Partial and Complete Observables for Canonical
 General Relativity}, {\bf 23}, 6155 (2006) (0507106).\hfill\break
 J.M.Pons, D.C.Salisbury and K.A.Sundermeyer, {\it Revisiting Observables
 in Generally Covariant Theories in the Light of Gauge Fixing
 Methods}, Phys.Rev. {\bf D80}, 084015 (2009) (0905.4564); {\it Observables
 in Classical Canonical Gravity: Folklore Demystified.}, arXiv
 1001.2726.

\bibitem{32}J.Stewart, {\it Advanced General Relativity} (Cambridge Univ.
Press, Cambridge, 1993).


\bibitem{33}T.D.Moyer, {\it Formulation for Observed and Computed
Values of Deep Space Network Data Types for Navigation} (John Wiley,
New York, 2003).


\bibitem{34}M.Soffel, S.A.Klioner, G.Petit, P.Wolf, S.M.Kopeikin,
P.Bretagnon, V.A.Brumberg, N.Capitaine, T.Damour, T.Fukushima,
B.Guinot, T.Huang, L.Lindegren, C.Ma, K.Nordtvedt, J.Ries,
P.K.Seidelmann, D.Vokroulicky', C.Will and Ch.Xu, {\it The IAU 2000
Resolutions for Astrometry, Celestial Mechanics and Metrology in the
Relativistic Framework: Explanatory Supplement} Astron.J.,
\textbf{126}, pp.2687-2706, (2003) (astro-ph/0303376).\hfill\break
 G.H.Kaplan, {\it The IAU Resolutions on Astronomical Reference
Systems, Time Scales and Earth Rotation Models}, U.S.Naval
Observatory circular No. 179 (2005) (astro-ph/0602086).


\bibitem{35} K.J.Johnstone and Chr.de Vegt, {\it Reference Frames in Astronomy},
Annu. Rev. Astron. Astrophys. {\bf 37}, 97 (1999).\hfill\break
 J.Kovalevski, I.I.Mueller and B.Kolaczek, {\it Reference Frames in
Astronomy and Geophysics} (Kluwer, Dordrecht, 1989).

\bibitem{36}M.Bartelmann, {\it The Dark Universe}, (0906.5036).

\bibitem{37}S.Jordan, {\it The GAIA Project: Technique, Performance
and Status}, Astron.Nachr. {\bf 329}, 875 (2008) (DOI
10.1002/asna.200811065).

\bibitem{38}S.G.Turyshev and V.T.Toth, {\it The Pioneer Anomaly}
(1001.3686).


\bibitem{39}M.Ross, {\it Dark Matter: the Evidence from Astronomy,
 Astrophysics and Cosmology} (arXiv: 1001.0316).\hfill\break
 K.Garret and G.Duda, {\it Dark Matter: A Primer} (arXiv: 1006.2483).\hfill\break
 E.Battaner and E.Florido, {\it The Rotation Curve of Spiral Galaxies
 and its Cosmological Implications}, Fund.Cosmic Phys. {\bf 21}, 1
 (2000).\hfill\break
 D.G.Banhatti, {\it Disk Galaxy Rotation Curves and Dark Matter
 Distribution}, Current Science {\bf 94}, 986 (2008).






\bibitem{42}M.Favata, {\it Post-Newtonian Corrections to the
Gravitational-Wave Memory for Quasi-Circular, Inspiralling Compact
Binaries}, (2009), arXiv 0812.0069.

\bibitem{43}J.L.Jaramillo, J.A.V.Kroon and E.Gourgoulhon, {\it From Geometry
to Numerics: Interdisciplinary Aspects in Mathematical and Numerical
Relativity}, Class. Quantum Grav. {\bf 25}, 093001 (2008).



\bibitem{44}I.Hinder, {\it The Current Status of Binary Black
Hole Simulations in Numerical Relativity}, (2010), arXiv 1001.5161.



\bibitem{45} G.Schaefer, {\it Post-Newtonian Methods: Analytic
Results on the Binary Problem}, (2009), to appear in the book "Mass
and Motion in General Relativity", proceedings of the CNRS School in
Orleans/France, eds. L. Blanchet, A. Spallicci, and B. Whiting
(0910.2857).



\bibitem{46}W.G.Dixon, {\it Description of Extended Bodies by
Multipole Moments in Special Relativity}, J.Math.Phys. {\bf 8}, 1591
(1967).


\bibitem{47}W.G.Dixon, {\it Mathisson's New Mechanics: its Aims and Realisation},
Acta Physica Polonica B Proc.Suppl. {\bf 1}, 27 (2008).


\bibitem{48}W.G.Dixon, {\it Extended Objects
in General Relativity: their Description and Motion}, in {\it
Isolated Gravitating Systems in General Relativity}, Proc.Int.School
of Phys. Enrico Fermi LXVII, ed. J.Ehlers (North-Holland, Amsterdam,
1979), p. 156; {\it A Covariant Multipole Formalism for Extended
Test Bodies in General Relativity}, Nuovo Cim. {\bf 34}, 317 (1964).



\bibitem{48}W.G.Dixon, {\it Extended Objects
in General Relativity: their Description and Motion}, in {\it
Isolated Gravitating Systems in General Relativity}, Proc.Int.School
of Phys. Enrico Fermi LXVII, ed. J.Ehlers (North-Holland, Amsterdam,
1979), p. 156; {\it A Covariant Multipole Formalism for Extended
Test Bodies in General Relativity}, Nuovo Cim. {\bf 34}, 317 (1964).











\end{thebibliography}
\end{document}